\documentclass[a4paper,11pt]{article}
\pdfoutput=1 
\usepackage{jcappub}
\usepackage{amsmath}
\usepackage{latexsym}
\usepackage{xspace}
\usepackage{color}
\usepackage{hyperref} 
\usepackage{bm}
\usepackage{relsize}
\usepackage{tabularx}
\usepackage{multirow}
\usepackage{amssymb}
\usepackage[table]{xcolor}
\usepackage{braket}
\usepackage[utf8]{inputenc}
\usepackage{slashed}
\usepackage{empheq}
\usepackage{cancel}
\usepackage{soul}
\usepackage{pdfpages}
\usepackage{booktabs}
\usepackage[most]{tcolorbox}
\usepackage{capt-of}
\usepackage{xfrac} 

\usepackage{floatrow}
\usepackage{graphicx}
\usepackage[label font=bf,labelformat=simple]{subfig}
\usepackage{upgreek}

\usepackage[compat=1.1.0]{tikz-feynman}

\hypersetup{colorlinks,breaklinks,
	linkcolor = orange,
	urlcolor  =  orange,
	citecolor =  orange,
	anchorcolor = orange}

\definecolor{myblue}{RGB}{166, 193, 222}
\definecolor{myblue2}{RGB}{68, 114, 196}
\definecolor{mygreen}{RGB}{175, 218, 162}
\definecolor{mygreen2}{RGB}{133, 210, 110}
\definecolor{myorange2}{RGB}{255, 192, 0}

\newcolumntype{P}[1]{>{\centering\arraybackslash}p{#1}}

\makeatletter
\gdef\@fpheader{}
\g@addto@macro\bfseries{\boldmath}
\makeatother

\usepackage{savesym}

\savesymbol{Ref}
\savesymbol{erf}

\newcommand{\ie}{\textsl{i.e.}~}

\newcommand{\eg}{\textsl{e.g.}\xspace}

\newcommand{\Tr}{\mathrm{Tr}}

\newcommand{\dd}{\mathrm{d}}
\newcommand{\ee}{e}

\newcommand{\boldmathsymbol}[1]{{\ensuremath{\boldsymbol{#1}}}}

\newcommand{\bmk}{\boldmathsymbol{k}}
\newcommand{\bmp}{\boldmathsymbol{p}}
\newcommand{\bmq}{\boldmathsymbol{q}}
\newcommand{\bmx}{\boldmathsymbol{x}}

\newcommand{\cs}{c_{_\mathrm{S}}}

\newcommand{\beq}{\begin{equation}}
\newcommand{\eeq}{\end{equation}}
\newcommand{\bea}{\begin{equation}\begin{aligned}}
\newcommand{\eea}{\end{aligned}\end{equation}}

\newlength{\wsingfig}
\newlength{\wdblefig}
\newlength{\wquadfig}
\newlength{\wtriplefig}
\newcommand{\Eq}[1]{Eq.~(\ref{#1})}
\newcommand{\Eqs}[1]{Eqs.~(\ref{#1})}
\newcommand{\Fig}[1]{Fig.~{\ref{#1}}}
\newcommand{\Figs}[1]{Figs.~{\ref{#1}}}

\newcommand{\Refs}[1]{Refs.~{\cite{#1}}}
\newcommand{\Sec}[1]{Sec.~\ref{#1}}

\newcommand{\bs}[1]{\boldsymbol{#1}}

\newcommand{\Os}{\delta \mathcal{O}_{S}}
\newcommand{\then}{\quad \Rightarrow\quad}
\newcommand{\exx}[1]{\langle\!\langle #1 \rangle\!\rangle}
\newcommand{\ex}[1]{\langle #1 \rangle}
\newcommand{\T}{\mathcal{T}}

\newcommand{\Mpl}{M_{\mathrm Pl}}

\newcommand{\phip}{\varphi_{+}}
\newcommand{\phim}{\varphi_{-}}

\renewcommand{\H}{\mathcal{H}}
\renewcommand{\O}{\mathcal{O}}

\def\pia{\pi_{a}}
\def\pir{\pi_{r}}
\def\dpir{\dot{\pi}_{r}}
\def\dpia{\dot{\pi}_{a}}

\def\phia{\varphi_{a}}
\def\phir{\varphi_{r}}
\def\dphir{\dot{\varphi}_{r}}
\def\dphia{\dot{\varphi}_{a}}

\def\gr{g^{00}}

\newcommand{\fini}{f}

\def\bfk{\textbf{k}}
\def\bfq{\textbf{q}}

\newcommand{\bfx}{{\mathbf{x}}}

\newcommand{\Al}{A_{||}}
\newcommand{\At}{A_{\perp}}

\newcommand{\at}{a_{\perp}}
\newcommand{\e}{\epsilon}

\newcommand{\gtt}[1]{\gamma^{tt}_{#1, \ell}}
\newcommand{\gts}[1]{\gamma^{ts}_{#1, \ell}}
\newcommand{\gss}[1]{\gamma^{ss}_{#1, \ell}}

\newcommand{\gtto}[1]{\gamma^{tt}_{#1, 0}}

\newcommand{\gtso}[1]{\gamma^{ts}_{#1, 0}}
\newcommand{\gsso}[1]{\gamma^{ss}_{#1, 0}}

\newcommand{\bt}[1]{\beta_{#1,\ell}}

\newcommand{\po}{\mathrm{P.O.}}

\newcommand{\stuck}{St\"uckelberg }
\restoresymbol{phys}{Ref}
\restoresymbol{phys}{erf}

\usepackage[english]{babel}
\usepackage{numprint}
\usepackage{indentfirst}
\usepackage[T1]{fontenc}
\usepackage{lmodern}
\usepackage{eurosym}
\usepackage{lipsum}
\usepackage{textcomp}
\usepackage{wasysym}
\usepackage{textgreek}
\usepackage{listings}
\usepackage{url}
\usepackage[margin=1in,includefoot]{geometry}

\usepackage{xfrac} 
\usepackage{amsbsy}
\usepackage{amsmath}
\usepackage{adjustbox}
\usepackage{mathtools}
\usepackage{gensymb}
\usepackage{breqn}
\usepackage{cancel}
\usepackage{multirow}
\usepackage{amssymb}
\usepackage{physics}
\numberwithin{equation}{section}

\usepackage{amstext} 
\usepackage{array}   
\newcolumntype{L}{>{$}l<{$}} 

\usepackage{latexsym}
\usepackage{graphicx}  
\usepackage{float} 
\usepackage{ifoddpage}
\graphicspath{ {images/} }

\usepackage{tikz}	
\usepackage{tikz-feynman}
\tikzfeynmanset{compat=1.0.0}

\subheader{}

\title{Lectures on Open Effective Field Theories}

\author[a]{Thomas Colas}
\affiliation[a]{Department of Applied Mathematics and Theoretical Physics, University of Cambridge, Wilberforce Road, Cambridge, CB3 0WA, UK}
\emailAdd{tc683@cam.ac.uk}

\abstract{Effective field theories offer a powerful method to unify diverse models under a small set of control parameters, allowing systematic expansions around well-established theories.
These techniques, developed in particle physics, were designed for experiments where the initial state — the vacuum before a scattering event — is as clean and isolated as possible. 
Besides colliders, realistic environments are often noisy and dissipative. 
The recognition of the limitations of traditional EFT techniques has, over the past decade, sparked intense progress at the interface of high-energy physics and condensed matter.
These considerations motivate a new approach to gravitation and cosmology, one that models the gravitational sector as evolving in the presence of an unobservable medium. 
Open Effective Field Theories provide a systematic and controllable field-theoretic framework for modeling dissipation and noise in gravitation and cosmology. 
These notes aim to introduce this versatile toolkit, enabling model-agnostic assessments of how unknown environments shape our observational probes.  \\

This set of lectures was prepared for the Summer School \textit{The Disordered Universe 2025}. 
}  

\date{}

\begin{document}
	
\sloppy
	
\maketitle


\setcounter{section}{-1}


\section{Introduction}\label{sec:intro}

Theoretical cosmology is a hybrid discipline, situated at the intersection of general relativity, quantum field theory, and statistical physics. The strength of the current cosmological paradigm lies in its ability to synthesize insights from all three. For example, the formation of heavy elements in the early universe --- known as Big Bang nucleosynthesis --- depends on a subtle interplay between the universe’s expansion rate, quantum decay processes, and both equilibrium and non-equilibrium statistical physics.

Contemporary open questions in cosmology --- such as the nature of dark matter, dark energy, or a possible early phase of inflation --- all share a common feature: the absence of direct experimental probes. As a result, these phenomena must be studied indirectly, through their gravitational and electromagnetic imprints. From our perspective, they act as an effective medium through which gravity and light propagate. To characterize this medium, we cannot afford to neglect any of the three theoretical pillars of cosmology: general relativity, quantum field theory, and statistical physics.

That said, every physicist has their own trajectory in acquiring knowledge, and naturally gravitates toward certain methods. Over the past 75 years, effective field theories (EFTs) --- developed primarily within particle physics --- have proven invaluable in systematically parameterizing our ignorance about unknown physics. The EFT of Inflation~\cite{Cheung:2007st} and the EFT of Dark Energy~\cite{Gubitosi:2012hu} provide universal frameworks for describing single-clock models of the early and late universe. While this particle-physics-inspired approach efficiently incorporates symmetry principles and scale hierarchies, it is less suited to phenomena that lie outside the realm of collider physics. At the same time, the universe is neither empty nor static; it is a setting for rich and complex phenomena, including dissipation and noise, decoherence and classicalization, entropy production, and non-equilibrium dynamics. 

The aim of these lecture notes is to provide readers with a background in particle physics a set of tools to extend EFT techniques to gravitational systems that exhibit such phenomena. For readers more familiar with general relativity, we hope the notes offer a gentle introduction to quantum field theory methods applied to cosmology.

\begin{itemize}
	\item In \textbf{Lecture~\ref{sec:SK}}, we introduce the \emph{Schwinger--Keldysh formalism}, an approach for handling non-equilibrium and open quantum systems.
	\item In \textbf{Lecture~\ref{sec:scalar}}, we construct a first example of an open effective theory for relativistic scalar fields.
	\item In \textbf{Lecture~\ref{sec:lec3}}, we apply this framework to the scalar sector of single-clock inflation.
	\item In \textbf{Lecture~\ref{sec:lec4}}, we extend the formalism to the simplest gauge theory: electromagnetism in a medium.
	\item Finally, in \textbf{Lecture~\ref{sec:lec5}}, we develop an open effective field theory for gravity in a medium, which serves as a foundation for exploring dissipative and stochastic effects in cosmology.
\end{itemize}

\paragraph{References.} Readers may find useful material in the following reviews \cite{2016RPPh...79i6001S, Liu:2018kfw, Burgess:2022rdo} and textbooks \cite{breuerTheoryOpenQuantum2002, kamenev_2011, Calzetta:2008iqa}. \\

\paragraph{Nomenclature.} There are a lot of variations in the terminology used by different communities to describe open and non-equilibrium dynamics. In these notes, we mostly follow the nomenclature from \cite{breuerTheoryOpenQuantum2002} used in the open systems community. Unitarity refers to the existence of a unitary evolution operator $\mathcal{U}(t,t_0)$ mapping an initial state $\rho(t_0)$ to a final state $\rho(t)$ through
\begin{align}\label{eq:unitevol}
	\rho(t_0) \rightarrow \rho(t) = \mathcal{U}(t,t_0)  \rho(t_0) \mathcal{U}^\dag(t,t_0),
\end{align}
with 
\begin{align}
	\mathcal{U}^\dag(t,t_0) \mathcal{U}(t,t_0)  = \mathcal{U}(t,t_0)    \mathcal{U}^\dag(t,t_0) =  \mathrm{Id},
\end{align}
and initial condition $\mathcal{U}(t_0,t_0) = \mathrm{Id}$. We further consider density matrices which are (i) normalised $\mathrm{Tr} \rho =1$, (ii) Hermitian $\rho^\dag = \rho$ and (iii) positive definite $\rho > 0$. As we will see at length throughout these lectures, not all physical evolution can be written under the form of \Eq{eq:unitevol}. Whenever some degrees of freedom are experimentally inaccessible, initially available information can get distributed in the unknown environment and eventually lost. Such evolution maps pure states to mixed states which can be computed by performing a Schwinger-Keldysh path integral. This path integral is controlled by an effective functional that we decompose into
\begin{align}\label{eq:Seffintro}
	S_{\mathrm{eff}}\left[ \varphi_+, \varphi_-\right] = S_{\mathrm{unit}}\left[\varphi_+\right] - S_{\mathrm{unit}}\left[\varphi_-\right] + S_{\mathrm{non-unit}}\left[\varphi_+,\varphi_- \right],
\end{align}
for a generic scalar $\varphi$, where $S_{\mathrm{unit}}$ alone leads to evolution of the form of \Eq{eq:unitevol} while $S_{\mathrm{non-unit}}$ contains terms mixing the $+$ and $-$ branches of the path integral. These terms are the ones responsible for driving pure states into mixed states. Later on, we will encounter a convenient basis known as the Keldysh basis, in which the field $\varphi$ is decomposed into a retarded $\varphi_r$ and advanced component $\varphi_a$ through
\begin{align}
	\varphi_r = \frac{\varphi_+ + \varphi_-}{2}, \qquad \varphi_a = \varphi_+ - \varphi_-.
\end{align}
We will often call \textit{stochastic} the operators $\mathcal{O}(\varphi_a^{2p})$ that are even in powers of $\varphi_a$. These contributions never come from the unitary part of the functional are related to the noise sourcing the open systems. The operators  $\mathcal{O}(\varphi_a^{2p + 1})$ that are odd in powers of $\varphi_a$ can always be decomposed into a unitary contribution of the form $S_{\mathrm{unit}}\left[\varphi_+\right] - S_{\mathrm{unit}}\left[\varphi_-\right]$ and a remaining part, intrinsically non-unitary, that we refer to as being dissipative. This terminology slightly departs from the one used in dissipative hydrodynamics. In particular, dissipation does not relate to the eventual conservation laws that the system may have.\\

\noindent \textit{Note}: These lecture notes are meant to be improved. Please report typos, mistakes or poorly written sections to \href{mailto:tc683@cam.ac.uk}{tc683@cam.ac.uk}. \\

\section{Lecture 1: Schwinger-Keldysh formalism}\label{sec:SK}

Primordial cosmology relies on the correspondence between late-time observables and early-time cosmological correlators. While the former involves classical statistics of cosmological tracers (temperature anisotropies of the CMB, density contrast of the LSS, $\cdots$), the latter features quantum expectation values of quantum field theoretic operators. Schematically, 
\begin{align} \label{eq:corresp}
	\langle \prod_{i = 1}^n \delta(\bmk_i) \rangle \quad \leftrightarrow \quad \langle \prod_{i = 1}^n \widehat{\phi} (\bmk_i)\rangle.
\end{align}
While many lecture notes (e.g. \cite{Baumann:2009ds}) and textbooks (e.g. \cite{Dodelson:2003ft}) develop the physical origin of the relation between the right and left hand sides of the above equation, our goal in this Section consists in reviewing the perturbative computation of the right-hand side only - that is the computation of cosmological correlators through the Schwinger-Keldysh formalism\footnote{For an investigation of the left-hand only, see \cite{Colas:2019ret} and references therein.}. The presentation closely follow \cite{Chen:2017ryl}.


\subsection{Cosmological correlators}

Let us consider a homogeneous and isotropic universe described by a FLRW background metric
\begin{align}
	\dd s^2 = a^2(\eta) \left( - \dd \eta^2 + \dd \bmx^2 \right).
\end{align}
The scale factor $a(\eta)$ expressed in terms of the conformal time $\eta$ controls the expansion of the universe. During inflation, approximating the universe by a de Sitter geometry, we have 
\begin{align}
	a(\eta) = - \frac{1}{H \eta}, \qquad \eta \in \left] - \infty, \eta_0 \right],
\end{align}  
where the Hubble parameter $H$ is a constant (up to slow-roll corrections). The conformal time $\eta_0$ corresponds to the end of inflation and the beginning of the hot Big-Bang, which is often taken in practice to $\eta_0 \rightarrow 0$.

On the top of the background geometry, we consider a massless scalar denoted $\phi(\eta, \bmx)$ with a speed of sound $c_s^2$. Its linear action is given by
\begin{align}\label{eq:zetaaction}
	S^{(2)}_\phi &= - \frac{1}{2}\int \dd \eta \mathrm{d}^3 \bs{x}   a^2(\eta)  \left[c_s^{-2}\phi^{\prime 2} -\left(\partial_i \phi\right)^2\right].
\end{align}
One can Fourier transform the field variable
\begin{align}\label{eq:Fourier}
		\phi_\bmk(\eta) = \int \dd^3 \bmx  \phi(\eta, \bmx)\ee^{i \bmk.\bmx}.
\end{align}	
in order to take advantage of the spatial homogeneity. 
Following the canonical quantisation prescription, field variables are promoted to quantum operators obeying the equal-time commutation relations 
\begin{align}
	[\widehat{\phi}_{\bs{k}}(\eta),\widehat{\Pi}_{\bs{q}}(\eta) ] = i \delta(\boldsymbol{k}+\bs{q}),
\end{align}
with $\widehat{\Pi}_{\bs{q}}$ the conjugate momentum.
Making use of the linear evolution, one can relate the field operators at time $\eta$ to $\widehat{a}_{\bs{k}}$ and $\widehat{a}^{\dag}_{-\bs{k}}$, the creation and annihilation operators of the Bunch-Davies vacuum $\ket{\Omega}$ in the asymptotic past through the mode-function decomposition
\begin{align}\label{eq:modefctdecomp}
	\widehat{\phi}_{\bs{k}}(\eta) &= \phi_{k}(\eta) \widehat{a}_{\bs{k}} +  \phi^{*}_{k}(\eta) \widehat{a}^{\dag}_{-\bs{k}} 
\end{align}
The mode-functions obey the classical equation of motion. Explicitly, rescaling $\phi$ by the scale factor $a$, the equation of motion for $\phi$ takes the familiar form of a parametric oscillator 
\begin{align}\label{eq:MSeom}
	(a\phi_{k})'' + \left(c_s^2 k^2 - \frac{a''}{a} \right) (a\phi_{k})=0.
\end{align}
Choosing the outgoing branch
\begin{align}\label{eq:BDmode}
	\phi_{k}(\eta)=\frac{H}{\sqrt{2 c_s k^3}} \left(1+ i c_s k \eta \right)\ee^{-i c_s k \eta}.
\end{align}
The Hilbert space of the $\phi$ field is constructed out of the tensor product of each mode's Fock space, that is $\mathcal{H} = \otimes_\bmk \mathcal{H}_\bmk$ with
\begin{align}\label{eq:Hspace}
	\mathcal{H}_\bmk = \left\{ \ket{\Omega} , \widehat{a}^\dag_{\bs{k}} \ket{\Omega}, \widehat{a}^\dag_{\bs{k}} \widehat{a}^\dag_{\bs{k}} \ket{\Omega}, \cdots \right\},
\end{align}
the Bunch-Davies vacuum being defined from $\widehat{a}_{\bs{k}} \ket{\Omega} = 0$.

We are now in position to define the observables presented on the RHS of \Eq{eq:corresp} we aim to compute. Cosmological correlators are defined as the expectation value of the Bunch-Davies vacuum of equal-time product of local operators at the future conformal boundary of dS,
\begin{align}\label{eq:def}
	\langle \prod_{i = 1}^n \widehat{\phi}(\bmk_i)\rangle \equiv \lim_{\eta_{0} \rightarrow 0} \langle \Omega | \prod_{i = 1}^n  \widehat{\phi}_{\bmk_i}(\eta_0)|\Omega \rangle.
\end{align}
Note that the terminology \textit{observable} may be misleading. Contrarily to scattering amplitudes, cosmological correlators are not field redefinition invariant. Instead, correlators dictate the summary statistics of the field $\phi$ at time $\eta_0$ (its mean, variance, skewness, kurtosis, $\cdots$), which might be more familiar in the context of statistical field theory. The knowledge of the infinite tower of cosmological correlators is equivalent to the knowledge of the full Probability Distribution Function (PDF) for $\phi$.\footnote{Which is \textit{not} the complete characterization of the quantum state $ \widehat{\rho}_\Omega = |\Omega \rangle \langle \Omega |$, but only the diagonal density matrix element in the field basis $P_\Omega(\phi_\bmk, \eta_0) = \langle \phi_\bmk, \eta_0 |\widehat{\rho}_\Omega | \phi_\bmk, \eta_0 \rangle$. Off-diagonal density matrix elements $\langle \widetilde{\phi}_\bmk, \eta_0 |\widehat{\rho}_\Omega  | \phi_\bmk, \eta_0 \rangle$ or, equivalently, correlators of the conjugate momentum $\widehat{\Pi}_\bmk$ are needed to fully characterize the state \cite{Colas:2024ysu}.} 


\subsection{In-in formalism}

The \textit{in-in formalism} \cite{Weinberg:2005vy, Weinberg:2006ac, Chen:2017ryl} is a well-established framework aiming at computing perturbatively cosmological correlators. Let us review its basic structure.

\subsubsection{Weinberg formula}

\paragraph{From Schr\"odinger to Heisenberg.}

In Schr\"odinger picture, the state $\ket{\Psi(\eta)}$ 
obey the Schr\"odinger equation 
\begin{align}\label{eq:eom:Schrodinger}
	\frac{\dd \ket{\Psi(\eta)} }{\dd \eta} = - i \widehat{H}({\eta})\ket{\Psi(\eta)}, 
\end{align}
where $\widehat{H}({\eta})$ is the Hamiltonian. This equation can be solved formally to obtain the quantum state of cosmological perturbations at the end of inflation
\begin{align}\label{eq:evolgen}
	\ket{\Psi(\eta_0)} = \widehat{\mathcal{U}}(\eta_0,- \infty)  \ket{\Omega} ,
\end{align}
where 
we have introduced the evolution operator 
\begin{align}
	\label{eq:U:H}
	\widehat{\mathcal{U}}(\eta_0, - \infty)= \mathcal{T} \exp\left[{-i \int_{- \infty}^{\eta_0} \dd{\eta}'\;  \widehat{H}({\eta}')}\right] 
\end{align}
$\mathcal{T}$ ($\overline{\mathcal{T}}$) representing (anti-) time ordering of quantum operators.
This is the so-called \textit{Schr\"odinger picture}, where the state $\ket{\Psi(\eta)}$ evolve with $\widehat{\mathcal{U}}$ and observables $\widehat{\phi}_{\bmk_i}$ are time-independent. Using the unitarity of the evolution operator $\widehat{\mathcal{U}}^\dag\widehat{\mathcal{U}} = \widehat{\mathcal{U}} \widehat{\mathcal{U}}^\dag= \mathbb{I}$ and the mapping from the Heisenberg to the Schr\"odinger picture
\begin{align}
 	\widehat{\mathcal{U}}^\dag(\eta_0, - \infty) \left[\widehat{\phi}_{\bmk_i}(\eta_0) \right]	\widehat{\mathcal{U}}(\eta_0, - \infty) = \widehat{\phi}_{\bmk_i}
\end{align}  
we find the Schr\"odinger picture representation of cosmological correlators
\begin{align}
	\langle \prod_{i = 1}^n \widehat{\phi}_{\bmk_i}\rangle = \lim_{\eta_{0} \rightarrow 0} \langle \Psi (\eta_0) | \prod_{i = 1}^n  \widehat{\phi}_{\bmk_i}|\Psi (\eta_0) \rangle  .
\end{align}

\paragraph{Interaction picture.}

To conveniently compute these quantities in perturbation theory, we divide the Hamiltonian into a free part and an interaction part,
\begin{align}
	\widehat{H}(\eta)= \widehat{H}_0(\eta) +  g \widehat{H}_{\mathrm{int}} (\eta)
\end{align}
and introduce the free evolution operator $\widehat{\mathcal{U}}_0$, defined as in \Eq{eq:U:H} where $\widehat{H}$ is replaced by $\widehat{H}_0$. This approach is known as the \textit{interaction picture} which provides an in between the Schr\"odinger and Heisenberg pictures. In this picture, quantum states evolve with the interaction Hamiltonian $g\widehat{H}_{\mathrm{int}}$ and operators evolve with the free Hamiltonian $\widehat{H}_0$. The link between the Schr\"odinger and the interaction picture is given by 
\begin{align}
	\label{eq:rho:Hint}
	| \widetilde{\Psi} (\eta_0)\rangle = \widehat{\mathcal{U}}^{\dag}_0(\eta_0,-\infty) 	\ket{ \Psi(\eta_0) } \, , 
\end{align}
where tildes denote quantities evaluated in the interaction picture. From \Eq{eq:eom:Schrodinger} it is easy to show that the state evolves according to
\begin{align}\label{eq:eom:IntPict}
	\frac{\dd | \widetilde{\Psi} (\eta)\rangle}{\dd \eta}=
	-ig  \widetilde{H}_{\mathrm{int}}(\eta)| \widetilde{\Psi} (\eta)\rangle \, ,
\end{align}
where
\begin{align}\label{eq:Hintintpic}
	\widetilde{H}_{ \mathrm{int}} (\eta) = \widehat{\mathcal{U}}^{\dag}_0(\eta,-\infty) \widehat{H}_{\mathrm{int}}(\eta) \widehat{\mathcal{U}}_0(\eta,-\infty)
\end{align}
and we have used that the evolution operator is Hermitian, \ie $\widehat{\mathcal{U}}_0 \widehat{\mathcal{U}}_0^\dag = \widehat{\mathcal{U}}_0^\dag \widehat{\mathcal{U}}_0 = \mathrm{Id}$, since $\widehat{H}_0=\widehat{H}_0^\dagger$. 

\paragraph{In-in correlators.}

Given an interaction Hamiltonian $g \widehat{H}_{\mathrm{int}}$, cosmological correlators can be written in the interaction picture 
\begin{align}\label{eq:ininstart}
	\langle \prod_{i = 1}^n \widehat{\phi} (\bmk_i)\rangle &= \lim_{\eta_{0} \rightarrow 0} \bra{\Omega} \left[\overline{\mathcal{T}} \ee^{i g \int_{-\infty(1-i\epsilon)}^{\eta_0} \dd{\eta}' \;  \widetilde{H}_{\mathrm{int}}({\eta}')}\right] \prod_{i = 1}^n  \widetilde{\phi}_{\bmk_i}(\eta_0) \left[ \mathcal{T}\ee^{-i g \int_{-\infty(1+i\epsilon)}^{\eta_0} \dd{\eta}'\; \widetilde{H}_{\mathrm{int}}({\eta}') }\right]\ket{\mathrm{\Omega}}.
\end{align}
The $i \epsilon$ deformation is here to ensure the projection of the adiabatic vacuum of the interacting theory onto the vacuum of the free theory in the asymptotic past, see \Refs{Adshead:2009cb,Kaya:2018jdo, Albayrak:2023hie} for in-depth discussions. In practice, the correlation functions of the theory are computed perturbatively, at a given order $n$ in $\widehat{H}_{\mathrm{int}}$, leading to
\begin{align}\label{eq:pertinin}
	\langle \prod_{i = 1}^n \widehat{\phi} (\bmk_i)\rangle &= \lim_{\eta_{0} \rightarrow 0}  (ig)^n  \int_{-\infty}^{\eta_0}\dd \eta_n\int_{-\infty}^{\eta_n}\dd \eta_{n-1} \cdots \int_{-\infty}^{\eta_{2}}\dd \eta_1 \Bigg. \\
	& \bra{\Omega}\left[\widetilde{H}_{\mathrm{int}}(\eta_1),\left[\widetilde{H}_{\mathrm{int}}(\eta_{2}), \cdots \left[\widetilde{H}_{\mathrm{int}}(\eta_n),\prod_{i = 1}^n  \widetilde{\phi}_{\bmk_i}(\eta_0) \right]\cdots \right] \right]\ket{\Omega} + \mathcal{O}(g^{n+1}) \nonumber
\end{align}
where the appropriate $i \epsilon$ prescription must be used depending on which branch of \Eq{eq:ininstart} provides the time integration. This approach has been used in a variety of problems in primordial cosmology to compute cosmological correlators of scalar and tensor perturbations.

\subsubsection{Path integral representation}

While the operator formalism has the advantage to make computations explicit, it relies in practice in performing a large number of commutation relations and Wick contractions, which makes it rapidly unpractical. The path integral representation aims bypassing these steps, often at the price of losing transparency. Here, we aim at making the connection between the operator and path integral formalisms explicit. A detailed derivation can be found in \cite{Boyanovsky:2015xoa}.

Let us reconsider the Heisenberg's picture of our cosmological correlators given in \Eq{eq:def}. An equivalent representation of the quantum state $\ket{\Omega}$ is given in terms of the \textit{density matrix} $\widehat{\rho}_\Omega \equiv \ket{\Omega} \bra{\Omega}$ \cite{breuerTheoryOpenQuantum2002}. The advantage of working with a density matrix, which accomadotates both \textit{pure} and \textit{mixed} states, will become transparent in the next section \ref{subsec:heavy}. In terms of the density matrix, \Eq{eq:def} is restated 
\begin{align}
		\langle \prod_{i = 1}^n \widehat{\phi}(\bmk_i)\rangle = \lim_{\eta_{0} \rightarrow 0} \mathrm{Tr}_{\mathcal{H}} \left[\prod_{i = 1}^n \widehat{\phi}_{\bmk_i}(\eta_0) \widehat{\rho}_\Omega  \right]
\end{align}
where the trace is over the full Hilbert space defined above \Eq{eq:Hspace}. One can explicitly compute this trace in a given basis. For instance, considering that the theory only contains one scalar degree of freedom $\phi$, one can express this expectation value in the field basis constructed out of the eigenstates of the field operators in the Heisenberg picture
\begin{align}
	\widehat{\phi}_\bmk(\eta_0) |\phi_\bmk ,\eta_0  \rangle = \phi_\bmk(\eta_0) |\phi_\bmk ,\eta_0 \rangle.
\end{align}
In this basis, the expectation value becomes
\begin{align}
	\langle \prod_{i = 1}^n \widehat{\phi}(\bmk_i)\rangle &= \lim_{\eta_{0} \rightarrow 0} \int \dd \phi_\bmk \langle \phi_\bmk, \eta_0 |  \prod_{i = 1}^n \widehat{\phi}_{\bmk_i}(\eta_0) \widehat{\rho}_\Omega|\phi_\bmk, \eta_0\rangle \label{eq:obsref}\\
	&= \lim_{\eta_{0} \rightarrow 0} \int \dd \phi_\bmk \int \dd  \widetilde{\phi}_\bmq \langle \phi_\bmk, \eta_0 |  \prod_{i = 1}^n \widehat{\phi}_{\bmk_i}(\eta_0)  |\widetilde{\phi}_\bmq , \eta_0\rangle \langle \widetilde{\phi}_\bmq , \eta_0|\widehat{\rho}_\Omega|\phi_\bmk, \eta_0 \rangle
\end{align}
where we used in the second line a resolution of the identity
\begin{align}
	\mathbb{I} =  \int \dd \widetilde{\phi}_\bmq |\widetilde{\phi}_\bmq , \eta_0\rangle \langle \widetilde{\phi}_\bmq , \eta_0|.
\end{align}

Cosmological correlators being diagonal in the field basis, 
\begin{align}
	\langle \phi_\bmk ,  \eta_0|  \prod_{i = 1}^n \widehat{\phi}_{\bmk_i}(\eta_0)  |\widetilde{\phi}_\bmq, \eta_0 \rangle = \prod_{i = 1}^n \phi_{\bmk_i}(\eta_0) \delta^{(3)} (\bmk - \bmq) \delta(\phi - \widetilde{\phi} ) .
\end{align}
all the information of interest is contained in the diagonal element of density matrix in the field basis 
\begin{align}\label{eq:PDFdef}
	P_\Omega[\phi_\bmk(\eta_0)] \equiv \langle \phi_\bmk, \eta_0 | \widehat{\rho}_\Omega | \phi_\bmk, \eta_0 \rangle.
\end{align}
This objects define a Probability Distribution Function (PDF) from which one extracts expectation values in the usual way,
\begin{align}\label{eq:exppdf}
	\langle \prod_{i = 1}^n \widehat{\phi}_{\bmk_i}\rangle &= \lim_{\eta_{0} \rightarrow 0}  \int \dd \phi_\bmk \prod_{i = 1}^n \phi_{\bmk_i}(\eta_0) P_\Omega[\phi_\bmk( \eta_0)]. 
\end{align}
When the state is pure ($\widehat{\rho}_\Omega^2 = \widehat{\rho}_\Omega$), this definition matches the wavefunction approach $P_\Omega[\phi_\bmk(\eta_0)] = |\Psi[\phi_\bmk(\eta_0)]|^2$.

\begin{figure}[tbp]
 	\centering
 	\includegraphics[width=0.9\textwidth]{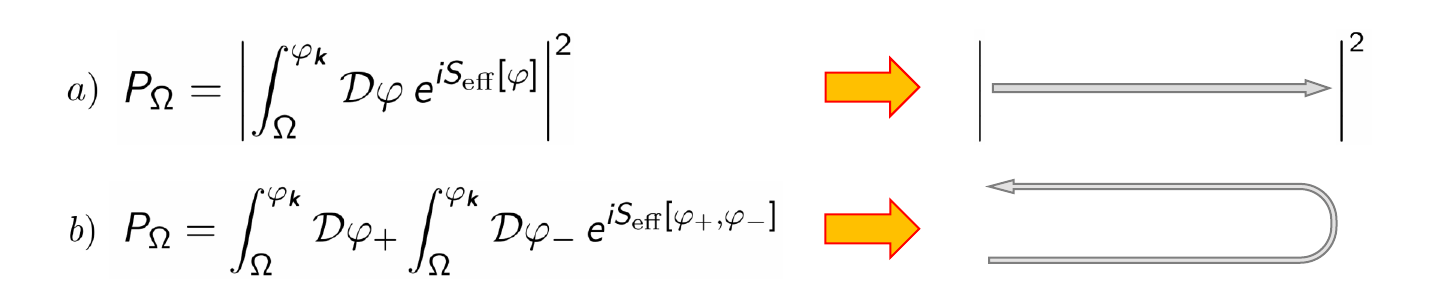}
 	\caption{Comparison between a) the single-branch path integral contour used in the computation of the wavefunction, and b) the double-branch path integral contour used in the computation of the density matrix. While the two approaches are equivalent in the case of pure states (see e.g. \cite{Donath:2024utn}), the in-in contour also accomodates mixed states, which have no single-branch analogue. }
 	\label{fig:contourunit}
 \end{figure}

The purpose in life of a path integral is to prepare the statistics sampled by an experiment. While single-branch path integrals sample a wavefunction, 
\begin{align}
	\Psi[\phi_\bmk(\eta_0)] &= \langle \phi_\bmk ,\eta_0 | \Omega \rangle = \langle \phi_\bmk | \Psi(\eta_0) \rangle = \langle \phi_\bmk | \widehat{\mathcal{U}}(\eta_0,- \infty) | \Omega \rangle \\
	&= \int_{\Omega}^{\phi_\bmk} \mathcal{D} \varphi_+ \ee^{i S[\varphi_+]},
\end{align}
a density matrix necessitates the use of double-branch path integrals,
\begin{align}
	P_\Omega[\phi_\bmk(\eta_0)]& = \langle \phi_\bmk, \eta_0 | \Omega \rangle \langle \Omega | \phi_\bmk, \eta_0 \rangle \\
	&= \int_{\Omega}^{\phi_\bmk} \mathcal{D} \varphi_+  \int_{\Omega}^{\phi_\bmk} \mathcal{D} \varphi_- \ee^{i \left(S[\varphi_+] - S[\varphi_-]\right)}.\label{eq:unitth}
\end{align} 
Such a path integral contour is presented in \Fig{fig:contourunit}. Known as the \textit{Schwinger-Keldysh}, \textit{in-in} or \textit{closed-time} path contour, it can be intuitively understood in the following manner. Starting from some initial state $|\Omega\rangle$, the path integral evolves the dynamics forward up to $\eta_0$, then backward to initial state $\langle \Omega |$ again. Compared to the familar \textit{in-out} formalism used in particle physics which specifies the state both in the asymptotic past and asymptotic future, the Schwinger-Keldysh formalism only relies on initial conditions, leaving the state at time $\eta_0$ unconstrained. It allows one to access transient dynamics which can be really far-from-equilibrium, which is the primary reason for introducing such a formalism.

The second important feature of the in-in contour is the apparent doubling of the degrees of freedom of the theory, one for each branch of the path integral: $\varphi_+$ for the forward branch and $\varphi_-$ for the backward branch. It is important to bear in mind that this is nothing but a trick to capture describe statistical and quantum fluctuations in the formalism. Physically, there is only one dynamical degree of freedom, a fact we can make manifest by computing dispersion relations and propagators. 

At last, note that the exponential in \Eq{eq:unitth} factorises between the $+$ branch contribution and the $-$ branch contribution, $S[\varphi_+] - S[\varphi_-]$. This is a peculiarity of \textit{unitary/closed} theories in which information is conserved \cite{Burgess:2024heo}. In \Sec{subsec:heavy}, we will discuss what happens when $\phi$ exchange information with hidden sectors, leading to the emergence of \textit{mixed states}. At the level of the path integral, this will correspond to investigate the appearance of terms mixing the branches of the path integral, leading to $S_{\mathrm{eff}} = S[\varphi_+] - S[\varphi_-] + F[\varphi_+, \varphi_-]$. The generic construction of these effective functionals describing pure and mixed states in a single way is the object of Lecture 2.

\subsubsection{In-in diagrammatics}\label{subsubsec:diagram}

We now aim at computing \Eq{eq:exppdf} in perturbation theory. To do so, we spell a set of Feynman rules and associated diagrammatic. 

\begin{figure}[tbp]
	\centering
	\includegraphics[width=1.1\textwidth]{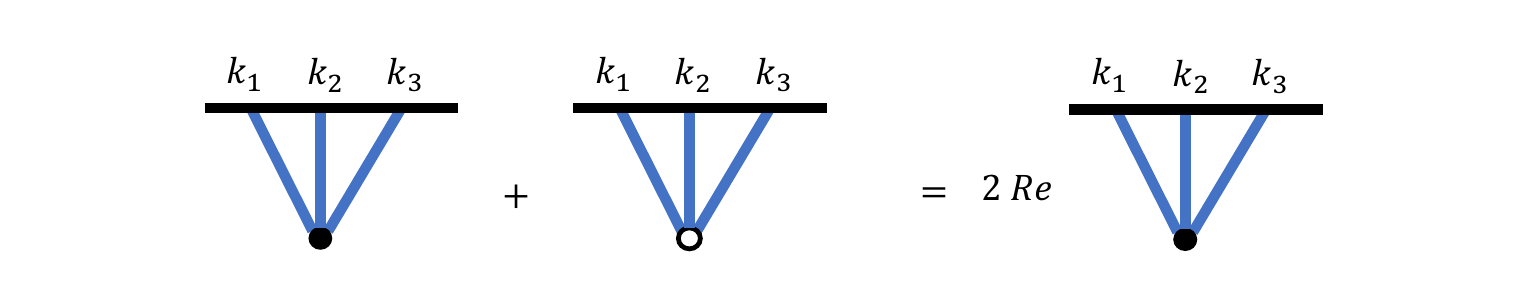}
	\caption{In-in diagrammatics of tree-level three-point correlator. The ``$+$'' and ``$-$'' diagram are related by complex conjugations for parity even interactions.
	}
	\label{fig:bispectrumpm}
\end{figure}

Invariance by spatial translation (homogeneity) implies that correlators must contain at least one Dirac delta for momentum conservation. While disconnected diagrams, proportional to two or more Dirac, can always be substracted by removing all products of lower $n$-point functions, connected correlators, proportional to a single Dirac, are the main objects of interest at a given order $n$. We introduce the notation: 
\begin{align}
		\langle \prod_{i = 1}^n \widehat{\phi}(\bmk_i)\rangle &\equiv (2\pi)^3 \delta\Big(\sum_{i = 1}^n \bmk_i\Big) B_n(\{\bmk\}). 
\end{align}
To compute this quantity in perturbation theory, we separate the free action from the interactions, 
\begin{align}
	S[\varphi] = S_{0}[\varphi] + S_{\mathrm{int}}[\varphi]
\end{align}
where $S_{\mathrm{int}}[\varphi]$ is at least cubic in $\varphi$. While the $S_{0}[\varphi]$ part can be exactly solved to obtain the propagators of the theory, we perform a systematic expansion in $S_{\mathrm{int}}$, leading to 
\begin{align}
	B_n(\{\bmk\})& = \int_{\Omega}^{\phi_\bmk} \mathcal{D} \varphi_+  \int_{\Omega}^{\phi_\bmk} \mathcal{D} \varphi_- \bigg\{ \varphi_+^n \Big( 1 + i S_{\mathrm{int}}[\varphi_+] -  i S_{\mathrm{int}}[\varphi_-] - \frac{1}{2} S^2_{\mathrm{int}}[\varphi_+]  - \frac{1}{2} S^2_{\mathrm{int}}[\varphi_-] \nonumber \\
	&\qquad  +  \frac{1}{2} S_{\mathrm{int}}[\varphi_+]  S_{\mathrm{int}}[\varphi_-] + \cdots \Big) \bigg\}  \ee^{i \left(S_0[\varphi_+] - S_0[\varphi_-]\right)} - \text{(disconned diagrams)}
\end{align}
where we inserted $\varphi_+^n$ to compute the associated $n$-point function \cite{Chen:2017ryl}.\footnote{One could equally insert $\varphi_-^n$, given the boundary conditions considered.}
In the operator language, the contribution from the ``$+$'' branch corresponds to acting on the left with $\widetilde{H}_{\mathrm{int}}$ in \Eq{eq:pertinin}, while the contribution from the ``$-$'' acting on the right. In the end, it amounts to evaluate these various correlators in the free theory using Wick contractions:
\begin{align}
	B_n(\{\bmk\})& = \exx{ \varphi_+^n }^{(0)} + i \exx{ \varphi_+^n S_{\mathrm{int}}[\varphi_+] }^{(0)}  - i \exx{ \varphi_+^n S_{\mathrm{int}}[\varphi_-] }^{(0)} - \frac{1}{2} \exx{ \varphi_+^n S^2_{\mathrm{int}}[\varphi_+] }^{(0)} \nonumber \\
	&\qquad  - \frac{1}{2} \exx{ \varphi_+^n S^2_{\mathrm{int}}[\varphi_-] }^{(0)} + \exx{ \varphi_+^n S_{\mathrm{int}}[\varphi_+] S_{\mathrm{int}}[\varphi_-]}^{(0)} + \cdots - \text{(disconned diagrams)},
\end{align}
where $\exx{ \cdots }^{(0)}$ stands for correlators evaluated in the free theory. Reproducing this computation everytime we have to evaluate a correlator is tedious, this is why people develop a set of rules to perform these steps in a systematic manner:

\begin{tcolorbox}[%
	enhanced, 
	breakable,
	skin first=enhanced,
	skin middle=enhanced,
	skin last=enhanced,
	before upper={\parindent15pt},
	]{}
	
	\vspace{0.05in}
	
\paragraph{Feynman rules.} 
\begin{enumerate}
	\item To compute $B_n$ in perturbation theory, draw a diagram with $V$ vertices, $I$ internal lines, each connecting two vertices, and $n$ external lines connecting a vertex to the future boundary $\eta_0 \rightarrow 0$ represented by a horizontal line at the top of the diagram, see e.g. \Fig{fig:Tpm}. Times run from bottom at $\eta \rightarrow - \infty$ to top at $\eta \rightarrow \eta_0$.
	\item Each vertex can either be a ``$+$'' vertex, coming from $S_{\mathrm{int}}[\varphi_+]$ or a ``$-$'' vertex, coming from $S_{\mathrm{int}}[\varphi_-]$. The final expression for $B_n$ is obtained by summing over the $2^V$ to label the $V$ vertices.
	\item To each of the $n$ external lines, associate a spatial momentum $\bmk_i$. Internal momenta are associated to $p_m$, with $m = 1, \cdots,I$.
	\item Internal lines represent four different types of bulk-to-bulk propagators:\footnote{We follow the notations in \cite{Chen:2017ryl}. Notice that the four bulk-to-bulk propagators are not linearly independent due to the identity $D_{++}+D_{--} = D_{+-} + D_{-+}$.}
	\begin{align}
		\label{eq_dSDmp}
		&~D_{-+}(k;\eta_1,\eta_2) = \phi_k(\eta_1)\phi^*_k(\eta_2) ,\quad ~D_{+-}(k;\eta_1,\eta_2) = \big[D_{-+}(k;\eta_1,\eta_2)\big]^* ,\\
		\label{eq_dSDpp}
		&~D_{\pm\pm}(k;\eta_1,\eta_2) = D_{\mp\pm}(k;\eta_1,\eta_2)\theta(\eta_1-\eta_2) + D_{\pm\mp}(k;\eta_1,\eta_2)\theta(\eta_2-\eta_1), \bigg.
	\end{align}
	depending whether  $+$ or $-$ vertices are connected by these propagators. $\phi_k(\eta_1)$ is the Bunch-Davies mode function found in \Eq{eq:BDmode}. External lines are associated to the bulk-to-boundary propagators
	\begin{align}
		\label{eq_flatK}
		K_+(k,\eta) = \phi^*_k(\eta)\phi_k(\eta_0) ,\qquad K_-(k,\eta)= \big[ K_+(k,\eta) \big]^*.
	\end{align}
	\item Vertices controlled by a coupling constant $g$ are associated to 
	\begin{align}
		\pm i g \int^{\eta_0}_{-\infty(1 \mp i \epsilon)} \frac{\dd \eta_A}{(H \eta_A)^4} , \qquad \text{for}~A = 1 , \cdots V, 
	\end{align}
	where the $\mp i \epsilon$ are there to project onto the interacting vacuum in the asymptotic past in the $+$ and $-$ contour respectively \cite{Kaya:2018jdo} and the $(H \eta_A)^{-4}$ comes from the volume factor $\sqrt{-g}$. The $\pm$ in front depend whether it is a $+$ or a $-$ vertex that is considered. 
	Moreover, each vertex comes with a delta conserving spatial momenta, a consequence of spatial homogeneity. All internal momenta $\bmp_m$ should be integrated over. Using the topological identiy $I - V + 1 = L$, there should remain in the end $L$ internal momenta integrals to perform, corresponding to the $L$ loops. 
	\item Combinatorial factors may need to be added, when permutations generate the same contribution. Check \cite{Chen:2017ryl} for a detailed discussion.
\end{enumerate}
In the absence of parity odd operators, there exists a simple relation between diagrams with $V$ ``$+$'' vertices and $\bar{V}$ ``$-$'' vertices denoted $B^{V \tilde{V}}_n(\{\bmk\})$, and the same diagram with exchanged ``$+$'' and ``$-$'' vertices, $B^{\bar{V} V}_n(\{\bmk\}) = (B^{V \bar{V}}_n(\{\bmk\}))^*$, see \Fig{fig:bispectrumpm}.
\end{tcolorbox} 

Let's illustrate these diagrammatic rules through a simple example. As a first example, we consider the leading cubic interactions in single-clock models of inflation \cite{Cheung:2007st}
\begin{align}
	S_{\mathrm{int}} = \int \dd^3 \bmx  \int \frac{\dd \eta}{(H \eta)^4} (- H \eta^3)\left[ -\frac{\lambda_1}{3!} \varphi^{\prime 3} - \lambda_2 \varphi' (\partial_i \varphi)^2 \right].
\end{align}
We explicitly treat the $\lambda_1$ interaction. The associated contact bispectrum, represented in \Fig{fig:bispectrumpm}, is given by 
\begin{align}
	B_3^{\lambda_1} = 2 \Re \left[ - i \frac{\lambda_1}{3!} \times 3! \times \int_{-\infty(1 - i \epsilon)}^{\eta_0} \frac{\dd \eta}{(H \eta)^4} (- H\eta)^3 K'_+(k_1,\eta) K'_+(k_2,\eta) K'_+(k_3,\eta) \right].
\end{align}
Let us discuss term by term the various contributions. First, the real part comes from the fact that the ``$-$'' diagram is nothing but the complex conjugate of the ``$+$'', as shown in \Fig{fig:bispectrumpm}. The $- \lambda_1/3!$ is the vertex contribution, while the $3!$ is the combinatorial factors, coming from the Indistinguishability between $k_1$, $k_2$ and $k_3$. At least, the integrand contains the $(-H\eta)^3$ factor from the vertex and the bulk-to-boundary propagators for the three field insertions. The latter feature a conformal time derivative coming from the fact we considered $\varphi^{\prime 3}$ and not $\varphi^3$. Injecting the expression of $\phi_k(\eta)$ found in \eqref{eq:BDmode} into the bulk-to-boundary propagator expression \eqref{eq_flatK}, we obtain 
\begin{align}
	B_3^{\lambda_1} &=  2 \Re \left[ i \lambda_1  \int_{-\infty(1 - i \epsilon)}^{\eta_0} \frac{\dd \eta}{(H \eta)} \prod_{i=1}^3 \frac{H^2}{2 c_s k_i^3} c_s^2 k_i^2 \eta \ee^{i c_s k_i \eta} \right] \\
	&=  2 \Re \left[ i\frac{ \lambda_1 H^5 c_s^3}{8 k_1 k_2 k_3} \int_{-\infty(1 - i \epsilon)}^{0} \dd \eta \eta^2 \ee^{i E_T \eta} \right] \\
	&= - \frac{\lambda_1 H^5 c_s^3}{2 k_1 k_2 k_3 E_T^3},
\end{align}
where in the last line we defined the total ``energy'' $E_T \equiv c_s (k_1 + k_2 + k_3)$ and took the $\eta_0 \rightarrow 0$ limit. The contribution from the $\lambda_2$ interaction can be computed through similar methods, though leading to a much more involved results \cite{Pajer:FTC}. In the \textit{Problem Set} part \ref{subsec:prob1} of this Lecture, the reader can practice in-in diagrammatics through simpler examples.


\subsection{Integrating out heavy fields}\label{subsec:heavy}

\textit{Foreword.} This section provides partial solutions to Problem 2 of the problem set in Section~\ref{subsec:prob1}, based on \cite{Colas:2025XXX}. Readers who wish to practice the methods introduced in Lecture 1 are encouraged to attempt the problem set first, before consulting this section.
\\

To illustrate the difference between in-out and in-in formalism, we make a short detour in flat spacetime, where the metric reads:
\begin{equation}
	\mathrm ds^2 = -\mathrm d t^2+ \mathrm d \bm x^2, \qquad \quad t \in ] - \infty, 0],\quad  \bm x \in \mathbb{R}^3.
\end{equation}
This ``half-Minkowski'' patch introduces subtle differences compared to the ``full-Minkowksi'' case where $t \in ] - \infty, \infty]$, which have been the object of recent investigations \cite{Salcedo:2022aal, Green:2024cmx}. As we will see, the $t\rightarrow0$ limit introduces a finite-time boundary, leading to time-translation symmetry breaking, particle production and emergent dissipative and stochastic effects.

\paragraph{Model.} Following \cite{Colas:2025XXX}, we consider the following two-field toy model with the Lagrangian, 
\begin{equation}
	\label{eq_flatL}
	\mathcal L[\varphi,\sigma] =  -\frac12 (\partial_\mu \varphi)^2 - \frac12 (\partial_\mu \sigma)^2 - \frac12 M^2\sigma^2 + \frac{1}{2\Lambda} \dot\varphi^2\sigma,
\end{equation}
with $\varphi$ being a massless scalar and $\sigma$ a massive scalar with mass $M$. Here $\Lambda$ is the cutoff scale for the dim-5 operator $\dot\varphi^2\sigma$, and we denote derivatives with respect to physical time $t$ by overdots. After quantization, scalar fields can be expanded in terms of creation/annihilation operators. For instance,
\begin{align}
	\sigma(t,\bm x) = \int \frac{\mathrm d^3\bm k}{(2\pi)^3}\,e^{i\bm k\cdot\bm x}\Big[u_\sigma(k,t)a_{_\sigma,\bm k} + u_\sigma^*(k,t)a_{_\sigma,-\bm k}^\dagger\Big],
\end{align}
where $u_\sigma(k,t)$ is the mode function of $\sigma$ determined by solving the Klein-Gordon equation with appropriate initial conditions. Assuming the Bunch-Davis (BD) vacuum, the mode function is given by:
\begin{align}
	\label{eq_flatusigma}
	u_\sigma(k,t) = \frac{e^{-i E_kt}}{\sqrt{2E_k}},\qquad E_k \equiv \sqrt{M^2+k^2}.
\end{align}
Similarly, the energy of massless $\varphi$ is $E_k=k$, so the mode function is:
\begin{align}
	\label{eq_flatuphi}
	u_\varphi(k,t) = \frac{e^{-i k t}}{\sqrt{2k}}.
\end{align}

\begin{figure}[tbp]
	\centering
	\includegraphics[width=0.8\textwidth]{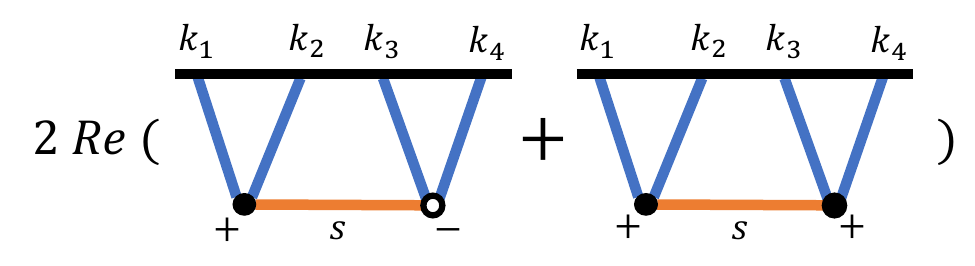}
	\caption{In-in diagrammatics of tree-level $s$-channel four-point correlator with a massive exchange. The massless scalar $\varphi$ is represented in \textit{blue} and the massive scalar $\sigma$ in \textit{orange}. 
	}
	\label{fig:Tpm}
\end{figure}

\paragraph{Correlator.} Below we will compute the $s$-channel four-point correlator:
\begin{align}
	\label{eq_flatI}
	B_4^s \equiv \langle \varphi_{\bm k_1}\varphi_{\bm k_2}\varphi_{\bm k_3}\varphi_{\bm k_4} \rangle_s',
\end{align}
where we use a prime to denote the momentum conservation factor $(2\pi)^3\delta(\bm k_1+\bm k_2+\bm k_3+\bm k_4)$ has been stripped off.
Focusing on this process, we study the diagrams presented in \Fig{fig:Tpm}. Our goal is to characterise and physically interpret the influence of the heavy field $\sigma$ while focusing on the massless field $\varphi$.
We proceed with the calculation via Schwinger-Keldysh formalism. Following \Sec{subsubsec:diagram}, we double all the field content, and the equal-time correlator (defined at $t=0$) can be expressed as the path integral:\footnote{
	Here the spacetime integral should be understood as $\int \dd^4x \equiv \int_{-\infty}^0\dd t\int \dd^3\bm x$.}
\begin{align}
	\label{eq_flatPathInt}
	\exx{ \mathcal O[\varphi,\sigma]} = \int_{\text{BD}}^\varphi \mathcal D\varphi_+\int_{\text{BD}}^\varphi \mathcal D\varphi_-\int_{\text{BD}}^\sigma \mathcal D\sigma_+\int_{\text{BD}}^\sigma \mathcal D\sigma_-\,
	\mathcal O[\varphi,\sigma]e^{i \int \dd^4x \, ( \mathcal L[\varphi_+,\sigma_+] - \mathcal L[\varphi_-,\sigma_-] )}.
\end{align}
By perturbation expansion of the interaction term in the Lagrangian \eqref{eq_flatL}, we can compute the momentum-space four-point correlator $B_4^s$ \eqref{eq_flatI} as an ``in-in'' integral:
\begin{align}
	\label{eq_flat4ptinin}
	B_4^s =&~\frac{1}{\Lambda^2} \sum_{\mathsf a,\mathsf b=\pm} (-\mathsf a\mathsf b) \int_{-\infty}^0 \mathrm dt_1\mathrm dt_2\, \partial_{t_1}K_{\mathsf a}^\varphi(k_1,t_1)\times \partial_{t_1}K_{\mathsf a}^\varphi(k_2,t_1)\nonumber\\
	&\times D_{\mathsf a\mathsf b}^\sigma(s;t_1,t_2)\times \partial_{t_2}K_{\mathsf b}^\varphi(k_3,t_2)\times \partial_{t_2}K_{\mathsf b}^\varphi(k_4,t_2).
\end{align}
Here $D_{\mathsf a\mathsf b}^\sigma(k;t_1,t_2)$ are the bulk-to-bulk propagators of $\sigma$:
\begin{subequations}
	\label{eq_flatD}
	\begin{align}
		\label{eq_flatDmp}
		&~D_{-+}^\sigma(k;t_1,t_2) = u_\sigma(k,t_1)u_\sigma^*(k,t_2) = \frac{e^{-i E_k (t_1-t_2)}}{2E_k},\\
		\label{eq_flatDpm}
		&~D_{+-}^\sigma(k;t_1,t_2) = \big[D_{-+}^\sigma(k;t_1,t_2)\big]^* = \frac{e^{+i E_k (t_1-t_2)}}{2E_k},\\
		\label{eq_flatDpp}
		&~D_{\pm\pm}^\sigma(k;t_1,t_2) = D_{\mp\pm}^\sigma(k;t_1,t_2)\theta(t_1-t_2) + D_{\pm\mp}^\sigma(k;t_1,t_2)\theta(t_2-t_1), \bigg.
	\end{align}
\end{subequations}
and
$K_\pm^\varphi$ are the bulk-to-boundary propagators of $\varphi$:
\begin{align}
	\label{eq_flatKv2}
	K_+^\varphi(k,t) = u^*_\varphi(k,t)u_\varphi(k,0) = \frac{e^{i kt}}{2k},\qquad K_-^\varphi(k,t)= \big[ K_+^\varphi(k,t) \big]^* = \frac{e^{-i kt}}{2k}.
\end{align}
We then plug the propagators \eqref{eq_flatDpm}-\eqref{eq_flatKv2} into \Eq{eq_flat4ptinin}, and it is straightforward to compute this ``in-in'' integral. The result is:
\begin{align}
	\label{eq_flatIresult}
	B_4^s = \frac{k_T+E_s}{8\Lambda^2E_sk_T(k_{12}+E_s)(k_{34}+E_s)},
\end{align}
where we have defined the total energy $k_T\equiv k_{1234}$ with the shorthand $k_{ij\cdots} \equiv k_i+k_j+\cdots$.

\paragraph{Discussion.}

\begin{figure}[tbp]
	\centering
	\includegraphics[width=1\textwidth]{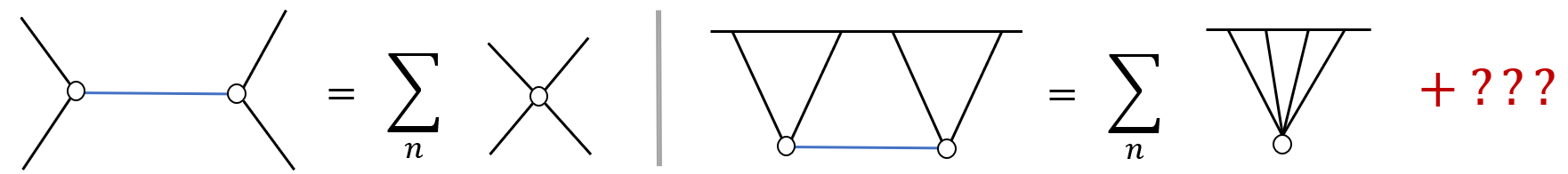}
	\caption{Illustration of the difference between amplitudes and correlators. In the in–out formalism, integrating out the heavy field produces a local effective action, while in the in–in framework it also gives rise to dissipative and stochastic operators involving odd powers of $1/M$.}
	\label{fig:diff}
\end{figure}

Let us consider the heavy mass limit of \Eq{eq_flatIresult}. As noted in \cite{Salcedo:2022aal, Green:2024cmx, DuasoPueyo:2025lmq}, it is impossible to recover all terms in \Eq{eq_flatIresult} in the traditional $\Box/M^2$ expansion of the heavy propagator. To see this, we expand \Eq{eq_flatIresult} in $M\gg k_i$ for all momenta (recall that $E_s = \sqrt{M^2+s^2}$):
\begin{align}\label{eq:expM}
	B_4^s = \frac{1}{8\Lambda^2  k_T } \frac{1}{M^2} - \frac{(k_{12} k_{34} + s^2)}{8\Lambda^2 k_T }  \frac{1}{M^4} + \frac{k_{12} k_{34} }{8\Lambda^2 } \frac{1}{M^5}+ \mathcal{O}\left( \frac{1}{M^6}\right),
\end{align}
where we find some odd powers of $1/M$ that cannot be produced from the $\Box/M^2$ expansion. This departs from the common expectation from scattering amplitudes, illustrated in \Fig{fig:diff}.\footnote{Indeed, the scattering amplitudes result can be recovered by looking at the $k_T$ pole \cite{Arkani-Hamed:2017fdk, Goodhew:2020hob, Pajer:2020wxk, Salcedo:2022aal, Cespedes:2025dnq}
\begin{align}
	\lim_{k_T \rightarrow 0}	B_4^s  \propto \Re \left\{ \frac{\mathcal{A}(k_1^\mu, k_2^\mu,k_3^\mu, k_4^\mu)}{k_T} \right\},
\end{align}
where $k_i^\mu = (k_i, \bmk_i)$ is a set of null momenta which characterize the incoming particles in the flat-space scattering process associated with the same graph. $\mathcal{A}(k_1^\mu, k_2^\mu,k_3^\mu, k_4^\mu)$ represents the corresponding 4-point scattering amplitude. The general relation is given by 
\begin{align}
	\lim_{k_T \rightarrow 0}	B_n({\bmk})  \propto \Re \left\{ \frac{i^{n+1}}{(i k_T)^{\Delta}} \mathcal{A}(k_1^\mu, \cdots, k_n^\mu)\right\}
\end{align}
where $\Delta_T$ quantifies
the degree of divergence of the total energy singularity, see \cite{Cespedes:2025dnq} for more details.} Then, what is the physical origin of the $1/M^5$ term appearing in \Eq{eq:expM}?

	\subsubsection{Keldysh basis perspective}
	
	This puzzle is first considered through the perspective of the \textit{Keldysh basis}, which provides a physical interpretation of the origin of the odd powers of $1/M$. Finite-time QFT is conveniently organized in the Keldysh basis of fields, which is a linear combination of $+/-$ fields on the two Schwinger contours \cite{kamenev_2011}. The retarded and advanced fields are respectively defined as:
	\begin{align}
		\label{eq_sigmaRA}
		\sigma_r = \frac{\sigma_+ + \sigma_-}{2},\qquad \sigma_a = \sigma_+ - \sigma_-,
	\end{align}
	and the propagators can be easily derived from the bulk-to-bulk propagators \eqref{eq_flatD}:
	\begin{subequations}
		\label{eq_GRAK}
		\begin{align}
			\label{eq_GR}
			-i G_\sigma^R(k;t_1,t_2) \equiv &~\exx{ \sigma_r(t_1,\bm k) \sigma_a(t_2,-\bm k)}  = D^\sigma_{++}(k;t_1,t_2)-D^\sigma_{+-}(k;t_1,t_2), \bigg.\\
			\label{eq_GA}
			-i G_\sigma^A(k;t_1,t_2) \equiv &~\exx{  \sigma_a(t_1,\bm k) \sigma_r(t_2,-\bm k) } = D^\sigma_{++}(k;t_1,t_2)-D^\sigma_{-+}(k;t_1,t_2),\\
			\label{eq_GK}
			-i G_\sigma^K(k;t_1,t_2) \equiv &~\exx{  \sigma_r(t_1,\bm k) \sigma_r(t_2,-\bm k) }
			= \frac12 \big[D_{-+}^\sigma(k;t_1,t_2)+D_{+-}^\sigma(k;t_1,t_2) \big],\\
			\label{eq_G0}
			&~\exx{  \sigma_a(t_1,\bm k) \sigma_a(t_2,-\bm k) } = 0.
		\end{align}
	\end{subequations}
	As will be discussed in depth in Lecture \ref{sec:scalar}, both the \emph{retarded propagator} $G^R$ and the \emph{advanced propagator} $G^A$ are the Green's functions satisfying:
	\begin{equation}
		\big[\partial_{t_1}^2+(k^2+M^2)\big]G^{R/A}_\sigma(k;t_1,t_2) = \delta(t_1-t_2),
	\end{equation}
	but are only supported on $t_1>t_2$ and $t_1<t_2$, respectively. On the other hand, the \emph{Keldysh propagator} $G^K$ is not a Green's function, satisfying the sourceless Klein-Gordon equation. We will come back to the physical interpretation of the propagators in the next lecture.
	
	For later convenience, we further define the \emph{principal-value propagator} $G^P$ and the \emph{Pauli-Jordan propagator}\footnote{We here use \textit{propagator} in a lose sense - referring to the fact these functions encode how information propagates in the environment $\sigma$.} $G^\Delta$ as a linear combination of retarded/advanced propagators:
	\begin{subequations}
		\label{eq_GHD}
		\begin{align}
			\label{eq_GH}
			&G_\sigma^P(k;t_1,t_2) = \frac12\big[G_\sigma^R(k;t_1,t_2) + G_\sigma^A(k;t_1,t_2)\big]
			=\frac i2\big[D^\sigma_{++}(k;t_1,t_2)-D^\sigma_{--}(k;t_1,t_2)\big],
			\\
			\label{eq_GD}
			&G_\sigma^\Delta(k;t_1,t_2) = G_\sigma^R(k;t_1,t_2)-G_\sigma^A(k;t_1,t_2) = i \big[ D_{-+}^\sigma(k;t_1,t_2) - D_{+-}^\sigma(k;t_1,t_2)\big]~.
		\end{align}
	\end{subequations}
	In terms of the mode function $u_\sigma(k,t)$ and its complex conjugate, we can write down the three linearly independent propagators:
	\begin{subequations}
		\label{eq_flatGHDK}
		\begin{align}
			\label{eq_flatGH}
			&G_\sigma^P(k;t_1,t_2) = \frac i2  \big[ u_\sigma(k,t_1) u^*_\sigma(k,t_2) - u^*_\sigma(k,t_1) u_\sigma(k,t_2) \big] \text{sn}(t_1-t_2),\\
			\label{eq_flatGD}
			&G_\sigma^\Delta(k;t_1,t_2) = i \big[ u_\sigma(k,t_1) u^*_\sigma(k,t_2) - u^*_\sigma(k,t_1) u_\sigma(k,t_2) \big],\\
			\label{eq_flatGK}
			&G_\sigma^K(k;t_1,t_2) =  \frac i2 \big[ u_\sigma(k,t_1) u^*_\sigma(k,t_2)+u^*_\sigma(k,t_1) u_\sigma(k,t_2) \big],
		\end{align}
	\end{subequations}
	where $\mathrm{sn}(x)\equiv\theta(x)-\theta(-x)$ denotes the sign function. 
	Finally, we insert the explicit expression for the mode function \eqref{eq_flatusigma} and obtain:
	\begin{subequations}
		\label{eq_flatGHDKex}
		\begin{align}
			\label{eq_flatGHex}
			&G_\sigma^P(k;t_1,t_2) = \frac{\sin E_k(t_1-t_2)}{2E_k} \text{sn}(t_1-t_2),\\
			\label{eq_flatGDex}
			&G_\sigma^\Delta(k;t_1,t_2) = \frac{\sin E_k(t_1-t_2)}{E_k},\\
			\label{eq_flatGKex}
			&G_\sigma^K(k;t_1,t_2) =  \frac{i \cos E_k(t_1-t_2)}{2E_k}.
		\end{align}
	\end{subequations}
	We summarize some basic properties of the propagators $G^P$, $G^\Delta$, and $G^K$ in Tab.\ \ref{tab_prop}, including that they are either real or (pure) imaginary; either symmetric or asymmetric under $t_1\leftrightarrow t_2$; and either factorised or nested in time order. By factorised, we mean that it can be written by a finite sum of factorized terms, and by nested, we mean that time integrals involving this propagator exhibit nested time integrals due to $\text{sn}(t_1-t_2)$. These characteristics can be easily verified from the explicit expressions \eqref{eq_flatGHDKex}, but they are actually fundamentally defined properties from \eqref{eq_flatGHDK}, and thus remain valid in arbitrary spacetime.
	\begin{table}[t] 
		\centering
		\caption{Basic properties of propagators in the Keldysh basis}
		\vspace{2mm}
		\begin{tabular}{ccccc}
			\toprule[1.5pt]
			Propagators &Real/Imaginary & Symmetric/Asymmetric & Factorised/Nested \\ \hline
			$G^P(k;t_1,t_2)$ & R & S & N\\
			$G^\Delta(k;t_1,t_2)$ & R & A & F\\
			$ G^K(k;t_1,t_2)$ & I & S & F\\
			\bottomrule[1.5pt] 
		\end{tabular}
		\label{tab_prop}
	\end{table}
	
	To gain some insights from the Keldysh basis, we turn to the new basis of propagators \eqref{eq_flatGHDK} for $\sigma$ in the ``in-in'' integral \eqref{eq_flat4ptinin}, and extract the contribution from each one. More explicitly, the four ``in-in'' propagators \eqref{eq_flatD} can be expressed in this basis:
	\begin{subequations}
		\label{eq_sub}
		\begin{align}
			\label{eq_sub1}
			&D^\sigma_{\mp\pm}(k;t_1,t_2) = -i G_\sigma^K(k;t_1,t_2) \mp \frac i2 G_\sigma^\Delta(k;t_1,t_2),\\
			\label{eq_sub2}
			&D^\sigma_{\pm\pm}(k;t_1,t_2) = -i G_\sigma^K(k;t_1,t_2) \mp i G_\sigma^P(k;t_1,t_2).
		\end{align}
	\end{subequations}
	Therefore, we can substitute the bulk-to-bulk propagator $D^\sigma_{\mathsf a\mathsf b}$ with \Eq{eq_sub} in the integral \eqref{eq_flat4ptinin} to obtain the corresponding contributions.
	In particular, the principal-value propagator $G^P$ contributes to $\pm\pm$ branches:
	\begin{align}
		\label{eq_flatIH}
		\mathcal I^P =
		&~\frac{1}{\Lambda^2} \sum_{\mathsf a=\pm} (-1)\int_{-\infty}^0 \mathrm dt_1\mathrm dt_2\, \partial_{t_1}K_{\mathsf a}^\varphi(k_1,t_1)\times \partial_{t_1}K_{\mathsf a}^\varphi(k_2,t_1)\nonumber\\
		&\times \big[- \mathsf a i G_\sigma^P(s;t_1,t_2) \big]\times \partial_{t_2}K_{\mathsf a}^\varphi(k_3,t_2)\times \partial_{t_2}K_{\mathsf a}^\varphi(k_4,t_2)\nonumber\\
		=&-\frac{1}{16\Lambda^2k_T}\Big(\frac{1}{k_{12}^2-E_s^2}+\frac{1}{k_{34}^2-E_s^2}\Big),
	\end{align}
	and the Pauli-Jordan propagator $G^\Delta$ contributes to $\pm\mp$ branches:
	\begin{align}
		\label{eq_flatID}
		\mathcal I^\Delta =
		&~\frac{1}{\Lambda^2} \sum_{\mathsf a=\pm} \int_{-\infty}^0 \mathrm dt_1\mathrm dt_2\, \partial_{t_1}K_{\mathsf a}^\varphi(k_1,t_1)\times \partial_{t_1}K_{\mathsf a}^\varphi(k_2,t_1)\nonumber\\
		&\times \Big[   \frac{\mathsf ai}2 G_\sigma^\Delta(s;t_1,t_2) \Big]\times \partial_{t_2}K_{-\mathsf a}^\varphi(k_3,t_2)\times \partial_{t_2}K_{-\mathsf a}^\varphi(k_4,t_2)\nonumber\\
		=&-\frac{k_T}{16\Lambda^2(k_{12}^2-E_s^2)(k_{34}^2-E_s^2)}. 
	\end{align}
	Finally, the Keldysh propagator $G^K$ contributes to all SK branches:
	\begin{align}
		\label{eq_flatIK}
		\mathcal I^K =
		&~\frac{1}{\Lambda^2}\sum_{\mathsf a,\mathsf b=\pm} (-\mathsf a\mathsf b)\int_{-\infty}^0 \mathrm dt_1\mathrm dt_2\, \partial_{t_1}K_{\mathsf a}^\varphi(k_1,t_1)\times \partial_{t_1}K_{\mathsf a}^\varphi(k_2,t_1)\nonumber\\
		&\times \big[ -i G_\sigma^K(s;t_1,t_2) \big]\times \partial_{t_2}K_{\mathsf b}^\varphi(k_3,t_2)\times \partial_{t_2}K_{\mathsf b}^\varphi(k_4,t_2)\nonumber\\
		=&~ \frac{ k_{12}k_{34}}{8\Lambda^2E_s(k_{12}^2-E_s^2)(k_{34}^2-E_s^2)}.
	\end{align}
	Indeed, one can easily recover the full correlator \eqref{eq_flatIresult} by summing up \Eqs{eq_flatIH}-\eqref{eq_flatIK}.
	
	\paragraph{Physical interpretation.} There are two main distinctions among the above three contributions \eqref{eq_flatIH}-\eqref{eq_flatIK}.
	First, it is only $\mathcal I^P$ that possesses the \emph{total energy pole} (or the $k_T$ pole), namely it diverges when $k_T\to 0$.
	This is because the (residue of) total energy pole corresponds to a unitary scattering amplitude (in flat spacetime) \cite{Arkani-Hamed:2017fdk, Goodhew:2020hob, Pajer:2020wxk, Salcedo:2022aal, Cespedes:2025dnq}, and thus it could only appear in a unitary theory. As we will see in \Sec{subsec:flatEFT}, when integrating out $\sigma$, the principal-value propagator encodes all the information about the unitary single-field EFT for $\varphi$. Conversely, the Pauli-Jordan and Keldysh propagators capture dissipation and noise, which are non-unitary effects. It explains why $\mathcal I^P$ is unique in possessing this $k_T$ pole.
	This can be also understood from a technical perspective: both $G^\Delta$ and $G^K$ are factorised in time, see Table \ref{tab_prop}, so they cannot maintain singularity in $k_T$ which is not factorisable. We also observe that $\mathcal I^\Delta$ vanishes at $k_T=0$, and it would be interesting to explore the underlying reason and whether this is a universal feature.
	Aside from the total energy pole, we find all these three parts have \emph{partial energy poles} ($k_{12}+E_s=0$ or $k_{34}+E_s=0$) and \emph{folded poles} ($k_{12}-E_s=0$ or $k_{34}-E_s=0$), while the latter will be canceled when the three parts add together to the full correlator \eqref{eq_flatIresult} due to the BD initial condition.
	
	Second, let us expand each contribution in the heavy mass limit $M\to \infty$:
	\begin{align}
		&\mathcal I^P = \frac{1}{8\Lambda^2k_T} \frac{1}{M^2}
		+ \frac{k_{12}^2+k_{34}^2-2s^2}{16\Lambda^2k_T}\frac{1}{M^4}
		+ \mathcal O\Big(\frac{1}{M^6}\Big),\\
		&\mathcal I^\Delta = -\frac{k_T}{16\Lambda^2} \frac{1}{M^4} + \mathcal O\Big(\frac{1}{M^6}\Big),\\
		&\mathcal I^K = \frac{k_{12}k_{34}}{8\Lambda^2} \frac{1}{M^5} + \mathcal O\Big(\frac{1}{M^7}\Big).
	\end{align}
	We can find that both $\mathcal I^P$ and $\mathcal I^\Delta$ contain even powers in $1/M$, while $\mathcal I^K$ only contains odd powers. 
	This can be traced back to the fact that $G^P$ and $G^\Delta$ vanish in the coincident time limit while $G^K$ does not, see \Eq{eq_flatGHDKex}.
	Hence, $G^P$ and $G^\Delta$ are built from the $\sin$ function, while $G^K$ follows from the $\cos$ function.
	Since the traditional (unitary and local) EFT is the $\Box/M^2$ expansion of the heavy propagator, it could only have even powers of $1/M$ and each term contributes to a contact graph possessing the total energy pole. That is, the traditional EFT reproduces the contribution from the principal-value propagator \eqref{eq_flatIH}. Corrections to this include even powers of $1/M$ from the Pauli-Jordan propagator \eqref{eq_flatID} that vanishes at $k_T=0$, and odd powers of $1/M$ from the Keldysh propagator \eqref{eq_flatIK}.

	\subsubsection{Top-down open EFT}\label{subsec:flatEFT}

We now construct an EFT for the massless field $\varphi$. This Section aims to convince the reader that in order to fully recover the results found above, the EFT has to be \textit{open}, that is cannot be written in the form $S_{\mathrm{eff}}[\varphi_+, \varphi_-] = S_{\mathrm{unit}}[\varphi_+] - S_{\mathrm{unit}}[\varphi_-]$. This serves as a motivation for the construction of generic open EFTs in the Schwinger-Keldysh contour.

We define the retarded and advanced fields for $\varphi$ 
\begin{equation}
	\varphi_r = \frac{\varphi_+ + \varphi_-}{2},\qquad \varphi_a = \varphi_+ - \varphi_-.
\end{equation}
Under this basis, the path integral \eqref{eq_flatPathInt} becomes:
\begin{align}
	\exx{ \mathcal O[\varphi,\sigma] } = \int_{\text{BD}}^\varphi \mathcal D\varphi_r\int_{\text{BD}}^0 \mathcal D\varphi_a\int_{\text{BD}}^\sigma \mathcal D\sigma_r\int_{\text{BD}}^0 \mathcal D\sigma_a\,
	\mathcal O[\varphi,\sigma]e^{i S_0^\varphi[\varphi_r,\varphi_a]+i S_0^\sigma[\sigma_r,\sigma_a]}e^{i S_{\text{int}}[\varphi_r,\varphi_a,\sigma_r,\sigma_a]},
\end{align}
where the free action reads:\footnote{Here the propagators are written in position space and are the Fourier transform of the form in \Eq{eq_flatGHDK}:
	\begin{equation}
		G(x,y) = \int \frac{\dd^3\bm k}{(2\pi)^3} G(k;t_1,t_2) e^{i \bm k\cdot (\bm x-\bm y)},
	\end{equation}
	where we have omitted the superscript ($H$, $\Delta$, $K$) and subscript ($\varphi$, $\sigma$) for the propagator. The retarded and advanced propagators in position space are defined similarly.
}
\begin{align}
	S_0^\varphi[\varphi_r,\varphi_a] =& -\frac12 \int \dd^4x\,
	\begin{pmatrix}
		\varphi_r & \varphi_a
	\end{pmatrix}
	\begin{pmatrix}
		G^K_\varphi & G^R_\varphi \\ G^A_\varphi & 0
	\end{pmatrix}^{-1}
	\begin{pmatrix}
		\varphi_r \\ \varphi_a
	\end{pmatrix},\\
	S_0^\sigma[\sigma_r,\sigma_a] =& -\frac12\int \dd^4x \,
	\begin{pmatrix}
		\sigma_r & \sigma_a
	\end{pmatrix}
	\begin{pmatrix}
		G^K_\sigma & G^R_\sigma \\ G^A_\sigma & 0
	\end{pmatrix}^{-1}
	\begin{pmatrix}
		\sigma_r \\ \sigma_a
	\end{pmatrix},
\end{align}
and the interaction part is:
\begin{align}
	S_{\text{int}}[\varphi_r,\varphi_a,\sigma_r,\sigma_a] =&~\frac{1}{2\Lambda} \int \dd^4x \, ( \dot\varphi_+^2\sigma_+ - \dot\varphi_-^2\sigma_-)\nonumber\\
	=&~\frac{1}{\Lambda} \int \dd^4x \, \Big( \dot\varphi_r\dot\varphi_a\sigma_r + \frac12 \dot\varphi_r^2\sigma_a + \frac18 \dot\varphi_a^2\sigma_a \Big).
\end{align}
We will develop in Lecture \ref{sec:scalar} the physical description of the various terms appearing in this functional. For the moment, let us simply take it as the initial theory rewritten in a funny basis.

We now aim to integrate out the field $\sigma$ and to obtain the single field open EFT for $\varphi$ only.
In practice, we follow a procedure similar to the one described in \cite{Proukakis:2024pua}. We first define the \emph{influence functional} $S_{\text{IF}}$ \cite{FEYNMAN1963118}:
\begin{align}
	\ee^{iS_{\mathrm{IF}} \left[ \pi_r, \pi_a\right]} = \int_{\mathrm{BD}}^\sigma \mathcal{D}\sigma_r \int_{\mathrm{BD}}^0 \mathcal{D}\sigma_a\,  \ee^{iS^\sigma_{0}\left[ \sigma_r, \sigma_a\right]} \ee^{iS_{\mathrm{int}}\left[ \varphi_r, \varphi_a, \sigma_r, \sigma_a\right]},
\end{align}
and then expand $S_{\mathrm{int}}$ to the second order in $1/\Lambda$ to obtain:
\begin{align}
	\ee^{iS_{\mathrm{IF}}}  = 1 + i \langle S_{\mathrm{int}} \rangle_\sigma - \frac{1}{2} \langle S^2_{\mathrm{int}}
	\rangle_\sigma + \cdots .
\end{align}
The theory being linear in $\sigma$ and having removed the tadpole contribution, we assume for the moment that $\langle S_{\mathrm{int}} \rangle_\sigma = 0$. We then identify the influence functional in the leading order:
\begin{equation}
	S_{\mathrm{IF}} \simeq \frac i2 \langle S^2_{\mathrm{int}} \rangle_\sigma.
\end{equation}
Therefore, the second-order effective action is simply obtained by replacing internal $\sigma$ legs by its propagators, leading to:
\begin{align}
	\label{eq_flatIF}
	&S_{\mathrm{IF}}[\varphi_r,\varphi_a] = ~\frac{1}{2\Lambda^2} \int \dd^4x \int \dd^4y \bigg\{ \dot\varphi_r(x) \dot\varphi_a(x)  \times G^{R}_\sigma(x,y) \times
	\left[ \frac12\dot\varphi_r^2(y) + \frac{1}{8}\dot\varphi_a^2(y)\right] \\
	+&\left[ \frac12\dot\varphi_r^2(x)+\frac18\dot\varphi_a^2(x)\right] \times G^{A}_\sigma(x,y) \times \dot\varphi_r(y)\dot\varphi_a(y) + \dot\varphi_r(x) \dot\varphi_a(x)  \times G^{K}_\sigma(x,y) \times \dot\varphi_r(y)\dot\varphi_a(y)
	\bigg\}. \nonumber
\end{align}
Notice that when doing the contraction, the two vertices give $1/\Lambda^2$.
We emphasize that the last term in \Eq{eq_flatIF} is manifestly non-unitary: it contains an even number of advanced fields, and thus if going back to the original $+/-$ basis, it can never be written in a factorised form $S_{\mathrm{eff}}[\varphi_+, \varphi_-] = S_{\mathrm{unit}}[\varphi_+] - S_{\mathrm{unit}}[\varphi_-]$ \cite{Salcedo:2024nex}.

The first two terms in \Eq{eq_flatIF} are more subtle. It is more intuitive to express $G^{R/A}$ in terms of $G^{H/\Delta}$ using \Eq{eq_GHD}, where the influence functional becomes:
\begin{align}
	\label{eq_flatIFHD}
	&S_{\mathrm{IF}}[\varphi_r,\varphi_a] = ~\frac{1}{2\Lambda^2} \int \dd^4x \int \dd^4y \bigg\{ \dot\varphi_r(x) \dot\varphi_a(x)  \times 2G^P_\sigma(x,y) \times
	\left[ \frac12\dot\varphi_r^2(y) + \frac{1}{8}\dot\varphi_a^2(y)\right] \\
	+&\dot\varphi_r(x) \dot\varphi_a(x)  \times G^{\Delta}_\sigma(x,y) \times
	\left[ \frac12\dot\varphi_r^2(y) + \frac{1}{8}\dot\varphi_a^2(y)\right] + \dot\varphi_r(x) \dot\varphi_a(x)  \times G^{K}_\sigma(x,y) \times \dot\varphi_r(y)\dot\varphi_a(y)
	\bigg\},\nonumber
\end{align}
where we have used $G^P_\sigma(y,x) = G^P_\sigma(x,y)$ and $G^\Delta_\sigma(y,x) = -G^\Delta_\sigma(x,y)$ to simplify the expression. Now the three terms in the non-local EFT action \eqref{eq_flatIFHD} have clear physical meanings:
\begin{enumerate}
	\item The term of principal-value propagator $G^P_\sigma$ is a unitary EFT. Written back in the $+/-$ basis, this term can be expressed as $S_{\text{unit}}[\varphi_+]-S_{\text{unit}}[\varphi_-]$ with
	\begin{equation}
		S_{\text{unit}}[\varphi_\pm] = \frac{1}{8\Lambda^2}\int \dd^4x \int \dd^4y\, \dot\varphi^2_\pm(x)\dot\varphi^2_\pm(y) G_\sigma^P(x,y).
	\end{equation}
	If we further expand $G^P_\sigma$ in the heavy mass limit $M\to \infty$ and integrate over $y$, we will recover the traditional (unitary and local) EFT.
	\item As a contrast, the term of Pauli-Jordan propagator $G^\Delta_\sigma$ is a non-unitary contribution which we interpret as the \emph{dissipation}. It cannot be decomposed into separable contributions in the $+/-$ basis and encodes energy exchange between $\varphi$ and $\sigma$ \cite{Calzetta:2008iqa}. The asymmetric property of $G^\Delta_\sigma$ discussed in Table \ref{tab_prop} may relate to the fact that dissipation generally breaks time translation symmetry, creating an effective arrow of time.\footnote{Conversely, the symmetry property of $G^P_\sigma$ makes it a natural candidate to control the unitary/Hamiltonian evolution which is time-reversal symmetric.}
	\item Finally, the term of Keldysh propagator $G^K_\sigma$ is the \emph{noise} term that is also non-unitary. Physically, it originates from the fluctuations of the $\sigma$ medium sourcing the $\varphi$ dynamics. This can be seen from the fact that $G^K_\sigma$ controls the amplitude of the perturbations in the $\sigma$ field (that is the power spectrum), which makes it a natural candidate to encode the environmental noise onto the system. 
\end{enumerate} 

\begin{tcolorbox}[%
	enhanced, 
	breakable,
	skin first=enhanced,
	skin middle=enhanced,
	skin last=enhanced,
	before upper={\parindent15pt},
	]{}
	
	\vspace{0.05in}
	
	\paragraph{Summary.}
	
	\Eq{eq_flatIFHD} represents the second-order effects of $\sigma$ on the dynamics of $\varphi$. It can be decomposed into three distinctive effects:
	\begin{enumerate}
		\item  The line controlled by $G_\sigma^P(x,y)$ is \textit{unitary} and corresponds to the generation of an effective (non-local) vertex in the Lagrangian of $\varphi$, sometimes called the \textit{Lamb shift}.
		\item The line controlled by $G^\Delta_\sigma(x,y)$ is \textit{non-unitary} and corresponds to the \textit{dissipative} evolution of $\varphi$ through the $\sigma$ medium.
		\item The line controlled by $G_\sigma^K(x,y)$ is \textit{non-unitary} and corresponds to the \textit{noise} generated by fluctuations of the $\sigma$ medium backreacting on the evolution of $\varphi$.
	\end{enumerate}
	
\end{tcolorbox}

This example illustrates how integrating out certain degrees of freedom ($\sigma$) along the Schwinger–Keldysh contour can induce non-unitary effects on the remaining fields ($\phi$) — effects that lie beyond the scope of standard EFT treatments. Our goal now is to understand how to systematize these observations.


\clearpage

\subsection{\textit{Problem set}}\label{subsec:prob1}

\paragraph{\textit{Exercise 1.}} \textit{Flat space in-in diagrams}\\

Consider a massless scalar field in flat spacetime where the metric is given by
\begin{equation}
	\mathrm ds^2 = -\mathrm d t^2+ \mathrm d \bm x^2, \qquad \quad t \in ] - \infty, 0],\quad  \bm x \in \mathbb{R}^3.
\end{equation}
The action is given by 
\begin{align}
	S = \int \dd^4 x  \left\{\frac{1}{2}\left[\dot{\varphi}^2 - (\partial_i \varphi)^2 \right] + \mathcal{L}_{\mathrm{int}} \right\}.
\end{align}
\begin{enumerate}
	\item Write down the mode function of the field.
	\item Write down the propagators.
	\item Compute $B_3(k_1, k_2,k_3)$ at order $\lambda$ for $\mathcal{L}_{\mathrm{int}} = -(\lambda/3!) \varphi^3$.
	\item Compute $B_3(k_1, k_2,k_3)$ at order $\lambda_t$ for $\mathcal{L}_{\mathrm{int}} = (\lambda_t/2) \dot{\varphi}^2 \varphi$.
	\item Compute $B_3(k_1, k_2,k_3)$ at order $\lambda_s$ for $\mathcal{L}_{\mathrm{int}} = - (\lambda_s/2) (\partial_i \varphi)^2 \varphi$.
	\item Comment the common structure of these correlators. It might be useful to consider $B_n(k_1,\cdots,k_n)$ at order $\lambda_n$ for $\mathcal{L}_{\mathrm{int}} = -(\lambda_n/n!) \varphi^n$.
\end{enumerate}

\begin{center}
	\noindent\rule{8cm}{0.4pt}
\end{center}

\paragraph{\textit{Exercise 2.}} \textit{Integrate out heavy fields} \\

Consider the following two-field toy model with the Lagrangian, 
\begin{equation}
	\mathcal L[\varphi,\sigma] =  -\frac12 (\partial_\mu \varphi)^2 - \frac12 (\partial_\mu \sigma)^2 - \frac12 M^2\sigma^2 + \frac{1}{2\Lambda} \dot\varphi^2\sigma,
\end{equation}
with $\varphi$ being a massless scalar and $\sigma$ a massive scalar with mass $M$. 

\begin{enumerate}
	\item Draw the four diagrams contributing to the $s$-channel massless trispectrum. Why is there in practice only two diagrams to compute? 
	\item Compute $B^s_4(k_1, k_2,k_3,k_4)$ at order $1/\Lambda^2$. Comment the pole structure of the result. 
	\item The expansion of $B^s_4$ in powers of $M \gg k_i$ reads
	\begin{align}
		B^s_4(k_1, k_2,k_3,k_4) = 	\frac{1}{8\Lambda^2  k_T } \frac{1}{M^2} - \frac{(k_{12} k_{34} + s^2)}{8\Lambda^2 k_T }  \frac{1}{M^4} + \frac{k_{12} k_{34} }{8\Lambda^2 } \frac{1}{M^5}+ \mathcal{O}\left( \frac{1}{M^6}\right).
	\end{align}
	The associated $2-2$ scattering amplitude is 
	\begin{align}
		\mathcal{A}^s_4 = \frac{- i }{\Lambda^2} \frac{1}{(s - M^2)}.
	\end{align}
	Comment the difference with the correlator result in the heavy mass expansion. 
	\item It is well known that $\mathcal{A}^s_4$ can be reproduced order by order in $1/M^2$ by considering a single-field EFT containing an infinite tower of higher-order operators,
	\begin{align}
		S_{\mathrm{EFT}} \supset \frac{1}{\Lambda^2 M^2} \int \dd^4 x \sum_{n=0}^\infty  \phi^2 \left(\frac{\Box}{M^2}\right)^{n} \phi^2.
	\end{align}
	In the Schwinger-Keldysh contour, is such a tower sufficient to reproduce the above result? 
	\item Propose a strategy to investigate the difference of a single heavy exchange in amplitudes and correlators, and for characterizing the corresponding effective field theories. 
\end{enumerate}
Some answers to the last questions can be found in \Sec{subsec:heavy}.


\section{Lecture 2: Open scalar theory}\label{sec:scalar}

\begin{figure}[h]
	\centering
	\includegraphics[width=0.35\textwidth]{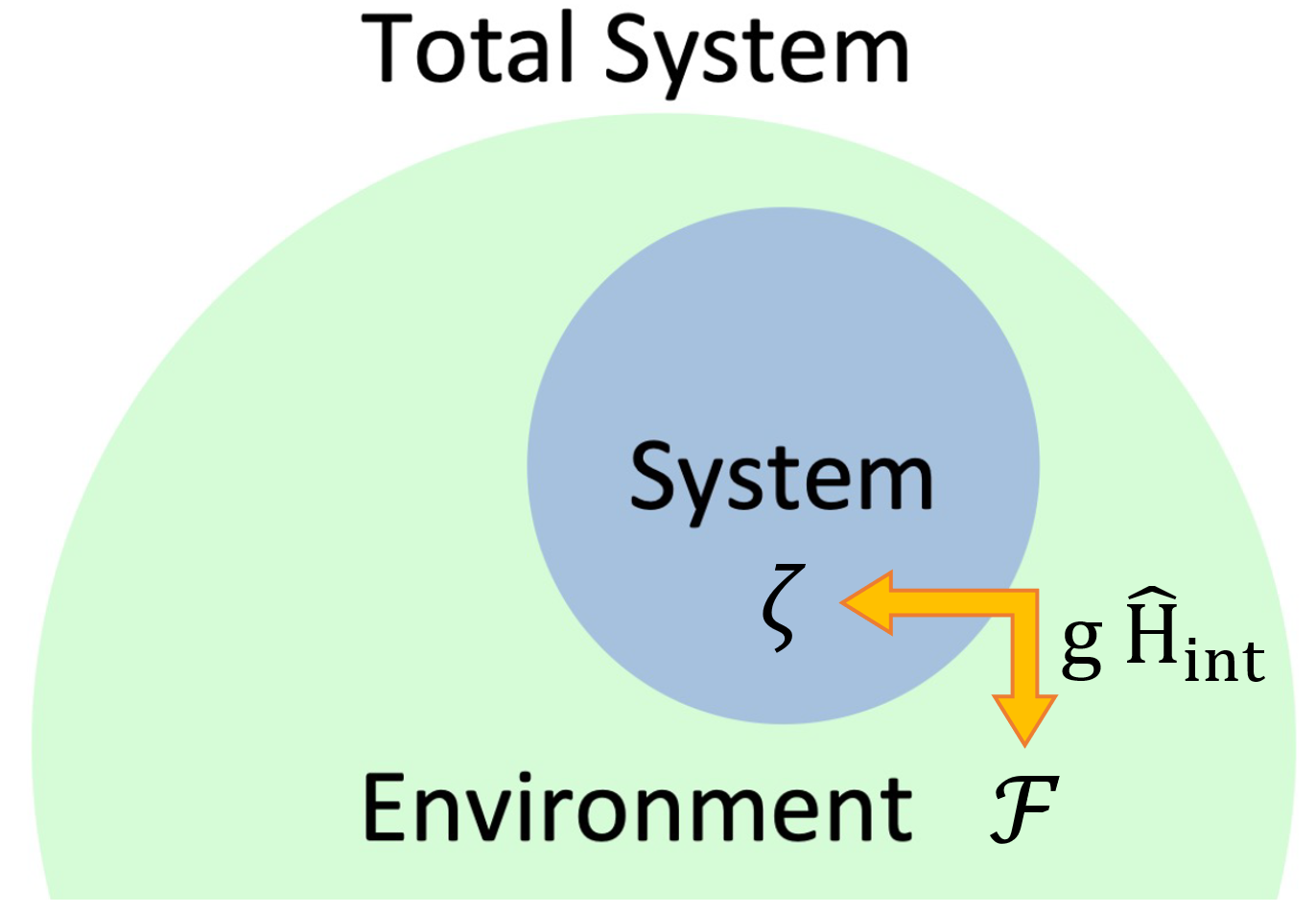}
	\caption{Schematic setup of open systems. The system, made of the degree of freedom $\zeta$, is embedded in the environment made of $\mathcal{F}$. Their interaction is specified by $g \widehat{H}_{\mathrm{int}}$. From the point of view of the system, this interaction renormalizes its energy level, generates energy loss through dissipation and information exchanges through noise. Figure adapted from \cite{manzanoShortIntroductionLindblad2020}.}
	\label{fig:setup}
\end{figure}

Open systems primary aim at describing physical systems exchanging energy and information with their surrounding environments. The system is made of the degrees of freedom we experimentally access - for instance the curvature perturbations $\zeta$ in the early universe. On the contrary, the environment is experimentally inaccessible and poorly specified. It characterises unobservable degrees of freedom that may have played a role in the description of the system but that we do not physically access. For instance, the cosmological collider considers the impact of heavy fields and higher-spin particles on $\zeta$ which can constitute a cosmological environment, denoted $\mathcal{F}$. \Fig{fig:setup} summarizes the setup considered.

Open systems theory provides a toolbox of effective methods to describe dynamical evolution of systems losing energy and information \cite{breuerTheoryOpenQuantum2002, Colas:2022hlq, Colas:2023wxa, Colas:2024lse}. These techniques originate from the XIX$^{\mathrm{th}}$ century with the study of particles of pollen immersed into water by Lord Brown \cite{doi:10.1080/14786442808674769}. The investigation and theoretical modelling of \textit{Brownian motion} lead to the discovery of atoms by Jean Perrin in 1905 \cite{Perrin1913-PERLA-7} following the pionerring work of Albert Einstein \cite{einstein1956investigations}.
\begin{figure}[tbp]
	\centering
	\includegraphics[width=1\textwidth]{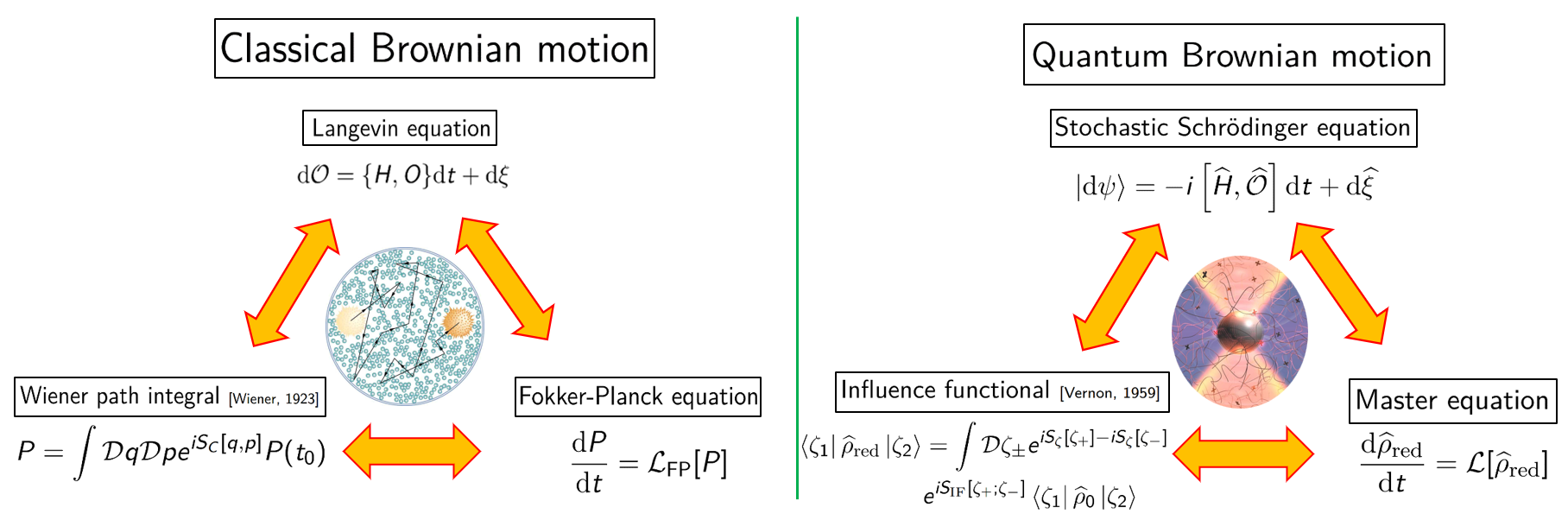}
	\caption{Open Classical and Quantum Systems. The effect of the surrounding environment is encoded through a set of stochastic variables called noises. It generates an effective dynamics which renormalizes the free evolution, dissipates energy into the environment and source the system's dynamics through noise. These effects are captured at the classical level by the Langevin equation, the Fokker-Planck equation or the Martin-Siggia-Rose path integral. These techniques have quantum analogues where they are replaced by the stochastic Schr\"odinger equation, the master equation and the influence functional respectively.}
	\label{fig:trinity}
\end{figure} 	
To illustrate the scope of open system methods, let us briefly discuss the case of a Brownian particle immersed in a bath. We consider a scalar variable $\varphi$ and aim at describing its dynamics in the presence of an environment.
The effect of the surrounding medium is encoded through a set of stochastic variables known as noises, which promote the deterministic equations of motion to stochastic differential equations such as the Langevin equation
\begin{align}\label{eq:langevinintro}
	\ddot{\varphi}_r + \gamma \dphir + c_s^2 k^2 \phir = \xi.
\end{align}
The new variable $\xi$ has to be understood in a statistical sense, for instance obyeing the Gaussian statistics
\begin{equation}\label{eq:noisecontactintro}
\ex{\xi(x)}=0, \qquad	\langle\xi(x) \xi(y)\rangle=2\beta \delta(x-y).
\end{equation}
In flat space, the late-time dynamics generated by \Eq{eq:langevinintro} is controlled by the equilibrium between the drift term $\gamma \dphir$ which slows down the Brownian particle and the source term $\xi$ which generates its erratic jumps. This is the well-known Brownian motion of a random walker.

There exists several ways to describe this dynamics, which can be mapped one to another \cite{breuerTheoryOpenQuantum2002}. Averaging over many stochastic realisations of the Langevin equation, we derive a dynamical equation for the probability distribution of being at a given position at a given time, known as a Fokker-Planck equation. This probability distribution has a path integral formulation known as the Martin-Siggia-Rose (MSR) path integral \cite{PhysRevA.8.423}. These techniques rely on the same physics and represent different aspects of a same problem. For instance, the Langevin equation focuses on the equations of motion and is well-suited for numerical simulations. The MSR path integral allows us to describe relativistic settings in a manifestly covariant formalism while the Fokker-Planck equation has been widely studied for its ability to implement resummations. As shown in the Right panel of \Fig{fig:trinity}, there exists an exact same language in the quantum framework where the Fokker-Planck equation is supplemented by master equations, the MSR path integral by the influence functional and the Langevin equation by a stochastic unravelling \cite{breuerTheoryOpenQuantum2002}. 

Since the main focus of these notes is relativistic QFT, the path integral formulation of open dynamics will be particularly suited. In particular, we aim at understanding the rules obeyed by the influence functional that do not depend on the microphysical details of the environment. Indeed, one can always try to model the environment in order to deduce its impact on the system. We will qualify this approach as the \textit{top-down} approach\footnote{Examples of top-down approaches in cosmological open quantum systems can be found in \cite{Colas:2021llj, Colas:2022kfu, Colas:2024xjy, Burgess:2024eng}.}. For instance, Caldeira-Leggett model \cite{Caldeira:1981rx, Caldeira:1982iu, Caldeira:1982uj} is a simple model coupling a harmonic oscillator $\varphi$ to a set of $N$ harmonic oscillators $\sigma_n$ ($n = 1, \cdots, N$) through
\begin{align}
	S[\varphi,\sigma_n] = \int \dd t \int \frac{\dd\bmk^3}{(2\pi)^3} \bigg\{\frac{1}{2}\left[\dot{\varphi}^2 - k^2 \varphi^2 \right] + \sum_{n = 1}^{N} \frac{1}{2} \left[ \dot{\sigma}^2_n - \left(k^2 + m_n^2\right)\sigma_n^2\right] + \sum_{n = 1}^{N} g_n \varphi \sigma_n \bigg\},
\end{align}
where $m_n$ are the masses of the environment oscillators and $g_n$ the coupling between the system oscillator $\varphi$ and the set of environment oscillators $\sigma_n$. By choosing ${m_n^2, g_n}$ such that the $N$ environmental oscillators form a thermal bath, we indeed recover the Langevin equation given in \eqref{eq:langevinintro} when deriving the dynamics of $\varphi$ - a clear derivation can be found in \cite{Lau:2024mqm}. While having a microphysical construction is often insightful, it also often obscures the fact that many properties of the systems dynamics do not depend on the microphysical details of the environmental model. Instead, these properties are imposed by physical principles such as symmetries, locality and unitarity. This is for instance the main message of Lindblad theorem \cite{Lindblad:1975ef} which states that any Markovian (memoryless) open systems evolves according to a rigid framework, the Lindblad or Gorini-Kossakowski-Sudarshan-Lindblad (GKSL) equation. By developping \textit{bottom-up} approach agnostic of the microphysical details of the surrounding environment, we aim at finding similar types of constraints. This will be the strategy followed in these notes.


\subsection{Effective functional}

The Schwinger-Keldysh formalism \cite{Schwinger:1960qe, Keldysh:1964ud} aims at computing expectation values of operators in far-from-equilibrium systems and/or open systems. In this Section, we follow the EFT philosophy and aim at providing the most general description of the system while remaining as agnostic as possible about the environment. We assume separation of scales between system and environment such that interactions mediated by the environment can be modeled  by a finite number of local, possibly dissipative, interactions of the system. We will therefore refrain from a detailed description of the environment, trading it for a self-consistent description of the system interactions. This is similar in spirit to the Lindblad description of open quantum system \cite{Lindblad:1975ef, 10.1063/1.522979} where the focus is made on the derivation of a well-defined open dynamics, avoiding microphysical modeling of the environment. For concreteness, we assume that the environment is homogeneous, isotropic, and invariant under time translation. Moreover, we neglect the backreaction of system on the environment. There are there three steps in constructing local EFTs: i) choosing the degrees of freedom; ii) determining the symmetries of the theory, iii) keeping a finite number of operators in a radiatively stable power counting scheme.

Here, we consider the latter situation where a system of interest interacts with an unspecified environment. The setup assumes a unitary $\{$system $+$ environment$\}$ evolution. Starting in a pure state and upon tracing out the environment, the system eventually ends up in a mixed state described by the density matrix $\widehat \rho(t)$. In this open quantum system, equal-time expectation values of fundamental operators are schematically computed by
\begin{align}\label{eq:expect}
	\ex{\widehat \O(t)}=\Tr[\widehat \rho(t)\widehat{\O}(t)]=\int \dd\phi \, \dd \phi' \, \rho_{\phi \phi'}(t) \bra{\phi'}\widehat \O(t)\ket{\phi}\,,
\end{align}
with 
\begin{align}\label{rhomatrix}
	\rho_{\phi\phi'}(t)\equiv \bra{\phi}\widehat \rho(t)\ket{\phi'}=\int^{\phi} \mathcal{D}\varphi_{+}\int^{\phi'}\mathcal{D} \varphi_{-}\,e^{iS_{\mathrm{eff}}[\varphi_{+},\varphi_{-}]}\,.
\end{align}
$S_{\mathrm{eff}}$ is a functional of the fields that we will refer to as \textit{open functional}. This is the sum of terms describing the unitary dynamics of the system plus an influence functional describing effects mediated by the environment \cite{FEYNMAN1963118}. Notice that $ \dd\phi$ denotes an average over boundary conditions at some time-slice $ t$, while $ \mathcal{D}\phip$ and $ \mathcal{D}\phim$ denote path integrals over histories \cite{Boyanovsky:2015xoa}. The initial conditions, which we left implicit in the formulae, are specified in terms of an initial density matrix, which we will always to be a pure state in the infinite past. For the moment, we will be interested in operators that are diagonal in the field basis, namely $\bra{\phi}\mathcal{O}\ket{\phi'}\propto \delta(\phi-\phi')$. This is the case for the product of fields $\phi$ at different spacetime points, but it is not the case if one includes their momentum conjugate. Hence we will be interested in the \textit{diagonal elements} of the density matrix $\rho_{\phi \phi}(t)$ appearing in \Eq{eq:expect} for which $\phi' = \phi$.

\begin{figure}[tbp]
	\centering
	\includegraphics[width=1\textwidth]{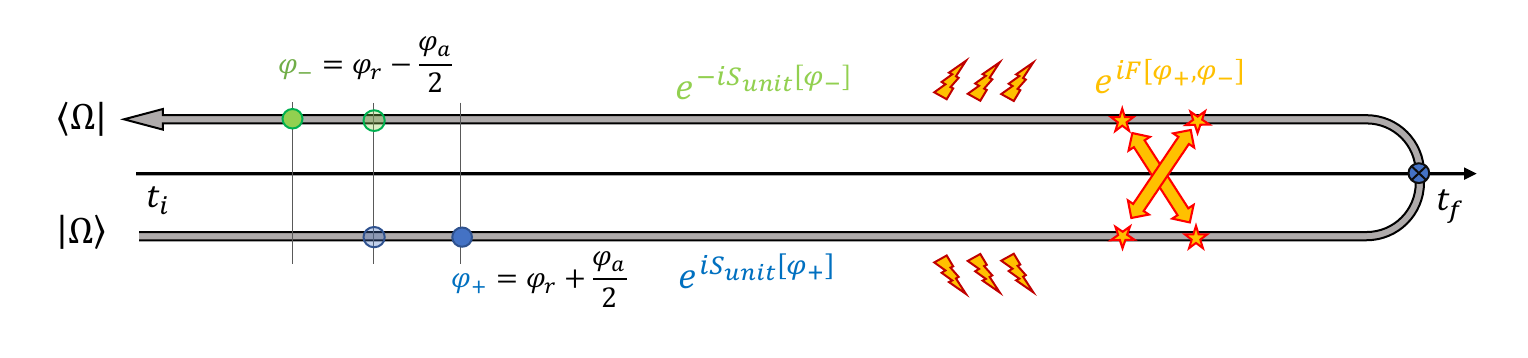}
	\caption{Illustration of the Schwinger-Keldysh path integral, where time is running from left to right in both contours and the arrow represent path ordering (time ordering in $\ket{\Omega}$ and anti-time-ordering in $\bra{\Omega}$). We consider an initially pure state $\ket{\Omega}\bra{\Omega}$ at initial time $t_i$. The unitary time evolution, which preserves the purity of the state, is the same with opposite sign on each branch of the path integral, captured by $S_{\mathrm{unit}}$. Dissipative and stochastic effects are then captured by $F$ which has no unitary counterpart and cannot be captured through a single-branch contour.}
	\label{fig:contour}
\end{figure}

In our case, $S_{\mathrm{eff}}$ is a functional that integrates an integrand over the whole of space, for example $\mathbb{R}^3$, and over some interval of time. Sometimes it might be convenient to break up the functional $ S_{\mathrm{eff}}$ into a part that describes unitary evolution on a normalizable pure state and the rest, that is
\begin{align}\label{eq:splitting}
	S_{\mathrm{eff}}[\varphi_{+},\varphi_{-}]&= S_{\mathrm{unit}}[\varphi_{+}]-S_{\mathrm{unit}}[\varphi_{-}] + F[\varphi_{+},\varphi_{-}]\,.
\end{align}
The decomposition is summarized in \Fig{fig:contour}. $F$ is often called the Feynman-Vernon influence functional and encodes the effects of the environment on the system \cite{FEYNMAN1963118}. We will not derive $F$ from an explicit model of the environment. Instead, our goal here is to model $F$ in the most generic way possible by assuming a set of symmetries and, most importantly, a separation of scales that ensures locality in time and space in the system sector. \\

\subsubsection{Non-equilibrium constraints}\label{subsec:NEQconst}

By definition, $\widehat\rho$ obeys the following constraints \cite{breuerTheoryOpenQuantum2002}
\begin{align}\label{eq:NEQ}
	\Tr \widehat\rho&=1\,, & \widehat{\rho}^{\dagger}&=\widehat\rho\,, & \widehat\rho&\geq0\,,
\end{align}
where the last is a shorthand notation for $ \bra{\psi}\widehat \rho \ket{\psi}\geq 0$ for all $ \ket{\psi}\in \H$. These conditions are crucial to ensure that the quantum mechanical formalism makes meaningful statistical predictions. These three defining properties of a density matrix impose additional constraints on the functional $ S_{\mathrm{eff}}$ \cite{Calzetta:2008iqa, Liu:2018kfw} that are respectively
\begin{align}
	S_{\mathrm{eff}} \left[\varphi_+,\varphi_+\right] &= 0\,,  \label{eq:norm1}\\
	S_{\mathrm{eff}} \left[\varphi_+,\varphi_-\right] &= - 	S^*_{\mathrm{eff}} \left[\varphi_-,\varphi_+\right] \,, \label{eq:herm1}\\
	\Im S_{\mathrm{eff}} \left[\varphi_+,\varphi_-\right] &\geq 0 \label{eq:im1}\,.
\end{align}
The derivation of these constraints is slightly cumbersome and may be skipped by the reader:

\begin{tcolorbox}[%
	enhanced, 
	breakable,
	skin first=enhanced,
	skin middle=enhanced,
	skin last=enhanced,
	before upper={\parindent15pt},
	]{}
	
	\vspace{0.05in}
	
	\paragraph{Derivation of $S_{\mathrm{eff}}$ constraints.}
	
	The derivation follows from \cite{Glorioso:2016gsa}. Let us consider a UV evolution operator $\widehat{\mathcal{U}}(t,t_0)$ under which the UV state evolves according to
	\begin{align}
		\widehat{\rho}(t) = \widehat{\mathcal{U}}(t,t_0)  \ket{\Omega} \bra{\Omega}\widehat{\mathcal{U}}^\dag(t,t_0),
	\end{align}
	starting from an initial vacuum state $\ket{\Omega}$. The reduced density matrix is obtained by tracing out the environmental degrees of freedom  denoted $\sigma$ such that 
	\begin{align}
		\widehat{\rho}_{\mathrm{red}} (t) &= \mathrm{Tr}_\sigma  \left[  \widehat{\mathcal{U}}(t,t_0)  \ket{\Omega} \bra{\Omega}\widehat{\mathcal{U}}^\dag(t,t_0) \right],\\
		&=\int \dd \sigma \bra{\sigma}  \widehat{\mathcal{U}}(t,t_0)  \ket{\Omega} \bra{\Omega} \widehat{\mathcal{U}}^\dag(t,t_0) \ket{\sigma}.
	\end{align}
	Let us consider the field-basis matrix element of reduced density matrix
	\begin{align}
		\rho_{\varphi \varphi'} (t) &\equiv \bra{\varphi} \widehat{\rho}_{\mathrm{red}} (t)\ket{\varphi'} \\
		&= \int \dd \sigma  \bra{\varphi} \otimes \bra{\sigma}  \widehat{\mathcal{U}}(t,t_0)  \ket{\Omega} \bra{\Omega} \widehat{\mathcal{U}}^\dag(t,t_0)  \ket{\varphi'} \otimes \ket{\sigma}\\
		&=\int \dd \varphi_i \dd \varphi_i^\prime \int \dd \sigma \int \dd \sigma_i \dd \sigma_i^\prime \bra{\varphi} \otimes \bra{\sigma}  \widehat{\mathcal{U}}(t,t_0) \ket{\varphi_i} \otimes \ket{\sigma_i} \nonumber \\
		&\qquad \qquad  \rho^{(0)}_{\varphi_i \varphi_i^\prime} \rho^{(0)}_{\sigma_i \sigma_i^\prime}\bra{\varphi_i^\prime} \otimes \bra{\sigma_i^\prime} \widehat{\mathcal{U}}^\dag(t,t_0)\ket{\varphi'} \otimes \ket{\sigma},
	\end{align}
	where $\ket{\varphi}$, $\ket{\sigma}$ are eigenstates of the position operators $\widehat{\varphi}$, $\widehat{\sigma}$ and we used four representations of the identity, two on each side of the vacuum density matrix. The initial matrix elements are
	\begin{align}
		\rho^{(0)}_{\varphi_i \varphi_i^\prime} &\equiv \langle \varphi_i|\Omega_\varphi\rangle \langle\Omega_\varphi|\varphi_i^\prime\rangle, \\
		\rho^{(0)}_{\sigma_i \sigma_i^\prime} &\equiv \langle\sigma_i|\Omega_\sigma\rangle \langle\Omega_\sigma|\sigma_i^\prime\rangle,
	\end{align}
	where we consider $\ket{\Omega} = \ket{\Omega_\varphi} \otimes \ket{\Omega_\sigma}$. The path integral representation of the evolution operator is 
	\begin{align}
		\bra{\varphi} \otimes \bra{\sigma}  \widehat{\mathcal{U}}(t,t_0) \ket{\varphi_i} \otimes \ket{\sigma_i} &= \int_{\varphi_i}^\varphi \mathcal{D}\varphi_+ \int_{\sigma_i}^\sigma \mathcal{D}\sigma_+  \ee^{i S_0\left[\varphi_+,\sigma_+ \right]}, \\
		\bra{\varphi_i^\prime} \otimes \bra{\sigma_i^\prime} \widehat{\mathcal{U}}^\dag(t,t_0)\ket{\varphi'} \otimes \ket{\sigma}  &= \int_{\varphi_i^\prime}^{\varphi'} \mathcal{D}\varphi_- \int_{\sigma_i^\prime}^\sigma \mathcal{D}\sigma_-  \ee^{-i S_0\left[\varphi_-,\sigma_- \right]},
	\end{align}
	such that we obtain 
	\begin{align}
		\rho_{\varphi \varphi'} (t) = \int \dd \varphi_i \dd \varphi_i^\prime \int_{\varphi_i}^\varphi \mathcal{D}\varphi_+ \int_{\varphi_i^\prime}^{\varphi'} \mathcal{D}\varphi_- \ee^{i S_{\mathrm{eff}}\left[ \varphi_+, \varphi_-\right]} \rho^{(0)}_{\varphi_i \varphi_i^\prime},
	\end{align}
	with the influence functional 
	\begin{align}
		\ee^{i S_{\mathrm{eff}}\left[ \varphi_+, \varphi_-\right]} = \int \dd \sigma \int \dd \sigma_i \dd \sigma_i^\prime  \int_{\sigma_i}^\sigma \mathcal{D}\sigma_+ \int_{\sigma_i^\prime}^\sigma \mathcal{D}\sigma_-  \ee^{i S_0\left[\varphi_+,\sigma_+ \right]-i S_0\left[\varphi_-,\sigma_- \right]} \rho^{(0)}_{\sigma_i \sigma_i^\prime}.
	\end{align}
	
	The central step of the proof is to consider $\sigma$ evolves as if $\varphi$ is a background (external source) see Appendix A of \cite{Glorioso:2016gsa}. In this case, we can consider the sourced evolution 
	\begin{align}
		\int_{\sigma_i}^\sigma \mathcal{D}\sigma_+  \ee^{i S_0\left[\varphi_+,\sigma_+ \right]} &= \bra{\sigma}   \widehat{\mathcal{U}}(t,t_0;\{\varphi_+\})\ket{\sigma_i}, \\ 
		\int_{\sigma_i^\prime}^\sigma \mathcal{D}\sigma_-  \ee^{-i S_0\left[\varphi_-,\sigma_- \right]} &= \left[\bra{\sigma}   \widehat{\mathcal{U}}(t,t_0;\{\varphi_-\})\ket{\sigma_i^\prime}\right]^\dag,
	\end{align}
	where $\widehat{\mathcal{U}}(t,t_0;\{\varphi_+\})$ is the sourced evolution acting on $\mathcal{H}_\sigma$ only. One can the reconsider the influence functional as being 
	\begin{align}
		\ee^{i S_{\mathrm{eff}}\left[ \varphi_+, \varphi_-\right]} &= \int \dd\sigma \bra{\sigma} \widehat{\mathcal{U}}(t,t_0;\{\varphi_+\}) \overbrace{\left[\int \dd \sigma_i  \ket{\sigma_i} \bra{\sigma_i} \right]}^{\mathbb{I}}\ket{\Omega_\sigma} \\
		&\bra{\Omega_\sigma} \underbrace{\left[ \int \dd \sigma_i^\prime  \ket{\sigma_i^\prime}  \bra{\sigma_i^\prime} \right]}_{\mathbb{I}}\widehat{\mathcal{U}}^\dag(t,t_0;\{\varphi_-\}) \ket{\sigma} \nonumber 
	\end{align}
	which finally reduces to 
	\begin{align}
		\ee^{i S_{\mathrm{eff}}\left[ \varphi_+, \varphi_-\right]} &= \bra{\Omega_\sigma} \widehat{\mathcal{U}}^\dag(t,t_0;\{\varphi_-\})
		\overbrace{\left[\int \dd \sigma \ket{\sigma} \bra{\sigma} \right]}^{\mathbb{I}}\widehat{\mathcal{U}}(t,t_0;\{\varphi_+\}) \ket{\Omega_\sigma},
	\end{align}
	that is 
	\begin{align}\label{eq:rate}
		\ee^{i S_{\mathrm{eff}}\left[ \varphi_+, \varphi_-\right]}  &= \langle \Omega^{ \{\varphi_- \} }_\sigma (t) | \Omega^{ \{\varphi_+ \} }_\sigma (t) \rangle 
	\end{align}
	\noindent where we defined the sourced-evolved states
	\begin{align}
		\ket{ \Omega^{ \{\varphi_+ \} }_\sigma (t)} &=                \widehat{\mathcal{U}}(t,t_0;\{\varphi_+\}) \ket{\Omega_\sigma}, \\
		\bra{ \Omega^{ \{\varphi_- \} }_\sigma (t)}&= \bra{\Omega_\sigma}\widehat{\mathcal{U}}^\dag(t,t_0;\{\varphi_-\}).
	\end{align}
	In this sense, the influence functional can be interpreted as a probability of obtaining a configuration $(\varphi_+,\varphi_-)$ given the unitary evolution of the UV theory and taking into consideration the lack of knowledge about the environment. The transition rate is physical if
	\begin{align}
		||\langle\Omega^{ \{\varphi_- \} }_\sigma (t) | \Omega^{ \{\varphi_+ \} }_\sigma (t) \rangle ||^2 \leq 1
	\end{align}
	such that 
	\begin{align}
		|\ee^{i S_{\mathrm{eff}}\left[ \varphi_+, \varphi_-\right]}| \leq 1 \qquad \Rightarrow \qquad \Im S_{\mathrm{eff}} \left[\varphi_+,\varphi_-\right] &\geq 0. 
	\end{align}
	Then, from \Eq{eq:rate}, one can easily see that 
	\begin{align}
		\ee^{-i S^*_{\mathrm{eff}}\left[ \varphi_+, \varphi_-\right]} = \langle\Omega^{ \{\varphi_+ \} }_\sigma (t) | \Omega^{ \{\varphi_- \} }_\sigma (t) \rangle = \ee^{i S_{\mathrm{eff}}\left[ \varphi_-, \varphi_+\right]}
	\end{align}
	from which we deduce
	\begin{align}
		S_{\mathrm{eff}} \left[\varphi_+,\varphi_-\right] &= - 	S^*_{\mathrm{eff}} \left[\varphi_-,\varphi_+\right].
	\end{align}
	Lastly, the causality structure which is obvious in the unitary theory, $S_0[\varphi_+] - S_0[\varphi_-] = 0$ if $\varphi_+ = \varphi_-$, is much less straightforward in the non-unitary case yet still holds as
	\begin{align}
		\ee^{i S_{\mathrm{eff}}\left[ \varphi_+, \varphi_+\right]}  &= \langle\Omega^{ \{\varphi_+ \} }_\sigma (t) | \Omega^{ \{\varphi_+ \} }_\sigma (t) \rangle  =1 \qquad \Rightarrow \qquad S_{\mathrm{eff}}\left[ \varphi_+, \varphi_+\right] = 0.
	\end{align}
	Note that an alternative derivation may be possible using the classical effective functional, that is the saddle-point evaluation of the path integral. 
	
\end{tcolorbox}

We often work in the Keldysh basis 
\begin{align}
	\varphi_r = \frac{\varphi_+ + \varphi_-}{2},\qquad \varphi_a = \varphi_+ - \varphi_- \qquad \Leftrightarrow \qquad \varphi_+ = \varphi_r + \frac{\varphi_a}{2}, \qquad \varphi_- = \varphi_r - \frac{\varphi_a}{2} .\label{eq:RAbasis}
\end{align}
The retarded and advanced components $\varphi_r$ and $\varphi_a$, which respectively correspond to the mean and the difference of the field inserted along each branch of the path integral, turn out to conveniently organise the perturbative expansion \cite{Kamenev:2009jj}. In this basis, the above non-equilibrium constraints become 
\begin{align}
	S_{\mathrm{eff}} \left[\varphi_r,\varphi_a = 0\right] &= 0 \,,\label{eq:norm} \\
	S_{\mathrm{eff}} \left[\varphi_r,\varphi_a\right] &= - 	S^*_{\mathrm{eff}} \left[\varphi_r,-\varphi_a\right] \,,\label{eq:herm} \\
	\Im S_{\mathrm{eff}} \left[\varphi_r,\varphi_a\right] &\geq 0. \label{eq:pos}
\end{align} 
While this basis is convenient to make manifest the causality structure of the theory \cite{kamenev_2011} and to understand the structure of non-unitary operators \cite{Salcedo:2024smn}, it renders much harder the identification of a unitary subset, for which the $+/-$ basis remains the best option.

Despite their simple looking, the constraints \eqref{eq:norm1}, \eqref{eq:herm1} and \eqref{eq:im1} are imposing important model independent and non-perturbative restrictions on $S_{\mathrm{eff}}$: 
\begin{enumerate}
	\item Following \eqref{eq:norm}, $S_{\mathrm{eff}}$ starts linear in $\varphi_a$;
	\item Following \eqref{eq:herm}, odd powers of $\varphi_a$ are purely real and even powers of $\varphi_a$ purely imaginary;
	\item Following \eqref{eq:pos}, some EFT coefficients have to obey a positivity bound.
\end{enumerate}
If the theory is unitary, that is, if $ F[\varphi_{+}, \varphi_{-}] = 0 $ in \eqref{eq:splitting}, it is straightforward to show that $ S_{\mathrm{eff}} $ contains only odd powers of $ \varphi_a $. However, the converse is not true, as there exist operators that are odd in $ \varphi_a $ but do not originate from a unitary theory. At last, even powers of $\varphi_a$ only appear in stochastic field theory and will shortly be related to noise variables.

A natural way to organise open EFTs is in powers of the advanced field
\begin{align}
	S_{\mathrm{eff}} = \sum_{n = 1}^{\infty} S_n \quad \text{with} \quad S_n = \mathcal{O}(\varphi^n_a)
\end{align}
The reason is that if the scalar field can be decomposed along a background profile, 
\begin{align}
	\varphi_{\pm} = \bar{\varphi}(t) + \delta \varphi_\pm
\end{align} 
the retarded component carries the background value of the field while the advanced components starts linear in perturbations, 
\begin{align}
	\varphi_r \equiv \frac{\varphi_{+} + \varphi_{-}}{2} = \bar{\varphi}(t) + \delta \varphi_r, \qquad \varphi_a \equiv \varphi_{+} - \varphi_{-} =  \delta \varphi_a
\end{align}
Hence, terms in $S_{n \geq 3}$ are at least cubic in perturbations. While in a vacuum QFT, $ \delta \varphi_r$ and $ \delta \varphi_a$ scale the same, the power counting of NEQ QFTs strongly depends on the occupation number of the state, $N_\varphi = \langle \widehat{a}^\dag_\varphi \widehat{a}_\varphi\rangle$ \cite{2016RPPh...79i6001S, Berges:2012ty, Rosenhaus:2025bgy, Radovskaya:2020lns}. At large occupation $N_\varphi \gg 1$, $\varphi_a \sim \mathcal{O}(\hbar) \ll \varphi_r \sim \mathcal{O}(1) $, operators contained in $S_{n \geq 3}$ become irrelevant and the influence functional reduces to its semiclassical limit, $S_{\mathrm{eff}} = S_1 + S_2$ known as the Martin-Siggia-Rose (MSR) formalism.\footnote{This is pretty much the same story as the difference between a vacuum and thermal QFT \cite{Rosenhaus:2025bgy}. The power counting is modified in the latter in which interactions become irrelevant. In this case, it is enough to work with the Matsubara propogator to explore the equilibrium properties of the thermal QFT.} We will come back to this point below when we will derive the equations of motion for the scalar field $\varphi_r$.


\subsubsection{Free functional} \label{subsec:BCs}

To explore the basic features of this formalism, let us first consider a single scalar degree of freedom $\varphi$ evolving in flat space according to 
\begin{align}\label{eq:BM}
	S_{\mathrm{eff}}[\phir,\phia] = \int \dd^4x \left(\dphir \dphia - c_s^2 \partial_i \phir \partial^i \phia + \gamma \dphir \phia + i \beta \phia^2\right)
\end{align}
where a dot represents a time derivative. First, note that the first two terms can be written in a factorized form in the original $+/-$ basis,
\begin{align}
	S_{\mathrm{unit}}[\varphi_+] - S_{\mathrm{unit}}[\varphi_-] \qquad \mathrm{with} \qquad S_{\mathrm{unit}}[\varphi] = \frac{1}{2} \int \dd^4x \left[\dot{\varphi}^2 - c_s^2 (\partial_i \varphi)^2 \right],
\end{align}
and hence represent the unitary evolution. 
Conversely, the last two terms have no analogue in the unitary case and encode the dissipative and diffusive effects of $F[\varphi_+,\varphi_-]$ characteristic of the open dynamics. Note that both $\gamma$ and $\beta$ should be real, in accordance with the conditions above. 

Let us first discuss the first non-unitary contribution $\gamma \dphir \phia$ appearing in \Eq{eq:BM}. This dissipative operator is crucial in describing the loss of energy of the system $\varphi$ into its surrounding environment. In the original $+/-$ basis, it contains the boundary term
\begin{align}
	S_{\mathrm{unit}}[\varphi_+] - S_{\mathrm{unit}}[\varphi_-] \qquad \mathrm{with} \qquad S_{\mathrm{unit}}[\varphi] = -\frac{1}{4} \int \dd^4 x \left(\frac{\dd}{\dd t} \left[\varphi^2\right] \right),
\end{align}
together with a mixing between the two branches of the path integral 
\begin{align}
	F[\varphi_+,\varphi_-] = - \frac{1}{2} \int \dd^4 x \left(\dot{\varphi}_+ \varphi_- - \varphi_+ \dot{\varphi}_- \right).
\end{align}
Upon specifying the boundary conditions of the diagonal density matrix element appearing in \Eq{eq:expect}
\begin{align}
	\varphi_+(\bmx, t_0) = \varphi_-(\bmx, t_0) = \phi(\bmx),
\end{align}
it is clear that the boundary term does not contribute to the diagonal elements of $\widehat \rho$. On the contrary, the mixing term $F[\varphi_+,\varphi_-] $ has no reason to vanish. This effect has no unitary counterpart and is precisely responsible for damping of $\varphi$ fluctuations due to dissipation. To make this fact manifest, we now need to derive the equations of motion.  

The latter are obtained by varying $S_{\mathrm{eff}}$ with respect to $\varphi_a$, then setting $\varphi_a = 0$. One could naively think that it implies only the functional $ S_1 \propto \mathcal{O}(\varphi_a)$ linear in $\varphi_a$ contributes to the equations of motion of $\varphi_r$. It turns out that this is not the case, because of the boundary conditions of the path integral. The boundary conditions for $  \varphi_{a} $ are those responsible for $  \varphi_{a} $ not propagating, despite the fact that $  \varphi_{a} $ appears in the quadratic action in a way that is very similar to $  \varphi_{r} $. If it was not for the boundary conditions, the theory would inevitably suffer from an instability.\footnote{To see this, imagine to incorrectly assume that $  \varphi_{a} $ has standard initial conditions (as opposed to satisfying a boundary condition problem) and integrate it out. Its equation of motion is given by $  \varphi_{a}=\Box \varphi_{r} $, which, once substituted back into the action would give terms as $  \varphi_{r}\Box^{2}\varphi_{r} $. This theory is the iconic example of a ghost, as seen from the fact that the propagator can be partial fractioned into two terms with opposite signs (this is best seen introducing a small mass, which can be sent to zero at the end). Another way to see this is that $  \varphi_{+} $ and $  \varphi_{-} $ have opposite sign kinetic terms (since we have $iS_{\mathrm{unit}}[\varphi_{+}]-iS_{\mathrm{unit}}[\varphi_{-}]  $) and one of the two would end being a ghost.} Because of the boundary conditions, $  \varphi_{a} $ does not propagate and the theory is healthy. However, something peculiar happens as a byproduct: the $  \varphi_{a} $ path integral does not admit a stationary phase! This is because, in the presence of fluctuations $  i \beta \varphi_{a}^{2} $, the stationary phase would be $  \varphi_{a}\sim \Box \varphi_{r} $. This relation is manifestly incompatible with the boundary conditions that $  \varphi_{r} $ is anything but $  \varphi_{a} $ vanishes at $  t=t_{0} $. In other words, \textit{if $  \varphi_{a} $ appears non-linearly its path integral does not admit a stationary phase}.

Things would change if $  \varphi_{a} $ appeared linearly. Then, its stationary phase would be some equation of motion for the \textit{other} fields, not involving $  \varphi_{a} $, and that would always be compatible with the vanishing of $  \varphi_{a} $ on the boundary. This is indeed the magic that ensues from the Hubbard-Stratonovich (HS) transformation \cite{Stratonovich1957,Hubbard1959}. Upon introducing an auxiliary noise field $ \xi   $, the open functional becomes \textit{linear} in $  \varphi_{a} $. It is only at this point that the path integral in $  \varphi_{a} $ is well approximated by a stationary phase and it is only now that $  \delta S/ \delta \varphi_{a} $ can be interpreted as an equation of motion, namely the Langevin equation. Explicitly, the key trick is the mathematical identity
\begin{equation}
	\exp\left(-\int d^{4}x \beta\phia^{2}\right)=\mathcal{N}_{0}\int\mathcal{D}\xi\;\exp\left[\int d^{4}x \left(-\frac{\xi^{2}}{4\beta}+i\xi\phia\right)\right],
\end{equation}
with $\mathcal{N}_{0}$ being the normalisation constant. Note that because of the positivity constraint \Eq{eq:pos}, convergence is always ensured.
After this transformation, the path integral in $  \varphi_{a} $ is well approximated by a stationary phase which generates the Langevin equation
\begin{align}\label{eq:langevin}
	\ddot{\varphi}_r + \gamma \dphir + c_s^2 k^2 \phir = \xi.
\end{align}
The new variable $\xi$, satisfying $\ex{\xi(x)}=0$, behaves as a Gaussian field with a prescribed two-point function
\begin{equation}\label{eq:noisecontact}
	\langle\xi(x) \xi(y)\rangle=2\beta \delta(x-y).
\end{equation}

We conclude that \Eq{eq:BM} provides a path integral representation of the Langevin equation introduced in \eqref{eq:langevinintro}. We stress that we are not assuming that the system or the environment are in thermodynamical equilibrium. In particular $\beta$ is not in general related to the inverse temperature.\footnote{In other words, Schwinger-Keldysh path integral considered prepares a density matrix that is not a thermal state.} If desired, one could impose additional constraints, known as Kubo–Martin–Schwinger (KMS) conditions (see \cite{Kubo:1966fyg, PhysRev.115.1342} and \cite{Hongo:2018ant} for a review), to specify our construction to the thermal case. We do not follow this approach because our goal is to develop a formalism for problems in cosmology and gravity where thermalization does not necessary take place.


\subsubsection{Interactions}

To get familiar with interactions in the Keldysh basis, let us first consider the simple toy model of a massive scalar with $  \lambda \varphi^{3} $ interaction
\begin{align}
	S_{\mathrm{eff}}[\phir,\phia] = S_{\mathrm{unit}}[\varphi_+] - S_{\mathrm{unit}}[\varphi_-] \quad \mathrm{with} \quad S_{\mathrm{unit}}[\varphi] = \int \dd^4x \left[ -\frac{1}{2}\left(\partial_{\mu}\varphi\right)^{2}-\frac{1}{2}m^{2}\varphi^{2} -\lambda \varphi^{3}\right], \label{eq:unitarySphi}
\end{align}
that is 
\begin{align}
	S_{\mathrm{eff}}[\phir,\phia] &=\int \dd^4x \left(-\partial_{\mu}\varphi_{r}\partial^{\mu}\varphi_{a}-m^{2}\varphi_{r}\varphi_{a}-3\lambda \varphi_{r}^{2} \varphi_{a}-\frac{\lambda}{4} \varphi_{a}^{3} \right)\,.\label{eq:unitaryIphi}
\end{align}
Note that the theory fulfills the NEQ constraints presented above. The unitary interaction $\lambda(\varphi_{+}^3 - \varphi_-^3)$ generates two types of vertices, $\varphi_r^2 \varphi_a$ and $\varphi_a^3$.

If we vary $S_{\mathrm{eff}}$ with respect to $\varphi_a$, the terms linear in $  \varphi_{a} $ precisely generate the classical equations of motion, with all the right coefficients. This type of interactions can be understood as classical non-linearities. One may wonder what is the role of the $\varphi_a^3$ term? It is not easy to see how they contribute to the Langevin equation because we have to perform more complex version of the Hubbard-Stratonovich transformation, as done in \cite{Salcedo:2024smn} for $\varphi_a^3$ (see also \cite{Chakrabarty:2019aeu}). If one manages to do so, it appears $\varphi_a^3$ can be understood as a white non-Gaussian noise, 
\begin{equation}\label{eq:noisecontactv2}
	\langle\xi(x) \xi(y) \xi(z)\rangle \propto \delta(x-y) \delta(y-z).
\end{equation}
We will come back to these terms below when we will compute some flat-space correlators, for which these contributions are crucial to recover the flat space vacuum results.

So far, we considered \textit{unitary interactions}. Note that if we detune the fine balance between $\varphi_r^2 \varphi_a$ and $\varphi_a^3$, for instance considering
\begin{align}
		S_{\mathrm{eff}}[\phir,\phia] \supset  \int \dd^4x\left[-3\lambda \varphi_{r}^{2} \varphi_{a}-\frac{\lambda}{4} (1 + \delta) \varphi_{a}^{3}\right],
\end{align}
the theory stops being unitary, $S_{\mathrm{eff}} \neq S_{\mathrm{unit}}[\varphi_+] - S_{\mathrm{unit}}[\varphi_-]$, as long as $\delta \neq 0$. Importantly, the theory still fulfills the constraints \eqref{eq:norm}, \eqref{eq:herm} and \eqref{eq:pos}. The theory can still make perfect sense if understood as an open theory. At last, we could also have considered some interactions of the form $\mathcal{O}(\varphi_r) \varphi_a^{2p}$ with $p$ an integer. These interactions couple the operator $\mathcal{O}(\varphi_r)$ to the noise variables. For instance, let us consider 
\begin{align}\label{eq:thfull}
	S_{\mathrm{eff}}[\phir,\phia] =\int \dd^4x \bigg[& \dphir \dphia - c_s^2 \partial_i \phir \partial^i \phia + \gamma \dphir \phia + i \beta \phia^2  \\
	-&3\lambda \varphi_{r}^{2} \varphi_{a}  + i \widetilde{\beta} \varphi_r\varphi_a^2 -\frac{\lambda}{4}(1 + \delta) \varphi_{a}^{3}  \bigg],
\end{align}
where the first line contains the linear operators discussed above and the second line some interactions.
The associated equations of motion are given by the non-linear Langevin equation \cite{Salcedo:2024smn}
\begin{align}
	\ddot{\varphi}_r + \gamma \dphir + c_s^2 k^2 \phir =   3 \lambda \varphi_r^2 + \left(1 +  \frac{\widetilde{\beta}}{\beta} \phir \right)\xi , 
\end{align}
with noise statistics
\begin{align}
	\langle\xi(x) \rangle &= 0,  \\
	\langle\xi(x) \xi(y) \rangle &= 2   \beta \delta(x-y) \\
	\langle\xi(x) \xi(y) \xi(z) \rangle &= 6 \lambda (1 + \delta) \delta(x-y)\delta(y-z).
\end{align}

It is interesting to understand if we can \textit{complete} an operator to obtain a unitary combination. A proposal to unitarize an operator is the following. Let us consider an operator $\mathcal{O}\left(\varphi_r,\varphi_a\right)$ made of field insertions of $\varphi_r$, $\varphi_a$ and their (at most single) derivatives. The combination of operators
\begin{align}\label{eq:unitarize}
	\mathcal{U}\left[\mathcal{O}\left(\varphi_r,\varphi_a\right)\right] = \mathcal{O}\left(\frac{\varphi_+}{2},\varphi_+\right) + \mathcal{O}(\frac{\varphi_-}{2},-\varphi_-) - \mathcal{O}\left(\frac{\varphi_-}{2},\varphi_-\right) - \mathcal{O}(\frac{\varphi_+}{2},-\varphi_+)
\end{align}
is unitary by construction. For operators such as $\varphi_a^2$ (the diffusion operators) that are intrinsically non-unitary, the above combination vanishes. Also note that some operators such as $\dot{\varphi}_r \varphi_a$ (the dissipation operators) are unitarized into total derivatives and so do not add any net contribution to the open effective functional. Lastly, some of the contributions obtained out of \Eq{eq:unitarize} eventually violate symmetries of the problem (\eg the fact that $\varphi_r$ non-linearly realises time-translations and boosts) and must then be discarded.


\subsection{Free theory}\label{subsec:freeth}

\subsubsection{Path ordering}\label{subsec:path}

Ultimately, the path integral aims at computing expectation values of observables. To achieve this task, we need to acknowledge what the path integral is actually computing. In this Section, we discuss the link between path integral and quantum operators. 
To make the following discussion more streamlined, it is useful to introduce the following shorthand notation for the path integral average over some function 
\begin{align}
	\exx{ \O[\varphi_+,\varphi_-]}\equiv \int \dd \phi \int^{\phi }_{\text{I.C.}} \mathcal{D}\varphi_+\int^{\phi}_{\text{I.C.}} \mathcal{D}\varphi_-\O[\varphi_+,\varphi_-] e^{iS_{\mathrm{eff}}[\varphi_+,\varphi_-]}\,.
\end{align}
Let us now focus on the closed-time contour of the Schwinger-Keldysh formalism, which has only two timefolds, one going forwards and one backwards in time. Along the ``$+$'' branch of the path integral, expectation values are time ordered (denoted $\T$), whereas along the ``$-$'' branch, expectation values are anti-time ordered (denoted $\bar \T$), \cite{Kamenev:2009jj}
\begin{align}
	\exx{\varphi_{+}(t)\varphi_{+}(t')}&=\ex{\T \widehat \phi(t)\widehat \phi(t')}\,,\\
	\exx{\varphi_{-}(t)\varphi_{-}(t')}&=\ex{\bar\T \widehat \phi(t)\widehat \phi(t')}\,.
\end{align}
When there is one insertion on each branch of the path integral, the expectation value is neither time or anti-time ordered and we obtain a Wightman function, 
\begin{align}
	\exx{\varphi_{+}(t)\varphi_{-}(t')}&=\ex{ \widehat \phi(t')\widehat \phi(t)}\,,\\
	\exx{\varphi_{-}(t)\varphi_{+}(t')}&=\ex{\widehat \phi(t)\widehat \phi(t')}\,.
\end{align}
Using these results repeatedly it is easy to check that
\begin{align}
	\exx{\varphi_{r}(t)\varphi_{a}(t')}&=\frac{1}{2} \exx{\left[ \varphi_{+}(t)+\varphi_{-}(t) \right]\left[ \varphi_{+}(t')-\varphi_{-}(t') \right] }\\
	&=\frac{1}{2} \left[\ex{\T \widehat \phi(t)\widehat \phi(t')}-\ex{\bar\T \widehat \phi(t)\widehat \phi(t')} + \ex{  \widehat \phi(t)\widehat \phi(t')}  - \ex{  \widehat \phi(t')\widehat \phi(t)} \right]\\
	&=\theta(t-t') \ex{[\widehat \phi(t),\widehat \phi(t')]}\,,
\end{align}
which is the definition of the retarded propagator. Similarly one finds the reversed time argument to correspond to the advanced propagator: 
\begin{align}
	\exx{\varphi_{a}(t)\varphi_{r}(t')}&=\theta(t'-t) \ex{[\widehat \phi(t'),\widehat \phi(t)]}.
\end{align}
The two-point function of the retarded propagator corresponds to the anticommutator of the field operator:
\begin{align}
	\exx{\varphi_{r}(t)\varphi_{r}(t')}&=\frac{1}{4}\exx{\left[\varphi_{+}(t)+\varphi_{-}(t)\right]\left[\varphi_{-}(t)+\varphi_{+}(t)\right]} \\
	&=\frac{1}{4}\left[\langle\mathcal{T}\widehat{\phi}(t)\widehat{\phi}(t')\rangle+\langle\widehat{\phi}(t)\widehat{\phi}(t')\rangle+\langle\widehat{\phi}(t')\widehat{\phi}(t)\rangle+\langle\bar{\mathcal{T}}\widehat{\phi}(t)\widehat{\phi}(t')\rangle\right]\\
	&=\frac{1}{2} \ex{\{\widehat \phi(t),\widehat \phi(t')\}},
\end{align}
where we have expanded the anti-time and time orderings in terms of theta functions and used $\theta(x)+\theta(-x)=1$. At last, a similar calculation gives
\begin{align}
	\exx{\varphi_{a}(t)\varphi_{a}(t')}&= \exx{\left[ \varphi_{+}(t)-\varphi_{-}(t) \right] \left[ \varphi_{+}(t')-\varphi_{-}(t') \right]}\\
	&=\ex{\T \widehat \phi(t)\widehat \phi(t')}+\ex{\bar\T \widehat \phi(t)\widehat \phi(t')}-\ex{  \widehat \phi(t)\widehat \phi(t')}-\ex{  \widehat \phi(t')\widehat \phi(t')}=0\,.
\end{align}
This illustrates the fact that $\varphi_{a}(t)$ cannot be considered as a propagating degree of freedom. These four two-point functions constitute the propagators of our theory, that we will now derive in an explicit model.

\subsubsection{Gaussian generating functional}

Let us consider the quadratic part of \Eq{eq:thfull}.
The form of \Eq{eq:thfull} allows us to write the action as a bilinear on the fields
\begin{equation}\label{eq:presc}
	S_{\text{eff}}^{(2)} = -\frac{1}{2}\int d^{4}x \left(\phir,\phia\right)\begin{pmatrix}
		0 & \widehat{D}_{A} \\ \widehat{D}_{R} & -2i\widehat{D}_{K}
	\end{pmatrix}\begin{pmatrix}
		\phir \\ \phia
	\end{pmatrix},
\end{equation}
the matrix being a second order differential operator acting to the right, which is made of
\begin{align}
	\widehat{D}_{A}& \equiv \partial_t^2 - \gamma \partial_t - c_s^2 \partial_i^2 ,\\
	\widehat{D}_{R}&\equiv \partial_t^2 + \gamma \partial_t - c_s^2 \partial_i^2 ,\\
	\widehat{D}_{K}&\equiv \beta.
\end{align}
The following path integral of the free theory computes the diagonal of the density matrix of the system
\begin{align}
	\rho_{\phi \phi}(t)= &\int_{\Omega}^\phi \mathcal{D}\phir  \int_{\Omega}^{0}\mathcal{D}\phia \exp\left\{i  S_{\text{eff}}^{(2)}[\phir,\phia]\right\}.
\end{align}
Upon the introduction of sources, we obtain the generating function
\begin{align}
	\mathcal{Z}[J_{r},J_{a}]=\int_{\Omega}^\phi \mathcal{D}\phir  \int_{\Omega}^{0}\mathcal{D}\phia \exp\Big\{-\frac{i}{2}  &\int d^{4}x \left(\phir,\phia\right)\begin{pmatrix}
		0 & \widehat{D}_{A} \\ \widehat{D}_{R} & -2i\widehat{D}_{K}
	\end{pmatrix}\begin{pmatrix}
		\phir \\ \phia
	\end{pmatrix} \nonumber\\
	&+\int d^{4}x \left(J_{a}\phir+J_{r}\phia\right)\Big\}
\end{align}
Completing the square for $\phia$ and $\phir$ allows us to factorise the dependence on the sources $J_{a}$ and $J_{r}$. This is done by introducing a shift in the path integral
\begin{equation}
	\begin{pmatrix}
		\phir \\ \phia
	\end{pmatrix}=\begin{pmatrix}
		\Pi_{r} \\ \Pi_{a}
	\end{pmatrix}+\int d^{4}y \begin{pmatrix}
		A_{11}(x,y) & A_{12}(x,y) \\  A_{21}(x,y) & 0
	\end{pmatrix}\begin{pmatrix}
		\phir \\ \phia
	\end{pmatrix}
\end{equation}   
Demanding that the terms linear in $\Pi_{r/a}$ vanish, we find\footnote{There exists a similar equation where we integrate by parts on $\widehat{D}_{A/R/K}$ which yields the same information.}
\begin{equation}
	- \frac{i}{2} \begin{pmatrix}
		0 & \widehat{D}_{A} \\ \widehat{D}_{R} & -2i\widehat{D}_{K}
	\end{pmatrix}\begin{pmatrix}
		A_{11}(x,y) & A_{12}(x,y) \\  A_{21}(x,y) & 0
	\end{pmatrix}+\frac{1}{2}\begin{pmatrix}
		\delta(x-y) & 0 \\  0 & \delta(x-y)
	\end{pmatrix}=\begin{pmatrix}
		0 & 0 \\  0 & 0
	\end{pmatrix}\,,
\end{equation}
where the Dirac delta $\delta(x-y)$ is a tensor density. These equations can be rewritten in a covariant form such that it becomes straightforward to recognise the advanced and retarded propagators
\begin{align}
	\widehat{D}_{R}(x)A_{12}(x,y)&=-i\delta(x-y)\;,\quad \;A_{12}(x,y)=-iG^{R}(x,y)\;,\quad \; G^{R}(x^{0}<y^{0})=0,\\
	\widehat{D}_{A}(x)A_{21}(x,y)&=-i\delta(x-y)\;,\quad \;A_{21}(x,y)=-iG^{A}(x,y)\;,\quad \; G^{A}(x^{0}>y^{0})=0.
\end{align}
A fundamental property of the retarded and advanced propagator is that they are mapped to each other under the exchange of the arguments:
\begin{equation}\label{eq:Rxy_Ayx}
	G^{R}(x,y)=G^{A}(y,x)\,.
\end{equation}
The Keldysh propagator is obtained from the matrix element $A_{11}(x,y)=-iG^{K}(x,y)$, which obeys the differential equation
\begin{equation}
	\widehat{D}_{R}(x)A_{11}(x,y)=2\widehat{D}_{K}(x)G^{A}(x,y).
\end{equation}
Notice that $G^{K}$ is thus not a Green's function of some equation of motion. We choose to symmetrise in $x\leftrightarrow y$ such that 
\begin{align}\label{eq:Keldyshreff}
	G^{K}(x,y)=&i\int d^{4}z  G^{R}(x,z)\widehat{D}_{K}(z)G^{A}(z,y) \nonumber \\
	&+i\int d^{4}z  G^{R}(y,z)\widehat{D}_{K}(z)G^{A}(z,x)
\end{align}
where we have used the property \eqref{eq:Rxy_Ayx} to write $A_{11}(x,y)$ in the most symmetric way\footnote{The generalisation to non-local noises is straightforward from here by upgrading the local $\widehat{D}_{K}(z)$ to a non-local $\widehat{D}_{K}(z_{1}-z_{2})$, where now we have the integrand to be $G^{R}(x,z_{1})\widehat{D}_{K}(z_{1}-z_{2})G^{A}(z_{2},y)$ with the appropriate factors of the square root of the metric.}. Note that causality is implemented in a natural way as the retarded propagator requires $x^{0}>z^{0}$ and the advanced propagator requires $z^{0}<y^{0}$.

This leaves the partition function to be 
\begin{equation}
	\mathcal{Z}[J_{r},J_{a}]= \exp\left\{ - \frac{i}{2}\int d^{4}x \int d^{4}y \left(J_{a}(x),J_{r}(x)\right)\begin{pmatrix}
		G^{K}(x,y) & G^{R}(x,y) \\ G^{A}(x,y) & 0
	\end{pmatrix}\begin{pmatrix}
		J_{a}(y) \\ J_{r}(y)
	\end{pmatrix}\right\}
\end{equation}
where we used the fact that $\mathcal{Z}[0,0] = 1$ in the closed time contour from trace normalisation \cite{kamenev_2011}. The two-point function reduces to
\begin{align}\label{eq:extract}
	&\exx{\phir(x_{1})\phir(x_{2})}=\left.\frac{\delta^{2}}{\delta J_{a}(x_{1})\delta J_{a}(x_{2})}\mathcal{Z}[J_{a},J_{r}]\right|_{J_{r,a}=0}=-i G^K(x_{1},x_{2}).
\end{align}
Evaluating the two-point function at coincident times in Fourier space provides an expression for the power spectrum of the system
\begin{align}
	&\exx{\phir(\bfk, \eta_{0})\phir(\bfk', \eta_{0})}=P_{k}(\eta_{0})(2\pi)^{3}\delta(\bfk+\bfk') \quad \mathrm{with} \quad P_{k}(\eta_{0}) = -i G^K(k; \eta_0,\eta_0).\bigg.
\end{align}      


\subsubsection{Propagators}\label{subsec:M4Pk}

Let us now derive the propagators in the explicit theory prescribed by \Eq{eq:presc}. The equations of motion for the propagators read  
\begin{align}
	\left(\partial_t^2 \pm \gamma \partial_t + c_s^2 k^2 \right) G^{R/A} (k; t_1, t_2) =  \delta(t_1-t_2)\,,			 
\end{align}
and, from \Eq{eq:Keldyshreff},
\begin{align}
	G^K (k;t_1,t_2) = 2 i \beta \int \dd t' G^{R} (k;t_1,t')G^{A} (k;t',t_2).
\end{align}
One must first solve for the retarded propagator to deduce the advanced and Keldysh ones.

The retarded and advanced propagators are easily found in frequency space, where 
\begin{align}
	G^{R/A} (k; \omega) &= - \frac{1}{\omega^2 \pm i \gamma \omega - c_s^2k^2}.
\end{align}
The dispersion relation 
\begin{align}\label{eq:disprelref}
	\omega^2 + i \gamma \omega - c_s^2 k^2 = 0
\end{align}
characterises the propagation in the system. Explicitly, it is the fact that \Eq{eq:disprelref} admits non-trivial solutions for the frequencies $\omega(k,\gamma)$ that ensures the existence of a dissipative degree of freedom. In this case, the solutions are 
\begin{align}\label{eq:soldispeg}
	\omega_\pm \equiv - i \frac{\gamma}{2} \pm E_k^{\gamma} \qquad \qquad \mathrm{and} \qquad \qquad 	E_k^{\gamma} \equiv \sqrt{c_s^2 k^2 - \frac{\gamma^2}{4}}.
\end{align}
which have both a real and imaginary part whenever $2c_s k > \gamma$. Note that in the limit $k \rightarrow 0$ with $\gamma$ finite, the dispersion relation is purely imaginary. We notice a gapless mode, corresponding to $\omega_+ \rightarrow 0$ for $k \rightarrow 0$ and a gapped mode associated to $\omega_- \rightarrow i \gamma$ for $k \rightarrow 0$. Importantly, for $\gamma > 0$ both modes have a negative imaginary part
\begin{align}
	\gamma > 0 \qquad  \Rightarrow \qquad \Im \omega_\pm < 0,
\end{align}
so that the retarded Green's function 
\begin{align}
	G^{R} (k; \omega) &= - \frac{1}{\omega^2 + i \gamma \omega - c_s^2k^2}= -\frac{1}{(\omega_- - \omega_+)} \left[ \frac{1}{\omega- \omega_-} - \frac{1}{\omega- \omega_+}   \right],
\end{align}
is manifestly convergent, that is
\begin{align}\label{eq:retardedM4dissip}
	G^{R} (k; t - t') &= \int \frac{\dd \omega}{2\pi} \ee^{i \omega (t-t')} G^{R} (k; \omega) = - \frac{\sin \left[E_k^{\gamma} (t-t')\right]}{E_k^{\gamma}} \ee^{- \frac{\gamma}{2}(t-t')}\theta\left(t-t'\right),
\end{align}
where we defined $E_k^{\gamma} = \sqrt{c_s^2 k^{2}-(\gamma/2)^{2}}$. Note that the causality property encoded in the above Heaviside theta function indeed implies analyticity of $G^R$ in the upper-half complex frequency plane.
Therefore, $\gamma > 0$ ensures causality and stability, via late-time convergence. 

Note that the noise term $ i \beta \varphi_a^2$ appearing in \Eq{eq:presc} does not affect the propagation of the dissipative degree of freedom \cite{kamenev_2011}. Indeed, the noise sources the dynamics of the system but does not directly transform the propagation of information within it (said it differently, the former relates to the Keldysh function whereas the latter to the Green's functions). The Keldysh component is also easily obtained in frequency space where
\begin{align}
	G^{K} (k; \omega) &= 2 G^{R} (k; \omega) \widehat{D}_K (k; \omega) G^{A} (k; \omega) = \frac{2 i  \beta}{\left(\omega^2-c_s^2k^2\right)^2+\gamma^2 \omega^2}.
\end{align}
We assume $E_k^{\gamma}$ to be real (\textit{damped regime}), keeping in mind that the \textit{overdamped regime} for which $E_k^{\gamma} \in i\mathbb{R}$ can be obtained by analytic continuation. One can analyse the pole structure and obtain the real-space propagator from 
\begin{align}
	G^{K} (k; \tau) &= 2 i \beta \int \frac{\dd \omega}{2\pi} \frac{\ee^{-i\omega \tau}}{\prod_{i=1}^{4}(\omega-\omega_i)}
\end{align}
with the poles satsfying $\omega_2 = - \omega_1$, $\omega_3 = - \omega^*_1$, $\omega_4 = \omega^*_1$ and 
\begin{align}
	\omega_1 = E_k^{\gamma} + i \frac{\gamma}{2}.
\end{align}
A long but straightforward derivation leads to 
\begin{align}\label{eq:distribM4dissip}
	G^{K} (k; \tau) &= i \frac{\ee^{-\frac{\gamma}{2}\tau}}{c_s^2 k^2} \bigg[\frac{2 \beta}{\gamma} \cos\left(E_k^{\gamma} \tau\right) + \beta \frac{\sin\left(E_k^{\gamma} \tau\right)}{E_k^{\gamma}}\bigg]
\end{align}
In the coincident time limit, we obtain the dissipative power spectrum
\begin{align}
	P_k = \frac{2 \beta}{\gamma c_s^2k^2}.
\end{align} 
From the fact that the equal-time power spectrum must be non-negative we conclude that the $\beta$'s must be positive. Note that the vacuum contribution to the power spectrum, the usual $P_k=1/2E_k^{\gamma = 0}$ term, is absent because of the exponential decay in time caused by dissipation. Nevertheless, with hindsight, one could still recover the standard vacuum Minkowski power spectrum by taking both $\beta_1$ and $\gamma$ to zero while keeping their ratio fixed. This will be the object of the next Section.

This computation highlights several features of the formalism. While $G^R(k, \tau)$ encodes the dynamics but is oblivious to the state of the system, $G^K(k, \tau)$ captures the state of the environment by probing the statistics of fluctuations \cite{kamenev_2011}. The final outcome is an interplay between the dissipation of the system into its surrounding and the fluctuations of the environment getting imprinted onto the observable sector. Crucially, these two effects cannot be easily disentangle from one another. 


\subsubsection{$i \epsilon$ prescription}\label{subsubsec:iepsilon}

\textit{Foreword.} This section provides partial solutions to Problem 1 of the problem set in Section~\ref{subsec:pb2}. Readers who wish to practice the methods introduced in Lecture 2 are encouraged to attempt the problem set first, before consulting this section.
\\

We make a quick detour to highlight how to recover the standard vacuum QFT propagators. In the case, it is required to introduce an $\epsilon$ prescription in \eqref{eq:unitaryIphi} to set up the correct initial conditions.
This prescription is \cite{2016RPPh...79i6001S}
\begin{equation}
	S_{\mathrm{eff}}[\phir,\phia] =\int \dd^4x \left[-\partial^{\mu}\varphi_{r}\partial_{\mu}\varphi_{a}-2\epsilon\dot{\varphi}_{r} \varphi_{a} -(m^{2}+\epsilon^{2})\varphi_{r}\varphi_{a}+i\epsilon \fini\varphi_{a}^{2} \right]\,.\label{eq:epsIphi}
\end{equation}
The first term featuring $\epsilon$ corresponds to an arbitrarily small dissipation that enforces the convergence of the time integrals as $t\to-\infty$ for the retarded Green's function. The $\epsilon$ correction to the mass term is present to ensure that the left-hand side Langevin equation is deformed as
\begin{equation}
	\left[(\partial_{0}+\epsilon)^{2}+m^{2}\right]\varphi=\xi+...
\end{equation}
which yields the small imaginary displacement of the real-axis pole in the frequency plane for the computation of the retarded and advanced propagators. At last, the $i\epsilon \fini\varphi_{a}^{2}$ term corresponds to the particle population of the state at initial time on which we are computing the in-in propagators. It directly relates to the initial occupation of the state through $\fini = 2E_k (1 + 2 n_0)$ with\footnote{The normalization $2 E_k$ should be double checked by confronting to the expected results for the thermal propagators. Here, it has been chosen following \cite{2016RPPh...79i6001S}, in order to recover the vacuum QFT results.}  
\begin{align}
	n_0 = \frac{1}{e^{E_k/T}-1},
\end{align}
for a bosonic system initially prepared in a thermal state at temperature $T$, with $E_k = \sqrt{k^2 + m^2}$ in this case. The vacuum limit is recover when taking $T\rightarrow 0$, that is $	n_0 \rightarrow 0$ and $f \rightarrow 2 E_k$. The Hubbard-Stratonovich trick in this case works just as above, that is
\begin{equation}
	\text{exp}\left[-\int d^{4}x\;\epsilon \fini\varphi_{a}^{2}\right]=\int[\mathcal{D}\xi]\,\text{exp}\left[\int d^{4}x\left(-\frac{\xi^{2}(x)}{4\epsilon \fini}+i\varphi_{a}\xi\right)\right].
\end{equation}

\Eq{eq:epsIphi} generates the equations of motion for the propagators  
\begin{align}\label{eq:Minkret}
	\left[(\partial_0 \pm \epsilon)^2 + k^2 + m^2 \right] G^{R/A} (k; t_1, t_2) =  \delta(t_1-t_2)			 
\end{align}
and
\begin{align}
	G^K (k;t_1,t_2) = - i f \epsilon \int \dd t' G^{R} (k;t_1,t') G^{A} (k;t',t_2).
\end{align}
One can benefit from the flat space frequency decomposition to solve \Eq{eq:Minkret} in temporal Fourier space
\begin{align}
	G^{R/A} (k; \omega) &= - \frac{1}{(\omega \mp i \epsilon)^2 - E_k^2}= -\frac{1}{2k} \left[ \frac{1}{\omega- (E_k\pm i\epsilon)} - \frac{1}{\omega- (-E_k\pm i\epsilon)}   \right]
\end{align}
The real-time Green functions are given by
\begin{align}\label{eq:retardedM4}
	G^{R/A} (k; \tau) &= \int \frac{\dd \omega}{2\pi} \ee^{i \omega \tau} G^{R/A} (k; \omega) =\pm \frac{\sin \left[E_k \tau\right]}{E_k}\theta\left(\pm \tau\right)
\end{align}
where $\tau \equiv t_1-t_2 $ and taking the limit $\epsilon \rightarrow 0$. We can then compute the Keldysh function either in Fourier or real space. For instance, in real space, it writes
\begin{align}
	G^K (k;t_1,t_2) &= - i \frac{f}{E_k^2} \epsilon \int \dd t'  \left\{\sin[E_k(t_1-t')] \ee^{\epsilon (t_1-t')} \theta(t'-t_1) \right\}\\
	&\qquad \qquad \qquad \qquad \times \left\{\sin[E_k(t'-t_2)] \ee^{-\epsilon (t'-t_2)} \theta(t'-t_2) \right\} \nonumber 
\end{align}
which leads to 
\begin{align}\label{eq:distribM4}
	G^K (k;\tau)  = i \frac{f(E_k)}{4E_k^2}\cos\left[E_k \tau\right].
\end{align}
As mentioned above, the $f$ factor characterises the occupation of the system's state. The vacuum result is indeed recovered for $f = 2 E_k$ which we assume to be the case below. Note that the $i \epsilon$ prescription in the Schwinger-Keldysh formalism relies on a subtle balance between fluctuation and dissipation controlled by the $\epsilon$ prescription. Indeed, $\epsilon$ appearing in $\widehat{D}_R$ ensures the convergence of the integrals. Taking $\epsilon \rightarrow 0$ makes the field interact for longer and longer in the asymptotic past. Then, the amplitude of the fluctuations controlled by $\widehat{D}_K$ must be rescaled by $\epsilon$ accordingly such that the contributions equilibrate despite longer interactions and a finite result is reached. This provides a mathematical trick to setup initial conditions through weak interactions between the system of interest and a fictitious bath with the desired properties.

If the $\epsilon$ prescription is so crucial to impose initial conditions and recover vacuum expectation values, the reader may wonder why it is so rarely discussed in the non-equilibrium literature. The reason is that whenever the open system experiences a small but finite dissipation $\gamma \dot{\varphi}_r \varphi_a$, initial conditions get erased after a finite time. Asymptotic observables do not depend on the initial occupation of the state and information about it has been lost throughout the dynamics. At a mathematical level, one can see that $\gamma$ induces a pole in the frequency plane that guarantees the convergence of the integral contour used to compute the retarded and advanced Green's function, such that the small $\epsilon$ deformation becomes unnecessary in practice.


\subsection{Interactions and diagrammatics}\label{subsec:flatinter}

Beyond the Gaussian statistics, higher-point functions are computed following the usual perturbative approach. Once the propagators are known, we can derive a new set of Feynman rules from which we construct correlators order by order in perturbation theory. In this section, after reviewing the standard in-in treatment of interactions, we study the structure of the bispectrum (three-point function in Fourier space) in flat space where analytic results are easily obtained. 


\subsubsection{Feynman rules}

\begin{figure}[h]
	\centering
	\includegraphics[width=0.8\textwidth]{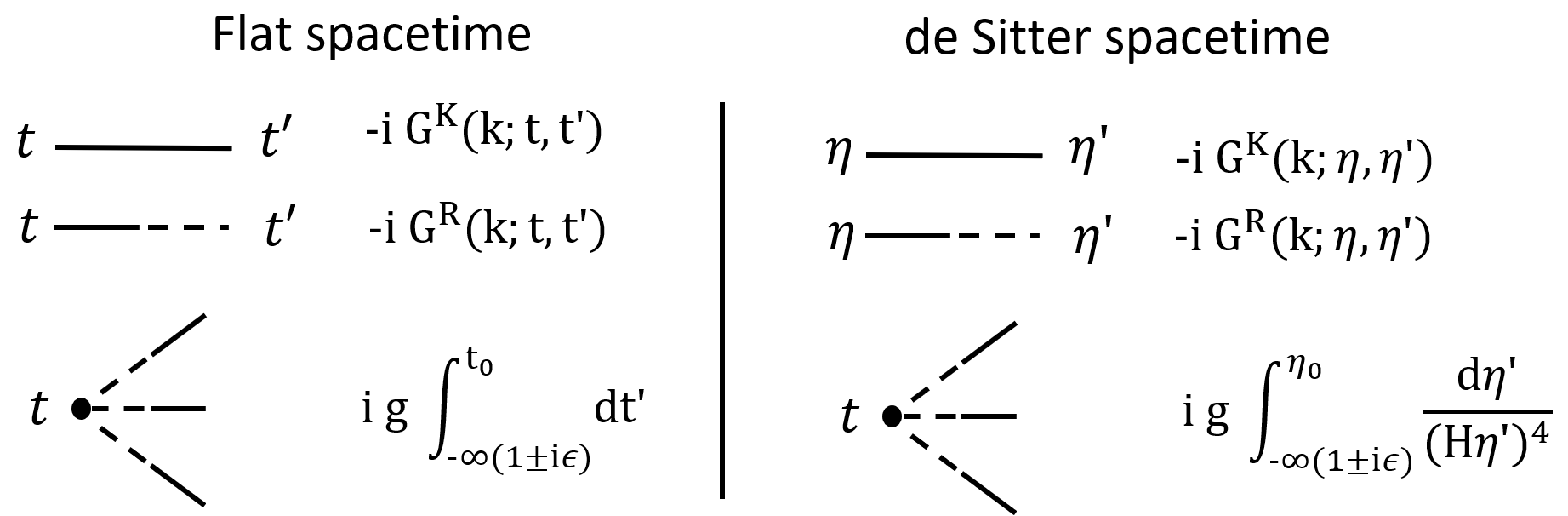}
	\caption{Feynman rules in the Keldysh basis. }
	\label{fig:rules}
\end{figure}

Interactions are treated as in the familiar in-in approach presented in \Sec{subsubsec:diagram}. This provides a comforting unified treatment for the cases of an open and closed system. The only small difference from some references is that we find it convenient to work in the Keldysh basis, $\varphi_{r,a}$ instead of the $\varphi_\pm$ basis. Expectation values of $\widehat{Q}(\eta) \equiv \widehat{\varphi}(\eta, \bmx_1) \cdots \widehat{\varphi}(\eta, \bmx_n)$ are defined through 
\begin{align}\label{eq:expvalQ}
	\langle \widehat{Q} \rangle =  \int \dd \phi \int_{\mathrm{I.C.}}^{\phi} \mathcal{D}\phir \int_{\mathrm{I.C.}}^{0} \mathcal{D}\phia  \left[\phir(\eta, \bmx_1) \cdots \phir(\eta, \bmx_n) \right] \ee^{i S_{\mathrm{eff}}\left[ \phir, \phia\right]}
\end{align}
where initial conditions lie on the boundaries of the path integral. Once the generating functional is known,  $\langle \widehat{Q} \rangle$ is extracted out of functional derivatives\footnote{Note that one needs to vary with respect to the \textit{advanced} source to obtained a \textit{retarded} field insertion. This comes from the NEQ constraint which imposes $\varphi_r J_a + \varphi_a J_r$ \cite{Liu:2018kfw}.}
\begin{align}
	\langle \widehat{Q} \rangle =  \left.\frac{\delta}{\delta J_{a}(\eta, \bmx_{1})} \cdots \frac{\delta}{\delta J_{a}(\eta, \bmx_{n})}\mathcal{Z}[J_{r},J_{a}]\right|_{J_{r,a}=0}
\end{align}
just as we did in \Eq{eq:extract} for the two-point function. The Feynman rules are derived as in \cite{Chen:2017ryl} and lead to \Fig{fig:rules}. Compared to the previous rules, the main differences are the following:
\begin{itemize}
	\item There are two propagators that are $-iG^K(k; \eta, \eta')$ (continuous line) which connects $\phir(k;\eta)$ to $\phir(k;\eta')$ and $-iG^R(k; \eta, \eta')$ (continuous-to-dashed line), which connects $\phir(k;\eta)$ to $\phia(k;\eta')$. Notice that the latter is directional, with the continuous line being attached to the $\phir(k;\eta)$ insertion and the dashed part to the $\phia(k;\eta')$. This leads to certain properties of the Feynman diagrams known as causality flows \cite{Radovskaya:2020lns, Ema:2024hkj}. There is no propagator connecting $\phia(k;\eta)$ to $\phia(k;\eta')$ which is a consequence of the causality structure of the closed time contour \cite{kamenev_2011}.
	\item Then, diagrams evaluation follows the exact same rules as in \cite{Chen:2017ryl}. As seen from \Eq{eq:expvalQ}, external legs connecting to the conformal boundary $\eta_0 \rightarrow 0^-$ are continuous, corresponding to $\phir$ insertions. 
\end{itemize}
An example is given in \Fig{fig:Btree} for a contact bispectrum. One can easily be convinced of these Feynman rules by recovering some known results as we do in \textit{Problem Set} \ref{subsec:pb2}.

\subsubsection{Flat-space dissipative bispectrum}\label{subsec:flatB}

Before turning our attention to primordial cosmology in Lecture \ref{sec:lec3}, it is instructive to discuss the generic structure of the contact three-point functions in Minkowski in the presence of dissipation. We derive these results following the Feynman rules enumerated in \Fig{fig:rules}, the propagators being given in \Eqs{eq:retardedM4dissip} and \eqref{eq:distribM4dissip}. For the sake of clarity, we temporarily set $c_s = 1$. 

\begin{figure}[h]
	\centering
	\includegraphics[width=0.5\textwidth]{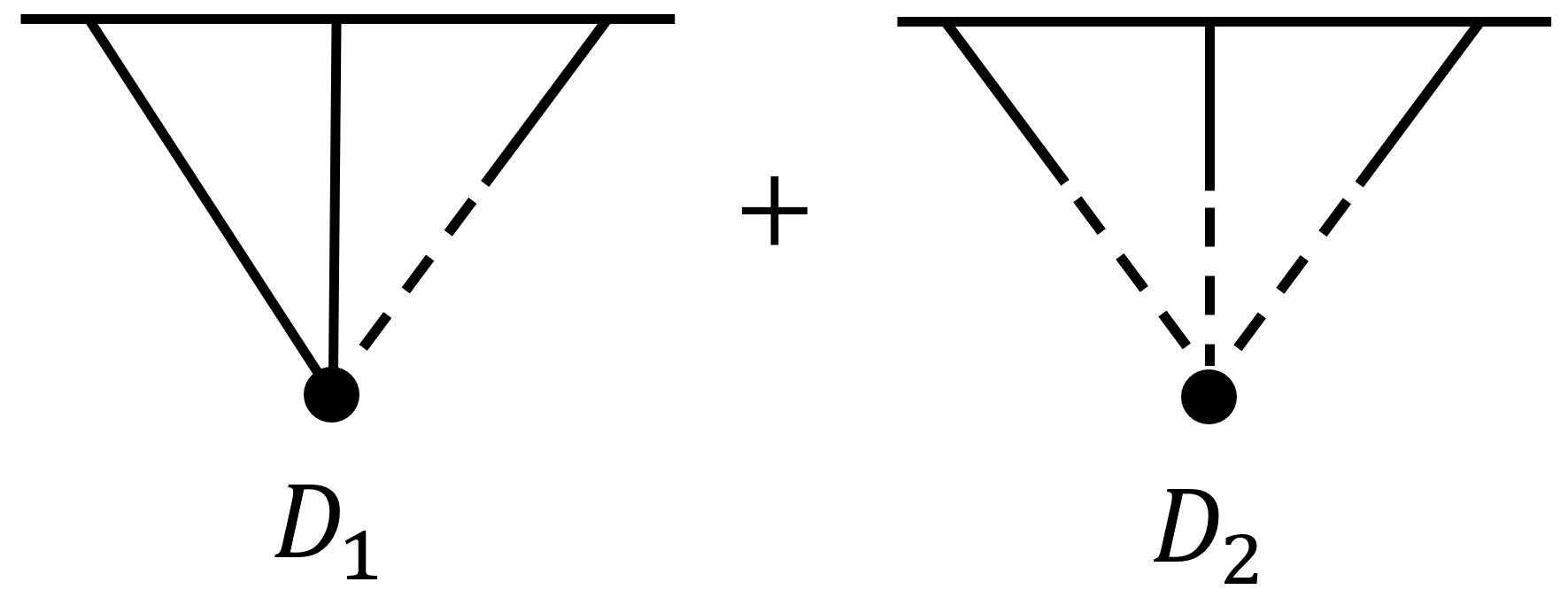}
	\caption{The two diagrams to compute the interaction given in \Eq{eq:cubicdot}. \textit{Left}: diagram corresponding to the vertex $\dot{\varphi}_r^2 \dot{\varphi}_a$, resulting in \Eqs{eq:D1dot}. \textit{Right}: diagram corresponding to the vertex $\dot{\varphi}_a^3$, resulting in \eqref{eq:D2dissip}.}
	\label{fig:Btree}
\end{figure}

\paragraph{$\dot{\varphi}^3$ interactions} Let us first consider 
\begin{align}\label{eq:cubicdot}
	\mathcal{L}_{\mathrm{int}} = -\frac{\alpha}{3!}\left(\dot{\varphi}^3_+ - \dot{\varphi}^3_-\right) = -\frac{\alpha}{2} \left(\dot{\varphi}_r^2 \dot{\varphi}_a + \frac{1}{12} \dot{\varphi}_a^3\right).
\end{align}
where the minus sign in front comes from the Lorentzian signature, assuming $\alpha>0$. We aim at computing the bispectrum 
\begin{align}\label{eq:bispecdef}
	\langle \varphi_{\bmk_1}  \varphi_{\bmk_2}  \varphi_{\bmk_3} \rangle \equiv (2\pi)^3 \delta(\bmk_1 + \bmk_2 + \bmk_3) B_\varphi(k_1,k_2,k_3),
\end{align}
for which we have two contact diagrams to consider presented in \Fig{fig:Btree}, $B_\varphi(k_1,k_2,k_3) = D^{\dot{\varphi}^3}_1 + D^{\dot{\varphi}^3}_2$. 

The first one leads to
\begin{align}\label{eq:D1dot}
	D^{\dot{\varphi}^3}_1 &=  \frac{\alpha}{2} \int_{-\infty(1\pm i \epsilon)}^{t_0} \dd t \left[\partial_t G^K(k_1; t_0,t)\right] \left[\partial_t G^K(k_2; t_0,t)\right] \left[\partial_tG^R(k_3; t_0,t)\right] + 5~\mathrm{perms}. 
\end{align}
Upon injecting \Eqs{eq:retardedM4dissip} and \eqref{eq:distribM4dissip} in \Eq{eq:D1dot} and  summing over the six possible permutations, we observe that $D^{\dot{\varphi}^3}_1 = 0$. The second diagram is made of the $\phia$ components only, and reads 
\begin{align}\label{eq:D2dissip}
	D^{\dot{\varphi}^3}_2 = \frac{\alpha}{24} \int_{-\infty(1\pm i \epsilon)}^{t_0} \dd t \left[\partial_t G^R(k_1; t_0,t) \right] \left[\partial_t G^R(k_2; t_0,t)\right] \left[ \partial_t G^R(k_3; t_0,t)\right] + 5~\mathrm{perms}.
\end{align}
Under the same procedure, this diagram leads to a non-zero contribution to the bispectrum such that
\begin{align}\label{eq:D2resultdissip}
	B_\varphi(k_1,k_2,k_3) &=  \frac{\alpha \gamma}{2} \frac{  \mathrm{Poly}_6\left(e_1^\gamma, e_2^\gamma,e_3^\gamma \right)}{ \mathrm{Sing}_\gamma} ,
\end{align}
where we defined the singularity structure
\begin{align}\label{eq:singdissip}
	\mathrm{Sing}_\gamma =& \left| E_1^{\gamma} + E_2^{\gamma} + E_3^{\gamma} + \frac{3}{2}i \gamma\right|^2 \left| -E_1^{\gamma} + E_2^{\gamma} + E_3^{\gamma} + \frac{3}{2}i \gamma\right|^2 \nonumber\\
	&\times \left| E_1^{\gamma} - E_2^{\gamma} + E_3^{\gamma} + \frac{3}{2}i \gamma\right|^2 \left| E_1^{\gamma} + E_2^{\gamma} - E_3^{\gamma} + \frac{3}{2}i \gamma\right|^2, 
\end{align}
remembering that $E_k^{\gamma} = \sqrt{c_s^2 k^2 - \gamma^2/4}$. This singularity structure captures most of the specificities of the non-unitary dynamics. It emerges from time integrals of the form
\begin{align}\label{eq:M4int}
	\int_{-\infty(1\pm i \epsilon)}^{t_0} \dd t \ee^{ \pm i E_1^{\gamma} (t_0 - t) } \ee^{\pm i E_2^{\gamma} (t_0 - t) } \ee^{\pm i E_3^{\gamma} (t_0 - t) } \ee^{- \frac{3}{2}\gamma(t_0-t)}\,,
\end{align}
which follow from the structure of the propagators given in \Eqs{eq:retardedM4dissip} and \eqref{eq:distribM4dissip}.
Physically, it represents $3 \leftrightarrow 0$ and $2 \leftrightarrow 1$ interactions mediated by the environment. Fluctuations generate folded singularities while dissipation displaces the pole and regularises the divergence. Consequently, the singularity is not located in the physical plane and the bispectrum remains under perturbative control over the whole kinematical space. As we will see below, this singularity structure is generic and does not depend on the details on the interactions. The details of the particular interaction are imprinted into $\mathrm{Poly}_n$, which is a $n^{\mathrm{th}}$-order polynomial of the elementary symmetric polynomials
\begin{align}\label{eq:var}
	e_1^\gamma = E_1^{\gamma} + E_2^{\gamma} + E_3^{\gamma}, \quad e_2^\gamma = E_1^{\gamma} E_2^{\gamma} + E_2^{\gamma}  E_3^{\gamma} + E_1^{\gamma} E_3^{\gamma} \quad  e_3^\gamma = E_1^{\gamma} E_2^{\gamma} E_3^{\gamma}.
\end{align}
For this specific case, 
\begin{align}
	\mathrm{Poly}_6\left(e_1^\gamma, e_2^\gamma,e_3^\gamma \right) &= 243\gamma^6 - 792 \gamma^4 e_2^\gamma + 396 \gamma^4 \left(e_1^\gamma\right)^2 + 576 \gamma^2 \left(e_2^\gamma\right)^2 -1088 \gamma^2 e_2^\gamma\left(e_1^\gamma\right)^2 \nonumber \\
	&+272 \gamma^2 \left(e_1^\gamma\right)^4 + 1024 \gamma^2e_1^\gamma e_3^\gamma + 512 \left(e_1^\gamma\right)^2 \left(e_2^\gamma\right)^2 - 384 \left(e_1^\gamma\right)^4 e_2^\gamma \\
	&-1024 e_1^\gamma e_2^\gamma e_3^\gamma + 64 \left(e_1^\gamma\right)^6 + 512 e_3^\gamma \left(e_1^\gamma\right)^3 + 768 \left(e_3^\gamma\right)^2. \nonumber
\end{align}
Note that in a unitary shift symmetric theory, the operators $\dot{\varphi}^3$ and $(\partial_i \varphi)^2 \dot{\varphi}$ do not generate any contact bispectrum in Minkowski due to the time reversal symmetry $t\rightarrow-t$ and $\varphi \rightarrow -\varphi$ (one can check explicitly that there is zero contribution to the bispectrum, each diagram vanishing independently). Dissipation spontaneously breaks this symmetry and we observe that the diagrams now lead to a non-zero contribution to the bispectrum. 

\begin{figure}[tbp]
	\centering
	\includegraphics[width=0.7\textwidth]{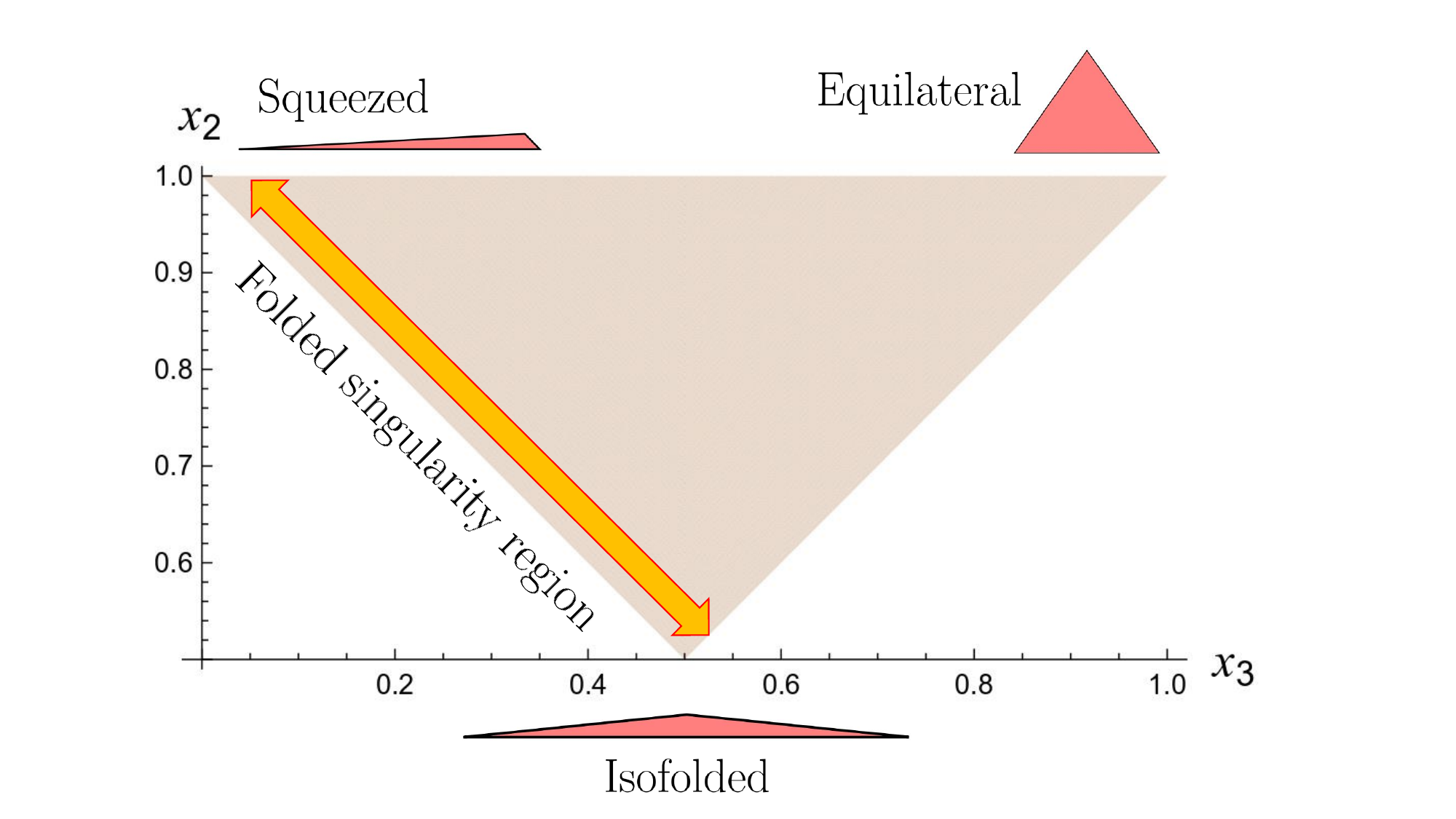}
	\caption{The shapes of accessible triangles fulfilling the momentum conservation $\delta(\bmk_1 + \bmk_2 + \bmk_3)$. The triangles are parameterised along $x_2 \equiv k_2/k_1$ and $x_3 \equiv k_3/k_1$ such that $x_3 < x_2 < 1$ and $x_2 + x_3 >1$, the two conditions to construct closed triangles. A region of particular interest for the scope of this article is the folded region where $x_2 + x_3 \simeq 1$, which interpolates between the squeezed and isofolded points. The singularity structure discussed in \Eq{eq:singdissip} peaks close to the folded region, and provides a smoking gun of dissipative dynamics.}
	\label{fig:triangles}
\end{figure} 

In \cite{Salcedo:2024smn}, we found that other cubic interactions follow the same structure. The generic structure is:
\begin{tcolorbox}[%
	enhanced, 
	breakable,
	skin first=enhanced,
	skin middle=enhanced,
	skin last=enhanced,
	before upper={\parindent15pt},
	]{}
	
	\paragraph{Dissipative bispectrum in Minkowski}
	\begin{align}\label{eq:gen}
		B_\varphi(k_1,k_2,k_3) =f(\mathrm{EFT}) \frac{  \mathrm{Poly}_n\left(e_1^\gamma, e_2^\gamma,e_3^\gamma \right)}{ \mathrm{Sing}_\gamma} 
	\end{align}
	where $f(\mathrm{EFT})$ a rational function of the EFT coefficients (and possibly the kinematics for spatial derivative interactions), $\mathrm{Poly}_n$ are polynomials of the variables given in \Eq{eq:var} and $\mathrm{Sing}_\gamma$ is the singularity structure expressed in \Eq{eq:singdissip}.
	
\end{tcolorbox}
The simplicity of the structure, which originates from integrals of the form of \Eq{eq:M4int}, might suggest the future development of bootstrap techniques for this kind of local dissipative dynamics. It also suggests the physics is well captured from the interpretation of $\mathrm{Sing}_\gamma$ controlling the amplitude of $3 \leftrightarrow 0$ and $2 \leftrightarrow 1$ interactions mediated by the environment.

\begin{figure}[tbp]
	\centering 
	\makebox[\textwidth][c]{\includegraphics[width=.5\textwidth]{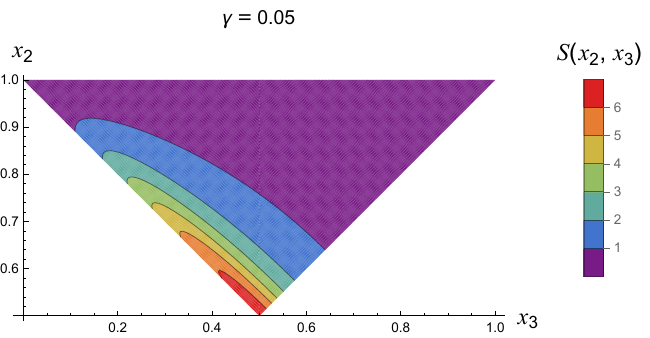}
		\includegraphics[width=.5\textwidth]{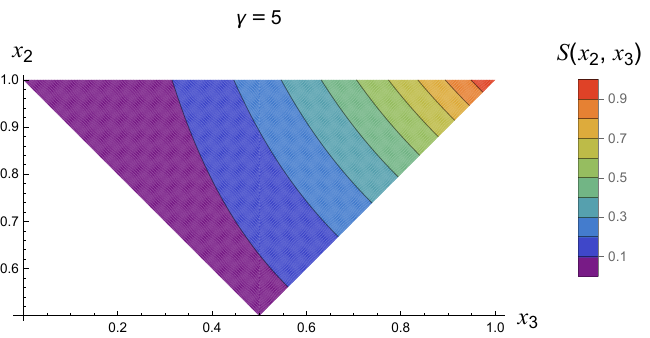}}
	\caption{\label{fig:shapeMink} Shape function of the contact bispectrum generated by $\dot{\varphi}_a^3$ in Minkowski given in \Eq{eq:D2resultdissip}. \textit{Left:} In the small dissipation regime, the singularity structure $\mathrm{Sing}_\gamma$ given in \Eq{eq:singdissip} becomes small in the folded region $x_2 + x_3 \simeq 1$ which enhances the bispectrum. \textit{Right:} In large dissipation regime, the imaginary contributions from the dissipation in dominates $\mathrm{Sing}_\gamma$ such that no bispectrum enhancement is observed in the $x_2 + x_3 \simeq 1$ folded region.}
\end{figure}

\paragraph{Shape function and non-Gaussianities phenomenology.} It is instructive to consider the \textit{shape} of the bispectrum, an object that informs the kinematic configuration that maximizes the signal. The momentum conservation in $\delta(\bmk_1 + \bmk_2 + \bmk_3)$ in \Eq{eq:bispecdef} forces the momenta to form a close triangle. As different inflationary models predict maximal signals in different triangular configurations (see e.g. \cite{Babich:2004gb, Chen:2006nt}), the shape function
\begin{align}\label{eq:shaperef}
	S(x_2,x_3) \equiv (x_2 x_3)^2 \frac{B(k_1, x_2 k_1, x_3 k_1)}{B(k_1,k_1,k_1)}
\end{align}
is an informative probe of the mechanism generating primordial non-Gaussianities. The variables $x_2 \equiv k_2/k_1$ and $x_3 \equiv k_3/k_1$ control the shape of the triangles and are restricted by $\delta(\bmk_1 + \bmk_2 + \bmk_3)$ to the region $\max(x_3, 1- x_3) \leq x_2 \leq 1$. In \Fig{fig:triangles}, we present the main shapes of interest. It appears that the singularity structure $\mathrm{Sing}_\gamma$ presented in \Eq{eq:singdissip} exhibits two different behaviour depending the magnitude of the dissipation coefficient $\gamma$. In the strong dissipation regime (\textit{right} panel of \Fig{fig:shapeMink}), the $ \frac{3}{2}i \gamma$ appearing in \Eq{eq:singdissip} always dominates the bispectrum contribution such that the signal peaks in the equilateral shape where $x_2 \simeq x_3 \simeq 1$. On the contrary, in the small dissipation regime, $\mathrm{Sing}_\gamma$ can become small in the folded region where $x_2 + x_3 \simeq 1$ such that the signal predominantly peaks near the isofolded configuration where $x_2 \simeq x_3 \simeq 1/2$ (\textit{left} panel of \Fig{fig:shapeMink}).

\begin{figure}[tbp]
	\begin{minipage}{6in}
		\centering
		\raisebox{-0.5\height}{\includegraphics[width=.45\textwidth]{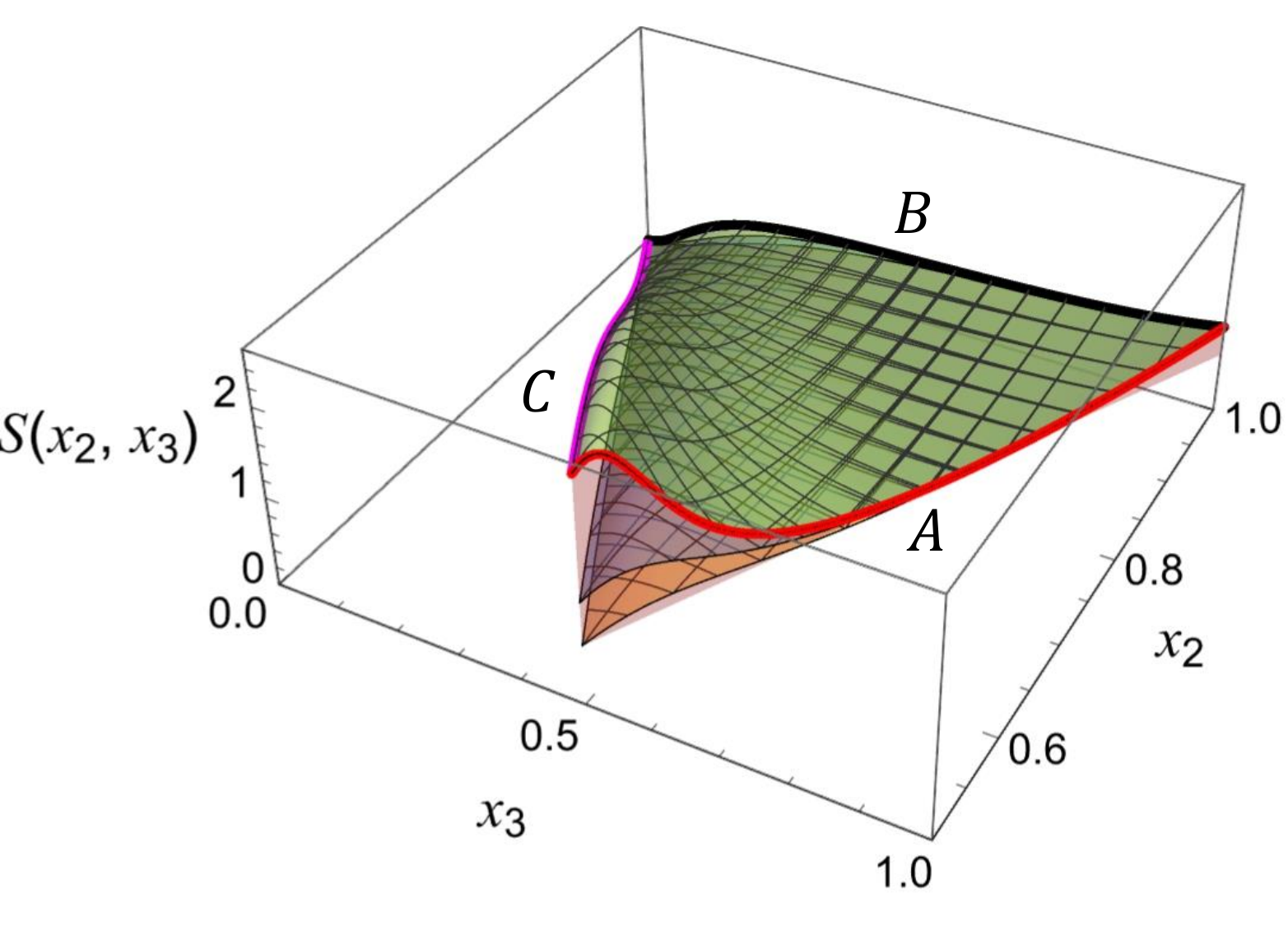}}
		\raisebox{-0.5\height}{\includegraphics[width=.45\textwidth]{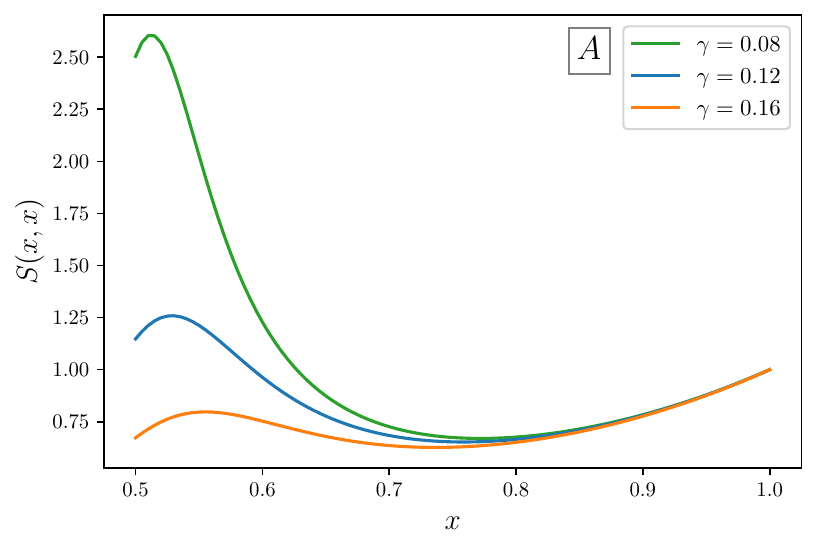}}
		\centering
		\raisebox{-0.5\height}{\includegraphics[width=.45\textwidth]{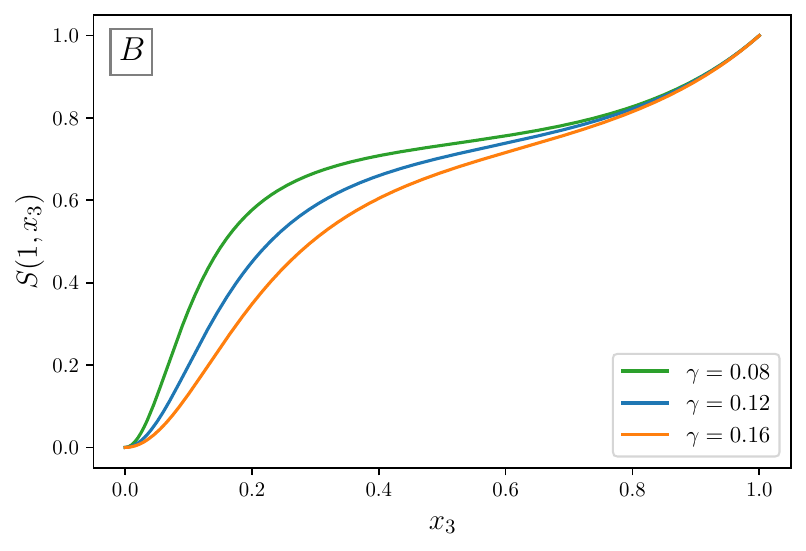}}
		\raisebox{-0.5\height}{\includegraphics[width=.45\textwidth]{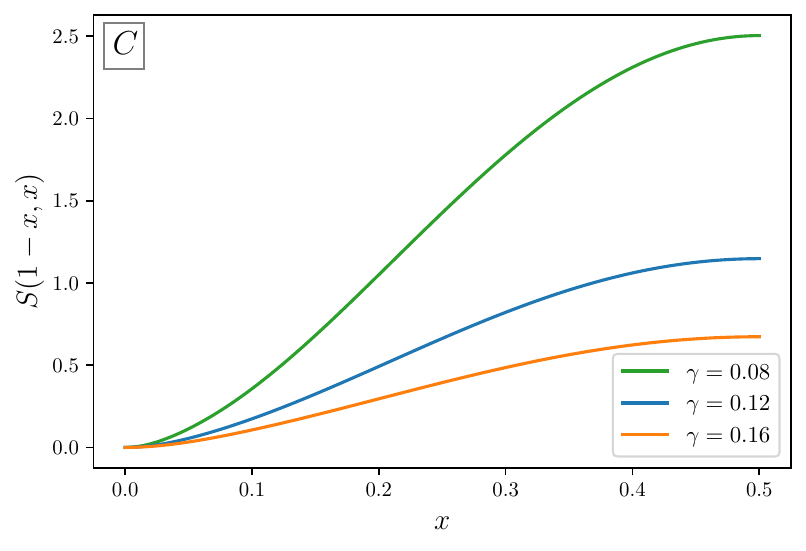}}
	\end{minipage}
	\caption{\label{fig:foldedMink} \textit{Top left:} $3\mathrm{d}$ shape function of the contact bispectrum generated by $\dot{\varphi}_a^3$ in Minkowski given in \Eq{eq:D2resultdissip} for three different values of the dissipation parameter $\gamma \in [0.08; 0.12; 0.16]$. We observe the equilateral-to-folded transition of the shape function as the dissipation parameter decreases. \textit{Top right:} $2\mathrm{d}$ cut along the direction $x_2 = x_3 = x$ appearing in red in the $3\mathrm{d}$ plot. The singularity is resolved such that the bispectrum remains well defined for any triangular configuration and any value of the dissipation parameter $\gamma$. \textit{Bottom left:} $2\mathrm{d}$ cut along the direction $x_2 = 1$ appearing in black in the $3\mathrm{d}$ plot. Consistency relations ensure the signal vanishes in the squeezed limit $x_3 \ll 1$. \textit{Bottom right:} $2\mathrm{d}$ cut along the direction $x_2 = 1 - x_3$ appearing in purple in the $3\mathrm{d}$ plot. Consistency relations are again observed in the squeezed limit $x_3 \ll 1$.}
\end{figure}

This type of singularities have already been encountered in cosmology, mostly in the context of non-Bunch-Davies initial states \cite{Holman:2007na, Chen:2006nt, Meerburg:2009ys, Agullo:2010ws, Ashoorioon:2010xg, Agarwal:2012mq, Ashoorioon:2013eia,  Albrecht:2014aga, Green:2020whw}. The main difference with the current investigation is that, due to the presence of the dissipative environment, the would-be folded singularity is regularised, \ie $\mathrm{Sing}_\gamma \neq 0$ whenever $\gamma \neq 0$. This clearly appears in \Fig{fig:foldedMink} where the peak of the shape function as one approaches the folded singularity $x_2 + x_3 \simeq 1$ is plotted on the \textit{top-right} panel. The resolution of the singularity is a useful feature of the formalism as it allows one to keep perturbative control over all configurations. In particular, one does not have to introduce an artificial cutoff to handle the dissipative interactions. 

\clearpage

\subsection{\textit{Problem set}}\label{subsec:pb2}

\paragraph{\textit{Exercise 1.}} \textit{A condensed matter approach to the $i \epsilon$ prescription} \\

Consider a massless scalar field $\varphi$ with free functional in the Keldysh basis given by 
\begin{equation}\label{eq:freeieps}
	S_{\mathrm{eff}} =\int \dd^4x \left(\dot{\varphi}_r \dot{\varphi}_a  -\partial^{i}\varphi_{r}\partial_{i}\varphi_{a} -2\epsilon\dot{\varphi}_{r} \varphi_{a} -\epsilon^{2}\varphi_{r}\varphi_{a}+i\epsilon \fini\varphi_{a}^{2} \right)\,,
\end{equation}
with $\epsilon$ a small parameter and $f$ a function to be determined.
\begin{enumerate}
	\item Interpret the various terms appearing in $S_{\mathrm{eff}}$.
	\item We first find the retarded Green's function.
	\begin{enumerate}
		\item What is the equation of motion obeyed by the retarded Green's function $G^{R} (k; t_1, t_2)$?
		\item Solve it it frequency space. Draw the poles of $G^{R} (k; \omega)$ in the complex plane. What is the role of $\epsilon$?
		\item Give the expression of $G^{R} (k; t_1, t_2)$ in real space. Take the limit $\epsilon \rightarrow 0$ and comment. 
	\end{enumerate}
	\item We now find the Keldysh propagator.
	\begin{enumerate}
		\item Express the Keldysh propagator in terms of the retarded Green's function. 
		\item Give the expression of $G^{K} (k; t_1, t_2)$ in real space. What is the role of $\epsilon$? Take the limit $\epsilon \rightarrow 0$ and comment.
		\item Which value $f$ should take to recover the vacuum power spectrum of a massless scalar, $1/2k$? Can you guess which physical parameter is controlled by $f$?
	\end{enumerate} 
\end{enumerate}
You have recovered the vacuum propagators. The $i \epsilon$ prescription is a way to implement the system's initial conditions in a vacuum or a thermal state. More details can be found in \Sec{subsubsec:iepsilon}.

\begin{center}
	\noindent\rule{8cm}{0.4pt}
\end{center}

\paragraph{\textit{Exercise 2.}} \textit{Diagrammatics in the Keldysh basis} \\

Consider the free theory specified in \Eq{eq:freeieps}, completed by the unitary interaction 
\begin{align}
	S_{\mathrm{eff}} = - \frac{\lambda}{3!} \int \dd^4x \left( \varphi_+^3 - \varphi_-^3\right).
\end{align}
\begin{enumerate}
	\item Write down the interaction in the Keldysh basis and comment the structure of the operators. 
	\item Draw the two diagrams contributing to the contact bispectrum $B_3(k_1,k_2,k_3)$ at order $\lambda$. Compute these diagrams separately and comment their singularity structure. 
	\item Combine these results and recover the expression of $B_3(k_1,k_2,k_3)$ found in \textit{Exercise 1} of \Sec{subsec:prob1}.
\end{enumerate}

\section{Lecture 3: Open inflation}\label{sec:lec3}

Despite the variety of microphysical models compatible with the current cosmological observations, the EFT of Inflation (EFToI) \cite{Cheung:2007st} and its modern avatar through the cosmological bootstrap \cite{Baumann:2022jpr} have extracted a vast number of model-independent results. Solely based on symmetries (scale invariance, homogeneity and isotropy), principles (unitarity, locality) and scale hierarchies (slow roll expansion), cosmological correlators can still be constrained to a large extent. In this Section, after reviewing the original construction of the EFToI in \Sec{subsec:EFToI}, we develop its extension to capture local dissipation and noise \Sec{subsec:open}. As we will see in \Sec{subsec:pheno}, model-independent phenomenological implications can still be derived in this case, some of which constitute a set of smoking-gun signatures for dissipative and stochastic dynamics.


\subsection{The Effective Field Theory of Inflation}\label{subsec:EFToI}

We first briefly review the original construction presented in \cite{Cheung:2007st, Gubitosi:2012hu, Piazza:2013coa}. We consider a single scalar field $\phi(t,\bfx)$ evolving unitarily in a perturbed Friedmann–Lema\^itre–Robertson–Walker (FLRW) geometry. Both the scalar field and the metric are expanded around their background value 
\begin{align}
	\phi(t,\bfx) = \bar \phi(t) + \delta \phi(t, \bfx), \qquad\qquad g_{\mu\nu}(t,\bfx) = \bar{g}_{\mu\nu}(t) + \delta g_{\mu \nu}(t,\bfx),
\end{align}
and we aim at understanding the dynamics of the fluctuations. The theory is conveniently constructed in the unitary gauge, which is defined by the condition\footnote{In this subsection, we use $\doteq$ to denote an equality that holds in unitary gauge.} $\delta \phi \doteq 0$. In this case, all the perturbations are absorbed into the metric and the homogeneous scalar field $\phi(t, \bfx) \doteq \bar{\phi}(t)$ can be used as a clock. Indeed, one can parametrize the time slicing in terms of the homogeneous value of the scalar field $t = t(\phi(t,\bfx))$ and construct geometrical objects based on this time foliation of spacetime. For instance, the unit vector perpendicular to the foliation is
\begin{align}\label{eq:n_mu_def}
	n_\mu \equiv - \frac{\partial_\mu \phi}{\sqrt{-g^{\mu\nu} \partial_\mu \phi \partial_\nu \phi}} \doteq - \frac{\delta^0_{~\mu}}{\sqrt{-g^{00}}},
\end{align}
where the second equality holds in the unitary gauge\footnote{A related useful expression is $n^\mu=-g^{\mu 0}/\sqrt{-g^{00}}$ and $\sqrt{-g^{00}}\doteq n^0 \doteq -1/n_0$. Note that $n^\mu$ is a future-pointing time-like vector.}.

\paragraph{Effective action.} This formulation allows us to write down the most general (unitary) EFT compatible with the symmetries of the problem \cite{Cheung:2007st}. The presence of the inflaton background $\bar \phi(t)$ spontaneously breaks time-translation symmetry, such that the resulting action is made of terms that are invariant under spatial diffeomorphisms only. Of course allowed terms include $4d$ covariant operators such as $R$, which are \textit{a fortiori} $3d$ covariant. But one should also allow time-dependent functions ($\Lambda(t),~\cdots$), contractions with $n_\mu$ such as $g^{00}$, or geometrical objects constructed out of the foliation such as the extrinsic curvature 
\begin{align}\label{eq:extrinsiccurv}
	K_{\mu\nu} \equiv \left(\delta_{\mu}^{\sigma} + n_\mu n^\sigma \right) \nabla_\sigma n_\nu \doteq \delta_\mu^i \delta_\nu^j \Gamma_{ij}^0 (-g^{00})^{-1/2}\,.
\end{align}
The most generic action takes the form \cite{Cheung:2007st}
\begin{align}
	S = \int \dd^4 x \sqrt{-g} F(R_{\mu\nu\rho\sigma}, g^{00}, K_{\mu\nu}, \nabla_\mu; t),
\end{align}
where $F$ is an arbitrary function. Expanding the metric in powers of the perturbations we obtain \cite{Gubitosi:2012hu}
\begin{align}\label{eq:universal}
	S = \int \dd^4 x \sqrt{-g} \left[ \frac{M^2_{\mathrm{Pl}}}{2} R - \Lambda(t) - c(t) g^{00}\right] + S^{(2)},
\end{align}
where $R$ is the Ricci scalar and $\Lambda$ and $c$ are functions of time, and $S^{(2)}$ starts at second order in perturbations, and consequently does not affect the background Friedmann equations 
\begin{align}\label{eq:backgroundeq}
	3\Mpl^2 H^2 = \Lambda(t) + c(t), \qquad 2\Mpl^2 \dot{H} = -2c(t),
\end{align}
together with the continuity equation 
\begin{align}\label{eq:contiEFToI}
	\dot{\Lambda}(t) + \dot{c}(t) + 6Hc(t) = 0.
\end{align}
The first three terms in \Eq{eq:universal} constitute the \textit{universal part} of the EFT of Inflation. 
Details about $S^{(2)}$ can be found in \eg \cite{Creminelli:2017sry}. At lowest order in derivatives acting on the metric, it takes the form
\begin{align}
	S^{(2)} =  \int \dd^4 x \sqrt{-g} \left[ \sum_{n=2}^\infty M^4_n(t) (1 + g^{00})^n\right].
\end{align}

\paragraph{St\"uckelberg trick}

The space-diff invariant unitary-gauge action is the starting point to derive a fully diff-invariant theory for the Goldstone boson $\pi$ of time translations via the St\"uckelberg trick \cite{Cheung:2007st}. This degree of freedom can be made manifest by performing the time diffeomorphisms 
\begin{align}
	t \rightarrow t + \pi(t,\bfx), \qquad \bfx \rightarrow \bfx.
\end{align}
Under this transformation, objects that are not $4d$ diff invariant do transform, such as time-dependent constants
\begin{align}
	c(t) \rightarrow c(t + \pi) =  c(t) + \dot{c}(t) \pi + \mathcal{O}(\pi^2)
\end{align}
or the time component of the metric 
\begin{align}
	g^{00} &\rightarrow  \frac{\partial ( t + \pi) }{\partial x^\alpha} \frac{\partial ( t + \pi) }{\partial x^\beta} g^{\alpha \beta} =  g^{00} + 2 g^{0\mu} \partial_\mu \pi + g^{\mu \nu} \partial_\mu \pi \partial_\nu \pi, \\
	g^{0i} &\rightarrow  \frac{\partial ( t + \pi) }{\partial x^\alpha} g^{\alpha i} = g^{0i} +  g^{i\nu} \partial_\nu \pi.
\end{align}
Under this transformation, the effective action 
\begin{align}\label{eq:fullaction}
	 S = \int \dd^4 x \sqrt{-g} \bigg[ &\frac{M^2_{\mathrm{Pl}}}{2} R - \Lambda(t+\pi) - c(t+\pi) \left(g^{00} + 2 g^{0\mu} \partial_\mu \pi + g^{\mu \nu} \partial_\mu \pi \partial_\nu \pi \right) \nonumber \\
	  & \quad + \sum_{n=2}^\infty M^4_n(t + \pi) (1 + g^{00} + 2 g^{0\mu} \partial_\mu \pi + g^{\mu \nu} \partial_\mu \pi \partial_\nu \pi)^n\bigg],
\end{align}
is manifestly $4d$ diff invariant as long as $\pi$ non-linearly realises time-translations, that is when $t\rightarrow t - \epsilon$, $\pi$ transforms as
\begin{align}
	\pi(t) \rightarrow \pi'(t) = \pi(t+\epsilon)+\epsilon.
\end{align}

\paragraph{Decoupling.} 
\textit{Foreword.} This paragraph provides partial solutions to Problem 1 of the problem set in Section~\ref{subsec:prob3}. Readers who wish to practice on problems related to the EFT of Inflation are encouraged to attempt the problem set first, before consulting this paragraph.
\\

The main interest of reintroducing the dynamical scalar $\pi$ is that in a certain limit known as decoupling, one can safely neglect the fluctuations of the metric. It is easier to see this in the flat gauge where 
\begin{equation}
	\delta g_{ij}^{\rm scalar}=0\quad,\quad\delta g_{00}=-2\phi=-\delta g^{00}\quad,\quad \delta g_{0i}=a(t)\partial_{i}F=\delta g^{0i}.
\end{equation}
Let's consider the universal part and the minimal extension controlled by $M_2^4(t)$. It leads to the Einstein equations
\begin{align}
	\frac{\Mpl^{2}}{2}\delta G_{00}&+\left[\frac{\Lambda(t)}{2} - M_2^4(t)\right]  \delta g_{00} - \left[c(t) + 2M_2^4(t) \right]\dot{\pi}+3Hc(t)\pi=0, \label{eq:E1v2}\\
	\frac{\Mpl^{2}}{2}\delta G_{0i}&+\left[\frac{\Lambda(t)}{2}-\frac{c(t)}{2}\right]\delta g_{0i}-2c(t)\partial_{i}\pi=0, \label{eq:E2v2}
\end{align}
where $\delta G_{\mu\nu}$ is the perturbed Einstein tensor.
In this case, solving for the constraints, one finds
\begin{align}\label{eq:constraintsol}
	\phi = 2 \epsilon H \pi, \qquad \nabla^2 F = - \frac{a \epsilon H}{c_s^2} \left[\dot{\pi} + \left(3 c_s^2- 2\epsilon \right) H \pi \right], 
\end{align}
where we introduced the first slow-roll parameter $\epsilon \equiv -\dot{H}/H^2$ and the speed of sound $c_s^{-2} \equiv [c(t) + 2 M_2^4(t)]/c(t)$.

One can then perturb \Eq{eq:fullaction} at second order in perturbations, substitute $\phi$ and $\nabla^2F$ found in \Eq{eq:constraintsol} and compare the contributions from $\phi$ and $F$ to the self dynamics of $\pi$ itself. For instance, the kinetic term of the $\pi$ roughly scales as 
\begin{align}
 (\mathrm{I}): \qquad	c(t) (\partial_\mu \pi)^2 \quad \sim \quad\epsilon \Mpl^2 H^2 \dot{\pi}^2 \quad \sim \quad  \epsilon \Mpl^2 H^4 \pi^2
\end{align}
where we used the background expression of $c(t)$ found in \Eq{eq:backgroundeq} and the heuristic estimate of the time derivatives during inflation $\dot{\pi}\sim H\pi$ \cite{Salcedo:2024smn}. Comparing this expression with the leading mixing with gravity \cite{Cheung:2007st}
\begin{align}
(\mathrm{II}): \qquad	c(t) \delta g^{00} \dot{\pi} \quad \sim \quad  \epsilon \Mpl^2 H^2 \phi \dot{\pi} \quad \sim \quad \epsilon^2 \Mpl^2 H^4 \pi^2,
\end{align}
where in the last line we used \Eq{eq:constraintsol}, we find that $(\mathrm{II}) \ll (\mathrm{I})$ as long as $\epsilon \ll 1$. 
This explicit computation shows that the contribution of the fluctuations of the metric on the scalar dynamics of $\pi$ is suppressed by the slow-roll parameter $\epsilon \ll 1$. A more rigorous derivation of the decoupling limit may be found in \cite{Salcedo:2025ezu}, where we perform an analogue computation in the Keldysh basis, extending the results of \cite{Cheung:2007st} to include dissipation and noise. 

In this limit, the action dramatically simplifies to 
\begin{align}
	S_\pi = \int \dd^4 x \sqrt{-g} \left\{\frac{1}{2} \Mpl^2 R + \epsilon \Mpl^2 H^2 \left[ \dot{\pi}^2 - \frac{(\partial_i \pi)^2}{a^2} \right] + 2 M_2^4 \left[ \dot{\pi}^2 + \dot{\pi}^3 - \dot{\pi}\frac{(\partial_i \pi)^2}{a^2} \right] - \frac{4}{3} M_3^4 \dot{\pi}^3 + \cdots \right\}, 
\end{align}
where the ellipsis represents terms at least quartic in $\pi$. From a theory with two tensor modes and one scalar degree of freedom, we end up with a theory with a single shift symmetric scalar \cite{Finelli:2018upr}. This framework is the starting point of the study of \cite{Salcedo:2024smn} which aims at extending this framework to include dissition and noise.


\subsection{Opening the theory}\label{subsec:open}

The starting point to study an open system is a choice of the degrees of freedom that we want to describe. Given our focus on single-clock inflation, our system will consists of a scalar field $\pi$, the St\"uckelberg field introduced above. We aim to construct an EFT for $\pi$ in the presence of dissipative and stochastic effects induced by interactions with the unknown environment. The statistics of the system is characterized by its density matrix in the field basis
\begin{align}
	\rho_{\mathrm{red}}\left[\pi, \pi';\eta_0\right] = \int_{\Omega}^\pi \mathcal{D}\pi_+ \int_{\Omega}^{\pi'} \mathcal{D}\pi_- \ee^{i S_{\mathrm{eff}}\left[ \pi_+, \pi_-\right]}.\label{eq:densitymatrix}
\end{align}
Here $\Omega$ represents the choice of initial state, we assume to be the Bunch-Davies state. The \textit{open effective functional} $S_{\mathrm{eff}}\left[ \pi_+, \pi_-\right]$ admits a Hermitian and a non-Hermitian part
\begin{align}\label{eq:Seffref}
	S_{\mathrm{eff}}\left[ \pi_+, \pi_-\right] = S_\pi\left[\pi_+\right] - S_\pi\left[\pi_-\right] + S_{\mathrm{IF}}\left[\pi_+,\pi_- \right].
\end{align}
Following \cite{Salcedo:2024smn}, our goal here is to construct and study $S_{\mathrm{eff}}\left[ \pi_+, \pi_-\right]$ for single-clock inflation. 


\subsubsection{Symmetries and locality}

\paragraph{Time-translation symmetry breaking.}

Symmetries further restrict the number and structure of EFT operators. Let's consider a schematic microscopic theory $S_{\mathrm{UV}}[\pi,\chi]$ with $\chi$ a collective variable capturing the presence of an environment. Following the EFToI construction presented above \cite{Cheung:2007st, Finelli:2018upr}, $S_{\mathrm{UV}}[\pi,\chi]$ is invariant under shift symmetry of the $\pi$ field. Then, the Schwinger-Keldysh action $S_{\mathrm{UV}}[\pi_+,\chi_+] - S_{\mathrm{UV}}[\pi_-,\chi_-] $ for the closed system is invariant under $\mathrm{shift}_+ \times \mathrm{shift}_- \equiv \mathrm{shift}_r \times \mathrm{shift}_a $. Integrating out the $\chi$ field generates terms mixing the branches of the path integral. In general, these terms do not transform nicely under transformation along each branch of the path integral. This leads to a symmetry breaking pattern under which $S_{\mathrm{eff}} [\pi_+,\pi_-]$ becomes only invariant under the smaller diagonal subgroup, 
\begin{align}\label{eq:SSBEFToI}
\mathrm{shift}_+ \times \mathrm{shift}_-  \quad \equiv  \quad \mathrm{shift}_r \times \mathrm{shift}_a \quad	\rightarrow  \quad \mathrm{shift}_r.
\end{align}

\begin{figure}[tbp]
	\centering
	\includegraphics[width=1\textwidth]{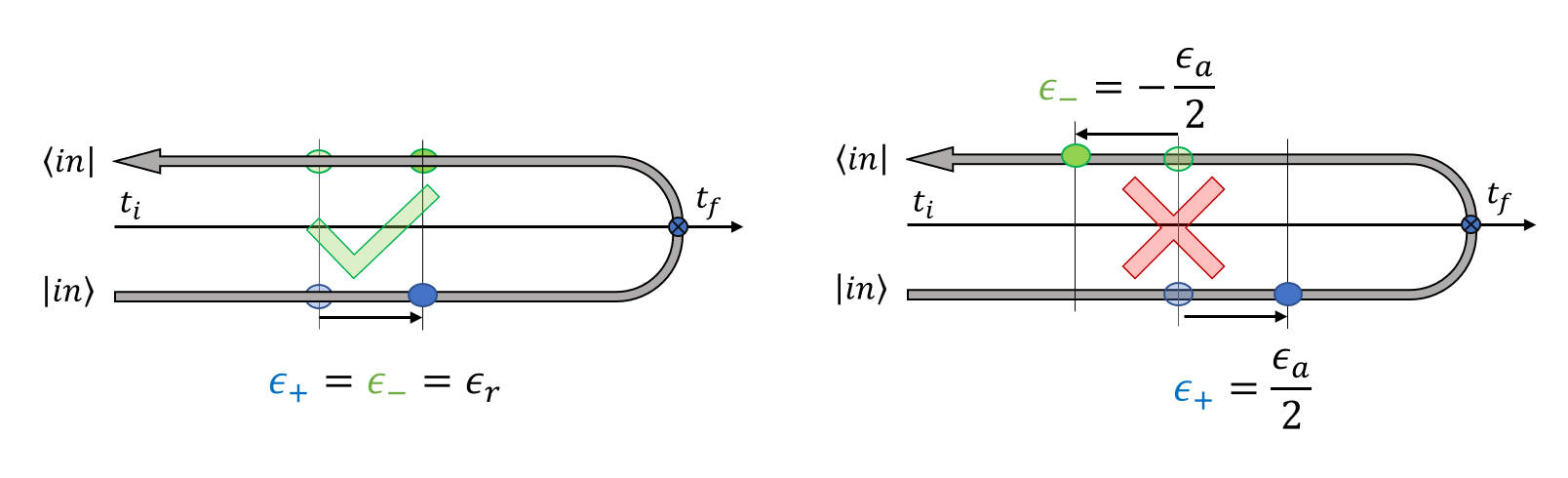}
	\caption{The $\epsilon_{r}$ and $\epsilon_{a}$ transformations on the closed time path, where time is running from left to right in both contours and the arrow represent path ordering (time ordering in $\ket{in}$ and anti-time-ordering in $\bra{in}$). The $\epsilon_{r}$ transformation translates the $``+"$ and $``-"$ variables in the same direction while $\epsilon_{a}$ transformation does it in the opposite directions. While the $\epsilon_{r}$ transformation is preserved, the $\epsilon_{a}$ one is explicitly broken due to dissipative effects \cite{Hongo:2018ant}.}
	\label{fig:sym}
\end{figure}

Explicitly, let us consider a UV-action that is invariant under independent time-translations, one transforming the $+$ branch of the path integral by $t - \epsilon_+$ and the other transforming the $-$ branch of the path integral by $t - \epsilon_-$. Under these transformation, $\pi_\pm$ \textit{non-linearly realise} the symmetry, 
\begin{align}\label{eq:SSB}
	\pi_+(t) \rightarrow \pi'_+(t) = \pi_+(t+\epsilon_+)+\epsilon_+,\quad\quad\pi_-(t) \rightarrow \pi'_-(t) = \pi_-(t+\epsilon_-)+\epsilon_-,  
\end{align}
that is they shift, while additional environment fields act linearly on the transformations. Upon tracing over the environment, the open effective functional $S_{\mathrm{eff}}\left[\pi_+,\pi_-\right]$ is not in general invariant under general $\epsilon_\pm$ translations. More precisely, the effective functional remains invariant under the translation (see the left-hand panel of \Fig{fig:sym}) \cite{Hongo:2018ant}
\begin{align}
	\epsilon_+ = \epsilon_- =\epsilon_{r}\,,
\end{align}
while the translations
\begin{align}
	\epsilon_+ =  - \epsilon_- =\frac{\epsilon_{a}}{2}
\end{align}
are explicitly broken (see the right-hand panel of \Fig{fig:sym}). In this way, out of two time-translational symmetries of the microscopic action, we are left with a single diagonal subgroup $\epsilon_+ = \epsilon_-$.

Let us consider the transformation of $\pi_r$ and $\pi_a$ under the diagonal subgroup of time translations and boosts. In the Keldysh basis, the $\epsilon_{r}$-transformations read \cite{Hongo:2018ant} (see left-hand panel of \Fig{fig:sym})\footnote{One can check that $S_{\mathrm{eff}}[\pi_r,\pi_a]$ is left invariant by the transformation given in \Eq{eq:translRsym}. To see it explicitly, one can consider the expansion
	\begin{align}\label{eq:translRsymexp}
		\pi_{r}(t,\bmx) &\rightarrow \pi_{r}(t,\bmx) + \epsilon_r\left[1+\dot{\pi}_{r}(t,\bmx)\right] + \mathcal{O}\left[(\epsilon_r)^2\right],\\
		\pi_{a}(t,\bmx) &\rightarrow  \pi_{a}(t,\bmx) + \epsilon_r\dot{\pi}_{a}(t,\bmx) + \mathcal{O}\left[(\epsilon_r)^2\right].
\end{align} } 
\begin{align}\label{eq:translRsym} 
	\pi_{r}(t,\bmx) \rightarrow \pi'_{r}(t,\bmx) &= \pi_{r}(t+\epsilon_{r},\bmx) + \epsilon_{r},\\
	\pi_{a}(t,\bmx) \rightarrow \pi'_{a}(t,\bmx) &= \pi_{a}(t+\epsilon_{r},\bmx),
\end{align} 
whereas the $\Lambda_{r}$-transformations follow 
\begin{align}\label{eq:LorentzRsym}
	\pi_{r}(t,\bmx) \rightarrow \pi'_{r}(t,\bmx) &= \pi_{r}\left(\Lambda^0_{r~\mu}x^\mu,\Lambda^i_{r~\mu}x^\mu \right)+\Lambda^0_{r~\mu}x^\mu - t, \\
	\pi_{a}(t,\bmx) \rightarrow \pi'_{a}(t,\bmx) &= \pi_{a}\left(\Lambda^0_{r~\mu}x^\mu,\Lambda^i_{r~\mu}x^\mu \right),
\end{align}
where we introduced $\Lambda^\mu_{r~\nu} \in \mathrm{SO}(1,3)$. The important point is that $\pi_r$ \textit{non-linearly realises time-translations and boosts} whereas $\pi_{a}$ \textit{transforms linearly}, just as ordinary matter \cite{Hongo:2018ant}. 


\paragraph{Locality.} 

In the most general case, tracing over the environment yields an unwieldy non-local effective functional that is intractable unless one knows the exact UV-completion. Instead, just like for standard EFTs, a dramatic simplification takes place in the presence of a \textit{separation of scales}. Here we focus on precisely this possibility: we envisage that the typical length and time scales characterising the environment are much shorter than the Hubble time and the Hubble radius at which we compute cosmological correlators. This hierarchy ensures that our open EFT is local in space and time, i.e. it features operators that are the product of fields at the same spacetime point and a finite but arbitrary number of derivatives thereof. Not all UV-models display such hierarchy. For example, the much studied models of inflation featuring the tachyonic production of gauge modes engendered by a $\phi F\tilde F$ coupling \cite{Anber:2009ua} would give a non-local open EFT for $\pi$ because the gauge fields are mostly produced around Hubble crossing. The fact that our open EFT is not useful to describe these models is nothing new or specific to open systems: the same would happen for a standard EFT if one tried to integrate out very light or massless fields. Instead, the construction that follows is able to match gauge and warm inflation models that features a hierarchy of scale such that the particle production can be contained within the sub-Hubble regime. Such a model was constructed and studied in detail in \cite{Creminelli:2023aly}, to which we match our EFT description.


\subsubsection{Open effective functional}

Now that we motivated a local open EFT, we want to write down all possible local operators compatible with the non-equilibrium constraints \eqref{eq:norm}, \eqref{eq:herm} and \eqref{eq:pos}, and retarded shift symmetry \eqref{eq:SSB}. We can do this by using invariant combinations as fundamental building blocks \cite{Akyuz:2023lsm}. The open effective functional can be constructed out of
\begin{align}\label{eq:Pmu}
	\text{building blocks:} \qquad \pi_{a}, \quad t+ \pi_{r},\quad \partial_\mu \pi_a, \quad   \partial_\mu (t + \pi_{r}).
\end{align} 
It will turn convenient to define $	P_\mu \equiv \partial_\mu (t + \pi_{r}) = \delta^0_{~\mu}+ \partial_\mu \pi_{r}$. Working at leading order in the derivative expansion, we now restrict ourselves to operators with at most one derivative per field. 
We now aim at writing the most generic local open effective functional. In addition to the derivative expansion, a useful way to organise the open effective functional is in powers of $\pi_a$ such that $ S_{\mathrm{eff}} = \int \dd^4 x \sqrt{-g }  \mathcal{L}_{\mathrm{eff}}$ with 
\begin{align}\label{eq:expref}
	\mathcal{L}_{\mathrm{eff}} = \sum_{n=1}^\infty \mathcal{L}_n\quad\mathrm{with}\quad \mathcal{L}_n = \mathcal{O}(\pi_{a}^n)
\end{align}
where we used the unitarity condition \eqref{eq:norm} to notice that $\mathcal{L}_{\mathrm{eff}}$ starts from the first order term in $\pi_{a}$. Restricting ourselves to cubic operators for practical applications, we focus on $\mathcal{L}_1$, $\mathcal{L}_2$ and $\mathcal{L}_3$ for which we illustrate the general procedure, the next order being at best quartic in $\pi_a$.

\paragraph{$\mathcal{L}_1$ functional.}

Let us illustrate the procedure by first considering $\mathcal{L}_1$. We aim at using the building blocks \eqref{eq:Pmu} to construct invariant combinations that are linear in $\pi_a$. The only option consists in multiplying $\pi_a$ and $P^\mu \partial_\mu \pi_{a} = \left(-\dot{\pi}_{a} + \partial^\mu \pi_{r} \partial_\mu \pi_{a}\right)$ by powers of
\begin{align}
	\left( P_\mu P^\mu  +1\right) = -2 \dot{\pi}_{r} + \left( \partial_\mu \pi_{r}\right)^2,
\end{align}
leading to \cite{Hongo:2018ant}\footnote{The sign in front of the $\gamma_n$ term is chosen for later convenience.}
\begin{align}
	\mathcal{L}_1 = & \sum_{n=0}^\infty \left( P_\mu P^\mu  +1\right)^{n} \left[ \gamma_n \pi_{a} -\alpha_n P^\mu \partial_\mu \pi_{a}  \right].
\end{align}
The EFT coefficients $\gamma_n$ and $\alpha_n$ are in general functions of $t + \pi_{r}$ which have to be real because of the conjugate condition \eqref{eq:herm}.  In the slow-roll regime, one can assume time-independence for the EFT coefficients at leading order in slow-roll \cite{Hongo:2018ant}. 

\begin{tcolorbox}[%
	enhanced, 
	breakable,
	skin first=enhanced,
	skin middle=enhanced,
	skin last=enhanced,
	before upper={\parindent15pt},
	]{}
	\paragraph{Background evolution.}
	
	Before describing the dynamics of the fluctuations, let us connect with the standard background evolution of the EFToI by discussing tadpole cancellation \cite{Cheung:2007st, Collins:2012nq}. As mentioned below \Eq{eq:expref}, the unitarity condition \eqref{eq:norm} imposes that $\mathcal{L}_{\mathrm{eff}}$ starts linear in $\pi_{a}$. Therefore, the only available tadpoles are \cite{LopezNacir:2011kk,Hongo:2018ant}
	\begin{align}\label{eq:tad1}
		S_{\mathrm{eff}}  \supset  \int \dd^4 x  \sqrt{-g} \left[ \gamma_0(t) \pi_a  + \alpha_0(t) \dot{\pi}_a \right],
	\end{align}
	which leads to the continuity equation 
	\begin{align}
		- \gamma_0 + \dot{\alpha}_0 + 3H \alpha_0 = 0.
	\end{align}
	We can compare this expression with \Eq{eq:contiEFToI} to identify 
	\begin{align}\label{eq:tadpole}
		\alpha_0(t) \equiv 2 c(t) \qquad \mathrm{and} \qquad \gamma_0 (t)\equiv  \dot{c}(t) - \dot{\Lambda}(t),
	\end{align}
	such that \eqref{eq:contiEFToI} and \Eqs{eq:tad1} are equivalent. The main physical outcome is that, as noticed in \cite{LopezNacir:2011kk}, there is no new tadpole for this class of local dissipative models of inflation. The background evolution is fixed by the slicing and probes the global energy density, which does not distinguish the contributions of the inflaton from those of the unknown environment.\footnote{It would be interesting to further investigate if there exists a slicing where system and environment are distinguishable from the background dynamics, for instance through different charges.} It is only at the level of the fluctuations that the distinction between system and environment becomes relevant, as we can disentangle observable degrees of freedom associated to the hydrodynamical direction $\pi$ and unobservable degrees of freedom that have been integrated out. We come back to this point in Lecture \ref{sec:lec5}.
\end{tcolorbox}

Now we have fixed the background dynamics, $\mathcal{L}_1$ takes the explicit form
\begin{align}
	\mathcal{L}_1 = - \alpha_0 \partial^\mu \pi_{r} \partial_\mu \pi_{a} &-\sum_{n=1}^\infty \alpha_n \left[-2 \dot{\pi}_{r} + \left( \partial_\mu \pi_{r}\right)^2\right]^{n}\left(-\dot{\pi}_{a} + \partial^\mu \pi_{r} \partial_\mu \pi_{a}\right) \nonumber \\
	+& \sum_{n=1}^\infty \gamma_n \left[-2 \dot{\pi}_{r} + \left( \partial_\mu \pi_{r}\right)^2\right]^n\pi_{a}.
\end{align}
Notice that only $\alpha_0$, $\alpha_1$ and $\gamma_1$ provide quadratic terms in $\pi$ relevant for the dispersion relation of the Goldstone mode \cite{Hongo:2018ant}. The $\alpha_0$ term is the usual kinetic term written in the Keldysh basis. The $\alpha_1$ term generates a non-trivial speed of sound accompanied by higher order operators controlling the appearance of equilateral non-Gaussianities \cite{Cheung:2007st}. 
In contrast to the $\alpha_0$ and $\alpha_1$ terms, the $\gamma_1$ term has no unitary counterpart and leads to a dissipative term in the $\pi_{r}$ equation of motion. Interestingly, the dissipation term $\dot{\pi}_{r}\pi_{a}$ is accompanied by a cubic interaction $\left( \partial_\mu \pi_{r}\right)^2 \pi_{a}$, as first noted in \cite{LopezNacir:2011kk}, such that the combination is invariant under Lorentz boosts. 

Let us explicitly consider the quadratic and cubic contributions. Indeed, in addition to the expansion in powers of $\pi_a$, one can classify
\begin{align}
	\mathcal{L}_{\mathrm{eff}} = \sum_{n,m=1}^\infty  \mathcal{L}^{(m)}_n \quad\mathrm{with}\quad \mathcal{L}^{(m)}_n = \mathcal{O}(\pi^m,~\pi_{a}^n,~ \pi_r^{m-n})
\end{align}
where $m$ labels the number of field operators. The quadratic contributions are
\begin{align}
	\mathcal{L}_1^{(2)} &= \left( \alpha_0 -2 \alpha_1\right) \dot{\pi}_{r} \dot{\pi}_{a} - \alpha_0 \partial_i \pi_{r}  \partial^i \pi_{a} - 2 \gamma_1  \dot{\pi}_{r} \pi_a
\end{align}
where $\alpha_0$ and $\alpha_1$ control the unitary kinetic term with a non-trivial speed of sound and $\gamma_1$ encodes the linear dissipation of the system onto its environment. At cubic order, $\mathcal{L}_1$ reads
\begin{align}\label{eq:cubicinter}
	\mathcal{L}_1^{(3)} =  &\left(4 \alpha_2 -3\alpha_1 \right) \dot{\pi}^2_{r} \dot{\pi}_{a} + \alpha_1 \left(\partial_i \pi_{r} \right)^2 \dot{\pi}_a + 2 \alpha_1  \dot{\pi}_{r}	\partial_i \pi_{r}  \partial^i \pi_{a} \Big.\nonumber \\
	&\qquad + \left(4 \gamma_2 -\gamma_1 \right)  \dot{\pi}^2_{r} \pi_{a}	+ \gamma_1 	\left(\partial_i \pi_{r} \right)^2 \pi_a 
\end{align} 
where the first line corresponds to parts of the unitary operators $\dot{\pi}^3$ and $(\partial_i \pi)^2 \dot{\pi}$ and the second line to the non-linear dissipation induced by the non-linearly realised symmetries, as discussed in \cite{LopezNacir:2011kk}.

\paragraph{$\mathcal{L}_2$ functional} 

Following \cite{Hongo:2018ant}, let us now construct $\mathcal{L}_2$, which is quadratic in $\pi_a$. Just as above, working with operators containing at most one derivative, in the slow-roll limit, we obtain  
\begin{align}\label{eq:diffquad}
	\mathcal{L}_2 &= i \Big[\beta_1 \pi_{a}^2 + \beta_2 \left( \partial_\mu \pi_{a}\right)^2 +\beta_3 \left(-\dot{\pi}_{a} + \partial^\mu \pi_{r} \partial_\mu \pi_{a}\right)\pi_{a} +\beta_4 \left(-\dot{\pi}_{a} + \partial^\mu \pi_{r} \partial_\mu \pi_{a}\right)^2 + \cdots \Big]
\end{align}
where the third and fourth terms are obtained from $P^\mu \partial_\mu \pi_{a} \pi_a$ and $(P^\mu \partial_\mu \pi_{a})^2$ respectively and the dots represent higher order terms obtained by multiplying the first four terms by arbitrary powers of $(P^\mu P_\mu + 1) = -2 \dot{\pi}_{r} + \left( \partial_\mu \pi_{r}\right)^2$. The action being at least quadratic in $\pi_a$, there is no tadpole contribution. The $i$ in front directly follows from the conjugate condition \eqref{eq:herm}. While the term proportional to $\beta_1$ is the standard noise term appearing in the Langevin equation, we observe the presence of derivative corrections such as $\left(\partial_\mu \pi_{a} \right)^2$ and $\dot{\pi}_{a}^2 $ in the $\beta_2$ and $\beta_4$ terms which make the noise scale-dependent. 

There exists a positivity condition on the $\beta$'s coefficients due to \Eq{eq:pos} which imposes $\Im S_{\mathrm{eff}} [\pi_r,\pi_a] \geq 0$. In flat space, making use of the derivative expansion which tells us that $\omega^2, k^2 \ll |\beta_1/\beta_{2,4}|$ (the quadratic term in $\beta_3$ can be written as a total derivative and removed), the authors of \cite{Hongo:2018ant} concluded that $\beta_1$ dominates in $\mathcal{L}_2$, such that the positivity constraint imposes 
\begin{align}
	\beta_1 > 0.
\end{align}
This positivity constraint on the noise kernel directly translates into consequences for the non-Gaussian signal if we multiply this operator by higher powers of $(P^\mu P_\mu + 1) = -2 \dot{\pi}_{r} + \left( \partial_\mu \pi_{r}\right)^2$. 

As above, let us explicitly consider the quadratic and cubic contributions for $\mathcal{L}_2$. The three quadratic noise are controlled by 
\begin{align}
	\mathcal{L}_2^{(2)} &= i \left[\beta_1 \pi_a^2 - \left(\beta_2 - \beta_4\right)\dot{\pi}_a^2 + \beta_2 \left(\partial_i \pi_{a} \right)^2\right]\,,\label{eq:piq2free}
\end{align}
where we removed the total derivative related to $\beta_3$ at quadratic order. At cubic order, we obtain 
\begin{align}\label{eq:cubicinter2}
	\mathcal{L}_2^{(3)} &= i \Big[-(\beta_3 -2 \beta_7)\dot{\pi}_{r} \dot{\pi}_{a} \pi_a + \beta_3 \partial_i \pi_{r}  \partial^i \pi_{a} \pi_a +2 (\beta_4+\beta_6 - \beta_8) \dot{\pi}_{r} \dot{\pi}_a^2 \nonumber \\
	& \qquad \quad - 2 \beta_4   \partial_i \pi_{r}  \partial^i \pi_{a} \dot{\pi}_{a} - 2\beta_5 \dot{\pi}_{r} \pi_a^2  - 2 \beta_6 \dot{\pi}_{r}(\partial_i\pi_a)^2\Big],
\end{align}
where we needed to introduce the Wilsonian coefficients $\beta_5, \beta_6, \beta_7$ and $\beta_8$ associated to the higher-order operators included in the dots of \Eq{eq:diffquad}, obtained from multiplying the first four terms in \Eq{eq:diffquad} by $(P^\mu P_\mu + 1) = -2 \dot{\pi}_{r} + \left( \partial_\mu \pi_{r}\right)^2$. The physical interpretation of these terms is discussed below in \Sec{subsubsec:unit}.

\paragraph{$\mathcal{L}_3$ functional}

One can carry on this construction to access $\mathcal{L}_3$. We have
\begin{align}
	\mathcal{L}_3 &= \delta_1 \pi_a^3 + \delta_2 (\partial_\mu \pi_a)^2 \pi_a + \delta_3 \left(-\dot{\pi}_{a} + \partial^\mu \pi_{r} \partial_\mu \pi_{a}\right) \pi_a^2 + \delta_4 \left(-\dot{\pi}_{a} + \partial^\mu \pi_{r} \partial_\mu \pi_{a}\right) (\partial_\nu \pi_a)^2 \Big.  \nonumber\\
	&\qquad \qquad + \delta_5 \left(-\dot{\pi}_{a} + \partial^\mu \pi_{r} \partial_\mu \pi_{a}\right)^2 \pi_a  + \delta_6 \left(-\dot{\pi}_{a} + \partial^\mu \pi_{r} \partial_\mu \pi_{a}\right)^3 + \cdots,
\end{align}
where the $\delta_3$ term originates from $(P^\mu \partial_\mu \pi_a) \pi_a^2$, the $\delta_4$ term from $(P^\mu \partial_\mu \pi_a) (\partial_\nu \pi_a)^2 $, the $\delta_5$ from $\left(P^\mu \partial_\mu \pi_a\right)^2 \pi_a$ and the $\delta_6$ term from $\left(P^\mu \partial_\mu \pi_a\right)^3$. As above, the dots represent higher order terms obtained by multiplying the first terms by arbitrary powers of $(P^\mu P_\mu + 1) = -2 \dot{\pi}_{r} + \left( \partial_\mu \pi_{r}\right)^2$. As we will see below, the interpretation of these terms is ambiguous, as they can either be associated to unitary or non-unitary operators depending on how they relate to contributions from $\mathcal{L}_1$. For this reason, we develop in \Sec{subsubsec:unit} a classification of these terms. 

As above, let us explicitly consider the quadratic and cubic contributions for $\mathcal{L}_3$. Since these contributions are at least cubic in $\pi_a$, there is no quadratic contribution. If we restrict ourselves to the cubic order, we obtain 
\begin{align}
	\mathcal{L}_3^{(3)} &=\delta_1 \pi_a^3 + (\delta_5- \delta_2)  \dot{\pi}_{a}^2 \pi_a  + \delta_2  (\partial_i \pi_{a})^2 \pi_a - \delta_4  (\partial_i \pi_{a})^2 \dot{\pi}_a + (\delta_4-\delta_6)\dot{\pi}_{a}^3 ,
\end{align}
the $\delta_3$ cubic contribution being a total derivative. 

\begin{tcolorbox}[%
	enhanced, 
	breakable,
	skin first=enhanced,
	skin middle=enhanced,
	skin last=enhanced,
	before upper={\parindent15pt},
	]{}
	\paragraph{Summary}
	
	Under the symmetry breaking pattern specified in \Eq{eq:SSBEFToI}, the most generic local second order open effective functional is
	\begin{align}\label{eq:L2ref}
		&\qquad \mathcal{L}^{(2)} =\left( \alpha_0 -2 \alpha_1\right) \dot{\pi}_{r} \dot{\pi}_{a} - \alpha_0 \partial_i \pi_{r}  \partial^i \pi_{a} \nonumber   \\
		&- 2 \gamma_1  \dot{\pi}_{r} \pi_a+i \left[\beta_1 \pi_a^2 - \left(\beta_2 - \beta_4\right)\dot{\pi}_a^2 + \beta_2 \left(\partial_i \pi_{a} \right)^2\right],
	\end{align}
	where the EFT coefficients are chosen to match the notations of \cite{Hongo:2018ant}. The first line corresponds to the usual unitary dynamics which the kinetic term and an effective speed of sound. The first term of the second line controlled by $\gamma_1$ corresponds to the dissipation due to the surrounding environment. At last, the $\beta_i$ coefficients control the diffusion (noise-induced) process.
	
	\indent At cubic order, the most generic open effective functional (up to total derivatives) writes
	\begin{align}
		\mathcal{L}^{(3)} &= \left(4 \alpha_2 -3\alpha_1 \right) \dot{\pi}^2_{r} \dot{\pi}_{a} + \alpha_1 \left(\partial_i \pi_{r} \right)^2 \dot{\pi}_a + 2 \alpha_1  \dot{\pi}_{r}	\partial_i \pi_{r}  \partial^i \pi_{a} \Big. \label{eq:L13} \\
		&\qquad \quad + \left(4 \gamma_2 -\gamma_1 \right)  \dot{\pi}^2_{r} \pi_{a}	+ \gamma_1 	\left(\partial_i \pi_{r} \right)^2 \pi_a \nonumber \\ 
		+&i \Big[-(\beta_3 -2 \beta_7)\dot{\pi}_{r} \dot{\pi}_{a} \pi_a + \beta_3 \partial_i \pi_{r}  \partial^i \pi_{a} \pi_a +2 (\beta_4+\beta_6 - \beta_8) \dot{\pi}_{r} \dot{\pi}_a^2 \label{eq:L23} \\
		& \qquad \quad - 2 \beta_4   \partial_i \pi_{r}  \partial^i \pi_{a} \dot{\pi}_{a} - 2\beta_5 \dot{\pi}_{r} \pi_a^2  - 2 \beta_6 \dot{\pi}_{r}(\partial_i\pi_a)^2\Big] \nonumber \\
		+&\delta_1 \pi_a^3 + (\delta_5- \delta_2)  \dot{\pi}_{a}^2 \pi_a  + \delta_2  (\partial_i \pi_{a})^2 \pi_a - \delta_4  (\partial_i \pi_{a})^2 \dot{\pi}_a + (\delta_4-\delta_6)\dot{\pi}_{a}^3, \label{eq:L33}
	\end{align}
	where \ref{eq:L13} originates from $\mathcal{L}_1^{(3)}$, \ref{eq:L23} originates from $\mathcal{L}_2^{(3)}$ and \ref{eq:L33} from $\mathcal{L}_3^{(3)}$.
	
\end{tcolorbox}

In de Sitter, in terms of the conformal time and the scale factor $a = -1/(H\eta)$, the open effective functional up to cubic order reads
\begin{align}
	&\quad S_{\mathrm{eff}}^{(2)} = \int \dd^4 x \Big\{\left( \alpha_0 -2 \alpha_1\right) a^2 \pi'_{r} \pi'_{a} - \alpha_0 a^2 \partial_i \pi_{r}  \partial^i \pi_{a}   \\
	-& 2 a^3 \gamma_1 \pi'_{r} \pi_a + i \left[\beta_1 a^4 \pi_a^2 - \left(\beta_2 - \beta_4\right) a^2\pi_a^{\prime2} + \beta_2 a^2 \left(\partial_i \pi_{a} \right)^2\right] \Big\}, \nonumber
\end{align}
and
\begin{align}
	S_{\mathrm{eff}}^{(3)} =  \int &\dd^4 x \Big\{\left(4 \alpha_2 -3\alpha_1 \right) a \pi_r^{\prime2} \pi'_{a} + \alpha_1 a \left(\partial_i \pi_{r} \right)^2 \pi'_a + 2 \alpha_1 a \pi'_{r}	\partial_i \pi_{r}  \partial^i \pi_{a} \\
	& \qquad \quad + \left(4 \gamma_2 -\gamma_1 \right)  a^2\pi_r^{\prime2} \pi_{a}	+ \gamma_1 a^2	\left(\partial_i \pi_{r} \right)^2 \pi_a \nonumber \Big. \\
	&+ i \Big[-(\beta_3-2\beta_7) a^2 \pi'_{r} \pi'_{a} \pi_a + \beta_3 a^2 \partial_i \pi_{r}  \partial^i \pi_{a} \pi_a +2 (\beta_4+\beta_6 - \beta_8) a \pi'_{r} \pi_a^{\prime2} \nonumber \\
	& \qquad \quad - 2 \beta_4  a \partial_i \pi_{r}  \partial^i \pi_{a} \pi'_{a} - 2\beta_5 a^3 \pi_r^{\prime} \pi_a^2  - 2 \beta_6 a \pi_r^{\prime} (\partial_i\pi_a)^2 \Big] \nonumber\\
	&+\delta_1 a^4 \pi_a^3 + (\delta_5- \delta_2) a^2 \pi_a^{\prime2} \pi_a  + \delta_2 a^2 (\partial_i \pi_{a})^2 \pi_a - \delta_4 a  (\partial_i \pi_{a})^2 \pi'_a + (\delta_4-\delta_6) a \pi_a^{\prime3} \Big\}. \nonumber
\end{align}


\subsubsection{Classification of the EFT operators}\label{subsubsec:unit}

The above construction exhibits a wide zoology of terms compared to its unitary counterpart: $5$ free parameters in $\mathcal{L}^{(2)}$ compared to only $1$ in the standard EFToI \cite{Cheung:2007st}; $13$ free parameters in $\mathcal{L}^{(3)}$ compared to only $1$ in \cite{Cheung:2007st}. While some operators describe faithful non-unitary effects generated by the presence of additional degrees of freedom, others are simply a consequence of writing unitary interactions in the Keldysh basis. In this section, we develop a procedure to distinguish unitary from non-unitary operators.

\paragraph{Recovering the EFToI} We expect the open effective functional $S_{\mathrm{eff}}$ to be able to reproduce in a certain limit the EFToI \cite{Cheung:2007st}. This limit defines the unitary direction of the parameter space of the theory. Let us first consider the quadratic terms of the EFToI which reads in the Keldysh basis
\begin{equation}
	\frac{1}{2}\left[\dot{\pi}_{+}^{2}-c_{s}^{2}(\partial_{i}\pi_{+})^{2}\right]-\frac{1}{2}\left[\dot{\pi}_{-}^{2}-c_{s}^{2}(\partial_{i}\pi_{-})^{2}\right]=\dot{\pi}_r\dot{\pi}_a-c_{s}^{2}\partial_{i}{\pir}\partial_{i}{\pia}.
\end{equation}
It can be matched with the first line of \Eq{eq:L2ref} for
\begin{equation}
	c_{s}^{2} = \frac{\alpha_0}{\alpha_{0}-2\alpha_1}.
\end{equation}
We can go to the next order in the EFToI and consider cubic operators. Starting with $\dot{\pi}^{3}$, the unitary interaction in the Keldysh basis reads
\begin{equation}\label{eq:Unitarypiq3}
	\dot{\pi}^{3}_{+}-\dot{\pi}^{3}_{-}=3\dot{\pi}_{r}^{2}\dot{\pi}_{a} + \frac{1}{4}\dot{\pi}^{3}_{a}.
\end{equation}
Comparing with the terms in \Eqs{eq:L13} and \eqref{eq:L33} which include $\dot{\pi}_{r}^{2}\dot{\pi}_{a}$ and  $\dot{\pi}^{3}_{a}$, the unitary combination in \Eq{eq:Unitarypiq3} imposes the relation among the EFT coefficients
\begin{equation}\label{eq:combin1}
	\delta_4-\delta_6= \frac{1}{12}\left(4 \alpha_2 -3\alpha_1 \right).
\end{equation}
A similar procedure follows for $(\partial_{i}\pi)^{2}\dot{\pi}$. In the Keldysh basis, this vertex reads
\begin{equation}
	(\partial_{i}\pi_{+})^{2}\dot{\pi}_{+}-(\partial_{i}\pi_{-})^{2}\dot{\pi}_{-}=(\partial_{i}\pir)^{2}\dot{\pi}_{a} + 2 \partial_{i}\pir\partial_{i}\pia\dot{\pi}_r+\frac{1}{4}\dot{\pi}_a(\partial_{i}\pir)^{2}.
\end{equation}
Comparing it to the operators appearing in \Eqs{eq:L13} and \eqref{eq:L33}, it specifies the unitary direction
\begin{equation}\label{eq:combin2}
	\delta_{4}=-\frac{1}{4}\alpha_1.
\end{equation}
One can then be reassured that in a certain limit, the current constructions reduces to the usual EFToI of \cite{Cheung:2007st}.

\paragraph{Unitary and orthogonal directions}

What about the other operators? The open effective functional has more Wilsonain coefficients than the EFToI. Does it imply that all the remaining operators are intrinsically related to non-unitary effects?  
The symmetry structure of the theory allows one to  answer this question is a systematic manner. In the limit where non-unitary effects are absent, \Eq{eq:Seffref} reduces to the unitary effective action 
\begin{align}\label{eq:Seffunit}
	S_{\mathrm{eff}}\left[ \pi_+, \pi_-\right] = S_\pi\left[\pi_+\right] - S_\pi\left[\pi_-\right].
\end{align}
This restriction is obtained by restoring the $\epsilon_a$ symmetry explicitly broken by the non-unitary effects (see \textit{Right} panel of \Fig{fig:sym}) \cite{Hongo:2018ant}. Indeed, in the unitary limit, the two branches of the path integral must transform equally under $\epsilon_\pm$                               
\begin{align}\label{eq:qsym}
	\pi_{\pm}(t,\bmx) \rightarrow \pi'_{\pm}(t,\bmx) &= \pi_{\pm}(t+\epsilon_{\pm},\bmx) + \epsilon_{\pm}.
\end{align}
One can impose this by acting on $S_{\mathrm{eff}}$ with the $\epsilon_{a}$ symmetry given by $\epsilon^0_+ =  - \epsilon^0_- = \epsilon_{a}/2$. Expressing \Eq{eq:qsym} in the Keldysh basis and expanding linear order in $\epsilon_a$ we obtain
\begin{align}
	\pi_{r}(t,\bmx) \rightarrow \pi'_{r}(t,\bmx) &= \pi_{r}(t,\bmx) + \frac{\epsilon_a}{2}\dot{\pi}_{a}(t,\bmx) + \mathcal{O}\left(\epsilon_a^2\right)\label{eq:translAsym} \\
	\pi_{a}(t,\bmx) \rightarrow \pi'_{a}(t,\bmx) &= \pi_{a}(t,\bmx) + \epsilon_a\left[1+\dot{\pi}_{r}(t,\bmx)\right] + \mathcal{O}\left(\epsilon_a^2\right). \label{eq:translAsym2}
\end{align} 
Unitary combinations of operators must leave $S_{\mathrm{eff}}$ invariant under the above transformation.

Let us illustrate this procedure with the kinetic terms $\dot{\pi}_r \dot{\pi}_a$ and $\partial_i\pi_r \partial^i \pi_a $ appearing in \Eq{eq:L2ref} that have been identified as being unitary through the comparison with the EFToI. Under the $\epsilon_{a}$ transformation \Eqs{eq:translAsym} and \eqref{eq:translAsym2}, we notice they lead to total derivatives. Hence, $S_{\mathrm{eff}}$ made of these terms is $\epsilon_a$-invariant, indicating they can be encountered in a unitary theory as one would expect. On the contrary, the quadratic dissipative term $\dot{\pi}_r \pi_a$ and diffusive terms $\pi_a^2$, $\dot{\pi}_a^2$ and $(\partial_i \pi_a)^2$ are not invariant. Consequently, they have no unitary counterpart and represent genuine non-unitary effects. 

For cubic interactions the effective action is invariant under $\epsilon_a$ only for the specific combinations identified in \Eqs{eq:combin1} and \eqref{eq:combin2}. Indeed, one can check the combinations $3\dot{\pi}_r^2\dot{\pi}_a + \dot{\pi}_a^3/4$ and $(\partial_i \pi_r)^2\dot{\pi}_a + 2 \dot{\pi}_r\partial_i \pi_r \partial^i \pi_a  +  (\partial_i \pi_a)^2\dot{\pi}_a/4$ are invariant. From these, one recovers the usual cubic interactions $\dot{\pi}^3$ and $(\partial_i \pi)^2\dot{\pi}$ expressed in the Keldysh basis. Any deviation from these fine-tuned combinations would be associated to non-unitary dynamics. In particular, notice that unitary combinations only involve \textit{odd} powers of $\pi_a$. Hence, any \textit{even} powers of $\pi_a$ always relate to diffusive/noise processes \cite{Hongo:2018ant}. 


\subsubsection{Scales and estimates}\label{subsec:energy}

The EFT coefficients are dimensionful quantities, such that $[\alpha_0] =[\alpha_1] = E^{4}$, $[\gamma_1] = E^{5}$, $[\beta_1] = E^{6}$ and $[\beta_2] =[\beta_4] = E^{4}$. One can then ask what are the relevant scales controlling the physics and the regime of validity of our EFT description. 

\subsubsection*{Energy scales and canonical normalisation} 

To treat the problems of the scales, we first canonically normalise the fields such that they have dimension of energy $E$. This canonical normalisation relates the original Wilsonian coefficients to quantities of physical interest such as the speed of sound $c_s$, the dissipation scale $\gamma$ or the fluctuations of the environment $\beta_1$. We first define the energy scale
\begin{align}
	f_\pi^4 \equiv \alpha_0 -2 \alpha_1\,.
\end{align}
From this we construct the canonically normalised field
\begin{equation}\label{eq:can}
	\pi \rightarrow  \pi/f_\pi^2.
\end{equation}
It follows that at leading order, the curvature perturbations are given by the relation
\begin{align}\label{eq:zetarel}
	\zeta =  - \frac{H}{f_\pi^2} \pi.
\end{align}
In terms of the canonically normalised variables, the quadratic action takes the form
\begin{align}\label{eq:canonorm} 
	&\quad S_{\mathrm{eff}}^{(2)} = \int \dd^4 x \Big\{ a^2 \pi'_{r} \pi'_{a} - c_{s}^{2} a^2 \partial_i \pir  \partial^i \pia  \\
	-&  a^3 \gamma \pi'_{r} \pia + i \left[\beta_{1} a^4 \pia^2 - \left(\beta_2 - \beta_4\right) a^2\pia^{\prime 2} + \beta_2 a^2 \left(\partial_i \pia \right)^2\right] \Big\}, \nonumber
\end{align}
where we used rescaled coefficients 
\begin{align}
	c_s^{2} \equiv \frac{\alpha^{(\text{old})}_{0}}{f_\pi^4}\;,\quad\; \gamma \equiv
	\frac{2\gamma^{(\text{old})}_{1}}{f_\pi^4}\;,\quad\;\beta_{i} \equiv \frac{\beta^{(\text{old})}_{i}}{f_{\pi}^{4}}\;\quad\mathrm{for}\quad i = 1 ~\mathrm{to}~8.
\end{align}
The dimensions of the parameters appearing above are 
\begin{align}
	[\pi] = E, \quad [f_\pi] = E, \quad [c_s] = E^0, \quad [\gamma] = E, \quad [\beta_1] = E^2, \quad [\beta_2] = [\beta_4] = E^0.
\end{align}
Expressed in terms of the canonically normalised variables, the cubic action becomes
\begin{align}\label{eq:canonormcub}
	S_{\mathrm{eff}}^{(3)} =   \frac{1}{f_\pi^2} \int & \dd^4 x \Big\{\Big[4 \alpha_2 -  \frac{3}{2} (c^2_s-1) \Big]  a \pir^{\prime2} \pi'_{a} +\frac{1}{2} (c^2_s-1) a \left[\left(\partial_i \pir \right)^2 \pi'_a + 2 \pi'_{r}	\partial_i \pir  \partial^i \pia \right] \\
	& \qquad \quad + \left(4 \gamma_2 - \frac{\gamma}{2}\right)  a^2\pir^{\prime2} \pia	+ \frac{\gamma}{2} a^2	\left(\partial_i \pir \right)^2 \pia \nonumber \Big. \\
	&+ i \Big[\left(2\beta_7-\beta_3\right) a^2 \pi'_{r} \pi'_{a} \pia + \beta_3 a^2 \partial_i \pir  \partial^i \pia \pia + 2(\beta_4+ \beta_6 - \beta_8) a \pi'_{r} \pia^{\prime2} \Big. \nonumber \\
	& \qquad \quad - 2 \beta_4  a \partial_i \pir  \partial^i \pia \pi'_{a} - 2\beta_5 a^3 \pir^{\prime} \pia^2  - 2 \beta_6 a \pir^{\prime} (\partial_i\pia)^2 \Big]  \Big. \nonumber\\
	&+\delta_1 a^4 \pia^3 + (\delta_5- \delta_2) a^2 \pia^{\prime2} \pia  + \delta_2 a^2 (\partial_i \pia)^2 \pia - \delta_4 a  (\partial_i \pia)^2 \pi'_a + (\delta_4-\delta_6) a \pia^{\prime3} \Big\},  \nonumber
\end{align}
where we defined the rescaled coefficients\footnote{Notice that $\alpha_2$ can be related to the EFToI \cite{Cheung:2007st} parameter $M_3^4 = - f_\pi^4 \alpha_2$.}
\begin{align}
	\alpha_2 \equiv \frac{\alpha^{(\text{old})}_2}{f_\pi^4}\;,\quad\; \gamma_2 \equiv \frac{\gamma^{(\text{old})}_{4}}{f_\pi^4} \;,\quad\; \delta_i \equiv \frac{\delta^{(\text{old})}_{i}}{f_\pi^4} \quad\mathrm{for}\quad i = 1 ~\mathrm{to}~6\,,
\end{align}
with dimensions
\begin{align}
	[\alpha_2] = E^{0}, &\quad [\gamma_2] = E, \quad [\beta_6] =  [\beta_8] = E^0, \quad [\beta_3] = [\beta_7] = E, \quad [\beta_5] = E^2, \nonumber \\
	& \qquad  [\delta_1] = E^3, \quad [\delta_5] = [\delta_2] = E, \quad [\delta_4] = [\delta_6] = E^0  . 
\end{align}
From now on, we work in this canonical basis and use the rescaled action given in \Eqs{eq:canonorm} and \eqref{eq:canonormcub}. 


\subsubsection*{Heuristic estimates} 

We are now in the position to carry out a rough estimate of the non-Gaussianities sourced by the cubic operators of \Eq{eq:canonormcub}. For simplicity, we set $\beta_2 = \beta_4 = 0$ and focus on the leading noise term of \Eq{eq:canonorm} controlled by $\beta_1$. The estimate relies on the following rules:
\begin{itemize}
	\item We can approximate $\pir$ from the amplitude of the primordial power spectrum $\Delta^2_{\zeta}$, accounting for the canonical normalisation and the leading-order relation with $\zeta$ such that
	\begin{equation}\label{eq:rule1}
		\pir\sim\frac{f_{\pi}^{2}}{H}\Delta_{\zeta}.
	\end{equation}
	\item  We can estimate spatial derivatives by the spatial momenta. The value of spatial derivatives in different directions is, on average, the same by isotropy:
	\begin{equation}\label{eq:rule2}
		\partial_{i}\pi_{r,a}\sim k\pi_{r,a}.
	\end{equation}
	Adiabatic perturbations of momentum $k$ freeze at a scale factor $a_*$ that depends on the dissipation parameter $\gamma$ \cite{LopezNacir:2011kk}. In \cite{Salcedo:2024smn}, we derived this freezing time by comparing the early and the late time limit of the power spectrum. This leads to the relation
	\begin{equation}\label{eq:rule3}
		c_s k\sim a_* H \sqrt{\frac{H+\gamma}{H}},
	\end{equation}
	where we use the expression $(H+\gamma)$ as a shorthand reminder of a quantity that scales to leading order in $\gamma\to\infty$ as $\gamma$ and to leading order in $\gamma \to 0$ as $H$. While freezing still occurs at wavelengths around (sound) horizon crossing at low dissipation, it is displaced to sub (sound) horizon wavelength at large dissipation.
	\item The characteristic frequencies of the retarded $\pir$ and advanced $\pia$ components are estimated to be \cite{Salcedo:2024smn}
	\begin{equation}\label{eq:rule4}
		\pi_{r,a}'\sim a H \pi_{r,a}.
	\end{equation}
	\item The retarded component $\pir$ evolves according to a dynamics sourced by the advanced component $\pia$ and controlled by the environment noise $\beta_1$ \cite{kamenev_2011}. This sourced dynamics implies that $\pir$ and $\pia$ are not of the same amplitude, their ratio being controlled by
	\begin{equation}\label{eq:rule5}
		\frac{\pir}{\pia} \sim \frac{\beta_1}{H(H+\gamma)}.
	\end{equation}
	This relation can be obtained from evaluating the equation of motion for $\pir$ at $a=a_{*}$ \cite{Salcedo:2024smn}.
\end{itemize}

These prescriptions imply some hierarchies among the quadratic operators. Comparing the kinetic terms $a^2 \pir'\pia'$ and $c_s^2 a^2 \partial_i \pir \partial^i \pia$ and the linear dissipation $a^3\gamma\pir' \pia$ with the noise $a^{4} \beta_1 \pia^2$, we observe that
\begin{align}\label{eq:quad1}
	\frac{a^2 \pir'\pia'}{a^{4} \beta_1 \pia^2} \sim \frac{H}{H+\gamma}, \qquad  \qquad  \frac{c_s^2 a^2 \partial_i \pir \partial^i \pia}{a^{4} \beta_1 \pia^2} \sim 1, \qquad \qquad \frac{a^3\gamma\pir' \pia}{a^{4} \beta_1 \pia^2} \sim \frac{\gamma}{H+\gamma}.
\end{align}
This illustrates the dynamical regimes of a driven-dissipative harmonic oscillator. At low dissipation, $a^3\gamma\pir' \pia$ is negligible and the system is mostly controlled by the other three operators. At large dissipation, we enter the overdamped regime in which $a^2 \pir'\pia'$ becomes subdominant compared to the other contributions. 

To estimate the size of non-Gaussianities, we can first approximate the ratio between the cubic operators in \Eq{eq:canonormcub} and the dominant quadratic terms in \Eq{eq:canonorm}. Based on the above estimate, a choice of operator that is valid both at large and small dissipation is $a^{4} \beta_1 \pia^2$ (or equivalently $c_s^2 a^2 \partial_i \pir \partial^i \pia$), leading to
\begin{equation}
	f_{\text{NL}}\Delta_{\zeta}\sim\frac{\mathcal{L}_{3}}{a^{4} \beta_1 \pia^2}\,.
\end{equation}
Like in the EFToI, one of the amplitudes of the unitary vertices in \Eq{eq:canonormcub} is controlled by the speed of sound $c_s$. The associated non-Gaussianities can be estimated through the above prescriptions, leading for instance to
\begin{equation}
	\mathcal{L}_{3} \supset \frac{(c^2_s-1)}{2f_{\pi}^{2}}  a \left(\partial_i \pir \right)^2 \pi'_a \quad \rightarrow \quad f_{\text{NL}}\sim \frac{(c^2_s-1)}{c_{s}^{2}} .
\end{equation}
This matches the usual expectation from \cite{Cheung:2007st}. We can then consider the two dissipative vertices $(\partial_{i}\pir)^{2}\pia$ and $\pir^{\prime 2}\pia$ controlled by $\gamma$ and related to the quadratic dissipation $\pir^{\prime}\pia$ through non-linearly realised boosts. Using the above prescriptions, we find for the first operator the linear-in-$\gamma$ scaling
\begin{equation}
	\mathcal{L}_{3} \supset \frac{\gamma}{f_{\pi}^{2}}a^2(\partial_{i}\pir)^{2}\pia \quad \rightarrow \quad f_{\text{NL}}\sim \frac{1}{c_s^2} \frac{\gamma}{H} ,
\end{equation}
in agreement with the results of \cite{LopezNacir:2011kk, Creminelli:2023aly}. The second vertex can be estimated by
\begin{equation}
	\mathcal{L}_{3} \supset \frac{\gamma}{f_{\pi}^{2}}a^2 \pir^{\prime 2}\pia \quad \rightarrow \quad f_{\text{NL}}\sim \frac{\gamma}{H+\gamma}.
\end{equation}
Noise terms quadratic in $\pia$ can also be estimated in the same manner, leading for instance to
\begin{equation}
	\mathcal{L}_{3} \supset \frac{i\beta_{5}}{f_{\pi}^{2}}a^{3}\pir'\pia^{2} \quad \rightarrow \quad  f_{\text{NL}}\sim \frac{\beta_{5}}{\beta_{1}}.
\end{equation}
Noticeably, this ratio of the noise amplitudes $\beta_1$ and $\beta_5$ is independent of the dissipation parameter $\gamma$ and leads to approximately constant $f_{\text{NL}}$ for any value of $\gamma/H$.  At last, cubic operators in $\pia$ also source a bispectrum signal such as
\begin{equation}\label{eq:5thestimate}
	\mathcal{L}_{3} \supset \frac{\delta_{1}}{f_{\pi}^{2}}a^{4}\pia^{3} \quad \rightarrow \quad f_{\text{NL}}\sim \frac{\delta_{1}}{\beta^2_{1}} (H+\gamma).
\end{equation}
The ratio of $\delta_{1}$ over $\beta_1^2$ controls the amplitude of the non-Gaussian noise compared to its Gaussian counterpart. Similar estimates can be obtained for all cubic operators of \Eq{eq:canonormcub}. This heuristic derivation correctly accounts for all contributions found in \cite{Salcedo:2024smn}, hence providing a valuable insight to the rich physics of the open EFT of inflation.


\subsection{Phenomenology of open inflation}\label{subsec:pheno}

Now the theory is established, we derive the phenomenology associated to this class of model. We first work out the power spectrum, before characterizing primordial non-Gaussianities expected from dissipation and noise. At last, we illustrate the scope of this construction by matching with an explicit model of gauge-inflation \cite{Creminelli:2023aly}.

\subsubsection{Power spectrum}

To derive the power spectrum, we follow the procedure introduced in \Sec{subsec:freeth}. Embedded in the inflating background, the retarded and Keldysh differential operators become
\begin{align}\label{eq:DRexp}
	\widehat{D}_{R} &= \frac{1}{H^2\eta^2} \left[\partial_\eta^2 - \frac{2 + \frac{\gamma}{H}}{\eta}\partial_\eta - c_s^2 \partial_i^2 \right]
\end{align}
and 
\begin{align}\label{eq:GKexp}
	\widehat{D}_K &=\frac{1}{H^2\eta^2}\left[\frac{\beta_1}{H^2\eta^2} - \left(\beta_2 - \beta_4\right) \left(\partial_\eta^2 -\frac{2}{\eta} \partial_\eta\right) + \beta_2 \partial_i^2  \right].
\end{align}
From now on, we set $c_s = 1$ for simplicity. The inclusion of $c_s$ is discussed in \cite{Salcedo:2024smn}.


\paragraph{Scaling dimensions} Before deriving the propagators of the theory, let us briefly comment on the two homogeneous solutions of $\widehat{D}_{R}$ in Fourier space (mode functions). Obeying the dynamical equation 
\begin{align}
	\left(\partial_{\eta}^2 - \frac{2 + \frac{\gamma}{H}}{\eta}\partial_{\eta} +  k^2  \right) \pi_k =0,
\end{align}
the two homogeneous solutions are given by 
\begin{align}
	\pi_k(\eta) \propto \eta^{\nu_\gamma}  H^{(1)}_{\nu_\gamma}(- k \eta) \qquad \mathrm{and} \qquad \propto \eta^{\nu_\gamma} H^{(2)}_{\nu_\gamma}(- k \eta) \,,
\end{align}
where 
\begin{align}
	\nu_\gamma \equiv \frac{3}{2}+ \frac{\gamma}{2H}\,,
\end{align}
and $H^{(1)}$ and $H^{(2)}$ are Hankel functions. Selecting positive frequency mode functions of the form $\ee^{- i k \eta}$ in the asymptotic past, $- k \eta \gg 1$, one can safely discard one of the two solutions. At late times, where $- k \eta \ll 1$, the dissipative mode functions acquire scaling solutions
\begin{align}
	\pi_k(z = - k \eta) &=  \mathcal{O}_+  z^{\Delta_+}  \left[1 +\mathcal{O}(z^2) \right] + \mathcal{O}_- z^{\Delta_-}  \left[1 +\mathcal{O}(z^2) \right]\,,
\end{align}
where $\mathcal{O}_+$ and $\mathcal{O}_-$ depend on $\gamma$ and $H$, and the scaling dimensions are
\begin{align}
	\Delta_+ = 0 \qquad \mathrm{and} \qquad \Delta_- = 3 + \frac{\gamma}{H}.
\end{align}
Comparing this result to the well-known relation for closed systems $\Delta_+ + \Delta_-= d$, involving representations related by a shadow transform, we note that dissipation acts in the same way as a continuation to a non-integer number of dimensions. 

These relations crucially depart from the unitary theory case. Yet, let us stress it would be misleading to derive the power spectrum simply by squaring these mode functions. Indeed, in an open theory, dissipation is inextricably linked to the fluctuations generated by the environment. Instead, the power spectrum is derived as follow.

\paragraph{Retarded Green function} \Eq{eq:DRexp} determines the equations of motion obeyed by the Green function  
\begin{align}\label{eq:Greenfct}
	\left(\partial_{\eta_1}^2 - \frac{2 + \frac{\gamma}{H}}{\eta_1}\partial_{\eta_1} + k^2  \right) G^{R} (k; \eta_1, \eta_2) = H^2 \eta_1^2 \delta(\eta_1-\eta_2).
\end{align}			
The normalisation is fixed by the continuity of the retarded Green function at coincident time and its first derivative discontinuity is controlled by the time-dependent prefactor of the $\partial_{\eta_1}^2$ term in \Eq{eq:Greenfct}, that is 
\begin{align}
	G^R(k; \eta_2,\eta_2) &= 0, \qquad \mathrm{and} \qquad
	\left. \partial_{\eta_1}G^R(k; \eta_1,\eta_2) \right|_{\eta_2 = \eta_1} = H^2 \eta^2_2.
\end{align}
This imposes
\begin{align}\label{eq:GRfinsimp}
	G^R(k; \eta_1,\eta_2) =  \frac{\pi}{2} H^2 (\eta_1 \eta_2)^{\frac{3}{2}}\left(\frac{\eta_1}{\eta_2} \right)^{\frac{\gamma}{2H}}  \Im\left[H^{(1)}_{\frac{3}{2}+ \frac{\gamma}{2H}}(-k \eta_1) H^{(2)}_{\frac{3}{2}+ \frac{\gamma}{2H}}(-k\eta_2)   \right] \theta(\eta_1-\eta_2). 
\end{align}
It turns out to be sometime convenient to express the retarded Green function in terms of Bessel functions of the first kind
\begin{align}\label{eq:GRfinsimpv2}
	G^R(k;\eta_1,\eta_2) = \frac{\pi}{2} \frac{H^2}{k^3} \left(\frac{z_1}{z_2}\right)^{\nu_\gamma} z_2^{3}\left[Y_{\nu_\gamma}(z_1)J_{\nu_\gamma}(z_2) -J_{\nu_\gamma}(z_1)Y_{\nu_\gamma}(z_2) \right] \theta(z_2-z_1)
\end{align}
where $z_i \equiv - k \eta_i$. 

\paragraph{Keldysh propagator} We now turn our attention to the computation of the Keldysh propagator of the theory from which the observables such as the power spectrum can be derived.
The Keldysh function is obtained from 
\begin{align}
	G^K(k; \eta_1,\eta_2) &= i\int \dd \eta' G^R(k; \eta_1,\eta') \widehat{D}_K(\eta' ) G^A(k; \eta',\eta_2) + (\eta_1 \leftrightarrow \eta_2) \\
	&= i \int \dd \eta' G^R(k; \eta_1,\eta') \widehat{D}_K(\eta' ) G^R(k; \eta_2,\eta') + (\eta_1 \leftrightarrow \eta_2).
\end{align}
Notice that this equation has precisely the structure one would expect for the power spectrum of a field obeying the Langevin equation with source fluctuations possessing a power spectrum $\widehat{D}_K(\eta' )$. Injecting \Eq{eq:GKexp} into the above expressions, we obtain three contributions
\begin{align}
	G^K_1(k; \eta_1,\eta_2) &= i \frac{\beta_1}{H^4}\int_{-\infty}^{\eta_2} \frac{\dd \eta'}{\eta^{\prime4}} G^R(k; \eta_1,\eta') G^R(k; \eta_2,\eta') + (\eta_1 \leftrightarrow \eta_2) \label{eq:GK1}\\
	G^K_2(k; \eta_1,\eta_2) &= i \frac{\beta_4 - \beta_2}{H^2}\int_{-\infty}^{\eta_2} \frac{\dd \eta'}{\eta^{\prime2}} G^R(k; \eta_1,\eta')\left( \partial_{\eta'}^2 - \frac{2}{\eta'}\partial_{\eta'}\right) G^R(k; \eta_2,\eta') + (\eta_1 \leftrightarrow \eta_2) \label{eq:GK2} \\
	G^K_3(k; \eta_1,\eta_2) &= i \frac{\beta_2 k^2}{H^2}\int_{-\infty}^{\eta_2} \frac{\dd \eta'}{\eta^{\prime2}} G^R(k; \eta_1,\eta') G^R(k; \eta_2,\eta') + (\eta_1 \leftrightarrow \eta_2) \,,\label{eq:GK3}
\end{align}
which correspond to the three noise terms appearing in \Eq{eq:GKexp}. Here we assumed $\eta_2 \leq \eta_1$ without loss of generality. The power spectrum is obtained from the coincident time limit $\eta_1 = \eta_2 = \eta_0$.

Let us compute the first contribution \Eq{eq:GK1}. Injecting \Eq{eq:GRfinsimpv2} in \Eq{eq:GK1}, we obtain the Keldysh propagator 
\begin{align}\label{eq:GKeq}
	G^K_1(k;\eta_1,\eta_2) = i\frac{\pi^2 \beta_1}{4 k^3}  (z_1 z_2)^{\nu_\gamma} \bigg\{& Y_{\nu_\gamma}(z_1) Y_{\nu_\gamma}(z_2) A^{(1)}_{\nu_\gamma}(z_2) + J_{\nu_\gamma}(z_1) J_{\nu_\gamma}(z_2) C^{(1)}_{\nu_\gamma}(z_2) \\
	-& \left[J_{\nu_\gamma}(z_1)Y_{\nu_\gamma}(z_2) + J_{\nu_\gamma}(z_2)Y_{\nu_\gamma}(z_1)\right]B^{(1)}_{\nu_\gamma}(z_2) \bigg\} + (1 \leftrightarrow 2)\nonumber
\end{align}
where 
\begin{align}
	A^{(1)}_{\nu_\gamma}(z) &\equiv \int_{z}^{\infty} \dd z' z^{\prime2 - 2{\nu_\gamma}} J^2_{\nu_\gamma}(z')  \label{eq:Fbetaref}\\
	B^{(1)}_{\nu_\gamma}(z) &\equiv \int_{z}^{\infty} \dd z' z^{\prime2 - 2{\nu_\gamma}} J_{\nu_\gamma}(z') Y_{\nu_\gamma}(z') \label{eq:Gbetaref} \\
	C^{(1)}_{\nu_\gamma}(z)&\equiv \int_{z}^{\infty} \dd z' z^{\prime2 - 2{\nu_\gamma}} Y^2_{\nu_\gamma}(z') \label{eq:Hbetaref}
\end{align} 
are complicated functions given explicitly at the end of the Lecture, in \Eqs{eq:Fbeta}, \eqref{eq:Gbeta} and \eqref{eq:Hbeta} respectively. 

\paragraph{Power spectrum} Considering that $\zeta = - H\pi/ f_\pi^2$ where $\pi$ is the canonically normalised field, the reduced power spectrum 
\begin{align}
	\Delta^2_\zeta(k) \equiv \frac{k^3}{2\pi^2}P_\zeta(k) \qquad \mathrm{with} \qquad \langle \zeta_\bmk \zeta_{-\bmk}\rangle = (2\pi)^3 \delta(\bmk + \bmk') P_\zeta(k)
\end{align}
is obtained in the coincident time limit of the Keldysh propagator given in \Eq{eq:GKeq}, such that
\begin{align}\label{eq:PKref}
	P_\zeta(k) = \frac{\pi^2\beta_1}{8 k^3}  \frac{H^2}{f_\pi^4} z^{2{\nu_\gamma}} \Big[& Y^2_{\nu_\gamma}(z)  A^{(1)}_{\nu_\gamma}(z) + J^2_{\nu_\gamma}(z) C^{(1)}_{\nu_\gamma}(z) -2J_{\nu_\gamma}(z)Y_{\nu_\gamma}(z) B^{(1)}_{\nu_\gamma}(z) \Big]\,.
\end{align}
In the super-Hubble regime $z\ll 1$, the power spectrum freezes and
\begin{align}
	\Delta^2_\zeta(k) &= \frac{1}{4} \frac{\beta_1}{H^2}   \frac{H^4}{f_\pi^4}  2^{2\nu_\gamma} \frac{\Gamma\left(\nu_\gamma-1\right)\Gamma\left(\nu_\gamma\right)^2}{\Gamma\left(\nu_\gamma - \frac{1}{2}\right)\Gamma\left(2\nu_\gamma - \frac{1}{2}\right)}\label{eq:PK1SH}
\end{align}
which is indeed dimensionless. Keeping in mind that $\nu_\gamma \equiv \frac{3}{2}+ \frac{\gamma}{2H}$, one can expand this result in the small and large dissipation regime leading to 
\begin{tcolorbox}[%
	enhanced, 
	breakable,
	skin first=enhanced,
	skin middle=enhanced,
	skin last=enhanced,
	before upper={\parindent15pt},
	]{}
	
	\paragraph{Dissipative power-spectrum}
	\begin{align}\label{eq:dissipPk}
		\Delta^2_\zeta(k)  &\propto  \begin{dcases}
			\frac{\beta_1}{H^2} \frac{H^4}{f_\pi^4}  + \mathcal{O}\left(\frac{\gamma}{H}\right), & \gamma \ll H,\\
			\frac{\beta_1}{H^2} \frac{H^4}{f_\pi^4}   \sqrt{\frac{H}{\gamma}} \left[1 + \mathcal{O}\left(\frac{H}{\gamma}\right) \right], & \gamma \gg H.
		\end{dcases}
	\end{align}
	The observational constraint $\Delta^2_\zeta = 10^{-9}$ is easily obtained by imposing hierarchies between the various scales of the problem. Note that if one further imposes thermal equilibrium of the environment such that the fluctuation-dissipation relation holds, the dynamical KMS symmetry imposes  $\beta_1 =2\pi \gamma T$ where $T$ is the environment temperature \cite{Hongo:2018ant}, such that in the large dissipation regime ($\gamma \gg H$)
	\begin{align}
		\Delta^2_\zeta \propto \frac{T}{H}\frac{H^4}{f_\pi^4} \sqrt{\frac{\gamma}{H}}
	\end{align}
	which reproduces the warm inflation expectation \cite{Berera:1995ie, Berera:1995wh, Berera:1999ws, Berera:2008ar, Ballesteros:2023dno, Montefalcone:2023pvh}.
\end{tcolorbox}

The two other contributions given in \Eqs{eq:GK2} and \eqref{eq:GK3} follow accordingly, the details of which can be found in Appendix C of \cite{Salcedo:2024smn}. For completeness, we here just state the result of computation. These two derivative noises also lead to equally valid scale invariant power spectra on super-Hubble scales
\begin{align}
	\Delta^2_\zeta(k) &\supset \begin{dcases}
		\frac{15}{32}(\beta_4- \beta_2)\frac{H^4}{f_\pi^4} 2^{2\nu_\gamma} \frac{\Gamma\left(\nu_\gamma-2\right)\Gamma\left(\nu_\gamma\right)^2}{\Gamma\left(\nu_\gamma - \frac{3}{2}\right)\Gamma\left(2\nu_\gamma - \frac{1}{2}\right)}, & \text{\Eq{eq:GK2}},\\
		\frac{3}{16} \beta_2\frac{H^4}{f_\pi^4} 2^{2\nu_\gamma} \frac{\Gamma\left(\nu_\gamma-2\right)\Gamma\left(\nu_\gamma\right)^2}{\Gamma\left(\nu_\gamma - \frac{3}{2}\right)\Gamma\left(2\nu_\gamma - \frac{3}{2}\right)}, & \text{\Eq{eq:GK3}},
	\end{dcases}
\end{align}
which expand in the large dissipation limit ($\gamma \gg H$) to 
\begin{align}
	\Delta^2_\zeta(k) &\supset \begin{dcases}
		(\beta_4- \beta_2)  \frac{H^4}{f_\pi^4}  \sqrt{\frac{H}{\gamma}}  \left[1 + \mathcal{O}\left(\frac{H}{\gamma}\right) \right], & \text{\Eq{eq:GK2}},\\
		\beta_2 \frac{H^4}{f_\pi^4}  \sqrt{\frac{\gamma}{H}} \left[1 + \mathcal{O}\left(\frac{H}{\gamma}\right) \right], & \text{\Eq{eq:GK3}}.
	\end{dcases}
\end{align}
The obtained contributions may again satisfy the observational constraint $\Delta^2_\zeta = 10^{-9}$ by imposing some hierarchy between the various scales (more stringent for the last contribution due to the $\sqrt{\gamma/H}$ enhancement in the spatial derivative case).

\subsubsection{Primordial non-Gaussianities}

The computation of primordial non-Gaussianities closely follow the flat space example given in \Sec{subsec:flatinter}. Remembering that the curvature perturbations are defined though $\zeta = - H \pi/f_\pi^2$ in the spatially flat gauge, the primordial bispectrum reads 
\begin{align}
	\langle \zeta_{\bmk_1}  \zeta_{\bmk_2}  \zeta_{\bmk_3} \rangle  = - \frac{H^3}{f_\pi^6}	\langle \pi_{\bmk_1}  \pi_{\bmk_2}  \pi_{\bmk_3} \rangle \equiv (2\pi)^3 \delta(\bmk_1 + \bmk_2 + \bmk_3) B(k_1,k_2,k_3).
\end{align} 
Quantities of interest to characterise the non-Gaussian signatures are the amplitude of the signal
\begin{align}
	f_{\mathrm{NL}}(k_1,k_2,k_3) \equiv \frac{5}{6} \frac{B(k_1,k_2,k_3)}{P(k_1) P(k_2) +  P(k_1) P(k_3)  + P(k_2) P(k_3) }\,,
\end{align}
discussed for specific configurations below, and the shape function, already defined in \Eq{eq:shaperef} which we reproduce here for clarity 
\begin{align}
	S(x_2,x_3) \equiv (x_2 x_3)^2 \frac{B(k_1, x_2 k_1, x_3 k_1)}{B(k_1,k_1,k_1)}\,.
\end{align}
Here $x_2 \equiv k_2/k_1$ and $x_3 \equiv k_3/k_1$ are restricted to the region $\max(x_3, 1- x_3) \leq x_2 \leq 1$. One can proceed just as in the flat space case presented in \Sec{subsec:flatB} to evaluate contact bispectra. The generic structure of the integrals is
\begin{align}\label{eq:Bexpref}
	&B(k_1,k_2,k_3) = (-i)^{n_K + n_R + 1}\frac{H^3}{f_\pi^6} \frac{g}{H^{4-n_d}} \int_{-\infty(1 \pm i \epsilon)}^{0^-} \frac{\dd \eta}{\eta^{4-n_d}} \nonumber \\
	&\qquad \widehat{\mathcal{D}}(\{\bmk_i\}, \partial_\eta)\left[ G^{K/R}(k_1, 0, \eta) G^{K/R}(k_2, 0, \eta) G^R(k_3, 0, \eta) + 5~\mathrm{perms}. \right]
\end{align}
where $n_K$ counts the number of Keldysh progagators, $n_R$ the number of retarded ones and $n_d$ the number of (spatial and temporal) derivatives. $\widehat{\mathcal{D}}(\{\bmk_i\}, \partial_\eta)$ is a differential operator schematically representing the $n_d^{\mathrm{th}}$ spatial and temporal derivatives acting on the propagators. Note that there is always at least one $G^R$ due to the at least linearity in $\pia$ inherited from \Eq{eq:norm}. The bulk-to-boundary propagators are 
\begin{align}\label{eq:GRB2b}
	G^R(k, 0, \eta) = -\frac{H^2}{2k^3}  z^{3} \left(\frac{z}{2} \right)^{ - \nu_\gamma}\Gamma(z) J_{\nu_\gamma}(z)
\end{align}
and 
\begin{align}
	G^K(k, 0,\eta) &= - i\frac{\pi}{4k^3} \beta_1 \left(2 z \right)^{\nu_\gamma} \Gamma(\nu_\gamma) \left[ Y_{\nu_\gamma}(z)A^{(1)}_{\nu_\gamma}(z)- J_{\nu_\gamma}(z)B^{(1)}_{\nu_\gamma}(z)\right]\,,
\end{align} 
where we expressed the quantities in terms of $z = - k \eta$ and  $\nu_\gamma = \frac{3}{2} + \frac{\gamma}{2H}$, and $A^{(1)}_{\nu_\gamma}(z)$ and $B^{(1)}_{\nu_\gamma}(z)$ are defined in \Eqs{eq:Fbetaref} and \eqref{eq:Gbetaref} respectively. It is also useful to consider 
\begin{align}
	\partial_\eta G^R(k, 0,\eta) = \frac{H^2}{2k^2} z^2 \left(\frac{z}{2} \right)^{-\nu_\gamma} \Gamma (\nu_\gamma) \left[ z J_{\nu_\gamma -1}(z)- (2 \nu_\gamma-3) J_{\nu_\gamma}(z)\right]
\end{align}
and
\begin{align}
	\partial_\eta G^K(k, 0,\eta)&= i\frac{\pi}{4k^2}\beta_1\left(2 z \right)^{\nu_\gamma-1} \Gamma(\nu_\gamma)  \bigg\{\left[-z Y_{\nu_\gamma+1}(z)+2\nu_\gamma Y_{\nu_\gamma}(z) + z Y_{\nu_\gamma-1}(z)\right]A^{(1)}_{\nu_\gamma}(z) \nonumber \\
	& \left[-z J_{\nu_\gamma+1}(z)+2\nu_\gamma J_{\nu_\gamma}(z) + z J_{\nu_\gamma-1}(z)\right]B^{(1)}_{\nu_\gamma}(z)\bigg\}.
\end{align}
It is now a matter of evaluating the time integral of \Eq{eq:Bexpref} injecting the above expressions for the propagators. This task is analytically hard in full generality, therefore, below we mostly rely on numerical integration and only derive analytical results in some specific regimes. 


\subsubsection*{Numerical results}

\begin{figure}[tbp]
	\sidesubfloat[]{\includegraphics[width=0.4\textwidth]{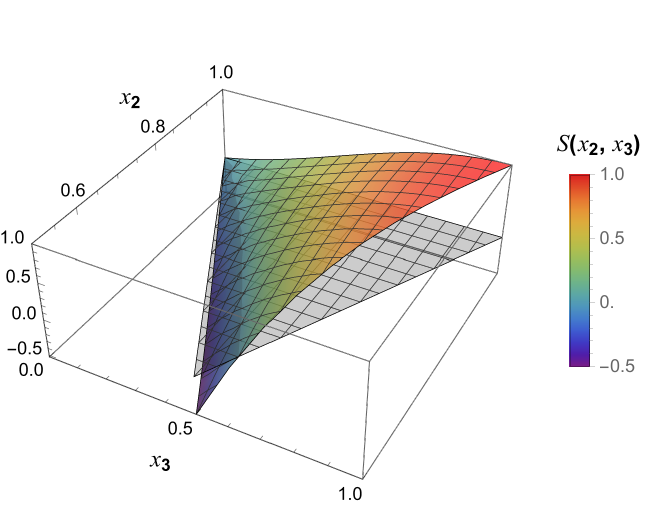}\label{fig:c}}
	\hfil
	\sidesubfloat[]{\includegraphics[width=0.4\textwidth]{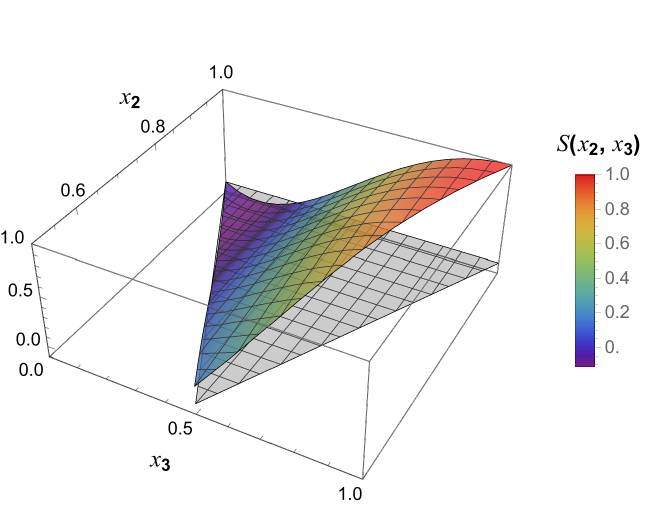}\label{fig:d}}
	
	\medskip
	\sidesubfloat[]{\includegraphics[width=0.4\textwidth]{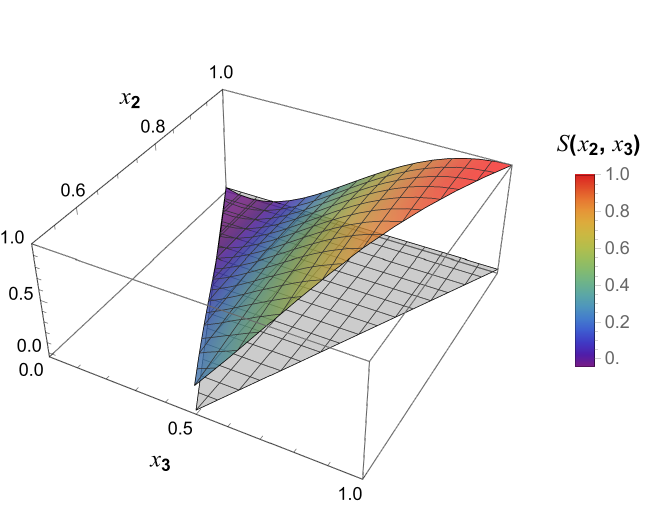}\label{fig:e}}
	\hfil
	\sidesubfloat[]{\includegraphics[width=0.4\textwidth]{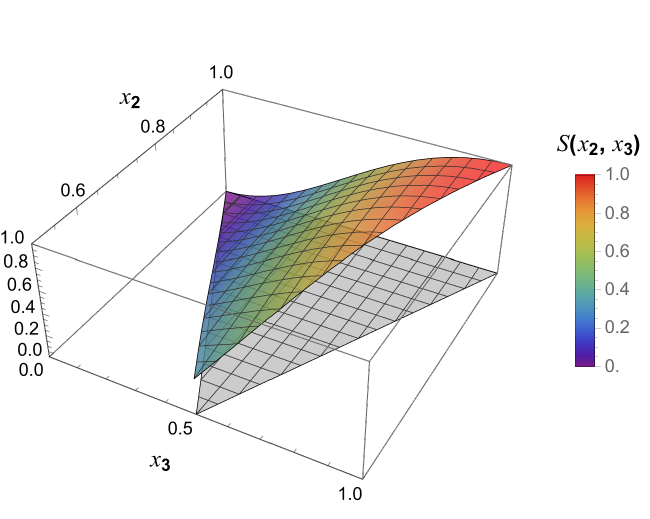}\label{fig:f}}
	\caption{Shapes of the bispectrum at large dissipation $\gamma = 5H$. In this regime, the signal peaks in the equilateral configuration $x_2 = x_3 = 1$. Consistency relations hold in the squeezed limit $x_3 \ll x_2 = 1$. \textbf{a.} $\left(\partial_i \pir \right)^2 \pia$ operator; \textbf{b.} $\dot{\pi}^2_{r} \pia$ operator; \textbf{c.} $\dot{\pi}_{r} \pia^2$ operator; \textbf{d.} $\pia^3$ operator.}
	\label{fig:shapes_high}
\end{figure}

\begin{figure}[tbp]
	\centering
	\sidesubfloat[]{\includegraphics[width=0.4\textwidth]{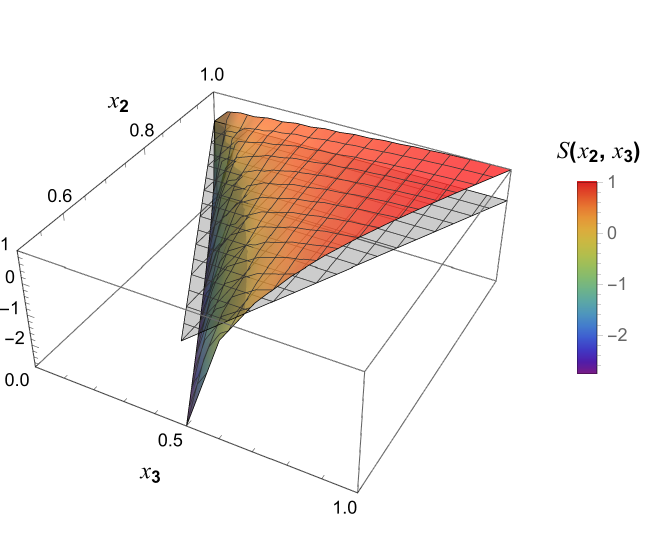}\label{fig:cbis}}
	\hfil
	\sidesubfloat[]{\includegraphics[width=0.4\textwidth]{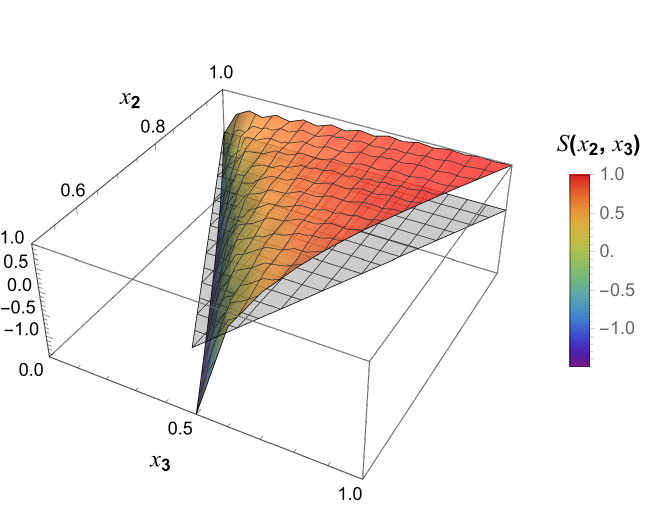}\label{fig:dbis}}
	
	\medskip
	\sidesubfloat[]{\includegraphics[width=0.4\textwidth]{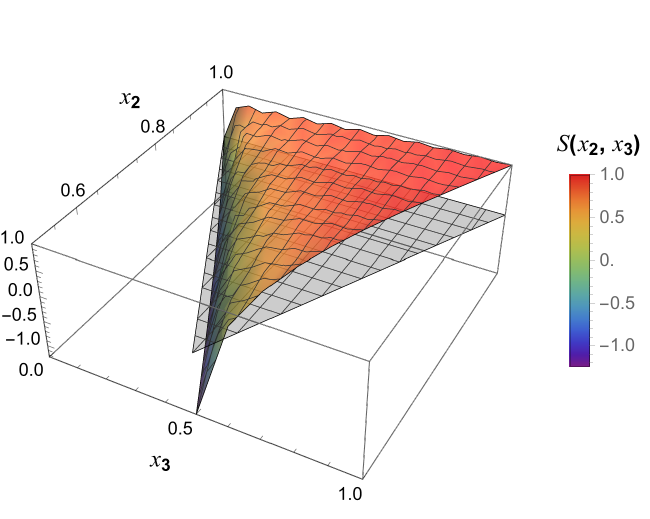}\label{fig:ebis}}
	\hfil
	\sidesubfloat[]{\includegraphics[width=0.4\textwidth]{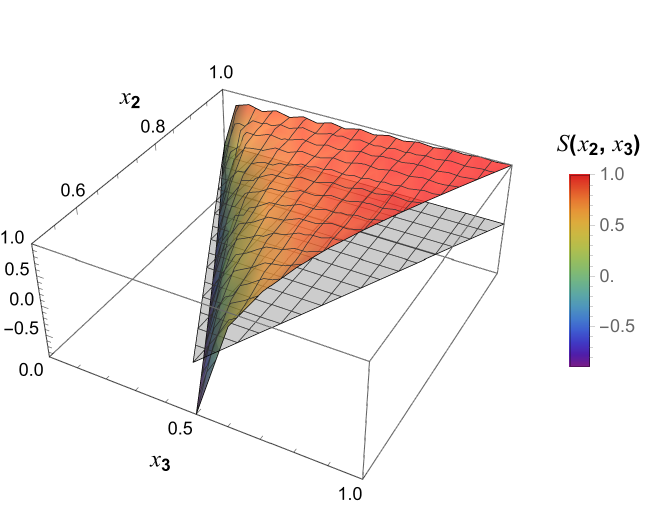}\label{fig:fbis}}
	\caption{Shapes of the bispectrum at low dissipation $\gamma= 0.001H$. In this regime, the signal peaks near folded configurations $x_2 + x_3 = 1$. Consistency relations still hold in the squeezed limit $x_3 \ll x_2 = 1$. The tiny oscillations are artefacts of the numerical integration over a finite range. 
		\textbf{a.} $\left(\partial_i \pir \right)^2 \pia$ operator; \textbf{b.} $\dot{\pi}^2_{r} \pia$ operator; \textbf{c.} $\dot{\pi}_{r} \pia^2$ operator; \textbf{d.} $\pia^3$ operator.}
	\label{fig:shapes_low}
\end{figure}

In this section, we summarise the main phenomenological implications of the contact bispectra generated by the cubic operators of \Eqs{eq:L13}, \eqref{eq:L23} and \eqref{eq:L33}.
Many of the sixteen cubic operators appearing in \Eqs{eq:L13}, \eqref{eq:L23} and \eqref{eq:L33} lead to similar signatures for the contact bispectrum. Hence, we only display the results for a subset of these operators.
The shapes of the contact bispectrum generated by these four operators are displayed in \Figs{fig:shapes_high} and \ref{fig:shapes_low} (the other operators essentially follow the same trend). Just as for the flat space case, different behaviours emerge in the large ($\gamma \gg H$) and small ($\gamma \ll H$) dissipation regime. While the former peaks in the equilateral configuration as already noted in \cite{LopezNacir:2011kk}, the latter reaches an extremum near the folded region. 

\begin{figure}[tbp]
	\centering
	\includegraphics[width=0.6\textwidth]{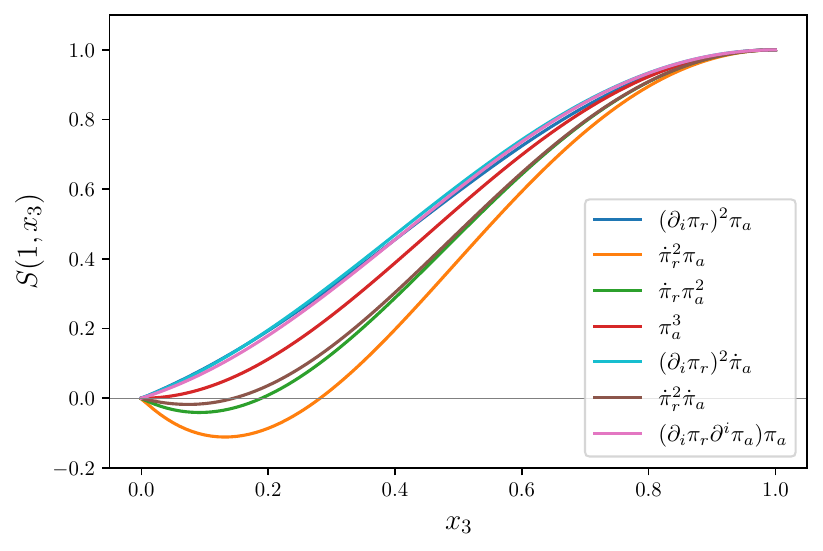}
	\caption{Shape function along the direction $x_2 = 1$ for different contributions to the contact bispectrum at large dissipation $\gamma = 5H$. Consistency relations ensure the signal vanishes in the squeezed limit $x_3 \ll 1$. }
	\label{fig:squeezed}
\end{figure}

This smoking gun of open dynamics might seem degenerate with other classes of models that also lead to a signal in the folded triangles such as non-Bunch Davies initial states \cite{Holman:2007na, Chen:2006nt, Meerburg:2009ys, Agullo:2010ws, Ashoorioon:2010xg, Agarwal:2012mq, Ashoorioon:2013eia,  Albrecht:2014aga, Green:2020whw}. A crucial difference, which appears in our numerical treatment and is confirmed analytically below, is that dissipation regularises the divergence by smoothing the peak and displacing it from the edge of the triangular configurations, leading to finite values of the bispectrum for any physical configuration.   
In particular, it implies no divergence in the squeezed limit of the bispectrum $k_1 \simeq k_2 \gg k_3$, which is displayed in \Fig{fig:squeezed}. Small values of $\gamma/H$ may eventually lead to an intermediate peak due to the regularised folded singularity, yet consistency relations hold \cite{Maldacena:2002vr, Creminelli:2012ed} and the squeezed limit goes to zero because of the symmetries of the theory. Notice that operators such as $\pia^3$ follow this trend despite what one may have naively thought in the absence of derivatives. This is because of the modified propagators compared to the free theory, which for instance ensure IR convergence.	

\begin{figure}[tbp]
	\begin{minipage}{6in}
		\centering
		\raisebox{-0.5\height}{\includegraphics[width=.48\textwidth]{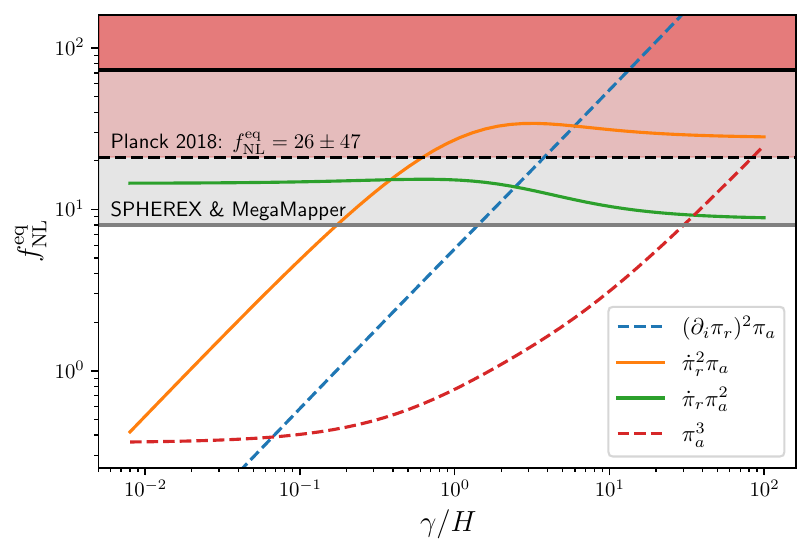}}
		\raisebox{-0.5\height}{\includegraphics[width=.48\textwidth]{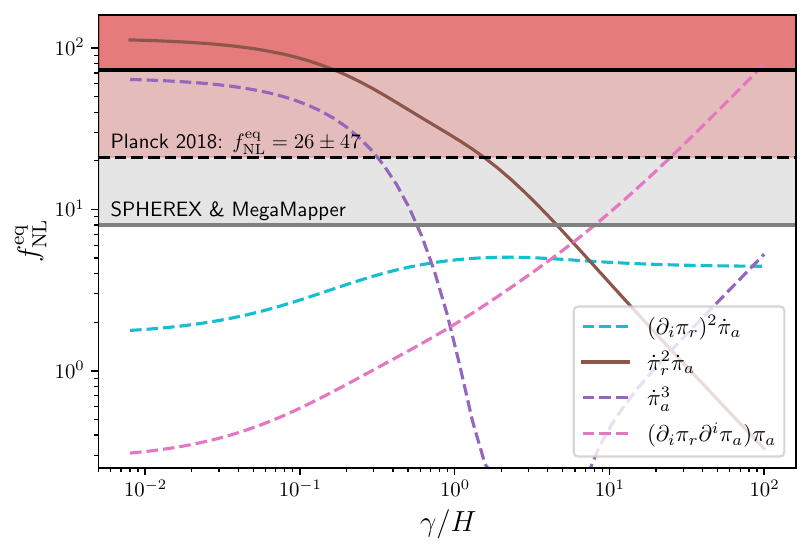}}
	\end{minipage}
	\caption{Amplitude in the equilateral configuration as a function of the dissipation parameter $\gamma/H$. Continuous (dashed) lines are positive (negative) values. \textit{Left:} The four cubic operators appearing in the matching with \cite{Creminelli:2023aly} performed in \Sec{sec:matching}. All the curves well reproduce the results from \cite{Creminelli:2023aly}. The otherwise arbitrary values of the EFT coefficients were chosen to qualitatively reproduce the results from \cite{Creminelli:2023aly}. \textit{Right:} New operators of the open EFToI that may lead to specific signatures on the non-Gaussian signal. Again, numerical values of the EFT coefficients are arbitrary, chosen to be comparable to the \textit{Left} panel. All scalings with $\gamma$ are consistent with the heuristic estimates of \Sec{subsec:energy}. }
	\label{fig:fnleq}
\end{figure}

The amplitude in the equilateral configuration is controlled by 
\begin{align}
	f_{\mathrm{NL}}^{\mathrm{eq}} \equiv \frac{10}{9} \frac{k^6}{(2\pi)^4}\frac{B(k,k,k)}{\Delta_\zeta^4}.
\end{align} 
In \Fig{fig:fnleq}, we display the dependence of $f_{\mathrm{NL}}^{\mathrm{eq}}$ on the dissipation parameter $\gamma/H$ from numerical integration of the bispectrum $B(k,k,k)$ for the operators considered above. One can use observational constraints $f_{\mathrm{NL}}^{\mathrm{eq}} = 26 \pm 47$ from \cite{Planck:2019kim} to place bounds on the EFT parameters in the large dissipation regime. For instance, a numerical fit of the $(\partial_i \pir)^2\pia$ contributions leads to $f_{\mathrm{NL}}^{\mathrm{eq}} \simeq - \gamma/(4H)$. It naively implies that $\gamma / H < 80$ at $68\%$ confidence. Of course, this is more of a proof of principle than a realistic estimate due to the cumulative effects of different cubic operators that cannot be disentangled one from another. Yet, it demonstrates how this class of model can be confronted to data. Such observational bounds could tighten thanks to future LSS experiments such as SPHEREX \cite{SPHEREx:2014bgr} or MegaMapper \cite{Cabass:2022epm, Braganca:2023pcp}, further constraining on this class of models.  

\subsubsection*{Analytical discussion}

An interesting question is the fate of folded singularities in an expanding background. In general, a folded singularity appears when two modes resonate with a third for an infinite amount of time. For this reason, the origin of the divergence can be derived from the early time oscillating behaviour of the propagators. Hence, let us consider the simplest cubic operator $\pia^3$ and the cubic bispectrum it generates
\begin{align}
	&B(k_1,k_2,k_3) = - 6 \frac{H^3}{f_\pi^6} \frac{\delta_1}{f_\pi^2}  \int_{-\infty(1 \pm i \epsilon)}^{0^-} \frac{\dd \eta}{H^4\eta^{4}} G^{R}(k_1, 0, \eta) G^{R}(k_2, 0, \eta) G^R(k_3, 0, \eta). 
\end{align}
Injecting \Eq{eq:GRB2b} into the above expression and expanding at early time in the sub-Hubble regime when $\eta \rightarrow - \infty$, we observe the time integral reduces to a collection of terms of the form
\begin{align}
	\left| \int \dd \eta \frac{\ee^{-i \left(\pm k_1 \pm k_2 \pm k_3 \right)\eta}}{\eta^{1 + \frac{3}{2}\frac{\gamma}{H}}}   \right| < \infty \qquad \mathrm{when} \qquad \frac{\gamma}{H} >0.
\end{align}
This is the analogue of the flat space case discussed in \Eq{eq:M4int} from which we concluded the resolution of the singularity in presence of non-vanishing dissipation. In the vanishing limit, we recover the log-folded singularity familiar to non-Bunch Davies initial states \cite{Holman:2007na, Chen:2006nt, Meerburg:2009ys, Agullo:2010ws, Ashoorioon:2010xg, Agarwal:2012mq, Ashoorioon:2013eia,  Albrecht:2014aga, Green:2020whw}. Contrarily to this dramatically uncontrolled divergence along the folded region of \Fig{fig:triangles}, dissipation tames the peak and displaces it from the edge of the physical region such that the folded singularity remains resolved and under perturbative control for any value of the kinematic variables.

One may wonder if there are implications of the above mentioned singularity on the squeezed limit of the bispectrum $k_1 \simeq k_2 \gg k_3$. As long as $\pir$ transforms as the usual pseudo-Goldstone boson, which non-linearly realises time-translation and boosts, the standard arguments \cite{Creminelli:2004yq, Cheung:2007sv, Creminelli:2012ed} should hold. As a consequence of the singularity being resolved, the curvature perturbation bispectrum still goes to zero in the squeezed limit with an eventual intermediate peak in the small dissipation regime, when $\gamma \ll H$.  

\subsubsection{Matching with a gauge-inflation model}\label{sec:matching}

For an EFT construction to be of interests, it has to encompass a class of physically motivated models. For this reason, we consider in this section the matching of the open EFT of inflation to a concrete ``ultra-violet'' model constructed and studied in \cite{Creminelli:2023aly}. This is not the first model of dissipative dynamics, but it has the distinguishing feature of leading to a \textit{local} low-energy dynamics around and below the Hubble scale. This is related to the fact that the window of instability for particle production involves wavelengths that are parametrically sub-Hubble. 
In addition to the inflaton field $\phi$, the model features a massive scalar field $\chi$ with a softly-broken $U(1)$ symmetry. The action of the model is given by
\begin{align}\label{eq:actionmatch}
	S= \int \dd^4 x &\sqrt{- g} \bigg[ \frac{1}{2} M_{\mathrm{Pl}}^2  R - \frac{1}{2} \left(\partial \phi \right)^2 - V(\phi) - \left|\partial \chi \right|^2 + M^2 \left| \chi \right|^2 \nonumber \\
	- &\frac{\partial_\mu \phi}{f}\left( \chi \partial^\mu \chi^* - \chi^* \partial^\mu \chi\right) - \frac{1}{2}m^2\left(\chi^2 + \chi^{*2}\right)\bigg].
\end{align}
The last term in \Eq{eq:actionmatch} breaks the $U(1)$ symmetry $\chi \rightarrow \ee^{i \alpha} \chi$ for $\alpha \in \mathbb{R}$. An important parameter for the dynamics of the system is $\rho \equiv \dot{\phi}_0(t)/f$, which controls the mixing of $\phi$ with $\chi$. Here $\dot{\phi}_0(t)$ is the time derivative of the inflaton background $\phi_0(t)$. The hierarchy of scales for which the treatment of \cite{Creminelli:2023aly} is valid is
\begin{align}
	f \gg \rho \gtrsim M \gg m \gg H.
\end{align}
The model exhibits a narrow instability band in the sub-Hubble regime, during which particle production occurs. The amount of particle production is controlled by the dimensionless parameter
\begin{align}
	\xi \simeq \frac{m^4}{8 H \rho M^2} \gtrsim \mathcal{O}(1).
\end{align}
The inflaton fluctuations experience a dissipative dynamics due to the presence of an environment of $\chi$ particles generated by the instability. The inflaton dynamics is effectively described in terms of a non-linear Langevin equation \cite{Creminelli:2023aly}
\begin{align}\label{eq:Paolo}
	\pi'' + \left( 2H + \gamma\right)a\pi' - \partial_i^2 \pi &\simeq \frac{\gamma}{2\rho f} \left[ \left( \partial_i \pi\right)^2 - 2 \pi \xi \pi^{\prime2}\right] - \frac{a^2m^2}{f} \left(1+2\pi\xi \frac{\pi'}{a \rho f} \right) \delta \mathcal{O}_S\,,
\end{align}
where we neglected the inflaton potential contribution $a^2 V'' \pi$ which is slow-roll suppressed. The dissipation parameter $\gamma$ (with units of mass) is determined by the following combination of microphysical parameters
\begin{align}\label{eq:gammaCrem}
	\gamma \simeq \frac{\xi m^4}{\pi M f^2} \ee^{2 \pi \xi}.
\end{align}
The effect of noise is captured in terms of its two- and three-point statistics
\begin{align}
	\langle  \delta \mathcal{O}_S(\bmk, \eta) \delta \mathcal{O}_S(\bmk',\eta')\rangle &\simeq (2\pi)^3 \delta(\bmk + \bmk') \delta(\eta - \eta')H^4 \eta^4 \nu_{\mathcal{O}}\,, \\
	\langle  \delta \mathcal{O}_S(\bmk, \eta) \delta \mathcal{O}_S(\bmk',\eta')  \delta \mathcal{O}_S(\bmk'',\eta'')\rangle &\simeq (2\pi)^3 \delta(\bmk + \bmk' +\bmk'') \delta(\eta - \eta') \delta(\eta - \eta'')H^8 \eta^8 \nu_{\mathcal{O}^3} \,,\label{eq:NGnoiseCrem}
\end{align}
with the amplitudes of the noises being controlled by 
\begin{align}
	\nu_{\mathcal{O}} \simeq \frac{M}{m} \frac{\ee^{4\pi \xi}}{4\pi^2} \qquad \mathrm{and} \qquad \nu_{\mathcal{O}^3} \simeq \frac{\ee^{6\pi \xi}}{\pi^2 m^2} .
\end{align}
The parameters at play are summarized in the first line of Table \ref{tab:matching}.

As we demonstrate below, the low-energy dynamics of this model is equivalently described in terms of
\begin{align}\label{eq:Seffmatch}
	\quad S_{\mathrm{eff}} &= \int \dd^4 x \Big[ a^2 \pi'_{r} \pi'_{a} - c_{s}^{2} a^2 \partial_i \pir  \partial^i \pia   -  a^3 \gamma \pi'_{r} \pia + i \beta_1 a^4 \pia^2 \nonumber \\
	+ &\frac{ \left(8\gamma_2 - \gamma\right)}{2f^{2}_{\pi}}  a^2\pir^{\prime2} \pia	+ \frac{\gamma}{2f_{\pi}^{2}} a^2	\left(\partial_i \pir \right)^2 \pia - 2i\frac{\beta_{5}}{f_\pi^2} a^3 \pir^{\prime} \pia^2 +\frac{\delta_1}{f_\pi^2} a^4 \pia^3  \Big],
\end{align}
upon matching the EFT coefficients to the microphysical parameters of the model through the relations of Table \ref{tab:matching}. 

\begin{table}
	\begin{center}
		\begin{tabular}{ |c|| P{1.2cm} P{1.2cm} P{1.2cm} P{1.2cm} P{1.2cm} P{1.2cm} P{1.2cm} |  }
			\hline 
			& \multicolumn{7}{|c|}{	$\Bigg.$
				\textit{Parameters:}} \\               
			\hline 
			$\Bigg.$UV completion \cite{Creminelli:2023aly} & $M$ & $m$ & $f$ & $\rho$ &  $\xi$ & $\gamma$ & $\nu_{\mathcal{O}}/\nu_{\mathcal{O}^3}$ \\
			\hline
			$\Bigg.$Open EFT \eqref{eq:Seffmatch} & $f^2_\pi $ & $c_s$ & $\gamma$ & $\gamma_{2}$ &  $\beta_1  $ & $\beta_{5}$ & $\delta_1 $ \\
			\hline
		\end{tabular}
		\vspace{0.2in}
		
		\begin{tabular}{ |P{4cm} P{4cm} |  }
			\hline 
			$\Bigg.$ \textit{Matching:} & $f^2_\pi = \rho f $ \\
			$\Bigg.$ $c_s = 1$ 	&	$\gamma = \frac{\xi m^4}{\pi M f^2} \ee^{2 \pi \xi}$ \\
			$\Bigg.$ $8\gamma_2 =\left(1-2 \pi \xi\right) \gamma$ & $\beta_1 = \frac{\nu_{\mathcal{O}}}{2 \rho f} \frac{m^4}{f^2} $ \\
			$\Bigg.$ $\beta_{5}=  2\pi\rho f\xi \beta_1$ & 		 
			$6\delta_1 = \rho f \frac{m^6}{f^3} \nu_{\mathcal{O}^3} $  \\
			\hline
		\end{tabular}
	\end{center}
	\caption{Matching of the EFT coefficients appearing in \Eq{eq:Seffmatch} to the microphysical parameters of \cite{Creminelli:2023aly}. The matching is obtained either at the level of the power spectrum and contact bispectra or by deriving the Langevin equation for the fluctuations.}
	\label{tab:matching}
\end{table}


\subsubsection*{Langevin equation}\label{subsec:Langevin}

The matching can be made by deriving a Langevin equation similar to \Eq{eq:Paolo} from the effective functional \eqref{eq:Seffmatch}. The comparison of the two expressions allows us to relate EFT coefficients to microphysical parameters. 

\paragraph{Linear dynamics.}

Let us first consider the linear part. The diagonal of the density matrix is given by a path integral over the $\pir$ component with boundary condition $\pir(\eta_{0})=\pi$ and the $\pia$ component with boundary condition $\pia(\eta_{0})=0$, leading to
\begin{equation}
	\rho[\pi,\pi]=\int_{\text{BD}}^{\pi}\mathcal{D}\pir\int_{\text{BD}}^{0}\mathcal{D}\pia e^{iS_{\text{eff}}[\pir,\pia]}.
\end{equation}
We start by considering the quadratic functional obtained by setting $\widehat{D}_{K}=0$, that is
\begin{equation}\label{eq:noiselessfree}
	S_{\text{eff}}[\pir,\pia] =  \int d^{4}x a^{4}\left(\frac{\pir'\pia'}{a^{2}} - c_{s}^{2}  \frac{\partial_i \pir  \partial^i \pia}{a^{2}} -   \gamma  \frac{\pir' \pia }{a}\right).
\end{equation}
Integrating by parts, we can write the influence functional as linear in $\pia$ (the boundary terms vanish because they are linear in $\pia(\eta_0)$) such that
\begin{equation}
	S_{\text{eff}}[\pir,\pia] =  \int d^{4}x a^{4}\left[ - \frac{\partial_{\eta}\left(a^{2}\pir'\right)}{a^{4}} + c_{s}^{2}  \frac{\partial_i^{2} \pir}{a^{2}} -   \gamma  \frac{\pir'}{a}\right] \pia.
\end{equation}
We can now consider the path integral over $\pia$ as enforcing the constraint
\begin{equation}
	\rho[\pi,\pi]=\int_{\text{BD}}^{\pi}\mathcal{D}\pir\;\delta\left[ - \frac{\partial_{\eta}\left(a^{2}\pir'\right)}{a^{4}} + c_{s}^{2}  \frac{\partial_i^{2} \pir}{a^{2}} -   \gamma  \frac{\pir'}{a}\right].
\end{equation}
This greatly simplifies the calculation of $n$-point functions of the $\pir$ component. If we want to compute the noise-less correlation functions of $\pir$, it is then enough to solve the deterministic equation of motion defined by the constrain, that is
\begin{equation}
	\pi_{\bfk}^{\prime \prime} + \left( 2H + \gamma\right)a \pi_{\bfk}^{\prime}+c_{s}^{2}k^{2}\pi_{\bfk}=0,
\end{equation}
which defines the differential operator $\widehat{D}_{R}$.
Comparing with \Eq{eq:Paolo}, we see that we were right to call $\gamma$ the same in both equations and that $c_{s}=1$. 

The derivation of the Langevin equation from the open effective functional relies on rewriting all terms to be linear in $\pia$. When we introduce the quadratic terms in $\pia$ like $ i {\beta}_{1}\pia^{2}$, this rewriting cannot be achieved by integration by parts. Rather, we have to introduce an auxiliary Gaussian field $\Os$ to rewrite quadratic terms in $\pia$ in a linear form. This procedure is known as the Hubbard–Stratonovich transformation \cite{Stratonovich1957,Hubbard1959}. We here focus on the term $ i {\beta}_{1}\pia^{2}$, which is the one usually  considered in the literature. This term can be written as the outcome of a path integral over a Gaussian field $\Os$
\begin{equation}\label{eq:HStrick}
	\exp\left(-\int d^{4}x a^{4}\beta_{1}\pia^{2}\right)=\mathcal{N}_{0}\int\mathcal{D}[\Os]\;\exp\left[\int d^{4}x a^{4}\left(-\frac{\Os^{2}}{4\beta_{1}}+i\Os\pia\right)\right],
\end{equation}
with $\mathcal{N}_{0}$ being the normalisation constant. In this way, the path integral that defines the density matrix of the system is linear in $\pia$
\begin{align}
	\rho[\pi,\pi]&=                \mathcal{N}_{0}\int\mathcal{D}[\Os]\int_{\text{BD}}^{\pi}\mathcal{D}\pir\int_{\text{BD}}^{0}\mathcal{D}\pia  \\
	& \qquad \times \exp\left\{\int d^{4}x a^{4}\left[-\frac{\Os^{2}}{4\beta_{1}}+i\left(-\widehat{D}_{R}\pir+\Os\right)\pia\right]\right\},\nonumber
\end{align}
from which we obtain the Langevin equation 
\begin{equation}\label{eq:LinearLangevin}
	\pi''_{\bfk} +\left( 2H + \gamma\right)a\pi_{\bfk}'+c_{s}^{2}k^{2}\pi_{\bfk}=a^2 \Os(\bfk, \eta).
\end{equation}
The new variable $\Os$ behaves as a Gaussian field, getting a non-vanishing two-point function under the path integral
\begin{equation}\label{eq:noisecontactv3}
	\langle\Os(\bfk,\eta)\Os(\bfk',\eta')\rangle=\frac{2\beta_{1}}{a^{4}(\eta)}\delta(\eta-\eta')(2\pi)^3\delta(\bfk+\bfk').
\end{equation}
A matching of $\beta_1$ with the microphysical parameters of \cite{Creminelli:2023aly} is possible by comparing the Gaussian statistics of the noise, from which we recover the above result. Alternatively, the linear Langevin equation \Eqs{eq:LinearLangevin} contains all the information needed to derive the power spectrum. One can indeed solve the Langevin equation using the convolution of the retarded Green function with the noise and then compute the power spectrum from the inhomogeneous solution
\begin{equation}\label{eq:GRxifree}
	\pi_{\bfk}(\eta)=\int_{-\infty}^{\eta}d\eta'a^{4}(\eta')G^R(k;\eta,\eta')\Os(\bfk,\eta').
\end{equation}
The two-point function of the noise given in \Eq{eq:noisecontactv3} sources the two-point function of the fluctuations through
\begin{align}
	&\langle\pi_{\bfk_{1}}(\eta_{1})\pi_{\bfk_{2}}(\eta_{2})\rangle=\int_{-\infty}^{\eta_{1}}d\eta_{1}'\int_{-\infty}^{\eta_{2}}d\eta_{2}'a^{4}(\eta'_{1})a^{4}(\eta'_{2})\nonumber \\
	&\qquad \qquad  G^{R}(k_1,\eta_{1},\eta'_{1})G^{R}(k_2, \eta_{2},\eta'_{2})\langle\Os(\bfk_{1},\eta'_{1})\Os(\bfk_{2},\eta'_{2})\rangle,
\end{align}
which obviously reproduces the above result 
\begin{align}
	\langle\pi_{\bfk_{1}}(\eta_{1})\pi_{\bfk_{2}}(\eta_{2})\rangle&=(2\pi)^3\delta(\bfk_{1}+\bfk_{2})\left[\beta_{1}\int_{-\infty}^{\eta_{1}} \frac{d\eta}{H^{4}\eta^{4}}G^{R}(k_1,\eta_{1},\eta)G^{R}(k_2, \eta_{2},\eta) +(\eta_{1}\leftrightarrow\eta_{2})\right].
\end{align}
As shown above, the matching between the previous results from  \Eq{eq:GK1} and \cite{Creminelli:2023aly} is obtained for
\begin{equation}
	\beta_{1}=\frac{\nu_{\mathcal{O}}}{2\rho f}\frac{m^{4}}{f^{2}}.
\end{equation}

\paragraph{Interacting theory.}

The Langevin equation formalism can be extended to include the interaction terms of the influence functional \eqref{eq:Seffref}.

The terms of \Eq{eq:Seffmatch} linear in $\pia$ are
\begin{align}
	S_{\text{eff}} \supset  \int \dd^4 x \Big[\frac{ \left(8\gamma_2 - \gamma\right)}{2f^{2}_{\pi}}    a^2\pir^{\prime2} \pia	+ \frac{\gamma}{2f_{\pi}^{2}} a^2	\left(\partial_i \pir \right)^2 \pia\Big].
\end{align}
These terms being already linear in $\pia$, their inclusion in the Langevin equation is straightforward, leading to
\begin{equation}
	\pi^{\prime \prime}+\left( 2H + \gamma\right)a\pi'-\partial_{i}^{2}\pi=a^2 \Os+\frac{ \left(8\gamma_2 - \gamma\right)}{2f^{2}_{\pi}}  \pi^{\prime2}+\frac{\gamma}{2f_{\pi}^{2}} 	\left(\partial_i \pi \right)^2.
\end{equation}
We can then compare with \Eq{eq:Paolo}, which leaves
\begin{equation}
	\rho f=f_{\pi}^{2}\;,\qquad \mathrm{and}\qquad \;8\gamma_2 =\left(1-2 \pi \xi\right) \gamma.
\end{equation}

The next term we need to consider is     
\begin{equation}
	S_{\text{eff}}\supset - 2i\frac{\beta_{5}}{f_\pi^2}  \int d^{4}x a^3 \pir^{\prime} \pia^2.
\end{equation}
This term can be included into the Langevin equation as a coupling between the noise and the system's variable $\pi$. To achieve this task, we first need to rewrite the interactions as being proportional to one of the quadratic terms in $\pia$. Assuming the quadratic noise to be much stronger than the noise-system coupling, we can work in perturbation theory. It leads to a modified Hubbard–Stratonovich trick that reads
\begin{align}
	&\exp\left[-\int d^{4}x a^{4}\left(\beta_{1}-2 \frac{\beta_{5}}{f_\pi^2}  \frac{\pir'}{a}\right)\pia^{2}\right]=
	\mathcal{N}(\pir')\int\mathcal{D}[\Os]\;\\
	&\qquad \qquad \times \exp\Bigg\{\int d^{4}x a^{4}\left[-\frac{\Os^{2}}{4\beta_{1}}+i\lambda(\pir')\Os\pia\right]\Bigg\}\nonumber
\end{align}
with
\begin{equation}
	\lambda^{2}(\pir)=1-2\frac{\beta_{5}}{f_\pi^2 \beta_{1}} \frac{\pir'}{a}.
\end{equation}
Note that the full Hubbard–Stratonovich trick implies that the normalisation of the path integral over $\Os$ depends on $\pir'$. Consequently, to follow this procedure we have to take the tree-level approximation where we drop the dependence on $\pir'$ from the normalisation of the integral and expand the square root in powers of $\lambda(\pir')$, leading to
\begin{align}
	&\exp\left[-\int d^{4}x a^{4}\left(\beta_{1}-2 \frac{\beta_{5}}{f_\pi^2}  \frac{\pir'}{a}\right)\pia^{2}\right]=\mathcal{N}\int\mathcal{D}[\Os]\; \\
	& \qquad \qquad \times \exp\Bigg[\int d^{4}x a^{4} \left(-\frac{\Os^{2}}{4\beta_{1}}+i\Os\pia-i\frac{\beta_{5}}{f_\pi^2 \beta_{1}} \frac{\pir'}{a}\Os\pia\right)\Bigg].\nonumber
\end{align}
This approach generates a perturbativity condition that is similar to demanding that the three-point function is smaller than the corresponding two-point signal, that is
\begin{align}
	\frac{2 \beta_{5}\pir' \pia^2}{f_{\pi}^{2} a \beta_{1} \pia^2}\ll1.
\end{align}
At last, we can include the noise-system coupling obtained on the right-hand side of the Langevin equation, leading to
\begin{equation}
	\pi^{\prime \prime}+\left( 2H + \gamma\right)a\pi'-\partial_{i}^{2}\pi=a^2 \Os- \frac{\beta_{5}}{f_\pi^2 \beta_{1}} a \pi'\Os. 
\end{equation}
Comparing with \Eq{eq:Paolo}, we recover
\begin{align}
	\beta_5 = 2\pi\rho f\xi \beta_1.
\end{align}

Finally, to recover the non-Gaussian statistics of the noise found in \cite{Creminelli:2023aly}, we consider 
\begin{equation}
	\quad S_{\mathrm{eff}} = \frac{\delta_1}{f_\pi^2} \int \dd^4 x  a^4 \pia^3.
\end{equation}
The terms that are cubic in $\pia$ can also be included at leading order into the Langevin equation by modifying the Hubbard–Stratonovich trick. These terms lead to quadratic corrections in the noise appearing on the right-hand side of the Langevin equation. As we will see below, these corrections mimic non-Gaussian statistics of the noise. Explicitly, the modification of the Hubbard–Stratonovich trick leads to 
\begin{align}
	&\exp\left[-\int d^{4}x a^{4}\left(\beta_{1}+i \frac{\delta_1}{f_\pi^2}\pia\right)\pia^{2}\right]=\mathcal{N}(\pia)\int\mathcal{D}[\Os] \\
	& \qquad \qquad \times \;\exp\left[\int d^{4}x a^{4}\left(-\frac{\Os^{2}}{4\beta_{1}+4i \frac{\delta_1}{f_\pi^2}\pia}+i\Os\pia\right)\right].\nonumber
\end{align}
In order to recover the Langevin equation, we again rely on perturbation theory, assuming the quadratic noise to be much stronger than the cubic operator in $\pia$. We expand at leading order in $\pia$ both the denominator and the normalisation constant, which leads to a term of the form $\Os^{2}\pia$, that is
\begin{align}
	&\exp\left[-\int d^{4}x a^{4}\left(\beta_{1}+i \frac{\delta_1}{f_\pi^2}\pia\right)\pia^{2}\right]\approx\mathcal{N}\int\mathcal{D}[\Os] \\
	&\qquad \qquad \times \;\exp\left[\int d^{4}x a^{4}\left(-\frac{\Os^{2}}{4\beta_{1}} + i\Os\pia + i\frac{\delta_{1}}{4f_\pi^2\beta_{1}^{2}} \Os^2 \pia\right) \right]. \nonumber 
\end{align}
Here, we again find a perturbativity condition
\begin{equation}
	\frac{\delta_{1}\pia^3}{f_{\pi}^{2}\beta_{1} \pia^2}\ll 1,
\end{equation}
that can be related to the heuristic estimate made in \Eq{eq:5thestimate}. 
The $\Os^{2}\pia$ term enters the Langevin equation as
\begin{equation}
	\pi^{\prime \prime}+\left( 2H + \gamma\right)a\pi'-\partial_{i}^{2}\pi=a^2 \Os+ \frac{ \delta_{1}}{4f_\pi^2\beta_{1}^{2}} a^2 \Os^{2}
\end{equation}
There is no direct matching with \Eq{eq:Paolo} yet, the connection can be made manifest if one introduces a field redefinition. Indeed, one can map the Gaussian noise $\Os$  to a noise with a non-Gaussian statistics through
\begin{equation}
	\Os^{\text{ng}}=\Os+\frac{\delta_{1}}{4f_\pi^2\beta_{1}^{2}}\Os^{2}\;,
\end{equation}
or equivalently in Fourier space
\begin{align}
	\delta\mathcal{O}_{S}^{\text{ng}}(\bfk, \eta)=\delta\mathcal{O}_{S}(\bfk,\eta)+\frac{\delta_{1}}{4f_\pi^2\beta_{1}^{2}}\int_{\bfq}\delta\mathcal{O}_{S}(\bfq, \eta)\delta\mathcal{O}_{S}(\bfk-\bfq, \eta).
\end{align}
Under this field redefinition, the non-Gaussian noise adquires a three-point function of the form
\begin{align}
	\langle\Os^{\text{ng}}(\bfk_{1},(\eta_{1})\Os^{\text{ng}}(\bfk_{2},\eta_{2})\Os^{\text{ng}}(\bfk_{3},\eta_{3})\rangle'&= \delta(\eta_{1}-\eta_{2})\delta(\eta_{2}-\eta_{3})\frac{24}{a^{4}(\eta_{1})a^{4}(\eta_{2})}  \frac{\delta_1}{f_\pi^2}.
\end{align}
where we used the notation $\langle \cdot \rangle = (2\pi)^3{\delta}(\bfk_{1}+\bfk_{2}+\bfk_{3}) \langle \cdot \rangle'$. Under this construction, we obtain the matching of the last parameter
\begin{equation}
	6\delta_1 = \rho f \frac{m^6}{f^3} \nu_{\mathcal{O}^3},
\end{equation}
which completes the reconstruction of the full non-linear Langevin equation of \cite{Creminelli:2023aly} from the path integral language of the open EFToI.


\clearpage

\subsection{\textit{Problem set}}\label{subsec:prob3}

\paragraph{\textit{Exercise 1.}} The decoupling limit of the EFT of inflation\\

We want to understand under which conditions 
\begin{align}\label{eq:Starget}
	S = \int \dd^4 x \sqrt{-g} \left[ \frac{M^2_{\mathrm{Pl}}}{2} R - \Lambda(t) - c(t) g^{00} + \sum_{n=2}^\infty M^4_n(t) (1 + g^{00})^n\right],
\end{align}
simplifies to 
\begin{align}\label{eq:Stargetpi}
	S_\pi = \int \dd^4 x \sqrt{-g} \left\{ \epsilon \Mpl^2 H^2 \left[ \dot{\pi}^2 - \frac{(\partial_i \pi)^2}{a^2} \right] + 2 M_2^4 \left[ \dot{\pi}^2 + \dot{\pi}^3 - \dot{\pi}\frac{(\partial_i \pi)^2}{a^2} \right] - \frac{4}{3} M_3^4 \dot{\pi}^3 + \cdots \right\}, 
\end{align}
It will be useful to consider the Friedmann equations
\begin{align}
	3\Mpl^2 H^2 = \Lambda(t) + c(t), \qquad 2\Mpl^2 \dot{H} = -2c(t),
\end{align}
and to define the first and second slow-roll parameters $\epsilon \equiv - \dot{H}/H^2 $ and $\eta \equiv \dot{\epsilon}/(\epsilon H)$, together with the speed of sound $c_s^{-2} \equiv [c(t) + 2 M_2^4(t)]/c(t)$.
\begin{enumerate}
	\item Perform a \stuck trick to \Eq{eq:Starget} to reintroduce $\pi$. 
	\item We expand the metric $g_{\mu\nu} = \bar{g}_{\mu\nu} + \delta g_{\mu\nu}$, where $\bar{g}^{\mu\nu} = \mathrm{diag}(-1, a^{2}, a^{2}, a^{2})$. Identify the four operators that generate mixing between $\pi$ and $\delta g_{\mu\nu}$ at quadratic order.
\end{enumerate}   
From now on, we work in the flat gauge where 
	\begin{equation}
		\delta g_{ij}^{\rm scalar}=0\quad,\quad\delta g_{00}=-2\phi=-\delta g^{00}\quad,\quad \delta g_{0i}=a(t)\partial_{i}F=\delta g^{0i}.
	\end{equation}
	The $00$ and $0i$ linear Einstein equations are given by
	\begin{align}
		\frac{\Mpl^{2}}{2}\delta G_{00}&+\left[\frac{\Lambda(t)}{2} - M_2^4(t)\right]  \delta g_{00} - \left[c(t) + 2M_2^4(t) \right]\dpir+3Hc(t)\pi=0, \label{eq:E1}\\
		\frac{\Mpl^{2}}{2}\delta G_{0i}&+\left[\frac{\Lambda(t)}{2}-\frac{c(t)}{2}\right]\delta g_{0i}-2c(t)\partial_{i}\pi=0, \label{eq:E2}
	\end{align}
	where $\delta G_{\mu\nu}$ is the perturbed Einstein tensor, reading
	 \begin{align}
		\delta G_{00} &= - \frac{2}{a(t)} H \nabla^2 F, \Bigg. \label{eq:G00}\\
		\delta G_{0i} &= \partial_i \left[ 2H\phi  - \left(2 \dot{H} + 3H^2 \right) a(t) F\right], \Bigg. \label{eq:G0i} 
	\end{align}
\begin{enumerate}\setcounter{enumi}{2}
	\item Solve for the constraints $\phi$ and $\nabla^2 F$ as a function of $\pi$ and $\dot{\pi}$. 
	\item Injecting the constraints in the linear mixing operators identified above, under which conditions these terms are they subdominant compared to the kinetic term of $\pi$? You may need the heuristic estimate $\dot{\pi} \sim H \pi$ during inflation.  
	\item What about the quadratic operators generated by the time derivatives of $\Lambda(t + \pi)$ and $c(t + \pi)$? Under which conditions are they subdominant compared to the kinetic term of $\pi$? 
	\item What are the operators that generate mixing between $\pi$ and $\delta g_{\mu\nu}$ at cubic order? Compare them with the cubic operators appearing in $\pi$. You may need the heuristic estimate $\cs k = a H$ during inflation.
	\item Under which condition do the decoupling limit hold during inflation? Give the expression of $\delta g^{00} = 1 + g^{00}$ in the decoupling limit. This constitutes the building block a the theory of a shift symmetric scalar in rigid de Sitter.
\end{enumerate}

\begin{center}
	\noindent\rule{8cm}{0.4pt}
\end{center}

\paragraph{\textit{Exercise 2.}} Constructing the effective functional of open inflation\\

To write a theory invariant under retarded shift symmetry, the building blocks at our disposal are $\pia$ and $\partial_\mu \pia$ together with $P_\mu = \partial_\mu (t + \pir) = \delta^0_{~\mu} + \partial_\mu\pir$, contracted with the background metric $\bar{g}^{\mu\nu} = \mathrm{diag}(-1, a^{-2}, a^{-2}, a^{-2})$. 
\begin{enumerate}
	\item Respecting the non-equilibrium constraints, write down the most general functional linear in the advanced field, $S_1$. 
	\begin{enumerate}
		\item Express the terms linear in perturbations. Derive the background continuity equation. By comparison with the usual $\dot{\Lambda}(t) + \dot{c}(t) + 6H c(t) = 0$, identify $c(t)$ and $\Lambda(t)$
		\item Express the terms quadratic in perturbations. Comment the physical origin of the various contributions. 
		\item Write down the cubic operators. Among those, which are the ones featured in the unitary EFT of inflation?  
	\end{enumerate}
	\item Write down the most general functional quadratic in the advanced field, $S_2$. How do you physically interpret these operators? What is the relation imposed by the non-equilibrium constraints on these operators. As above, decompose them in quadratic and cubic contributions. 
	\item At last, write down the most general functional cubic in the advanced field, $S_3$. Only keeping the terms cubic in perturbations, these constitutes the last contributions to the contact bispectrum. Among those, which are the ones featured in the unitary EFT of inflation?  
\end{enumerate}

\clearpage

\subsection*{\textit{Appendix. de Sitter Keldysh functions}}

This appendix gathers the outcome of the lenghty integrals appearing in the main text in the expression of \Eq{eq:GK1}. Indeed, to analytically access the Keldysh function given in \Eq{eq:GK1}, we have three integrals to compute
\begin{align}
	A^{(1)}_{\nu_\gamma}(z) &\equiv \int_{z}^{\infty} \dd z' z^{\prime2 - 2{\nu_\gamma}} J^2_{\nu_\gamma}(z'), \\
	B^{(1)}_{\nu_\gamma}(z) &\equiv \int_{z}^{\infty} \dd z' z^{\prime2 - 2{\nu_\gamma}} J_{\nu_\gamma}(z') Y_{\nu_\gamma}(z'), \\
	C^{(1)}_{\nu_\gamma}(z)&\equiv \int_{z}^{\infty} \dd z' z^{\prime2 - 2{\nu_\gamma}} Y^2_{\nu_\gamma}(z'),
\end{align}
that lead for the first contribution to 
\begin{align}
	A^{(1)}_{\nu_\gamma}(z) &= \frac{1}{4} \bigg[\frac{\Gamma ({\nu_\gamma} -1)}{\Gamma \left({\nu_\gamma} -\frac{1}{2}\right) \Gamma \left(2 {\nu_\gamma} -\frac{1}{2}\right)} \label{eq:Fbeta} \nonumber\\
	&\qquad \qquad -z^3 \Gamma \left({\nu_\gamma} +\frac{1}{2}\right) \, _2\tilde{F}_3\left(\frac{3}{2},{\nu_\gamma} +\frac{1}{2};\frac{5}{2},{\nu_\gamma} +1,2 {\nu_\gamma} +1;-z^2\right)\Bigg], 
\end{align}
for the second contribution to 
\begin{align} 
	B^{(1)}_{\nu_\gamma}(z) 
	&=\frac{z^{3-2 {\nu_\gamma} } \, _2F_3\left(\frac{1}{2},\frac{3}{2}-{\nu_\gamma} ;1-{\nu_\gamma} ,\frac{5}{2}-{\nu_\gamma} ,{\nu_\gamma} +1;-z^2\right)}{3 \pi  {\nu_\gamma} -2 \pi  {\nu_\gamma} ^2}  \label{eq:Gbeta} \nonumber \\
	+&\frac{z^3 \Gamma (-{\nu_\gamma} ) \, _2F_3\left(\frac{3}{2},{\nu_\gamma} +\frac{1}{2};\frac{5}{2},{\nu_\gamma} +1,2 {\nu_\gamma} +1;-z^2\right)}{3 \sqrt{\pi } \Gamma \left(\frac{1}{2}-{\nu_\gamma} \right) \Gamma (2 {\nu_\gamma} +1)}-\frac{\Gamma \left(\frac{3}{2}-{\nu_\gamma} \right)}{4 \Gamma (2-{\nu_\gamma} ) \Gamma \left(2 {\nu_\gamma} -\frac{1}{2}\right)},
\end{align}
and for the third contribution to
\begin{align}
	\label{eq:Hbeta}
	C^{(1)}_{\nu_\gamma}(z) &= \frac{2 \cot (\pi  {\nu_\gamma} ) z^{3-2 {\nu_\gamma} } \, _2F_3\left(\frac{1}{2},\frac{3}{2}-{\nu_\gamma} ;1-{\nu_\gamma} ,\frac{5}{2}-{\nu_\gamma} ,{\nu_\gamma} +1;-z^2\right)}{3 \pi  {\nu_\gamma} -2 \pi  {\nu_\gamma} ^2} \nonumber \\
	+&\frac{4^{{\nu_\gamma} } \Gamma ({\nu_\gamma} )^2 z^{3-4 {\nu_\gamma} } \, _2F_3\left(\frac{3}{2}-2 {\nu_\gamma} ,\frac{1}{2}-{\nu_\gamma} ;1-2 {\nu_\gamma} ,\frac{5}{2}-2 {\nu_\gamma} ,1-{\nu_\gamma} ;-z^2\right)}{\pi ^2 (4 {\nu_\gamma} -3)}\nonumber\\
	-&\frac{4^{-{\nu_\gamma} } z^3 \cot ^2(\pi  {\nu_\gamma} ) \, _2F_3\left(\frac{3}{2},{\nu_\gamma} +\frac{1}{2};\frac{5}{2},{\nu_\gamma} +1,2 {\nu_\gamma} +1;-z^2\right)}{3 \Gamma ({\nu_\gamma} +1)^2}\nonumber \\
	+&\frac{\Gamma ({\nu_\gamma} -1)}{4 \Gamma \left({\nu_\gamma} -\frac{1}{2}\right) \Gamma \left(2 {\nu_\gamma} -\frac{1}{2}\right)}-\frac{4^{3 {\nu_\gamma} -2} \sin (2 \pi  {\nu_\gamma} ) \Gamma (3-4 {\nu_\gamma} )}{\Gamma (1-{\nu_\gamma} ) \Gamma (2-{\nu_\gamma} )}.
\end{align}

\section{Lecture 4: Open electromagnetism}\label{sec:lec4}

While the open theory of inflation presented in the previous lecture is sufficient to derive the statistics of scalar perturbations seeding cosmological inhomogeneities during inflation, it does not predict the amplitude of primordial gravitational waves emitted during inflation. Moreover, it falls short to describe the mixing between the perturbations of the metric and the pseudo-Goldstone boson. To overcome these shortcomings, we need to establish a open theory of gravity in a medium. A necessary ingredient being diffeomorphism invariance, a preliminary step consists in studying gauge symmetries in the Schwinger-Keldysh contour, which is the goal of this Lecture. \\

\textit{References:} This section explores gauge symmetries in open effective theories, focusing on the simplest possible setting: an Abelian, linear theory in flat spacetime. For a broader discussion of symmetries in open quantum systems, see Section II.D of \cite{2016RPPh...79i6001S}. For a systematic approach based on the coset construction, see \cite{Akyuz:2023lsm}.

\subsection{Electromagnetism in a medium}

Let us consider the description of electromagnetism in a medium. At the microscopic level, the description relies quantum electrodynamics, where photons carry the electric $\mathbf{E}$ and magnetic $\mathbf{B}$ fields and protons and electrons describe the distribution of charged nuclei. This level of detail is not mandatory to capture the phenomenology of the problem at macroscopic scales. There, one can work with effective fields, such as the electric displacement field  $ \mathbf{D}=\epsilon \mathbf{E}$ where $ \epsilon$ is the electric permittivity and the magnetic field strength $ \mathbf{H}=\mathbf{B}/\mu$ where $ \mu$ is the permeability of the medium.\footnote{This approach is actually only valid for the simplest ``linear'' material, where the response of the material's own electric and magnetic fields is proportional to $ \mathbf{E}$ and $ \mathbf{B}$. More generally, the presence of a material can induce non-linearities.} The response of the material, that is how the medium reacts when we apply external forces on it, and how this reaction affects the propagation of the fields inside it, is encoded through effective parameters that are the permittivity $\epsilon$ or the permeability $\mu$. The interplay between the response of the medium and the presence of macroscopic sources, that are the currents $ \mathbf{J}_{free}$ and charges $ \rho_{free}$ that can move around in the material (as opposed to being fixed somewhere like around an atom or a molecule), lead to the emergence of new properties, such as a modified speed of propagation or a new dispersion relation. 

This physics is traditionally described at the level of the equations of motion, using the well-known Maxwell equations in a medium \cite{Jackson:1998nia, Griffiths_2017}
\begin{align}\label{maxmediumv0}
	\nabla \cdot \mathbf{D} &= \rho_{\text{free}} \qquad \text{and} \qquad \nabla \times \mathbf{H} - \frac{\partial \mathbf{D}}{\partial t} = \mathbf{J}_{\text{free}}.
\end{align}
The goal of this Lecture is to recover this standard effective description from the Schwinger-Keldysh path integral introduced previously. Revisiting this XIXth century problem will not bring us any new physics, but will eventually bring some light on the role played by gauge symmetries in the Schwinger-Keldysh contour.


\subsection{Retarded and advanced gauge transformation}

\begin{figure}[tbp]
	\centering
	\includegraphics[width=1\textwidth]{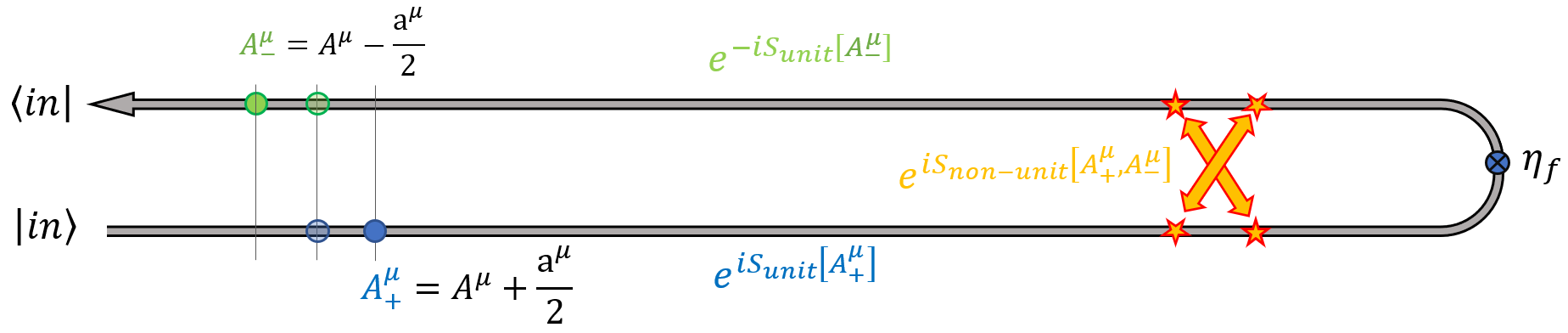}
	\caption{Schwinger-Keldysh contour considered. We aim at constructing the most general influence functional $S_{\mathrm{eff}}$ containing both unitary (the same along each branch of the path integral) and non-unitary (mixing between the branches of the path integral) effects.}
	\label{fig:contourEM}
\end{figure}

In \cite{Salcedo:2024nex}, we studied Abelian gauge theories within the Schwinger-Keldysh formalism. Specifically, we constructed the most general open effective field theory for electromagnetism in a dielectric medium. Here, we review our main findings, providing a road-map to the study of dynamical gravity. We begin by doubling the gauge fields, $A^\mu \to A^\mu_\pm$, and expressing the result in the Keldysh basis
\begin{align}
	\text{retarded: } A^{\mu}&=\frac{1}{2}\left( A_{+}^{\mu}+A_{-}^{\mu}\right) \,, & 
	\text{advanced: } a^{\mu}=A_{+}^{\mu}-A_{-}^{\mu}\,.
\end{align}
We aim at describing both the unitary and non-unitary effects that emerge from integrating out the unknown environment, as illustrated in \Fig{fig:contourEM}.

\paragraph{Gauge transformations in the Schwinger-Keldysh contour.} In the usual electromagnetism case, gauge invariance is there to introduce a redundancy such that one can use a vector field $A^\mu$ to only describe the presence of two propagating degrees of freedom: the two polarizations. Here, we have two ways to realize the gauge transformation:
\begin{enumerate}
	\item One can perform a \textit{retarded gauge transformation}, which acts on both $A_+$ and $A_-$ in the same way through $\epsilon_{+}=\epsilon_{-}=\e_r$. In this case, the retarded field $ A^{\mu}$ transforms as the usual gauge field, while $ a^{\mu}$ does not transform:
	\begin{align}\label{eq:gaugetrans}
		\text{retarded gauge transformation: } A^{\mu}&\to A^{\mu}+\partial^{\mu}\e_r\,, & a^{\mu}&\to a^{\mu}\,.
	\end{align}
	\item One can perform a \textit{advanced gauge transformation}, in which $A_+$ and $A_-$ transform in opposite directions,  $\epsilon_{+}=- \epsilon_{-}=\e_a/2$. In this case, the retarded and advanced components transform as 
	\begin{align}
		\text{advanced gauge transformation: } A^\mu \rightarrow A^\mu, \qquad a^\mu \rightarrow a^\mu + \partial^\mu \epsilon_a.
	\end{align}
\end{enumerate}

The main question of interest is to understand what happen when one impose the invariance of $S_{\mathrm{eff}}$ under these transformations. For the global symmetries $\e_r(t,\bmx) = \e_r$ and  $\e_a(t,\bmx) = \e_a$, we have seen that the presence of dissipative effects break the symmetry group to its diagonal subgroup \cite{Hongo:2018ant, Akyuz:2023lsm}
\begin{align}
	U(1)_r \times U(1)_a \xrightarrow{\text{open}} U(1)_r.
\end{align}
A natural starting point then consists in writing down a theory invariant under retarded gauge transformation. A simple implementation consists in building the effective functional out of the retarded field strength $F^{\mu\nu}=\partial^{\mu}A^{\nu}-\partial^{\nu}A^{\mu}\,$. 
Imposing the non-equilibrium constraints presented in Lecture \ref{sec:scalar} and assuming homogeneity and isotropy (having in mind applications to cosmology), we can write down the most generic functional invariant under retarded gauge transformation. 

\paragraph{Open E\&M effective functional.} Since E\&M is a free theory, we can now write all possible quadratic terms, starting at least linear in the advanced fields to fulfill the non-equilibrium constraints \eqref{eq:NEQ}.
Importantly, we work directly to all orders in derivatives. Indeed, to linear order (quadratic action) in flat space, we can formally specify an open functional in Fourier space that permits such construction. Let us first consider the terms linear in the advanced fields $ a^{0}$ and $ a^{i}$, which control the dispersion relations of the propagating degrees of freedom \cite{kamenev_2011, Salcedo:2024smn}. The most general open functional reads
\begin{align}
	S_1 = \int \frac{\dd \omega}{2\pi} \int \frac{\dd^3 \bmk}{(2\pi)^3} \left[a^{0}(\gamma_{ij}^{t}F^{ij}+\gamma^{t}_{i}F^{0i}+j_{0})+a^{i}\left(  \gamma^{s}_{ij}F^{0j}+\gamma^{s}_{ijl}F^{jl}+j_{i}\right)\,\right].
\end{align}
where  ``$t $'' stands for time and ``$s$'' for space. Recall that in our Fourier conventions $\partial_0=-i\omega$ and $\partial_i=i k_i$.
Assuming homogeneity and isotropy, we can write the EFT coefficients as
\begin{align}
	\gamma^{t}_{i}&=\gamma^{t}(\omega,k) ik_{i}\,, & \gamma^{t}_{ij}&=\gamma^{t}_{2}(\omega,k)k_{i}k_{j}+\gamma^{t}_{3}(\omega,k)\delta_{ij}\,, \\
	\gamma^{s}_{ij}&=\gamma_{1}(\omega,k)k_{i}k_{j}+\gamma_{2}(\omega,k)\delta_{ij}\,, & \gamma_{ijl}^{s}&=\gamma_{3}(\omega,k)\delta_{i[j}ik_{l]}+\gamma_{4}(\omega,k)\e^{ijl}\,.
\end{align}
where we introduced seven scalar functions of frequency and momentum. Also, we inserted an $i$ for every derivative for later convenience. 
The non-equilibrium constraints \eqref{eq:norm}, \eqref{eq:herm} and \eqref{eq:pos} impose a reality condition on the EFT coefficient of $S_1$ in real-space. This translates into conditions on the frequency-space coefficients, for instance \cite{Crossley:2015evo}
\begin{align}
	\gamma_2(\omega, k) = \Gamma - i \omega + \Gamma_{20}  \omega^2 + \Gamma_{02}  k^2 + i \Gamma_{30} \omega^3 + \cdots
\end{align}
that we normalized $\gamma_2(\omega, k) $ in this way for later convenience. It readily follows that the non-equilibrium constraints guarantee the $\Gamma$'s coefficients are all real.

Notice that one cannot construct an anti-symmetric two-index tensor and so $ \gamma^{t}_{ij}F^{ij}=0$. Also, notice that $ \gamma_{ijl}^{s}$ had to be anti-symmetric in $ jl$. As a consequence, we are left with
\begin{align}\label{OEM}
	S_1&= \int_{\omega,\bmk}  \left[ a^{0}(\gamma^{t} i k_{i}F^{0i}+j_{0}) +a_{i}\left( \gamma_{1}k^{i}k_{j}F^{0j}+ \gamma_{2}F^{0i}+\gamma_{3}ik_{j}F^{ij}+\gamma_{4}\e^{i}_{\,jl}F^{jl}+j_{i}\right)\right]\,, \Bigg.
\end{align}
where $\int_{\omega,\bmk}$ is a short-hand notation for $\int \frac{\dd \omega}{2\pi} \int \frac{\dd^3 \bmk}{(2\pi)^3}$. A few coefficients can be removed by rescalings and field redefinitions. First, one can set $\gamma^{t}=1$ (or to any other value) by rescaling $a^{0}$. Then, the term proportional to $\gamma_{1}$ can be removed by the field redefinition
\begin{align}
	a^{0}\to a^{0}-i\gamma_{1}k_{i}a^{i}\,.
\end{align}
It is a simple exercise to keep $\gamma^{t}$ and $\gamma_{1}$ to confirm they do not have any consequence for the dynamics. In the following, we simply set $\gamma_{t}=1$ and $ \gamma_{1}=0$. 
Moreover, we momentarily neglect the currents, as these can be easily re-inserted at the end of the calculation. As a result, we land on the following theory
\begin{align}\label{OEMnog}
	S_1=\int_{\omega,\bmk}  \left[ a^{0}  i k_{i}F^{0i}+a_{i}\left(  \gamma_{2}F^{0i}+\gamma_{3}ik_{j}F^{ij}+\gamma_{4}\e^{i}_{\,jl}F^{jl}\right)\right]\,.
\end{align}

Similarly, one can construct $S_2$ which controls the noise part of the theory. Invariance under retarded gauge transformation does not impose any constraint on the quadratic noise, $a^{\mu}$ being by construction invariant. We end up with a fairly generic functional 
\begin{equation}
	S_2= i \int_{\omega,\bmk} a^{\mu}N_{\mu\nu}a^{\nu},
\end{equation}
where $N_{\mu\nu}$ is any $4\times4$ positive definite matrix, the positivity following from the non-equilibrium constraints. 

\paragraph{Unitary and non-unitary cases.} We now investigate how this functional changes under retarded and advanced gauge transformation. Let us first consider the deterministic part $S_1$. Because of the use of the retarded field strength $F^{\mu\nu}$, it is manisfestly invariant under the retarded gauge transformation. On the contrary, it does transform under advanced gauge transformation, through\footnote{The operators proportional to $\gamma_3$ and $\gamma_4$ do not transform under advanced gauge transformations, the former being manifestly advanced gauge invariant and the latter being conserved by Bianchi identity, that is
	\begin{align}
		i \gamma_3 a_i k_j F^{ij} \qquad &\rightarrow \qquad- \gamma_3 \epsilon_a k_i k_j F^{ij} = 0,\\
		\gamma_4 a_i \widetilde{F}^{0i} \qquad &\rightarrow \qquad i \gamma_4 k_i\widetilde{F}^{0i} = 0,
	\end{align}
	where we defined $\widetilde{F}^{0i} \equiv \e^{i}_{\,jl}F^{jl}$.
}
\begin{align}
	\Delta S_1 = \int_{\omega, \bmk} \epsilon_a \left( i \omega + \gamma_2\right) i k_i F^{0i}.
\end{align}
Several remarks hold.\\

\underline{\textit{Remark 1}}: When $\gamma_2 = -i \omega$, the functional becomes invariant under both retarded and advanced gauge transformation. In this case, unitarity is restored, in the sense we can write
\begin{align}
	S_1(\gamma_2 \rightarrow -i \omega) = S_{\mathrm{unit}}[A^\mu_+] - S_{\mathrm{unit}}[A^\mu_-],
\end{align}
with 
\begin{align}
	S_{\mathrm{unit}}[A^\mu] = \frac{1}{4}\int \dd^4 x \left[F^{\mu\nu}F_{\mu\nu} + (\cs^2 - 1) F^{ij}F_{ij} + \theta F^{\mu\nu}\widetilde{F}_{\mu\nu} \right],
\end{align}
where $\widetilde{F}_{\mu\nu} = \frac{1}{2} \epsilon^{\mu\nu\rho\sigma}F_{\rho\sigma}$, the speed of sound $\cs^2$ is related to the $\gamma_3$ parameter and the birefringence coefficient $\theta$ is related to the $\gamma_4$ parameter. We conclude that the operators controlled by $\gamma_3$ and $\gamma_4$ are familiar unitary operators written in the Keldysh basis, and that the limit $\gamma_2 = -i \omega$ reproduces familiar results of Maxwell in a medium. \\

\underline{\textit{Remark 2}}: Conversely, when $\gamma_2 \neq -i \omega$, the functional is not invariant under advanced gauge transformation. It follows that the theory is open, which can be made manifest by observing that the current is not conserved. Explicitly, the on-shell equations of motion are obtained by varying $S_{\mathrm{eff}} = S_1 + S_2$ with respect to the advanced field
\begin{align}
	\frac{\delta S_1}{\delta a^0} &= 0 \quad \Rightarrow \quad i k_i F^{0i} =  j^0 + \xi^0, \label{eq:M1}\\
	\frac{\delta S_1}{\delta a^i} &= 0 \quad \Rightarrow \quad \gamma_{2}F^{0i}+\gamma_{3}ik_{j}F^{ij}+\gamma_{4}\epsilon^{i}_{\,jl}F^{jl} =  j^i + \xi^i, \label{eq:M2}
\end{align}
where we reintroduced the current $j^\mu$ and performed the Hubbard-Stratonovich trick on $S_2$ from which we obtained a set of stochastic sources encoded in $\xi^\mu$ \cite{Salcedo:2024nex}. 
This yields a modified Maxwell equation 
\begin{align}\label{eq:Gamma}
	\partial_\mu F^{\mu\nu} + \delta^\nu_{~i} \left[\Gamma F^{0i}+(\gamma_{3}+1)ik_{j}F^{ij}+\gamma_{4}\epsilon^{i}_{\,jl}F^{jl}\right] =  j^\nu + \xi^\nu ,
\end{align}
Combining \Eqs{eq:M1} and \eqref{eq:M2}, we obtain the \textit{non-standard current conservation}, 
\begin{align}\label{nonstand}
	- \gamma_2 (j_0 + \xi_0)+ i k^i (j_i + \xi_i) = 0\,,
\end{align}
Making explicit the dissipative term, $\gamma_2 = \Gamma - i \omega$, we obtain that the lack of conservation of the current is proportional to the dissipation coefficient through  
\begin{align}\label{eq:noiseconstEM}
	\boxed{\partial^\mu (j_\mu + \xi_\mu) = \Gamma (j_0 + \xi_0)}\,.
\end{align}
This relation holds out of equilibrium and relates the sourcing of the system dynamics through the noise $\xi_\mu$ and its damping through the dissipation $\Gamma$. Because of the non-trivial relation it imposes on the noise variables that were so far unconstrained, we dubbed this relation the \textit{noise constraint} in \cite{Salcedo:2024nex}.

\begin{figure}[tbp]
	\centering
	\includegraphics[width=0.5\textwidth]{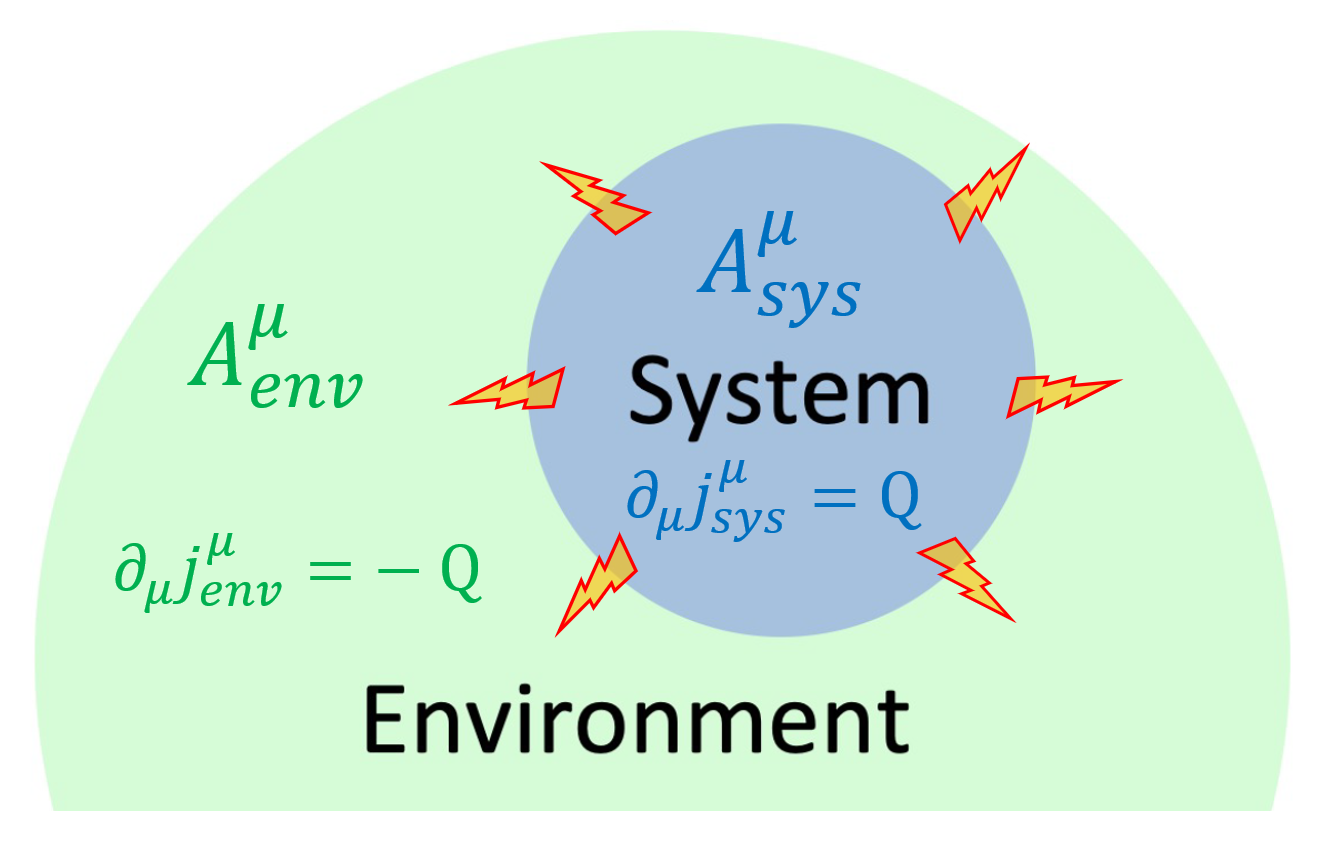}
	\caption{Illustration of the lack of conservation of the system's current. This effect is a mere consequence of the total charge conservation $\partial_\mu J^\mu = 0$. It simply illustrates the fact that charges initially located in the system sink in the environment.}
	\label{fig:sink}
\end{figure}

\begin{tcolorbox}[%
	enhanced, 
	breakable,
	skin first=enhanced,
	skin middle=enhanced,
	skin last=enhanced,
	before upper={\parindent15pt},
	]{}
	
	\vspace{0.05in}

\paragraph{Why is the current not conserved?} The fact that the current $(j+\xi)^\mu$ does not satisfy the standard conservation equation might appear concerning. Intuitively the electric charge should still be conserved even if we separate system and environment. Moreover, in standard electrodynamics, charge conservation is necessary for gauge invariance, and gauge invariance is necessary for unitarity. Here we want to show in a more transparent and intuitive way that the non-standard current conservation in \eqref{nonstand} is precisely what is expected from the fact that the full electric charge is conserved. 

Consider the standard Maxwell equation
\begin{align}
	\partial_\mu F^{\mu\nu} = J^\nu\,,
\end{align}
with $J^\mu $ the total conserved current $\partial_\mu J^\mu = 0$. By our assumption of separation of scales, the environment is gapped and so $J^\mu$ can be written as a fixed external current $j^\mu$ plus a function of the light fields in the Open EFT, namely the photons. Since $J^\mu$ is gauge invariant, its expectation value must be written in terms of $F_{\mu\nu}$. To lowest order in derivatives the simplest possibility is\footnote{To keep the discussion as transparent as possible here we focus just on the dissipation $\Gamma$ and neglect $\gamma_{3,4}$.} 
\begin{align}
	J^\mu=j^\mu - \Gamma \delta^\mu_{~i} F^{0i}+\dots\,,
\end{align}
where $\Gamma$ is a model dependent local function of time and space derivatives, $\Gamma=\Gamma(\omega,k^2)$ and the minus sign has been chosen for later convenience. Since part of the environment's current is now proportional to $F^{\mu\nu}$, it feels natural to re-write the Maxwell's equestions as
\begin{align}\label{eq:dissipMax}
	\partial_\mu F^{\mu\nu}+ \Gamma \delta^\nu_{~i} F^{0i} = j^\nu\,.
\end{align}
We recognize $\Gamma$ as the deviations from Maxwell's theory that we encountered in \eqref{eq:Gamma}, which plays the role of a friction in the equations of motion for $A_i$. The time-component imposes
\begin{align}
	\partial_\mu F^{\mu0}  = j^0\,.
\end{align}
Taking the gradient of \eqref{eq:dissipMax} and making use of the anti-symmetry of the field strength, we find
\begin{align}
	\partial_\mu j^\mu = \Gamma j^0  \,,
\end{align}
which is precisely the non-standard current conservation we found in \eqref{nonstand}. This illustrates how the conservation of the total current induces a non-trivial relation in the system when the latter is open, as shown in \Fig{fig:sink}.  

\end{tcolorbox}

\underline{\textit{Remark 3}}: One of the main findings of \cite{Salcedo:2024nex} is the relation between the lack of unitarity due to the openness of the dynamics and the transformation of the effective functional under advanced gauge transformation. It is crucial to stress that despite this fact, the advanced gauge invariance is not \textit{broken} but rather \textit{deformed}. To see this, we first focus on $S_1$ and re-write the above functional in the form
\begin{align}\label{openf}
	S_1=\int_{\omega, \bmk} a^{\mu}M_{\mu\nu}A^{\nu}\,,
\end{align}
with 
\begin{align}\label{eq:matrix1}
	M=\begin{pmatrix}  k^{2} & - \omega k_{i} \\ -i \gamma_{2} k_{i} & i\gamma_{2}\omega \delta_{ij}+\gamma_{3}(k^{2}\delta_{ij}-k_{i}k_{j})-2i\gamma_{4}\e_{ijl}k_{l} \end{pmatrix}\,.
\end{align}
A few comments apply. First, it is straightforward to check that this matrix has a vanishing determinant, as it should be since the retarded gauge invariance \eqref{eq:gaugetrans} must imply $ M_{\mu\nu} k^{\nu}=0$ where $ k^{\mu}=(\omega,\mathbf{k})$. 
Importantly, the presence of a ``right kernel" due to the retarded gauge invariance implies the existence of a ``left kernel" $v^\mu M_{\mu \nu} = 0$. In general, when the dynamics is open, the left kernel and the right kernel are spanned by different vectors since $M$ is non-Hermitian. For instance, an inspection of $M$ leads to the left kernel being spanned by 
\begin{align}
	v^\mu = (i\gamma_2, \bmk).\label{eq:vmu}
\end{align}
The difference between the left and right kernel is a reminder of the dissipative nature of the dynamics, since in the unitary case ($\gamma_2 = - i \omega$) one would find $v^\mu=k^\mu$.\footnote{In this case, $M$ becomes Hermitian and, being also real, symmetric. This also means that the left kernel of $M$ is spanned by the same vector $ k^{\mu}$ as the right kernel. Again this shows that there are two copies of the standard E\&M gauge group in the absence of dissipation.} This is to be expected because in the absence of dissipation one should recover the two independent gauge groups acting on each branch of the closed-time contour.

These observations are striking. Even if we did not impose any structure on the appearance of the advanced field $a^\mu$, the gauge invariance of the retarded sector enforces enough constraints to generate an \textit{advanced gauge invariance} in the advanced sector. 
\begin{tcolorbox}[%
	enhanced, 
	breakable,
	skin first=enhanced,
	skin middle=enhanced,
	skin last=enhanced,
	before upper={\parindent15pt},
	]{}
	
	\vspace{0.05in}
	
	\paragraph{Retarded and advanced gauge invariances:} $S_1$ remains unchanged under the transformations
	\begin{align}\label{eq:advgauge}
		A^\mu &\rightarrow A^\mu + \e_r k^\mu,& a^\mu \rightarrow a^\mu + \e_a v^\mu.
	\end{align}
	
\end{tcolorbox}
\noindent As we will see shortly, this new symmetry extends to the noise sector as well and plays a crucial role in reducing the number of advanced components needed to describe the problem. Indeed, the advanced gauge invariance illustrates the presence of a redundancy in the advanced sector - that is there are more advanced variables than needed to physically describe the problem. One then becomes allowed to gauge fix in the advanced sector, which is in some sense deeply reinsuring for the following reason.

\begin{tcolorbox}[%
	enhanced, 
	breakable,
	skin first=enhanced,
	skin middle=enhanced,
	skin last=enhanced,
	before upper={\parindent15pt},
	]{}
	
	\vspace{0.05in}

\paragraph{Why can we gauge fix in the advanced sector?}

Let's begin with a simple observation. When the number of advanced fields is larger than the number of retarded fields, an issue arises unless additional structure is taken into account. Indeed consider the toy model with one retarded field $\phi_r$ and two advanced fields $\phi_{a1}$ and $\phi_{a2}$,
\begin{align}\label{nabiggernr}
	S_{SK}=\int \, \dd^4x \left[\phi_{a1}F_1(\phi_r)+\phi_{a2}F_2(\phi_r) \right]\,,
\end{align}
where $F_{1,2}$ are some generic functionals of $\phi_r$. The classical deterministic equations of motions are $F_{1}(\phir)=0$ and $F_{2}(\phir)=0$. Unless $F_{1}$ and $F_2$ are related to each other in such a way to admit the same solutions, the classical equations of motion have no solutions whatsoever. 

Now let's see why this observation is relevant for a gauge theory with a symmetry group $G$. Because of the doubling of the path integral contour in the SK formalism, the group is naively doubled to $G_+\times G_-$. Now one expects that generic dissipative effects couple the two branches and break the anti-diagonal combination leaving only the retarded diagonal symmetry $G_r$ unbroken. If one fixes the retarded gauge, e.g. by setting to zero some retarded fields, the number of retarded fields generically decreases by one but the number of advanced fields remains unchanged. One hence worries about the problem pointed out above, namely that there are more classical equations than fields and there may be no solutions. Of course the theory does not change under gauge fixing, so it must be that non-trivial relations are present among the operators linear in the different advanced fields. In other words, additional structure must be accounted for. 

Accounting for this additional structure may at times render the construction of a theory more involved. In \cite{Salcedo:2024nex}, we discuss three different ways to deal with this issue, which give each different strategies to account for the additional structure:
\begin{itemize}
	\item One doubles the fields only \textit{after} having already fixed the gauge in the unitary theory so that the number of retarded and advanced fields matches by construction. In some sense in this case we are simply \textit{not} dealing with a gauge theory because we have fixed the gauge to begin with. A shortcoming of this approach is that the construction of the dissipative theory is gauge dependent from the start and to express the theory in a different gauge one needs to start back from the beginning.
	\item One fixes the retarded gauge but then notices that a new, \textit{deformed advanced gauge} is automatically present in the linear theory. Fixing this deformed advanced gauge brings the number of retarded and advanced fields to match again. We have not yet investigated what happens to this deformed advanced gauge transformation beyond the Abelian theory studied in \cite{Salcedo:2024nex}. It would be desirable to have an interpretation of the above transformation of $a^\mu$ in terms of differential geometry, just like we think of the standard gauge field as a connection on the principle fiber bundle.
	\item One never fixes the retarded gauge and proceeds with gauge invariant quantization, as often done in electromagnetism or in BRST quantization. 
\end{itemize}     
\end{tcolorbox}

Let us close the discussion on the deformed advanced gauge invariance by presenting a complementary perspective from \cite{Lau:2024mqm}. One may choose to recover manifest advanced gauge invariance by introducing a St\"uckelberg field $X_a$ that non-linearly realises advanced gauge transformation 
\begin{align}
	X_a \rightarrow X_a - \epsilon_a,
\end{align}
such that the combination $\mathcal{A}_a^\mu \equiv a^\mu + \partial^\mu X_a$    is manifestly advanced gauge invariant. The action constructed from the promotion of $a^\mu \rightarrow \mathcal{A}_a^\mu$ is then invariant under both retarded and advanced gauge transformations. Let us consider
\begin{align}
	S_1^{\mathrm{new}} &= \int_{\omega,\bmk}\left[ \mathcal{A}_a^{0}  i k_{i}F^{0i}+\mathcal{A}_{ai}\left(  \gamma_{2}F^{0i}+\gamma_{3}ik_{j}F^{ij}+\gamma_{4}\e^{i}_{\,jl}F^{jl}\right)\right], \\
	&= S_1^{\mathrm{old}} + \int_{\omega, \bmk} X_a \left(  i \omega + \gamma_2 \right) i k_i F^{0i}.
\end{align}
where $S_1^{\text{old}}$ was given in \eqref{OEMnog}. Now deriving the equation of motion for $X_a$ one finds that the on-shell relations impose the system must be closed through
\begin{align}
	\frac{\delta S_1}{\delta X_a} = 0 \quad \Rightarrow \quad \gamma_2 = - i \omega,
\end{align}
The constraint $\gamma=-i\omega$ naively appears to prevent any deviation from the unitary theory. The issue is that we are asking for a medium that dissipates photons but we are not coupling the system to an external current/source. To obtain a non-trivial result, we need to allow for an external current $j^\mu$, which will also contain noise contributions $\xi_\mu$. Applying the St\"uckelberg transformation to the coupling to an external current we find
\begin{align}\label{jpxi}
	S_1 \supset - \int_{\omega, \bmk} (j_\mu + \xi_\mu) a^\mu \rightarrow -\int_{\omega, \bmk} (j_\mu + \xi_\mu) \mathcal{A}_a^\mu
\end{align}
The new equation of motion for $X_a$ now becomes
\begin{align}\label{eomXa}
	\frac{\delta S_1}{\delta X_a} = 0 \quad \Rightarrow \quad \Gamma i k_i F^{0i} = i \omega (j_0+\xi_0) + i k_i (j^i+\xi^i) .
\end{align}
We can now make use of the on-shell equations of motion \Eqs{eq:M1} and \eqref{eq:M2} to recover the \textit{non-standard current conservation} found in \Eq{nonstand}. 
The present derivation employed the St\"uckelberg trick for explicitly broken advanced gauge transformations. However, this result was straightforwardly derived simply taking the gradient of the dissipative and stochastic Maxwell equations. Here we have emphasized the St\"uckelberg derivation to show the equivalence between our approach in \cite{Salcedo:2024nex} and that presented in \cite{Lau:2024mqm}. 

\underline{\textit{Remark 4}}: Ensuring retarded gauge invariance guarantees the theory only propagates two helicities. To see this, one can check what happens when the functional is not invariant under neither the advanced gauge transformation nor the retarded one. To study this in a concrete example, we add a mass term to the above effective action, i.e. we study the particular implementation of a \textit{dissipative Proca theory}. Explicitly, we consider
\begin{align}
	S_1 \supset - \int_{\omega, \bmk} m^2 a^\mu A_\mu.
\end{align}
As a consequence of this new term, $S_1$ is not invariant under both advanced and retarded gauge transformations.

One can straightforwardly study this theory as is. However, in certain situations one is interested in the simplification that may happen at high energies where the longitudinal degree of freedom is expected to decouple from the transverse one, a result know as the \textit{decoupling theorem} in the context of particle physics. In that case, it is useful \cite{Stueckelberg:1938hvi,Coleman:1969sm,Arkani-Hamed:2002bjr} to restore invariance under both gauge transformations, we perform a retarded St\"uckelberg trick to addition to the advanced St\"uckelberg trick we just discussed. More in detail, we introduce a St\"uckelberg field $X_r$ that non-linearly realises retarded gauge transformations 
\begin{align}
	X_r \rightarrow X_r - \epsilon_r.
\end{align}
It follows that the combination $\mathcal{A}_r^\mu \equiv A^\mu + \partial^\mu X_r$ is manifestly retarded gauge invariant, basically by construction. The functional constructed from the promotion of $a^\mu \rightarrow \mathcal{A}_a^\mu$ and  $A^\mu \rightarrow \mathcal{A}_{r}^\mu$ is then invariant under both retarded and advanced gauge transformations. Under this promotion, 
\begin{align}
	S^{\mathrm{new}}_1 \supset S^{\mathrm{old}}_1 - m^2\int \dd^4x  \left[\partial^\mu X_a \partial_\mu X_r - a^\mu \partial_\mu X_r - \partial^\mu X_a A_\mu \right],
\end{align}
where we wrote the expression in real space to improve the readibility. In the high energy limit $E \gg m$, we recover the familiar \textit{decoupling} limit where the mixing between the the scalar $(X_{r}, X_a)$ and the gauge vector $(A^\mu, a^\mu)$ becomes negligible. The take-home message is that breaking retarded gauge invariance triggers new degrees of freedom, in this case $X_r$, while the breaking of advanced gauge invariance caused by cross-branch interactions representing open effect does \textit{not}. Moreover, in electrodynamics, the presence or absence of the additional scalar degree of freedom $X_r$ is a model dependent choice. The theory with just the two photon polarizations exists and can be dissipative. 

\begin{tcolorbox}[%
	enhanced, 
	breakable,
	skin first=enhanced,
	skin middle=enhanced,
	skin last=enhanced,
	before upper={\parindent15pt},
	]{}
	
	\vspace{0.05in}

	\paragraph{Relation with dissipative hydrodynamics.} Readers may find useful to connect this construction with the eventually more familiar approach of dissipative hydrodynamics \cite{Crossley:2015evo} and quasi-hydrodynamics \cite{Baggioli:2023tlc}. To make the comparison easier, we adopt here the notations of these articles, where, compared to the above, $\mathcal{A}_a^\mu \rightarrow B_{r,a}^\mu$ and $X_{r,a} \rightarrow \varphi_{r,a}$. The symmetry content of the theory is 
	\begin{align}
		&\text{gauge symmetry}: \qquad U(1)_r \times U(1)_a \\
		&\text{global symmetry}: \qquad U(1)_r \times U(1)_a  \rightarrow  U(1)_r
	\end{align}
	where in the second, the breaking of the global symmetries to their diagonal subgroup entails the dissipative and stochastic nature of the theory. Because the global $U(1)_a$ symmetry is broken, the covariant block in the advanced sector is $B_a^\mu = a^\mu + \partial^\mu \varphi_a$, where $\varphi_a \rightarrow \varphi_a - \epsilon_a$ shifts under advanced gauge transformation. Note that a similar block $B_r^\mu = A^\mu + \partial^\mu \varphi_r$ exists in the retarded sector, but because $U(1)_r$ symmetry is preserved, $\varphi_r$ never appears in $S_{\mathrm{eff}}$. We could then use $B_a^\mu$ and $B_r^\mu$ to construct the effective functional, following the approach of \cite{Crossley:2015evo}. Instead, we find convenient to work in an \textit{advanced unitary gauge} in which $\varphi_a = 0$. In this gauge, the construction of $S_{\mathrm{eff}}$ boils down to write the most generic functional invariant under retarded gauge transformation. It is then enough to consider the contractions of $F^{\mu\nu}$ with $a^\mu$, as we did above. 
\end{tcolorbox}

\begin{figure}[tbp]
	\centering
	\includegraphics[width=1\textwidth]{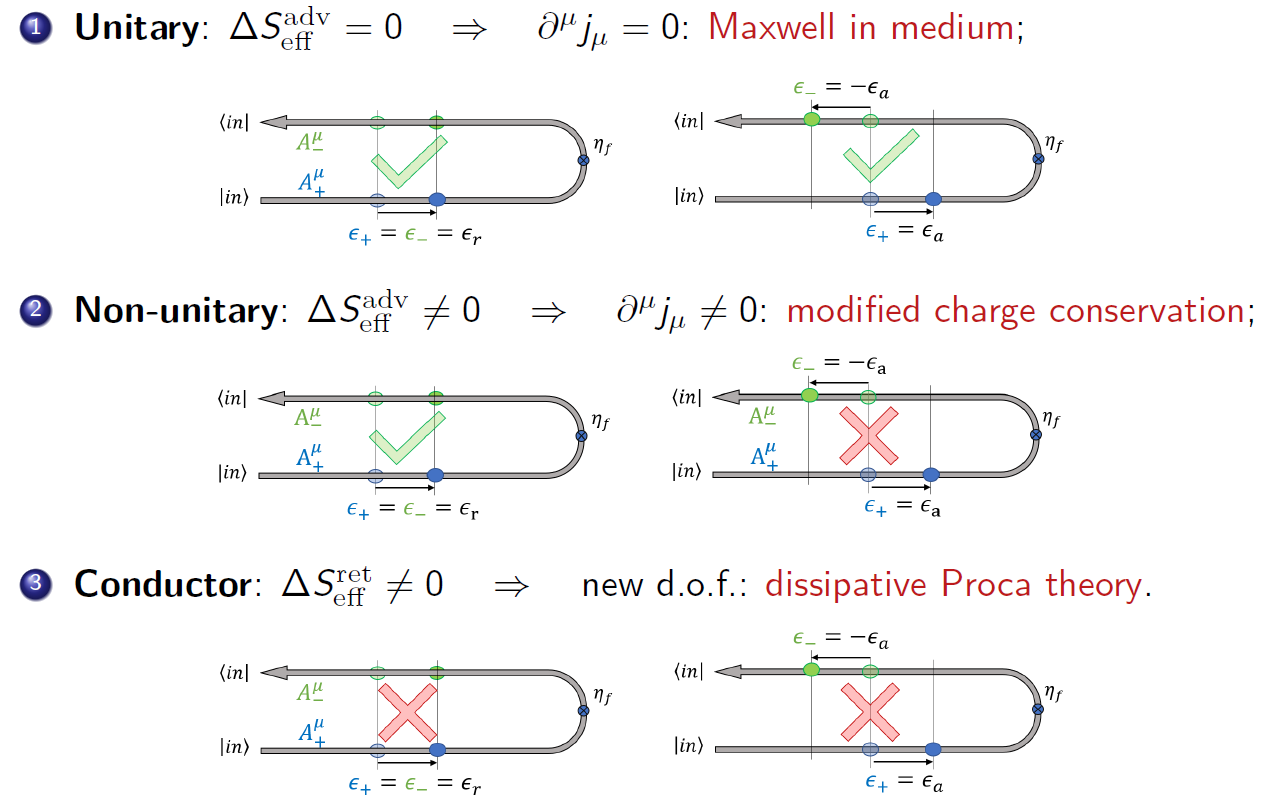}
	\caption{Consequences of the influence functional invariance under retarded and advanced transformations. i) When $S_{\mathrm{eff}}$ is invariant under both \textit{retarded} and \textit{advanced} transformations, the theory is \textit{closed} and the current is \textit{conserved in the standard way}, $\partial_\mu j^\mu = 0$. ii) When $S_{\mathrm{eff}}$ transforms under the \textit{advanced} transformation, the theory is \textit{open} and the current is \textit{conserved in a non-standard way}, $\partial_\mu j^\mu \neq 0$. iii) When $S_{\mathrm{eff}}$ transforms under both \textit{retarded} and \textit{advanced} transformations, \textit{new degrees of freedom} are present beyond $A_\mu$.}
	\label{fig:summary}
\end{figure}

The main findings associated with the retarded and advanced gauge transformations are summarized in \Fig{fig:summary}.

\subsection{Macroscopic dynamics and material properties}

We conclude this Lecture by the analysis of the dispersion relation of the propagating helicities and their relation with the familiar theory of electromagnetism in a medium. 

\subsubsection{Dispersion relations}\label{subsec:disprel}

Let us consider the number of propagating degrees of freedom and their dispersion relation in the (retarded \textit{and} advanced) Coulomb gauges. Although this is not a necessary assumption, the physics becomes quite transparent assuming we are studying a plane wave propagating in the $\hat z$ direction. First, we impose the Coulomb gauges by projecting $A^{\mu}$ and $a^{\mu}$ on the space perpendicular to $\hat z$. It amounts to drop their last component. For a wave moving in an arbitrary direction this would boil down to dividing $A^{\mu}$ and $a^{\mu}$ into their longitudinal and transverse components,
\begin{align}
	A^{i}=\Al^{i}+\At^{i}\,, \quad \text{with} \quad k_{i}\At^{i}=k^{j}\epsilon_{ijl}\Al^{j}=0\,,
\end{align}
and similarly for $a^\mu$. Then we derive a $ 3\times 3$ linear operator that describes the open dynamics of the 3 components in $ A^{\mu}$ and $ a^{\mu}$ that are not fixed by a gauge condition by dropping the longitudinal directions.  Inserting $A^{i}=\At^{i}$ and $a^{i}=\at^{i}$ in the open functional in \eqref{openf}, one finds
\begin{align}
	S= \int_{\omega, \bmk} \left[ k^{2} a^{0}A^{0} +\left(  i\gamma_{2}\omega+\gamma_{3}k^{2}\right)\at^{i}\At^{i}-2i\gamma_{4} \epsilon_{ijl}k^{i}\at^{j}\At^{l}\, \right].
\end{align}
This can also be written as the linear operator
\begin{align}\label{eq:matrix2}
	M_{\perp} =\begin{pmatrix}  k^{2} & 0 & 0  \\  0 &  i\gamma_{2}\omega+\gamma_{3}k^{2} &  -2i\gamma_{4} k\\ 0 &  2i\gamma_{4} k &  i\gamma_{2}\omega+\gamma_{3}k^{2}  \end{pmatrix}\,,
\end{align}
where the second and third rows and columns refer to the two independent components of $ \At$ and $ \at$, which are an orthonormal bases of the plane perpendicular to $ k^{i}$. The three eigenvalues are
\begin{align}\label{eq:disprelfull}
	(k^2\,,  i \gamma_{2} \omega+ \gamma_{3}  k^2 + 2 \gamma_{4} k, i \gamma_{2} \omega+ \gamma_{3}  k^2 - 2 \gamma_{4} k )\,.
\end{align}
This result is consistent with our expectation of having two propagating degrees of freedom. First, we can see that there is no choice of $ \omega$ such that the $00$-component of $ M_{\perp} $ vanishes. This indicates the existence of at least one \textit{constrained} degree of freedom, which is not propagating. This tells us that $ A^{0}$ is a constrained field, as expected. Second, we can diagonalize the remaining $2\times 2 $ block of $ M_\perp$ to find two eigenvalues. Demanding that they vanish gives us the dispersion relations for the two dynamical degrees of freedom (the two polarizations)
\begin{align}\label{eq:disprelfin}
	i \gamma_{2} \omega+ \gamma_{3}  k^2 \pm 2 \gamma_{4} k   =0\,.
\end{align}
To interpret this relation, let us first consider Maxwell theory in the vacuum. In this case, the dispersion relation reduces to $\omega^{2}=c^{2}k^{2}$ where we made the speed of light in the vacuum $c$ explicit. This also tells us that the constant part of $ \gamma_{3}$ is (minus) the square of the speed of light in the medium, that is $ \gamma_{3}=-v^{2}/c^2$. As we will see in \Sec{subsec:standardEM}, it is easy to relate this parameter to the refractive index $n=1/v$. The other leading order effect is a constant real part in $ \gamma_{2}$, that is $\gamma_{2}(\omega,k^2) \simeq \Gamma -i\omega$, which introduces the standard dissipation
\begin{align}
	\omega^{2} +i\Gamma \omega - v^{2}k^{2} = 0\then \omega=-i\frac{\Gamma}{2}\pm \sqrt{v^{2}k^{2}-(\Gamma/2)^{2}}\,.
\end{align}
For $ \Gamma >0$, this gives a stable system. At last, one can consider the effect of $\gamma_4$, leading to the dispersion relation
\begin{align}
	\omega^{2} +i\Gamma \omega - v^{2}k^{2} \pm 2 \gamma_4 k = 0\then \omega=-i\frac{\Gamma}{2}\pm \sqrt{v^{2}k^{2}-(\Gamma/2)^{2} \mp 2 \gamma_4 k}\,.
\end{align}
The two polarizations now have distinctive propagations, an effect known as birefringence. The discussion of the dispersion relations can be extended by accounting for higher-orders in derivatives. In general, $\gamma_{2,3,4}$ should be analytic functions of $ \omega$ and $ k^{2}$ around the origin by locality in time and space, assuming isotropy.


\subsubsection{Recovering electromagnetism in a medium}\label{subsec:standardEM}

E\&M in a medium is one of the most studied theories ever. Understanding how it compares with the above construction thus seems unavoidable. All results can be written in three equivalent forms: form notation with Hodge dual and exterior derivatives, covariant notation with Lorentz indices and finally in terms of the good old electric and magnetic vector fields. Here are some of the constitutive equations
\begin{align}
	&\text{Gauss law:} ~~~~~ d\star F= \star J \then \partial_{i} F^{i0} = \mu_0 J^0 \then \nabla \cdot \mathbf{E} = \frac{\rho}{\epsilon_0},\bigg. \\
	&\text{Ampere law:} ~~~~~ d\star F= \star J \then  \partial_{\mu} F^{\mu i} = \mu_0 J^i \then -\frac{1}{c^2} \frac{\partial \mathbf{E}}{\partial t} + \nabla \times \mathbf{B} = \mu_0 \mathbf{J}, \bigg.\\
	&\text{magnetic Gauss law:} ~~~~~ dF=0 \then \partial_{i}~\! ^\star\! F^{i0} = 0 \then \nabla \cdot \mathbf{B} = 0, \Bigg. \\
	&\text{Faraday induction:} ~~~~~ dF=0 \then  \partial_{\mu}~\! ^\star\! F^{\mu i} = 0 \then \frac{\partial \mathbf{B}}{\partial t} + \nabla \times \mathbf{E} = 0. \bigg.
\end{align}
The last two equations come from the Bianchi identity and do not involve any source. Consequently, they will not change in a material. 
Conversely, the first two equations come from the equations of motion, feature charges and current and change in the presence of a material.

A standard exercise consists in combining these linear equations in vacuum and solving them to find lightwaves propagating perpendicularly to $\mathbf{E}$ and $ \mathbf{B}$. In a material, the modifications of the the first two equations can be captured through the following manner. We introduce $ \mathbf{D}=\epsilon \mathbf{E}$ with $ \epsilon$ the electric permittivity and $ \mathbf{H}=\mathbf{B}/\mu$ with $ \mu$ the permeability.\footnote{This approach is actually only valid for the simplest ``linear'' material, where the response of the material's own electric and magnetic fields is proportional to $ \mathbf{E}$ and $ \mathbf{B}$. More generally, the presence of a material can induce non-linearities.} Then
\begin{align}\label{maxmedium}
	\nabla \cdot \mathbf{D} &= \rho_{\text{free}} \qquad \text{and} \qquad \nabla \times \mathbf{H} - \frac{\partial \mathbf{D}}{\partial t} = \mathbf{J}_{\text{free}}
\end{align}
where $ \mathbf{J}_{free}$ and $ \rho_{free}$ are the currents and charges that can move around in the material (as opposed to being fixed somewhere like around an atom or a molecule). In the absence of free currents we can easily solve these equations again and find the speed of light in the material $ v^{2}=1/(\epsilon \mu) $,
\begin{align}\label{eq:eomref}
	\mu \epsilon \ddot{\mathbf{E}}-\nabla^{2}\mathbf{E}=0\,.
\end{align}
The speed $ v$ is sometime related to the index of refraction $ n\equiv c/v$. The analysis of the dispersion relation obtained below \Eq{eq:disprelfin} revealed that the constant coefficient of the EFT parameter $\gamma_3$ directly relates to the speed of propagation in the medium through $\gamma_3 = - v^2$, that is $n = 1/\sqrt{-\gamma_3}$.

Higher derivatives tend to enrich the phenomenology while keeping the theory linear. One can simply work in frequency space in which all higher derivatives are collected into frequency and momentum dependence of the permittivity and permeability, $ \epsilon=\epsilon(\omega, k)$ and $ \mu=\mu(\omega, k)$. In Fourier space, both $ \epsilon(\omega, k)$ and $ \mu(\omega, k)$ can be complex. It turns out that the imaginary part of $ \epsilon(\omega, k)$ and $ \mu(\omega, k)$ leads to attenuation/dissipation, but also to a phase shift between the oscillations of $ \mathbf{E}$ and $ \mathbf{B}$.
From \Eq{eq:eomref}, the dispersion relation simply reads
\begin{align}
	\mu(\omega,k) \epsilon(\omega,k)\omega^{2} - k^{2}= 0\,.
\end{align}
Depending on the functions $ \epsilon(\omega,k)$ and $ \mu(\omega,k)$, there can be many different phenomena, such as normal propagation or absorption, in a way that is different for different frequencies and/or different wavenumbers. 

We now see how to derive this phenomenology from the Schwinger-Keldysh formalism developed in the previous sections. To make contact with electromagnetism in a medium, it is useful to start from \eqref{OEMnog} rephrased in terms of $ F^{0i}=E^{i}$ and $ F^{ij}=\epsilon^{ijl}B_{l}$. Adding the noise contributions, the constraint and equations of motion become
\begin{align}\label{eq:cEM}
	\frac{\delta S_{\mathrm{eff}}}{\delta a^{0}}=0\quad &\then \quad \nabla. \mathbf{E}= j_{0} + \xi_0, \\
	\frac{\delta S_{\mathrm{eff}}}{\delta a^{i}}=0\quad &\then \quad \gamma_{2}\mathbf{E} + \gamma_{3} \nabla \times \mathbf{B} - 2 \gamma_4 \mathbf{B} =\mathbf{j} + \boldmath{\xi}\,.
\end{align} 
Up to rescalings, we hence get equations similar to \eqref{maxmedium}, that is
\begin{align}
	\nabla \cdot \mathbf{D} &= \rho + \Xi \qquad \mathrm{and} \qquad \nabla \times \mathbf{H} - 2 \gamma_B \mathbf{H} + i \omega \mathbf{D}  = \mathbf{j} + \boldmath{\xi}\,.
\end{align}
To reach these expressions, we identified the permeability $\mu = 1/\gamma_3$, the permittivity $\epsilon = \gamma_2/(i\omega)$ and the birefringence index $\gamma_B = \gamma_4/\gamma_3$. We also used
the redefinitions $\rho = \epsilon j_0 $ and $\Xi = \epsilon \xi_0$. 
Apart from birefringence, the main difference with \Eq{maxmedium} comes from the stochastic contributions $\Xi$ and $\boldmath{\xi}$ which model the presence of random impurities in the medium through which light propagates.  Explicitly, \Eq{maxmedium} is recovered by setting $\gamma_B = \Xi = \boldmath{\xi} = 0$. 

In summary, the standard textbook treatment of electromagnetism in a medium is easily recovered from the open EFT construction. The effective functional $S_{\mathrm{eff}}$ generates modified Gauss and Ampère laws, accounting for the propagation of light in a dispersive ($\gamma_3$), dissipative ($\gamma_2$), anisotropic ($\gamma_4$) and random ($\xi_\mu$) medium.


\clearpage



\subsection{\textit{Problem set}}

\paragraph{\textit{Exercise 1.}} \textit{A constructive approach} \\

In this Exercise, we develop a classification of operators. Let us consider the functional linear in the advanced field
\begin{align}\label{openf2}
	S_1=\int_{\omega, \bmk} a^{\mu}M_{\mu\nu}A^{\nu}\,.
\end{align}
We consider the matrix $M$ in \Eq{openf2} to be a sum over operators
\begin{equation}\label{eq:matform}
	M_{\mu\nu}(\omega,\bmk)=\sum_{n}\mathcal{O}^{(n)}_{\mu\nu}(\omega,\bmk).
\end{equation}
Our goal is to construct and classify these operators.
\begin{enumerate}
	\item What is the restriction imposed by the retarded gauge invariance on $M_{\mu\nu}$? What about the restriction coming from unitarity (in contrast to non-unitary operators)?
	\item Let us first consider Lorentz invariant and unitary operators. The associated building blocks are the flat-space metric $\eta_{\mu\nu}$ and the four momentum $k^\mu$. 
	\begin{enumerate}
		\item At second order in derivatives, what are the two possible operators?
		\item Using the retarded gauge invariance, reduce them to one.
		\item Can you identify this operator? What is its physical interpretation?
	\end{enumerate}
	\item We now consider operators that are still unitary, but break Lorentz invariance to the Euclidean group SO(3), that is rotations and translations. Physically, they capture the fact that a homogeneous and isotropic material selects a preferred reference frame. Mathematically, we account for the existence of a preferred reference frame by introducing a timelike direction $n^{\mu}$, which we normalize by $n^\mu n_\mu=-1$.
	\begin{enumerate}
		\item At second order in derivatives, what are the three new operators one can construct?
		\item Using the retarded gauge invariance, reduce them to one.
	\end{enumerate}
 	The ratio between the EFT coefficients controlling this operator and the previous one uniquely determine the speed of sound. 
	\item There exists another Lorentz-breaking unitary operator linear in $k^\mu$ and $n^\mu$ which uses the Levi-Civita symbol $\epsilon_{\mu\nu\alpha\beta}$.
	\begin{enumerate}
		\item Construct this operator.
		\item Check its invariance under retarded gauge transformation.
		\item Check its unitarity.
	\end{enumerate}
	\item At last, let us discuss operators that are non-unitary and break Lorentz invariance. At lowest order in derivatives, the tensor structure is linear in $k^\mu$.
	\begin{enumerate}
		\item At linear order in derivatives, what are the three operators one can construct?
		\item Using retarded gauge invariance, reduce them to one.
		\item Check the non-unitarity of this operator. What is its left zero eigenvector?
	\end{enumerate}
\end{enumerate}
We have built the most general matrices compatible with retarded gauge invariance that includes at lowest-order in derivatives a kinetic term, non-trivial speed of sound, birefringence and dissipation. More details can be found in Section 4.5 of \cite{Salcedo:2024nex}.


\begin{center}
	\noindent\rule{8cm}{0.4pt}
\end{center}


\paragraph{\textit{Exercise 2.}} \textit{Covariant gauges} \\

The Coulomb gauge is not always the most convenient choice. There is a choice of gauge that retains Lorentz invariance and some gauge symmetry. These are the so-called covariant gauges. In the unitary theory they are given by introducing a new term into the action that makes the equation of motion invertible
\begin{equation}\label{eq:zetaterm}
	\mathcal{L}=-\frac{1}{4}F_{\mu\nu}F^{\mu\nu}-\frac{1}{2\zeta}\left(\partial_{\mu}A^{\mu}\right).
\end{equation}
In this Exercise, we generalize this set of covariant gauges to Open E\&M.

\begin{enumerate}
	\item Let us start from the open effective functional constructed above in \Eq{eq:matrix1}. 
	\begin{enumerate}
		\item Write \Eq{eq:zetaterm} in the Keldysh basis.
		\item Modify the kinematic matrix appearing in the matrix $M_{\mu\nu}$.
	\end{enumerate}
	\item We now investigate the properties of the kinematic matrix.
	\begin{enumerate}
		\item Compute the associated eigenvalues and compare them with \Eq{eq:disprelfull}. You should find that the two propagating modes are unchanged. The last two eigenvalues are then associated to a ghost and a constrained mode.
		\item Compute the determinant of $M_{\mu\nu}$. What is the difference compare with the above results? What practical consequence follows from $\zeta \neq 0$?
	\end{enumerate} 
\end{enumerate}
It follows that whenever $\zeta \neq 0$, the Gaussian path integral can be performed analytically, without any further requirement. The final result for the partition function is
\begin{equation}
	Z[J^{A},J^{a}]=Z[0,0]\text{exp}\left\{\frac{1}{2}\int\frac{d^{4}k}{(2\pi)^{4}}\left(J^{A}_{\mu}(-\omega, -\bmk)J^{a}_{\mu}(-\omega, -\bmk)\right)H^{\mu\nu}(\omega, \bmk)\begin{pmatrix}
		J^{A}_{\nu}(\omega, \bmk)\\
		J^{a}_{\nu}(\omega, \bmk)
	\end{pmatrix}\right\},
\end{equation}
where
\begin{equation}\label{eq:matrixH}
	H^{\mu\nu}=\begin{pmatrix}
		2[M(\omega, \bmk)^{-1}]^{\mu\alpha}[M(-\omega, -\bmk)^{-1}]^{\nu\beta}\left[N_{\alpha\beta}(\omega, \bmk)\right] & i[M(\omega, \bmk)^{-1}]^{\mu\nu}\\
		i[M(-\omega, -\bmk)^{-1}]^{\nu\mu}  & 0
	\end{pmatrix}.
\end{equation}
If we now provide an explicit expression for $M^{-1}$ and $N$ the noise kernel, we can extract the expression of the propagators in the covariant gauge. More details can be found in Section 4.6 of \cite{Salcedo:2024nex}.

\begin{center}
	\noindent\rule{8cm}{0.4pt}
\end{center}


\paragraph{\textit{Exercise 3.}} \textit{Topological operators} \\

Allowing for dissipation and noise enlarge the class of operators accessible from within the EFT. Do some of these new operators have particular properties? In the unitary theory, a well known non-trivial extension of Maxwell theory consists in adding a topological operator known as \textit{theta term}
\begin{align}
	S_{\theta_1} = \frac{\theta_1}{4} \int \dd^4x ^\star\!F^{\mu\nu} F_{\mu\nu} = \frac{\theta_1}{4} \int \dd^4x \bm{\mathrm{E}}.\bm{\mathrm{B}}
\end{align}
where 
\begin{align}
	^\star\! F^{\mu\nu} = \frac{1}{2} \epsilon^{\mu\nu\rho\sigma}F_{\rho\sigma}.
\end{align}
When $\theta_1$ is a constant, it is easy to check $S_{\theta_1}$ turns out to be a total derivative, 
\begin{align}
	S_{\theta_1} = \frac{\theta_1}{8} \int \dd^4x \partial_\mu\left(\epsilon^{\mu\nu\rho\sigma}A_\nu\partial_{\rho}A_\sigma\right),\label{eq:thetauni}
\end{align}
such that all the physical information is encoded in the spacetime boundary. Are there any non-unitary operators that exhibit similar properties? 
\begin{enumerate}
	\item Construct an operator linear in $A_\rho$ and $a_\sigma$ which uses the Levi-Civita symbol $\epsilon^{\mu\nu\rho\sigma}$. Prove its uniqueness. Is it a total derivative? What is its relation with \Eq{eq:thetauni}?
	\item Construct an analogue operator quadratic in the advanced field $a_\sigma$. Is it a total derivative? What are the NEQ constraints on this operator? What could be a physical interpretation for the presence of this operator?
\end{enumerate}
More details can be found in Section 4.7 of \cite{Salcedo:2024nex}.


\section{Lecture 5: Open gravity}\label{sec:lec5}


Now we have earned some understanding of gauge symmetries in the Schwinger-Keldysh contour, we can develop a systematic construction of gravity in a medium. This lecture summarizes the main steps of the construction carried out in \cite{Salcedo:2025ezu}. 


\subsection{Gravity in a medium}

\begin{figure}[tbp]
	\centering
	\includegraphics[width=1\textwidth]{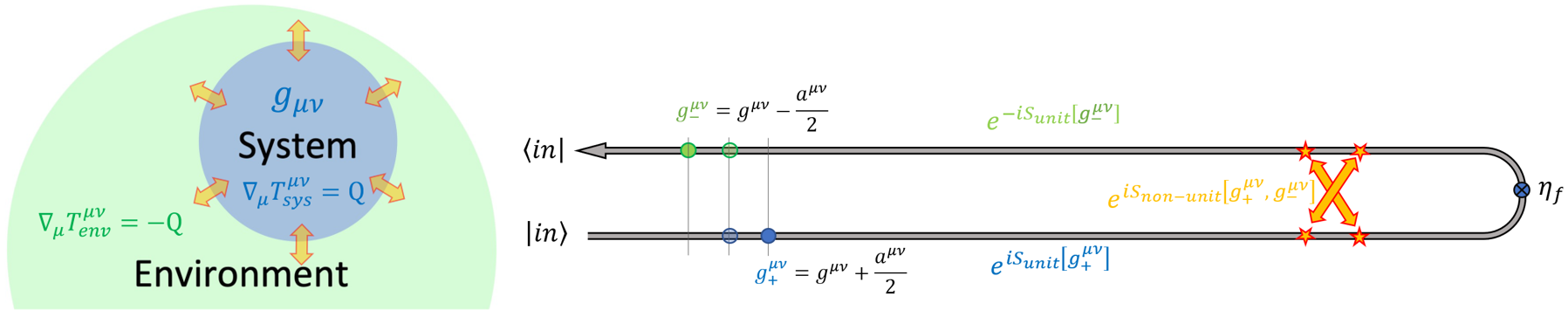}
	\caption{\textit{Left}: Illustration of the lack of conservation of the system's stress-energy tensor. This effect is a mere consequence of the total charge conservation ${\nabla}^\mu  {T}^{(\text{all})}_{\mu\nu} = 0$. It simply illustrates the fact that system and environment exchange energy and momentum. \textit{Right}: Schwinger-Keldysh contour considered to model dissipation and noise in open gravity.}
	\label{fig:ograv}
\end{figure}

Let us start from a full ``closed" theory, where the evolution is unitary and conservative. The Einstein equations take the usual form
\begin{align}
	\Mpl^2 {G}_{\mu\nu} = {T}^{(\text{all})}_{\mu\nu}  \qquad \Rightarrow \qquad {\nabla}^\mu  {G}_{\mu\nu}  = 0 = {\nabla}^\mu  {T}^{(\text{all})}_{\mu\nu}, 
\end{align}
where ${T}^{(\text{all})}_{\mu\nu} $ is the full energy-momentum tensor of the theory. We now separate ${T}^{(\text{all})}_{\mu\nu} $ into a system ${T}^{(\text{sys})}_{\mu\nu} $ and an environment ${T}^{(\text{end})}_{\mu\nu} $. Let us assume a hierarchy of scales between the characteristic time and length scales we want to study and those of the environment. In this case, all the degrees of freedom in the environment are assumed to be non-dynamical, and have hence been integrated out. 
As a consequence, ${T}^{(\text{all})}_{\mu\nu} $ should be substituted with its expectation value, which in turn can be written in terms of ``light'' degrees of freedom, namely the metric, and the system's degrees of freedom. As a result, Einstein's equations take the schematic form
\begin{align}\label{eq:mainidea}
	\Mpl^2 {G}_{\mu\nu} - \ex{{T}^{(\mathrm{env})}_{\mu\nu}}= {T}^{(\text{sys})}_{\mu\nu}  \qquad \Rightarrow \qquad  \Mpl^2 {G}_{\mu\nu} +\text{modifications}= {T}^{(\text{sys})}_{\mu\nu} \,,
\end{align}
where the ``modifications'' are terms built out of the metric and its derivatives that are fixed by diff-invariance. Since the Einstein tensor obeys the contracted Bianchi identities, $\nabla^\mu G_{\mu\nu} = 0$, but the modifications in general do not, one finds that $ \nabla^\mu {T}^{(\text{sys})}_{\mu\nu}\neq 0$.
This apparent non-conservation of ${T}^{(\text{sys})}_{\mu\nu}$, implied by our modified Einstein equations, is actually simply the statement that the \textit{full} energy-momentum tensor is conserved
\begin{align}
	\nabla^\mu{T}^{(\text{sys})}_{\mu\nu}=-\nabla^\mu{T}^{(\text{env})}_{\mu\nu}\,.
\end{align}
The situation of interest is described in \Fig{fig:ograv}, by analogy to open electromagnetism.

How can we model systematically these observations? Building on the learnings from the previous lectures, we will construct an effective functional including both unitary and non-unitary effects in the Schwinger-Keldysh contour (\textit{Right} panel of \Fig{fig:ograv}). As always, one needs to first define the field content of the theory, before spelling the symmetries. In \cite{Salcedo:2025ezu}, we considered the metric $g_{\mu\nu}$, together with a clock-field, aiming to reproduce the field content of the EFT of Inflation \cite{Cheung:2007st} and Dark Energy \cite{Gubitosi:2012hu}. Importantly, the clock field can be parametrized in three different ways: by a scalar field $\phi$, reabsorbed into the time variable $t$ in the so-called unitary gauge and reintroduced through a \stuck trick into the pseudo-Goldstone boson of time-translation symmetry breaking $\pi$. Understanding these various manipulations in the Schwinger-Keldysh contour will be a key aspect of open gravity.

Now comes the symmetry story.
To construct an \textit{open theory of gravity and a single clock}, we start with doubling the fields. This gives us two copies of diff invariance in each branch of the Schwinger-Keldysh path integral, which we denote as diff$_+$ and diff$_-$. It is convenient to work with diffs that act in the same (retarded) or opposite (advanced) direction in each branch, which we denote as diff$_r$ and diff$_a$ respectively. Since an open theory by definition includes couplings between fields in the $+$ and $-$ branch, the 4d advanced diffs are explicitly broken by dissipative effects. The 4d retarded diffs are broken to 3d spatial retarded diffs by the clock foliation. Schematically
\begin{align}
	( \text{4d-diff}_+ \times \text{4d-diff}_- )\simeq (\text{4d-diff}_a \times \text{4d-diff}_r) \quad\overset{\text{open}}{\longrightarrow} \quad(\text{4d-diff}_r ) \quad\overset{\text{clock}}{\longrightarrow}\quad (\text{3d-diff}_r)\,.
\end{align}
Our construction then mimics the one of the EFT of Inflation \cite{Cheung:2007st}. Let us highlight the general idea before going through into the explicit constuction in \Sec{subsec:SIF}. We first define a notion of \textit{unitary gauges}, that is we choose a convenient coordinate system such that the clock is unperturbed, $\phi_r=\bar{\phi}(t)$ and $\phi_a = 0$. 
The open theory of gravity plus a single clock is then defined as the most generic theory of an advanced and retarded metric that is invariant under 3d-diff$_r$, denoted $S_{\text{3d-diff}_r}[g_{\mu\nu},a_{\mu\nu} ]$. 

Having defined our theory, we can choose any convenient way to analyse it. One can reintroduce the scalar clocks by performing a retarded and advanced \stuck trick, which simultaneously generate the new fields $\pi_r$ and $\pi_a$ and make the theory retarded and advanced time-diff invariant. For single-clock cosmologies we have
\begin{align}
	S_{\text{3d-diff}_r}[g_{\mu\nu},a_{\mu\nu} ] \quad \overset{\text{St\"uck.}^2}{\longrightarrow} \quad S_{\text{4d-diff}_r\times \text{t-diff}_a} [g_{\mu\nu},a_{\mu\nu}, \pi_r,\pi_a] \quad\overset{\text{decoup.}}{\longrightarrow} \quad S[\pi_r,\pi_a]\,,
\end{align}
where the last step corresponds to the Open Effective Field Theory of Inflation (Open EFToI) in the decoupling limit \cite{Salcedo:2024nex} studied in Lecture \ref{sec:lec3}.


\subsection{Open functional in unitary gauges}\label{subsec:SIF}

Let us jump in the construction of the functional for an open theory of gravity in a medium. Just like any field is doubled in the Schwinger-Keldysh formalism, so is the metric. One then must consider two rank-2 tensors $g_{\mu\nu}^\pm$. Expressed in the Keldysh basis, the retarded and advanced metric read
\begin{align}\label{eq:retadvmetricsdef}
	g_{\mu\nu} &= \frac{\left(g_+\right)_{\mu\nu} + \left(g_-\right)_{\mu\nu}}{2}, \qquad \mathrm{and} \qquad 
	a^{\mu\nu} = \left(g_+\right)^{\mu\nu} - \left(g_-\right)^{\mu\nu} \,.
\end{align}
Since physical fields are associated to the retarded sector, we will later find it convenient to construct geometrical objects based on the retarded metric $g_{\mu\nu}$. 

In addition to the metric we have to consider the clock of the theory, which is a scalar field we will call $\phi$. Since we work with the Schwinger-Keldysh contour we have to double this field to distinguish insertion of $\phi$ in the plus and minus branch, so we must have $\phi_\pm(t,\bfx)$. The corresponding retarded and advanced fields are
\begin{align}
	\phi_r(t,\bfx)&\equiv\frac12 [\phi_+(t,\bfx)+\phi_-(t,\bfx)]\,, &
	\phi_a(t,\bfx)&\equiv \phi_+(t,\bfx)-\phi_-(t,\bfx)\,.
\end{align}
If generic dissipative effects are present, at this point the theory is invariant under 4d-diff$_r$ while we expect all advanced diffs to be broken
\begin{align}
	S=S_{\text{4d-diff}_r}[g_{\mu\nu},a^{\mu\nu},\phi_r,\phi_a]\,.
\end{align}

To construct the theory following the EFToI recipe \cite{Cheung:2007st}, it is convenient to fix retarded and advanced time diffs with some gauge prescriptions.\footnote{In these notes, we slightly simplify the construction compared to \cite{Salcedo:2025ezu} by fixing \textit{both} the retarded and advanced unitary gauge. While the two approaches are equivalent for the purpose of this discussion, readers interested in further developing the formalism are encouraged to go through \cite{Salcedo:2025ezu} for an extended discussion.} In the EFToI we would simply choose coordinates such that $\phi=\bar{\phi}(t)$. Here, we proceed as follows.

\begin{tcolorbox}[%
	enhanced, 
	breakable,
	skin first=enhanced,
	skin middle=enhanced,
	skin last=enhanced,
	before upper={\parindent15pt},
	]{}
	
	\vspace{0.05in}
	
	\paragraph{Building unitary gauges.} 

Since it is the retarded/symmetric combination of $\phi_\pm(t,\bfx)$ that contains the background of the field, the first natural option is to choose the retarded and advanced gauges such that
\begin{align}
	\phi_r(t,\bfx)\doteq \bar\phi(t)\, \qquad \phi_a(t,\bfx)\doteq 0.
\end{align}
We dub these gauges the \textit{clock retarded and advanced unitary gauges}. There is actually a second natural choice. To see this, let's go back to the fully 4d-diff$_r$ invariant theory and consider the following \textit{field redefinition}
\begin{align}
	\phi_+(t,\bfx)&\equiv \bar\phi(t_+(t,\bfx))\,, &\phi_-(t,\bfx)&\equiv \bar\phi(t_-(t,\bfx))\,,
\end{align}
where we have simply traded the two fields $\phi_\pm(t,\bfx)$ for the two fields $t_\pm (t,\bfx)$, which can always be done as long as $\bar \phi$ is a monotonic function. The name ``$t$'' for these new fields suggest the interpretation of maps of the spacetime to the doubled fluid space as suggested in \cite{Liu:2018kfw}. We stress however that $t_\pm(t,\bfx)$ are dynamical fields in the action, as opposed to coordinates that are integrated over. The second gauge that fixes retarded time diffs is then
\begin{align}\label{tr=t}
	\boxed{ t_r(t,\bfx)\equiv\frac12 \left[t_+(t,\bfx)+t_-(t,\bfx)\right]\doteq t, \qquad  t_a(t,\bfx)\equiv t_+(t,\bfx) - t_-(t,\bfx)\doteq 0}\,.
\end{align}
We dub this second set of gauges the \textit{(time) retarded and advanced unitary gauges}.

Before proceeding we point out that the existence of an advanced clock $t_a$ introduces a few subtleties. Here, we exploit the fact that in the specific case of a single-clock cosmology and at the classical stochastic order, one can remove the $t_a$ dependent terms with a field redefinition  \cite{Salcedo:2025ezu}. Building on the ideas developed for Open Electromagnetism in Lecture \ref{sec:lec4}, this is equivalent to work in the advanced unitary gauge defined above. In a general gauge, the coupling constants in the EFT of perturbations are functions of the fields $\phi_{\pm}(t, \bfx)$, that is of both the retarded and advanced clocks $t_r(t, \bmx)$ and $t_a(t, \bmx)$. Hence a generic coupling constant $\gamma$ will have the form $ \gamma(t_r; t_a)$. During inflation, the expansion in $t_r$ is known to be slow-roll suppressed \cite{Cheung:2007st}, however the dependence of the open functional on $t_{a}$ is \textit{a priori} arbitrary and un-restricted. In other words, in contrast to the original EFToI \cite{Cheung:2007st}, there are no constraints relating the derivatives of the coupling constants to the coupling constants of terms containing $t_{a}^{n}$ (see \cite{Finelli:2018upr}). 

\end{tcolorbox}

\paragraph{Building the action.} All perturbations are now absorbed in $g_{\mu\nu}$ and $a^{\mu\nu}$. Following the approach of \cite{Cheung:2007st}, we want our effective functional to be invariant under retarded spatial diffeomorphisms
\begin{align}
	S=S_{\text{3d-diff}_r}[g_{\mu\nu},a^{\mu\nu}] = \sum_{n=1}^{\infty} S_n \qquad \mathrm{with} \qquad S_n = \mathcal{O}(\mathrm{adv}^n), 
\end{align}
where in the second equality, we used the unitarity constraints \Eqs{eq:norm}, \eqref{eq:herm} and \eqref{eq:pos} which provide a convenient expansion scheme in powers of the advanced components \cite{Liu:2018kfw, Hongo:2018ant, Salcedo:2024smn}. Restricting ourselves to order $\O((a^{\mu\nu})^2)$ for practical applications, we focus on $S_1$ and $S_2$, for which we illustrate the general procedure. $S_1$ encodes how many and what degrees of freedom the EFT describes while $S_2$ models the noise.

A simple procedure to guarantee that the effective functional is invariant under retarded \textit{spatial} diffeomorphisms is to use geometric objects built out of $g_{\mu\nu}$. We can define a normal vector $n_\mu$ to the time foliation as in \eqref{eq:n_mu_def}, 
\begin{align}\label{eq:n_mu_defv2}
	n_\mu \equiv - \frac{\partial_\mu \phi_\pm}{\sqrt{-g^{\mu\nu} \partial_\mu \phi_\pm \partial_\nu \phi_\pm}}=- \frac{\partial_\mu \bar\phi(t_r)}{\sqrt{-g^{\mu\nu} \partial_\mu \bar\phi(t_r) \partial_\nu \bar\phi(t_r)}} \doteq - \frac{\delta^0_{~\mu}}{\sqrt{-g^{00}}}\,,
\end{align}
where the last equality is true in retarded unitary gauge $t_r=t$. From this we can define the induced metric $h_{\mu\nu} \equiv g_{\mu\nu} + n_\mu n_\nu$, the extrinsic curvature $K_{\mu\nu}$ in \eqref{eq:extrinsiccurv}, the covariant derivative%
$\nabla_\mu$ and the Riemann tensor $R_{\mu\nu\rho\sigma}$ are all built from the retarded metric $g_{\mu\nu}$. 

Hence, in retarded unitary gauge, the first contribution to the action takes the form 
\begin{align}\label{eq:S1try}
	S_1 = \int \dd^4 x \sqrt{-g}& \Big[M_{\mu\nu} (R_{\mu\nu\rho\sigma}, g^{00}, K_{\mu\nu}, \nabla_\mu; t) a^{\mu\nu} \Big] , 
\end{align}
where $M_{\mu\nu}$ can break retarded time diffeomorphisms but has to transform as a rank-$2$ cotensor under retarded spatial diffeomorphisms. Notice that it is the determinant of the retarded metric $g$ that appears in the volume measure $\sqrt{-g}$. We can also construct the noise functional 
\begin{align}\label{eq:S2try}
	S_2 &= i \int \dd^4 x \sqrt{-g} \Big[N_{\mu\nu\rho\sigma} (R_{\mu\nu\rho\sigma}, g^{00}, K_{\mu\nu}, \nabla_\mu; t) a^{\mu\nu}  a^{\rho\sigma} \Big],  
\end{align}
where $N_{\mu\nu\rho\sigma}$ has to transform as a rank-$4$ cotensor under retarded spatial 3d-diffs. As always, the Hubbard-Stratonovich trick \cite{Hubbard:1959ub,Stratonovich1957} replaces the quadratic terms in $a^{\mu\nu}$ with a path integral over an auxiliary field $\xi_{\mu\nu}$ 
\begin{align}\label{eq:HStrickOGR}
	&\text{exp}\left\{-\int \dd^{4}x\sqrt{-g}N_{\mu\nu\rho\sigma} a^{\mu\nu}  a^{\rho\sigma}\right\}  \nonumber \\
	&\qquad =\int[\mathcal{D}\xi_{\mu\nu}]\;\text{exp}\left\{-\int \dd^{4}x \sqrt{-g}\left[\frac{1}{4}(N^{-1})^{\mu\nu\rho\sigma}\xi_{\mu\nu}\xi_{\rho\sigma}+i\xi_{\mu\nu}a^{\mu\nu}\right]\right\},
\end{align}
leading to the stochastic Einstein-Langevin equations in the unitary gauges
\begin{align}
	\frac{\delta S_1}{\delta a^{\mu\nu}}  + \frac{\delta S_2}{\delta a^{\mu\nu}} = 0 \qquad \Rightarrow \qquad M_{\mu\nu}  = \xi_{\mu\nu}.
\end{align}
The equations of motion are now stochastic and their solutions should be averaged over the distribution of the noise. 


\paragraph{Construction of $S_1$.}

We now proceed to expanding the action $S_1$ up to second order in derivatives. 
Before we start, notice that any EFT operator can be multiplied by arbitrary powers of $g^{00}$, which transforms as a scalar under spatial retarded diffeomorphisms and does not change the order in derivatives. Just as in the original EFToI \cite{Cheung:2007st}, it is convenient to work with $(1+g^{00})$, as this expression vanishes on an FLRW background and starts explicitly at linear order. Hence, 
\begin{align}
	M_{\mu\nu} &= \sum_{\ell=0} (g^{00} +1)^\ell M_{\mu\nu, \ell}(R_{\mu\nu\rho\sigma}, K_{\mu\nu}, n_\mu, \nabla_\mu; t)   \, ,
\end{align}
where the index $\ell$ here is not a spacetime-index, but rather denotes the fact that $M_{\mu\nu,\ell}$ will in general have different EFT coefficients at different orders in $(1+g^{00})$. To construct the most generic $M_{\mu\nu, \ell} (R_{\mu\nu\rho\sigma}, K_{\mu\nu}, n_\mu, \nabla_\mu; t)$ up to second derivatives, we split it along the foliation according to
\begin{equation}\label{eq:decomp}
	M_{\mu\nu, \ell} =  n_\mu n_\nu M^{tt}_\ell + n_{(\mu} M^{ts}_{\nu),\ell} + M^{ss}_{\mu\nu, \ell} \, .
\end{equation}
Here $M^{tt}_\ell$, $M^{ts}_{\rho,\ell}$ and $M^{ss}_{\rho \sigma, \ell}$ denote the most generic scalar, vector and rank-2 tensor under spatial retarded diffs, which will again be constructed using $R_{\mu\nu\rho\sigma}$, $K_{\mu\nu}$ $n_\mu$, $\nabla_\mu$ and $g_{\mu\nu}$.
The rank-2 tensor $M^{ss}_{\mu\nu, \ell}$ can be further decomposed into: 
\begin{equation}\label{eq:decomp_Mmunu}
	M^{ss}_{\mu\nu, \ell} = M^{ss}_{\ell} g_{\mu\nu} + \Tilde{M}^{ss}_{\mu\nu, \ell} \, .
\end{equation}
Now all free indices in $M^{ts}_{\nu,\ell}$ and $\Tilde{M}^{ss}_{\mu\nu, \ell}$ should not be $\sim n_\mu$ or $\sim g_{\mu\nu}$, or else they are redundant with the other terms in \Eqs{eq:decomp} and \eqref{eq:decomp_Mmunu}. 
Since $M^{tt}_\ell$ and $M^{ss}_\ell$ must be scalars under spatial 3d-diffs, they are constructed from contractions of the tensors $R^{\mu\nu\rho\sigma}$, $K^{\mu\nu}$ and $g^{\mu\nu}$ with all possible combinations of $n_\mu$, $\nabla_\mu$, $g_{\mu\nu}$ and $K_{\mu\nu}$, up to second order in derivatives. Explicitly, we get
\begin{align}
	M^{tt}_\ell &= \gtt{1} + \gtt{2} K + \gtt{3}  K^2 + \gtt{4} K_{\alpha\beta} K^{\alpha\beta} + \gtt{5} \nabla^0 K + \gtt{6} R + \gtt{7} R^{00} \, , \bigg. \label{eq:Mtt} \\
	M^{ss}_{\ell} &= \gss{1} + \gss{2} K + \gss{3}  K^2 + \gss{4} K_{\alpha\beta} K^{\alpha\beta} + \gss{5} \nabla^0 K + \gss{6} R + \gss{7} R^{00} \, , \bigg. \label{eq:Mss_scalar} 
\end{align}
in the retarded unitary gauge. To obtain $M^{ts}_{\rho,\ell}$ and $\Tilde{M}^{ss}_{\rho \sigma, \ell}$, we have to construct the most generic vector and rank-2 tensor under retarded spatial 3d-diffs: we contract $R^{\mu\nu\rho\sigma}$ and $K^{\mu\nu}$ with all possible combinations of $K_{\mu\nu}$, $n_\mu$, $\nabla_\mu$ and $g_{\mu\nu}$, up to second order in derivatives, such that one and two indices remain uncontracted. Note that, as mentioned previously, the free indices may not be $n_\mu$ or $g_{\mu\nu}$. This leads to
\begin{align}
	M^{ts}_{\mu, \ell} &= \gts{1} n_\alpha g_{\mu\gamma} g_{\beta \delta} R^{\alpha \beta \gamma \delta} + \gts{2} \nabla_\mu g_{\alpha\beta} K^{\alpha\beta} + \gts{3} g_{\mu\alpha} \nabla_\beta K^{\alpha\beta} \bigg. \label{eq:Mts} \\
	&= \gts{1} R^{0}{}_\mu + \gts{2} \nabla_\mu K + \gts{3} \nabla_\beta K^{\beta}{}_\mu \, \bigg. , 
\end{align}  
and
\begin{align}
	\Tilde{M}^{ss}_{\mu\nu, \ell} &= \big(\gss{8} g_{\mu\alpha} g_{\nu\beta} + \gss{9} g_{\mu\alpha} g_{\nu\beta} n^\rho \nabla_\rho + \gss{10} g_{\mu\alpha} K_{\nu\beta} + \gss{11} g_{\alpha\beta} K_{\mu\nu} \big) K^{\alpha\beta} \bigg. \label{eq:Mss} \\
	+& \left[g_{\mu\alpha} g_{\nu \gamma} \left( \gss{12} g_{\beta\delta} + \gss{13} n_\beta n_\delta \right)  \right] R^{\alpha\beta\gamma\delta} \bigg. \nonumber \\
	&= \gss{8} K_{\mu\nu} + \gss{9} \nabla^0 K_{\mu\nu} + \gss{10} K_{\mu\alpha} K^{\alpha}{}_{\nu} + \gss{11} K K_{\mu\nu} + \gss{12} R_{\mu\nu} + \gss{13} R_\mu{}^0{}_\nu{}^0 \, \bigg.  ,
\end{align}  
where we have expressed the final results in the retarded unitary gauge. 

We may also construct parity-violating operators by considering additional terms constructed with the totally antisymmetric tensor $\epsilon_{\mu\nu\rho\sigma}$. Up to second order in derivatives, there are only two additional terms contributing to $S_1$:
\begin{align}\label{eq:bir1}
	M^{\mathrm{P.O.}}_{\mu\nu, \ell} &= \epsilon^{\alpha\beta\gamma\delta}  g_{\mu \alpha} n_\beta \left(\gamma^{\po}_{1,\ell}  \nabla_\gamma K_{\nu\delta} + \gamma^{\po}_{2,\ell} g^{\rho\sigma} n_\sigma R_{\gamma\delta\rho\nu}  \right) \, ,
\end{align}
which may be included in $M^{ss}_{\mu\nu, \ell}$. Any other potential term including the totally antisymmetric tensor either vanishes due to its symmetries, is of higher order in derivatives or of higher order in advanced variables.

Eventually, in the retarded unitary gauge the action $S_1$ reads
\begin{align}\label{eq:S1v1}
	S_1 &= \int \dd^4 x \sqrt{-g} \, \sum_{\ell=0} \, \left(g^{00} + 1 \right)^\ell \bigg\{ \, a^{00} \Big[ \gtt{1} + \gtt{2} K + \gtt{3}  K^2  + \gtt{4} K_{\alpha\beta} K^{\alpha\beta} \nonumber \\
	&\quad+ \gtt{5} \nabla^0 K + \gtt{6} R + \gtt{7} R^{00}  \Big] \bigg.  + a^{0\mu} \Big[ \gts{1} R^{0}{}_\mu + \gts{2} \nabla_\mu K + \gts{3} \nabla_\beta K^{\beta}{}_\mu \Big] \bigg.\nonumber \\
	& \quad + a^{\mu\nu} \Big[ g_{\mu\nu} \Big( \gss{1} + \gss{2} K + \gss{3}  K^2 + \gss{4} K_{\alpha\beta} K^{\alpha\beta} + \gss{5} \nabla^0 K+ \gss{6} R + \gss{7} R^{00} \Big)  \bigg.\nonumber \\ 
	&\quad  + \gss{8} K_{\mu\nu} + \gss{9} \nabla^0 K_{\mu\nu} + \gss{10} K_{\mu\alpha} K^{\alpha}{}_{\nu} \nonumber \bigg. + \gss{11} K K_{\mu\nu} + \gss{12} R_{\mu\nu} + \gss{13} R_\mu{}^0{}_\nu{}^0 \\
	&\quad  + \gamma^{\po}_{1,\ell} \epsilon_\mu{}^{\alpha\beta0} \nabla_\alpha K_{\beta\nu} + \gamma^{\po}_{2,\ell} \epsilon_\mu{}^{\alpha\beta0} R_{\alpha\beta}{}^0{}_\nu \Big]  \bigg\} .
\end{align}
We can absorb a few terms by rescaling and redefining the field $a^{\mu\nu}$. First, we identify the operators $\gsso{12} a^{\mu\nu} R_{\mu\nu}$ and $\gsso{6} a^{\mu\nu} g_{\mu\nu} R$ as the two terms appearing in the Einstein tensor. We assume throughout that these two EFT coefficients are non-zero. Consequently we can set $\gsso{12} = \Mpl^2/2$ by rescaling $a^{\mu\nu}$, redefine $\gsso{6} \rightarrow - \Mpl^2/4 + \gsso{6}  $ and rescale all other EFT coefficients by $\gamma_{i, \ell} \rightarrow (2\gsso{12}/\Mpl^2) \gamma_{i, \ell} $. Moreover we can redefine 
\begin{align}\label{eq:amunu_redef}
	a^{\mu\nu} &\rightarrow \Tilde{a}^{\mu\nu} = a^{\mu\nu} + \alpha_1(t) a^{0(\mu} g^{\nu)0} + \alpha_2(t) a^{00} g^{\mu\nu} + \alpha_3(t) a^{00} g^{\mu 0} g^{\nu 0} \, .
\end{align}
to absorb a few EFT operators. This field redefinition preserves boundary conditions of the advanced metric, does not mix the order in advanced fields and transforms covariantly under retarded spatial diffs, hence it is allowed. Under this redefinition $a^{00}$ and $a^{0\mu}$ change according to
\begin{align}
	a^{00} &\rightarrow a^{00} (1-\alpha_1-\tfrac{1}{4}\alpha_2+\tfrac{3}{4}\alpha_3) + \dots \, ,\\
	a^{0\mu} &\rightarrow a^{0\mu}(1-\tfrac{1}{2}\alpha_1) + a^{00} g^{0\mu}(\tfrac{1}{2}\alpha_1+\tfrac{1}{4}\alpha_2-\tfrac{3}{4}\alpha_3) + \dots \, ,
\end{align}
where we have only kept terms at order $\ell=0$ and dots denote terms at higher order in $(g^{00}+1)$. 
Indeed, plugging this redefinition into the action simply shifts the EFT coefficients in $M^{tt}_{\ell}$ and $M^{ts}_{\mu, \ell}$, while the coefficients in $M^{ss}_{\mu\nu, \ell}$ remain unaffected.
For instance, $\gtso{1}$ is shifted by
\begin{equation}
	\gtso{1} \rightarrow \gtso{1}(1-\tfrac{1}{2}\alpha_1) + \tfrac{\Mpl^2}{2} \alpha_1 \, ,
\end{equation}
where the last term originates from the operator $\gtso{12}=\Mpl^2/2$. Consequently $\gtso{1}$ can be set to zero by choosing $\alpha_1 = 2\gtso{1}/(\gtso{1}-\Mpl^2)$, as long as $\gtso{1} \neq \Mpl^2$. 

An appropriate choice of $\alpha_2$ and $\alpha_3$ may further remove two operators in the first three lines of \Eq{eq:S1v1} at a fixed order in $\ell$. 
However, for this procedure to consistently eliminate an operator, the terms in $M^{ts}_{\mu, \ell}$ and $M^{ss}_{\mu\nu, \ell}$ responsible for generating the shift in its coefficient must not all simultaneously be zero. If they were, the redefinition would have no effect and the corresponding operator could not be removed. To avoid this issue, a safe approach is to remove only those operators whose shift is generated by the two terms that also appear in the Einstein tensor.
Based on this criterion, we can also safely remove $\gtto{6}$ and $\gtto{7}$, such that the action at $\ell = 0$ takes the form: 
\begin{align}\label{eq:S1fin}
	S_{1,0} &= \int \dd^4 x \sqrt{-g} \, \bigg\{ \, a^{00} \Big[ \gtto{1} + \gtto{2} K + \gtto{3}  K^2 + \gtto{4} K_{\alpha\beta} K^{\alpha\beta} + \gtto{5} \nabla^0 K \Big]  \\
	& \quad +a^{0\mu} \Big[ \gtso{2} \nabla_\mu K + \gtso{3} \nabla_\beta K^{\beta}{}_\mu \Big]  + a^{\mu\nu} \Big[ g_{\mu\nu} \Big( \gsso{1} + \gsso{2} K + \gsso{3}  K^2 + \gsso{4} K_{\alpha\beta} K^{\alpha\beta}  \nonumber \Bigg.\\ 
	&\quad + \gsso{5} \nabla^0 K + \gsso{6} R + \gsso{7} R^{00} \Big)  + \gsso{8} K_{\mu\nu} + \gsso{9} \nabla^0 K_{\mu\nu} + \gsso{10} K_{\mu\alpha} K^{\alpha}{}_{\nu} \nonumber \\
	&\quad+ \gsso{11} K K_{\mu\nu} + \frac{\Mpl^2}{2} G_{\mu\nu} + \gsso{13} R_\mu{}^0{}_\nu{}^0+ \gamma^{\po}_{1,0} \epsilon_\mu{}^{\alpha\beta0} \nabla_\alpha K_{\beta\nu} + \gamma^{\po}_{2,0} \epsilon_\mu{}^{\alpha\beta0} R_{\alpha\beta}{}^0{}_\nu \Big] 
	\bigg\} \, , \nonumber
\end{align}
whereas all terms at $\ell \geq 1$ remain as displayed in \Eq{eq:S1v1}. 


\paragraph{Construction of $S_2$.}

The construction of $S_2$ proceeds in a manner analogous to that of $S_1$.  To keep things simple, we restrict the construction of $S_2$ to \textit{zeroth order} in derivatives in this section. Just as before, we begin by expanding each term in $S_2$ in powers of $(g^{00}+1)$: 
\begin{align}
	N_{\mu\nu\rho\sigma} &= \sum_{\ell=0}\left(g^{00} + 1 \right)^\ell N_{\mu\nu\rho\sigma, \ell}(R_{\mu\nu\rho\sigma}, K_{\mu\nu}, n_\mu, \nabla_\mu; t)  \, .
\end{align}
Restricting to zeroth order in derivatives, tensorial objects, such as vectors and tensors, may only be build out of $n_\mu$ and $g_{\mu\nu}$ multiplying arbitrary EFT coefficients, as all other objects are of higher order in derivatives (see e.g. $M^{ss}_{\mu, \ell}$ and $\Tilde{M}^{ss}_{\mu\nu, \ell}$ in \Eqs{eq:Mts} and \eqref{eq:Mss}). As a result, we obtain: 
\begin{align}
	N_{\mu\nu\rho\sigma, \ell} &= \bt{4} n_\mu n_\nu n_\rho n_\sigma + \bt{5} g_{\mu\nu} g_{\rho\sigma} + \bt{6} g_{\mu(\rho} g_{\sigma)\nu} \bigg. \\
	+& \frac{1}{2} \bt{7} \left(g_{\mu\nu}n_\rho n_\sigma + n_\mu n_\nu g_{\rho\sigma} \right) + \frac{1}{2} \bt{8} \left(g_{\mu(\rho} n_{\sigma)} n_\nu + g_{\nu(\rho} n_{\sigma)} n_\mu \right) \bigg. \, .
\end{align}
In the retarded unitary gauge, the noise functional thus takes the form:\footnote{Notations are chosen to match \cite{Salcedo:2025ezu} where three other EFT operators controlled by $\beta_{1-3,\ell}$ are considered.} 
\begin{align}\label{eq:S2v1}
	S_2 = i \int \dd^4 x \sqrt{-g} \sum_{\ell=0} \left(g^{00} + 1 \right)^\ell \Big[ &\beta_{4,\ell} \left( a^{00} \right)^2  + \beta_{5,\ell} (a^{\mu\nu} g_{\mu\nu})^2 + \beta_{6,\ell} a^{\mu\nu} g_{\mu\rho} g_{\nu\sigma} a^{\rho\sigma} 
	\nonumber \\
	+& \beta_{7,\ell} a^{\mu\nu} g_{\mu\nu} a^{00} + \beta_{8,\ell} a^{0\mu} a^{0\nu} g_{\mu\nu} \Big] . \bigg. 
\end{align}
Finally, one can trade these quadratic terms in the advanced variables by linear terms in said variables through the Hubbard–Stratonovich trick as in \Eq{eq:HStrickOGR} \cite{Hubbard:1959ub,Stratonovich1957}. We illustrate this procedure in \cite{Salcedo:2025ezu}.

\paragraph{An illustration.} To build physical intuition, we present here a subset of the theory where only a handful of EFT coefficients are considered. Let us consider
\begin{align}\label{eq:th2}
	S_{\rm eff}=\int\sqrt{-g}\bigg[&\frac{\Mpl^{2}}{2}G_{\mu\nu}a^{\mu\nu}+\frac{\Lambda(t)}{2}g_{\mu\nu}a^{\mu\nu}-c(t)a^{00}+\frac{c(t)}{2}g^{00}g_{\mu\nu}a^{\mu\nu}\nonumber \\
	-&\frac{M_{2}(t)}{4}\left(1+g^{00}\right)^{2}g_{\mu\nu}a^{\mu\nu}+M_{2}(t)\left(1+g^{00}\right)a^{00} \bigg.\nonumber\\
	+& \frac{\Gamma(t)}{3H}\left(1 + g^{00} \right)\left( a^{00} + g_{\mu\nu} a^{\mu\nu}\right)-\xi_{\mu\nu}a^{\mu\nu}\bigg]. 
\end{align}
The operators considered here play a particular role in the scalar dynamics. Indeed, $S_{\mathrm{eff}}$ encompasses the universal part of the EFToI controlled by the EFT coefficients $\Lambda(t)$ and $c(t)$ \cite{Salcedo:2025ezu}. It also contains a speed of sound controlled by $M_{2}(t)$ and a simple non-unitary extension made of the dissipative coefficient $\Gamma(t)$ and the noise contribution controlled by $\xi_{\mu\nu}$. 

At the background level, \Eq{eq:th2} reads
\begin{equation}
	S_{\rm eff}=\int \dd^4x \sqrt{-g}\left[\frac{\Mpl^{2}}{2}\bar{G}_{\mu\nu}+\frac{\Lambda(t)-c(t)}{2}\bar{g}_{\mu\nu}a^{\mu\nu}-c(t)a^{00}\right],
\end{equation}
from which we obtain the background Einstein's equations by varying with respect to $a^{\mu\nu}$, 
\begin{equation}
	\frac{\Mpl^{2}}{2}\bar{G}_{\mu\nu}+\frac{\Lambda(t)-c(t)}{2}\bar{g}_{\mu\nu}-c(t)\delta_{\mu}^{0}\delta_{\nu}^{0}=0.
\end{equation}
We recover the usual Friedmann equations
\begin{align}\label{eq:Fried}
	3\Mpl^{2}H^{2}=&\Lambda(t)+c(t),\qquad
	2\Mpl^{2}\dot{H}=-2c(t).
\end{align}
In \cite{Salcedo:2025ezu}, we reintroduced the scalar field perturbations by performing both a retarded and advanced \stuck trick. We then assumed a slow-roll hierarchy $|\dot{H}| \ll H^2$ holds and observed that the mixing between the \stuck field and the perturbations of the metric is negligeable compared to the self dynamics of the \stuck field itself. This illustrates how the quadratic dynamics of \cite{Salcedo:2024smn},
\begin{align}\label{eq:Sdecoup}
	S_{\pi}&\xrightarrow{(\text{decoupl.})} \int \dd^4x \sqrt{-g}\left\{- \dpir \dpia + \cs \bar{g}^{ij} \partial_i\pir \partial_j \pia + 2 \gamma \dpir \pia + \xi_{\pi} \pia\right\}, 
\end{align}
can indeed be understood as the decoupling limit of the theory \eqref{eq:th2} of open gravity. Interested reader may found details of the analysis in Section 6 of \cite{Salcedo:2025ezu}.


\subsection{Dissipative and stochastic gravitons} The central novel feature of our theory is its incorporation of dynamical gravity, which allows us to study gravitational effects, such as the generation and propagation of gravitational waves (GWs) during inflation. These phenomena are encoded in Transverse and Traceless (TT) metric perturbations. In this section, we derive the quadratic action that governs this sector, and we compute the corresponding propagators and power spectrum. Finally, we extract the tensor-to-scalar ratio, which enables us to confront the parameters of our theory with current and upcoming observational data.

\subsubsection*{The transverse traceless sector of linear perturbations} 

To derive the linear equation of motion for GWs, we expand \Eq{eq:S1v1} and \eqref{eq:S2v1} to quadratic order in perturbations and perform a scalar-vector-tensor decomposition of $\delta g_{\mu\nu}$ and $a^{\mu\nu}$ (recall that the latter is already of first order in perturbations) and restrict to the TT sector according to:
\begin{equation}
	g_{ij} = a^2(t) \left(\delta_{ij} + h_{ij}\right), \qquad a^{ij} = a^{-2}(t) h_{ij}^{a},
\end{equation}
with $h_{ij}$ and $h^a_{ij}$ transverse $\partial_i h_{ij}=0=\partial_i h^a_{ij}$ and traceless $\delta^{ij} h_{ij}=0= \delta^{ij}h^a{}_{ij}$. The index $a$ on $h^a_{ij}$ denotes that this is an advanced variable.

We begin by extracting the TT sector of $S_1$ in \Eq{eq:S1v1}. $a^{00}$ and $a^{0\mu}$ do not contain any transverse and traceless component, such that we obtain
\begin{align}
	&S_1^{(2)} = \int \dd^4 x \sqrt{-g} \,  a^{ij} \Big[ a^2 h_{ij} \Bar{M}^{ss}_{\ell=0}  + \gsso{8} \delta K_{ij} + \gsso{9} \nabla^0 \delta K_{ij} + \gsso{10} \delta \left( K_{i\alpha} K^{\alpha}{}_{j} \right) \nonumber \\
	&\quad+ \gsso{11} \Bar{K} \delta K_{ij} + \frac{\Mpl^2}{2} \delta R_{ij} + \gsso{13} \delta R_i{}^0{}_j{}^0 + \gamma^{\po}_{1,0} \epsilon_i{}^{\ell m 0} \delta \left( \nabla_\ell  K_{m j} \right) + \gamma^{\po}_{2,0} \epsilon_i{}^{\ell m 0} \delta R^0{}_j{}_{\ell m}
	\Big]  \, ,
\end{align}
where $\Bar{M}^{ss}_{\ell=0}$ denotes \Eq{eq:Mss_scalar} evaluated on the background $\Bar{g}_{\mu\nu}$. The perturbed tensors are given by
\begin{align}
	\delta R_{ij} &= -\frac{1}{2} \nabla^2 h_{ij} + \frac{a^2}{2} \left[ \ddot h_{ij} + 3H \dot h_{ij} + \left(6H^2 + 2\dot H\right) h_{ij} \right] , \label{eq:deltaRijTT} \\
	\delta R_i{}^0{}_j{}^0 &= -a^2 \left[ \frac{1}{2} \ddot{h}_{ij} + H \dot h_{ij} + \left(\dot{H}+H^2\right) h_{ij} \right] , \label{eq:deltaRi0j0TT} \\ 
	\delta R^0{}_j{}_{lm} &= \frac{a^2}{2} \left(\partial_l \dot{h}_{mj} - \partial_m \dot{h}_{lj} \right) , \label{eq:deltaRlm0jTT} \\
	\delta K_{ij} &= \frac{a^2}{2} \dot{h}_{ij} + a^2 H h_{ij} , \label{eq:deltaKijTT}
\end{align}
Plugging these into the action, along with the background values yields:
\begin{align}
	S_1^{(2)} = \frac{1}{2} \int \dd^4 x \sqrt{-g} \, &h_{ij}^{a} \bigg\{\left(\frac{\Mpl^2}{2}-\gsso{9}-\gsso{13}\right)\ddot{h}_{ij} \\
	&+\left[\gsso{8}+ \left(2\gsso{10}+3\gsso{11}+3\frac{\Mpl^2}{2}+2\gsso{13} \right)H\right] \dot{h}_{ij} \nonumber \\
	&- \frac{\Mpl^2}{2} \frac{\nabla^2}{a^2} h_{ij} + \frac{1}{a} \left(\gamma^{\po}_{1,0} + 2\gamma^{\po}_{2,0} \right) \Tilde{\epsilon}_{imn} \partial_m \dot{h}_{nj} \nonumber
	\bigg\}  \, .
\end{align}
To arrive at this result, we have used the spatial background equation, which imposes $\Bar{M}^{ss}_{\ell=0} = - \Tilde{\Bar{M}}^{ss}_{ii,\ell=0}$. Consequently, every term in $a^2 h_{ij} \Bar{M}^{ss}_{\ell=0}$ cancels with a corresponding $\sim h_{ij}$ term in the perturbed quantities in \Eqs{eq:deltaRijTT}, \eqref{eq:deltaRi0j0TT}, \eqref{eq:deltaRlm0jTT} and \eqref{eq:deltaKijTT}, such that no mass term for the graviton is produced. This is in agreement with our expectation that the graviton is massless, as long as we do not break retarded spatial diffs. Our theory only contains the two helicities of the graviton and the clock field. Note that we lowered the indices on all expressions in this equation, and that $\Tilde{\epsilon}_{imn}$ denotes the totally antisymmetric symbol, not tensor, the two being related via $\epsilon_{0ijk} = \sqrt{-g} \, \Tilde{\epsilon}_{ijk}$.

At quadratic order in perturbations, we further expect to obtain contributions from $S_2$. These contributions, being quadratic in advanced fields, encode stochastic noise terms generated by random fluctuations in the environment. The only contribution to the TT sector at this order in perturbations comes from
\begin{equation}
	S_2 \supset i \int \dd^4 x \sqrt{-g} \, \beta_{6,0} a^{\mu\nu} g_{\mu\rho} g_{\nu\sigma} a^{\rho\sigma} , 
\end{equation}
which results in
\begin{equation}
	S_2^{(2)} = i \int \dd^4 x \sqrt{-g} \, \beta_{6,0} h_{ij}^{a} h_{ij}^{a} .
\end{equation}
Note that we do not have to consider any higher action $S_{\geq 3}$ as these start at least at cubic order in perturbations.

The total quadratic action for the TT sector is therefore given by $S^{(2)} = S_1^{(2)} + S_2^{(2)}$. 
To simplify this expression we introduce
\begin{align}
	c_T^{-2} &\equiv 1 - \frac{2 \gsso{9}}{\Mpl^2} - \frac{2 \gsso{13}}{\Mpl^2} , \\
	\Gamma_T & \equiv c_T^2 \frac{2 \gsso{8}}{\Mpl^2} + c_T^2 H \left[\frac{4\gsso{10}}{\Mpl^2} + \frac{6\gsso{11}}{\Mpl^2} + \frac{4\gsso{13}}{\Mpl^2} +3\left(1-\frac{1}{c_T^2}\right)\right], \\
	\chi & \equiv c_T^2 \left(\frac{2 \gamma^{\po}_{1,0}}{\Mpl^2} + \frac{4 \gamma^{\po}_{2,0}}{\Mpl^2}\right),
\end{align}
with dimensions
\begin{equation}
	\left[c_T\right] = E^0, \qquad \left[\Gamma_T\right] = E^1, \qquad \left[\chi\right] = E^0 , \qquad \left[\beta_{6,0}\right] = E^4 . 
\end{equation}
The quadratic action becomes
\begin{align}
	S^{(2)} = \int \dd^4 x \sqrt{-g} \, &\frac{\Mpl^2}{4 c_T^2} \, h_{ij}^{a}  \bigg[\ddot{h}_{ij} - c_T^2 \frac{\nabla^2}{a^2} h_{ij} +\left(\Gamma_T + 3 H\right) \dot{h}_{ij} \label{eq:S_2_coefficients} \\
	&+ \frac{\chi}{a} \Tilde{\epsilon}_{imn} \partial_m \dot{h}_{nj}  + i c_T^2 \frac{4 \beta_{6,0}}{\Mpl^2} h_{ij}^{a}
	\bigg] . \nonumber
\end{align}
We briefly comment on the effects caused by the various operators that appear in $S^{(2)}$.

\paragraph{Noise $\beta_{6,0}$.} The coefficient $\beta_{6,0}$ characterizes stochastic fluctuations of gravitational waves sourced by the environmental sector. Upon performing the Hubbard-Stratonovich trick \cite{Hubbard:1959ub,Stratonovich1957} this quadratic term in $h_{ij}^a$ can be traded for a term linear in $h_{ij}^a$, at the expense of introducing an auxiliary field $\xi_{ij}$: 
\begin{align}
	&\text{exp}\left\{-\int \dd^{4}x\sqrt{-g}\left[ c_T^2 \frac{4 \beta_{6,0}}{\Mpl^2} h_{ij}^a  h_{ij}^a\right]\right\}  \nonumber \\
	&\qquad =\int[\mathcal{D}\xi_{ij}]\;\text{exp}\left\{-\int \dd^{4}x \sqrt{-g}\left[ \frac{\Mpl^2}{\beta_{6,0} c_T^2}\xi_{ij}\xi_{ij}+i\xi_{ij}h_{ij}^a\right]\right\}.
\end{align}
The field $\xi_{ij}$ is transverse and traceless and follows Gaussian statistics. After introducing $\xi_{ij}$, the equation of motion for $h_{ij}$ reads:
\begin{align}\label{eq:GWs_bf_eom}
	\ddot{h}_{ij} - c_T^2 \frac{\nabla^2}{a^2} h_{ij} +\left(\Gamma_T + 3 H\right) \dot{h}_{ij}  + \frac{\chi}{a} \Tilde{\epsilon}_{imn} \partial_m \dot{h}_{nj} = \xi_{ij}.
\end{align}


\paragraph{Speed of propagation $c_T^2$.} The presence of $c_T^2$ changes the speed of propagation for tensor modes. An operator that causes such an effect has already been identified in the closed theory as $\delta K_{\mu\nu} \delta K^{\mu\nu} - \delta K^2$ \cite{Creminelli:2014wna}. We can translate this unitary operator into the Schwinger-Keldysh basis by considering the difference of two copies of this term, one for each branch of the path integral
\begin{equation}\label{eq:K2}
	\int \dd^4x \left[ \sqrt{-g_+} \left( K[g_+]^2 - K_{\mu\nu}[g_+]^2 \right) - \sqrt{-g_-} \left( K[g_-]^2 - K_{\mu\nu}[g_-]^2 \right) \right] .
\end{equation}
We can rewrite the above using
\begin{equation}
	R_{\mu\nu} n^\mu n^\nu = K^2 - K_{\mu\nu} K^{\mu\nu} - \nabla_\mu \left(n^\mu \nabla_\nu n^\nu \right) + \nabla_\nu \left(n^\mu \nabla_\mu n^\nu \right) , 
\end{equation}
and dropping boundary terms. Expressing \eqref{eq:K2} in the Keldysh basis $g_\pm = g \pm a/2$ and expanding to first order in $a^{\mu\nu}$ yields 
\begin{equation}\label{eq:ct2_unitary_1}
	\int \dd^4x \sqrt{-g} \left(-\frac{1}{2} a^{\mu\nu} g_{\mu\nu} R_{\alpha\beta}n^\alpha n^\beta + n^\mu n^\nu \delta_a R_{\mu\nu} - 2 \frac{a^{0\mu}}{g^{00}} n^\nu R_{\mu\nu} - \frac{a^{00}}{g^{00}} n^\mu n^\nu R_{\mu\nu} \right) , 
\end{equation}
where the first term comes the expansion of the metric determinant and the last two terms originate from the expansion of $n^\mu$. Meanwhile the variation of the Ricci-tensor $\delta_a R_{\mu\nu}$ is given by the Palatini identity, one can reexpress it in terms of the extrinsic curvature through \cite{Salcedo:2025ezu}: 
\begin{equation}\label{eq:ct2_unitary_2}
	\int \dd^4x \sqrt{-g} \, n^\mu n^\nu \delta_a R_{\mu\nu} = \int \dd^4x \sqrt{-g} \, a^{\mu\nu} \left( -n^\rho \nabla_\rho K_{\mu\nu} - K K_{\mu\nu} + \frac{1}{2}g_{\mu\nu} K^2 + \frac{1}{2} g_{\mu\nu} n^\rho \nabla_\rho K \right) .
\end{equation}
From \Eqs{eq:ct2_unitary_1} and \eqref{eq:ct2_unitary_2} one can read of all the operators of the open theory that correspond to the unitary operator $K^2 - K_{\mu\nu}K^{\mu\nu}$ of the closed theory. Indeed, the first term in \Eq{eq:ct2_unitary_2} corresponds to the operator $\gsso{9}$. For simplicity, we will set $c_T = 1$ from now on.

\paragraph{Dissipation $\Gamma_T$.} The term proportional to $\Gamma_T$ captures the dissipation of GWs into the environment. It serves as an example of an operator that arises in an open theory. In a classical (closed) action the size of the term proportional to $\dot{h}_{ij}$ would be fixed by the Hubble parameter. The operator $\gsso{8}$, which generates dissipation at lowest order in derivative expansion, has previously been identified in \cite{Lau:2024mqm}.

\paragraph{Dissipative birefringence $\chi$.} When decomposing $h_{ij}$ into a polarization basis, the term $\chi$ acquires opposite signs for each polarization state due to the presence of the Levi-Civita symbol. Similar parity-violating operators also exist in the closed formulation of the EFToI, however these operators come with at least three derivatives \cite{Creminelli:2014wna}. This implies that the operator found here must correspond to a non-unitary effect. Indeed, just as it was the case for $\Gamma_T$, one can verify that it is not possible to construct a term in a closed action that yields a contribution of the form $\sim \epsilon_{ilm}\partial_l \dot{h}_{mj}$ in the equation of motion. For instance, a term like $\epsilon_{ilm}\partial_l \dot{h}_{mj} h_{ij}$ is a total derivative and thus does not contribute to the dynamics. Consequently we call the effect found here ``dissipative birefringence'', to distinguish it from the ``unitary birefringence'' described in \cite{Creminelli:2014wna}.

\subsubsection*{Tensor-to-scalar ratio}

We proceed to compute the propagators from the action in \Eq{eq:S_2_coefficients}, which will ultimately be used to determine the noise induced power spectrum of GWs and the corresponding tensor-to-scalar ratio.
The action simplifies when both $h_{ij}$ and $h^{a}_{ij}$ are expanded in a polarization basis:
\begin{align}
	h_{ij}(t,x) = \int_{\bfk} \sum_{s} e_{ij}^s(\hat{\bfk}) h^s(t,\bfk) e^{i\bfk \cdot \bfx}, \\
	h^{a}_{ij}(t,\bfk) = \int_{\bfk} \sum_{s} e_{ij}^s(\hat{\bfk}) h_a^s(t,\bfk) e^{i\bfk \cdot \bfx},
\end{align}
where the polarization tensors fulfill
\begin{align}
	e_{ii}^s(\hat{\bfk}) &= k^i e_{ij}^s(\hat{\bfk}) = 0 , &\qquad
	e_{ij}^s(\hat{\bfk}) &= e_{ji}^s(\hat{\bfk}) , &\qquad
	e_{ij}^s(\hat{\bfk}) e_{jk}^s(\hat{\bfk}) &= 0 , \Big. \\
	e_{ij}^s(\hat{\bfk}) e_{ij}^{s'}(\hat{\bfk})^* &= 2 \delta_{s s'} , &\qquad
	e_{ij}^s(\hat{\bfk})^* &= e_{ij}^s(-\hat{\bfk}) &\qquad i \Tilde{\epsilon}_{ijk} k_j e_{km}^s &= \frac{s}{2} k e_{im}^{s} . \label{epsid}
\end{align}
For the left and right circular polarization we have $s=\pm 2$. The quadratic action takes the form
\begin{align}
	S^{(2)} = &\frac{\Mpl^2}{2} \sum_{s} \int_{\bfk} \int \text{d}t \sqrt{-g} \, h_{a}^s(t, -\bfk) \bigg[ \ddot{h}^{s}(t,\bfk) +  \frac{\bfk^2}{a^2} h^{s}(t,\bfk) \label{eq:S2_proptimev1} \\ 
	&+\left(\Gamma_T+ 3 H + \frac{ks}{2a} \chi \right) \dot{h}^{s}(t,\bfk) + \frac{4 \beta_{6,0}}{\Mpl^2} h_{a}^s(t,\bfk)
	\bigg]. \nonumber 
\end{align}


\paragraph{Without birefringence.} For simplicity, we first restrict to the case $\chi=0$. Changing to conformal time $\dd t = a(t) \dd \eta$ and canonically normalizing 
\begin{equation}\label{eq:gws_can_norm}
	h^{s}_{c}\equiv \frac{\Mpl}{\sqrt{2}} h^s , \qquad \mathrm{and} \qquad h_{c,a}^{s}\equiv \frac{\Mpl}{\sqrt{2}} h^s_a ,
\end{equation}
results in 
\begin{align}
	S^{(2)} = \sum_{s} &\int_{\bfk} \int \text{d}\eta \, a(\eta)^2 \, h_{c,a}^{s}(\eta, -\bfk) \bigg[ h^{s}_{c}(\eta, \bfk)'' + \bfk^2 h^{s}_{c}(\eta, \bfk) \\
	&+a(\eta)\left(\Gamma_T+ 2 H\right) h^{s}_{c}(\eta, \bfk)' 
	+ i a(\eta)^2 \frac{4 \beta_{6,0}}{\Mpl^2} h_{c,a}^{s}(\eta, \bfk) \bigg] . \nonumber
\end{align}
This is just two copies of the action for the decoupled Goldstone $\pir$, which has been studied in detail in \cite{Salcedo:2024smn}. Following their approach, we can rewrite the action as bilinear in the fields
\begin{equation}
	S = \frac{1}{2} \sum_{s} \int_{\bfk} \int \dd \eta \, \left(h^s_c(\eta, -\bfk), h^s_{c,a}(\eta, -\bfk) \right) \begin{pmatrix}
		0 & \hat{D}_A \\
		\hat{D}_R & 2i\hat{D}_K 
	\end{pmatrix} 
	\begin{pmatrix}
		h^s_c(\eta, \bfk) \\
		h^s_{c,a}(\eta, \bfk)
	\end{pmatrix} , 
\end{equation}
with
\begin{align}
	\hat{D}_R &= a^2(\eta) \left[ \partial_\eta^2 + \left(2H + \Gamma_T\right) a(\eta) \partial_\eta + k^2 \right] \bigg. \,,\\
	\hat{D}_A &= a^2(\eta) \left[ \partial_\eta^2 + \left(2H - \Gamma_T \right) a(\eta) \partial_\eta + k^2 - 3a^2 H \Gamma_T \right] \,,\\
	\hat{D}_K &= a^4(\eta) \frac{4 \beta_{6,0}}{\Mpl^2} . 
\end{align}
The retarded Green's function obeys
\begin{equation}
	\hat{D}_R(\eta_1) G^R(k;\eta_1, \eta_2) = \delta(\eta_1-\eta_2) , 
\end{equation}
which results in \cite{Salcedo:2024smn}
\begin{equation}\label{eq:retprop}
	G^R(k;\eta_1,\eta_2) = \frac{\pi}{2} \frac{H^2}{k^3} \left( \frac{z_1}{z_2} \right)^{\nu_\Gamma} ( z_2)^3 \left[ Y_{\nu_\Gamma}(z_1) J_{\nu_\Gamma}(z_2)  - J_{\nu_\Gamma}(z_1) Y_{\nu_\Gamma}(z_2)\right] \theta(\eta_1 - \eta_2) ,
\end{equation}
in terms of Bessel functions of the first kind, with 
\begin{equation}
	\nu_\Gamma \equiv \frac{3}{2} + \frac{\Gamma_T}{2H} \qquad \mathrm{and} \qquad z_i \equiv -k\eta_i \, .
\end{equation}
This can also be rewritten in terms of Hankel functions as
\begin{equation}\label{eq:retpropHankelfct}
	G^R(k;\eta_1,\eta_2) = \frac{\pi}{2} H^2 (\eta_1 \eta_2)^{\tfrac{3}{2}} \left( \frac{\eta_1}{\eta_2} \right)^{\frac{\Gamma_T}{2H}} \Im \text{m} \left[ H^{(1)}_{\nu_\Gamma}\left(-k\eta_1 \right) H^{(2)}_{\nu_\Gamma}\left(-k\eta_2 \right) \right] \theta(\eta_1 - \eta_2) .
\end{equation}
The Keldysh propagator is given by 
\begin{align}
	G^K(k;\eta_1,\eta_2) = i \frac{4\beta_{6,0}}{\Mpl^2} \int \frac{\dd \eta'}{H^4 \eta'{}^4} G^R(k; \eta_1, \eta') G^R(k; \eta_2, \eta') + (\eta_1 \leftrightarrow \eta_2) .
\end{align}
The power spectrum is obtained in the coincident limit of the Keldysh propagator $P_T(k, \eta) = -i G^K(k;\eta, \eta)$.
The reduced GW power spectrum is
\begin{equation}
	\Delta_h^2(k) \equiv \frac{k^3}{2\pi^2} P_T(k), \qquad \mathrm{with} \qquad
	\langle h_{ij}(\bfk) h_{ij}(\bfk') \rangle = (2\pi)^3 \delta^3(\bfk+\bfk') P_T.
\end{equation}
In the super-Hubble regime $z \ll 1$ is thus given by 
\begin{equation}\label{eq:gws_ps_Gamma}
	\Delta_h^2(k) =  \frac{4 \beta_{6,0}}{\Mpl^4} 2^{2\nu_\Gamma} \frac{\Gamma(\nu_\Gamma-1 ) \Gamma(\nu_\Gamma)^2}{\Gamma(\nu_\Gamma-\tfrac{1}{2} )\Gamma(2\nu_\Gamma-\tfrac{1}{2})},
\end{equation}
where an additional factor of 2 accounts for both polarizations. 
\begin{figure}
	\centering
	\includegraphics[width=0.7\linewidth]{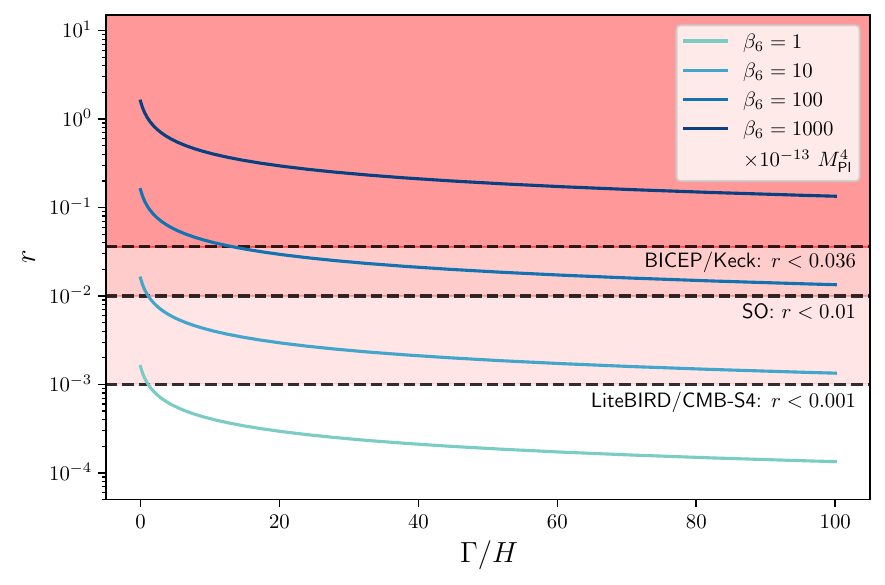}
	\caption{The tensor-to-scalar ratio $r$ as a function of $\Gamma_T/H$, for different choices of $\beta_{6,0}$. The observational constraint $r < 0.036$ \cite{BICEPKeck:2024stm} restricts the allowed region of parameters. Ongoing and future surveys, such as SO \cite{SimonsObservatory:2018koc}, CMB S4 \cite{CMB-S4:2020lpa} and LiteBIRD \cite{LiteBIRD:2022cnt} will put tighter constraint on this parameter.}
	\label{fig:gw_ps_nobf}
\end{figure}
The result is shown in \Fig{fig:gw_ps_nobf} as a function of $\Gamma_T/H$. As expected, the power spectrum is damped for large dissipation. 
Expanding in the weak $\Gamma_T \ll H$ and strong $\Gamma_T \gg H$ dissipation regime  yields: 
\begin{align}
	\Delta_h^2(k) &\propto \begin{dcases}
		\frac{\beta_{6,0}}{\Mpl^4}, &\mathrm{for} \quad \Gamma_T \ll H,\\
		\frac{\beta_{6,0}}{\Mpl^4} \sqrt{\frac{H}{\Gamma_T}} \left[ 1 + \O\left(\frac{H}{\Gamma_T}\right) \right], & \mathrm{for} \quad \Gamma_T \gg H .
	\end{dcases}
\end{align}
The tensor-to-scalar ratio is given by\footnote{For an application of quantum information to investigate the optimal inference of $r$, see \cite{Piotrak:2025zhy}.}
\begin{equation}
	r = \frac{\Delta_h^2(k)}{\Delta_\zeta^2(k)} .
\end{equation}
Using the observed $\Delta_\zeta^2 = 2.1 \times 10^{-9}$ this becomes:
\begin{equation}
	r =  \frac{\beta_{6,0}}{\Mpl^4} 2^{2\nu_\Gamma} \frac{\Gamma(\nu_\Gamma-1 ) \Gamma(\nu_\Gamma)^2}{\Gamma(\nu_\Gamma-\tfrac{1}{2} )\Gamma(2\nu_\Gamma-\tfrac{1}{2})} \times 1.9 \times10^{9} .
\end{equation}
This is plotted for different choices of $\beta_{6,0}$ against $\Gamma_T/H$ in \Fig{fig:gw_ps_nobf}. From this figure we can infer that $\beta_{6,0} \lesssim \left(10^{-3} \Mpl \right)^4 $ to be compatible with the observational bound of $r < 0.036$, for small values of $\Gamma_T/H$. For realizations near the observational bound, the noise sourcing the tensor sector is significant. It would be interesting to investigate if this can be achieved in concrete UV models scenarios. Also note that for a given UV model increasing the dissipation parameter $\Gamma_T$ might not lead to a decreased power spectrum - in realistic scenarios increasing the influence of the environment via $\Gamma_T$ also leads to an increase in the noise $\beta_{6,0}$, such that $\Delta_h$ might increase overall. This is for instance realized by the KMS condition $(4 \beta_{6,0}/\Mpl^4) = 2 \pi \Gamma_T T_{\mathrm{eq}}$ for environments that are in thermal equilibrium, if the temperature of the thermal bath $T_{\mathrm{eq}}$ remains fixed.  


\paragraph{Including birefringence.}  In the asymptotic past, $-k\eta \gg 1$, the dissipative birefringence term dominates the gravitational wave equation of motion, such that the resulting equation reads
\begin{equation}
	\left[ \partial_\eta^2 + \frac{s \chi}{2} k \partial_\eta + k^2 \right] h^s_k = 0 ,
\end{equation}
which has plane waves $e^{-i\omega \eta}$ as solutions with dispersion relations
\begin{equation}\label{eq:gws_bf_dr}
	\omega = -\frac{1}{4} i \chi s k \pm \frac{1}{2} k \sqrt{4-\chi^2} .
\end{equation}
It is clear that the system features an instability for $\chi \neq 0$, as the mode function of the $s=+2$ polarization (since we assumed $\chi\geq 0$) experiences an exponential enhancement as $\eta \to - \infty$, whereas the $s=-2$ polarization decays. Instability in one of the two polarizations also occurs in UV-models that include birefringent terms, such as Chern-Simons (CS) gravity. Within CS-gravity, the instability can be handled by either imposing a UV-cutoff or by coupling the $R \tilde{R}$-term to the inflaton field (or any other scalar field), and hence giving this coupling a time-dependence \cite{Alexander:2004wk, Alexander:2009tp, Creque-Sarbinowski:2023wmb, Dyda:2012rj, Alexander:2004us, Callister:2023tws}. Note that when the magnitude of $h_{ij}$ becomes comparable to the background metric, the perturbative approach breaks down and a non-perturbative treatment is necessary.  We stress, that one such modification is necessary to tame the divergence of the power spectrum of one of the two polarizations. For the moment, we leave the computation of the birefringent power spectra as an open question for the future. 

\clearpage


\subsection{\textit{Problem set}}

\paragraph{\textit{Exercise 1.}} \textit{Background evolution in open gravity} \\

The modified Friedman equations are obtained from 
\begin{align}\label{eq:S1tryback}
	\bar{S}_1 = \int \dd^4 x \sqrt{-g}& \, M_{\mu\nu} (\bar{R}_{\mu\nu\rho\sigma}, \bar{g}^{00}, \bar{K}_{\mu\nu}, \bar{\nabla}_\mu; t) a^{\mu\nu},  
\end{align}
where bars represent background values, leading to the background Einstein equations
\begin{align}
	\frac{\delta \bar{S}_1}{\delta a^{\mu\nu}} = 0 \qquad \Rightarrow \qquad M_{\mu\nu} (\bar{R}_{\mu\nu\rho\sigma}, \bar{g}^{00}, \bar{K}_{\mu\nu}, \bar{\nabla}_\mu; t) = 0.
\end{align}
\begin{enumerate}
	\item Derive the first Friedmann equation from the $00$-component of the background Einstein equations evaluated on a flat FLRW background.
	\item Similarly, derive the second Friedmann equation from the trace-part of the $ij$-components of the background Einstein equations.
	\item Redefine the EFT coefficients to reach the simple form
	\begin{align}
		3\Mpl^{2}H^2 &= \alpha_{1}+\alpha_{2}H \,, \Big. \label{eq:modifiedFriedmanfinal}\\
		2\Mpl^{2}\dot{H} &= \alpha_{3}+\alpha_{4}H\,. \label{eq:modifiedFriedmanfinal2}
	\end{align}
	Comment on how these expressions departs from the usual conservative single-fluid system.
	\item Based on the identification $\alpha_1 = 1$ and $\alpha_3 = - (\rho + p)$ and assuming $\alpha_2$ and $\alpha_4$ constant, derive the continuity equation for the fluid. Comment on how this expression depart from the usual conservative single-fluid system. 
	\item What is the acceleration equation $\ddot{a}/a$? Discuss the implications in terms of the late-universe acceleration.
\end{enumerate}


\begin{center}
	\noindent\rule{8cm}{0.4pt}
\end{center}


\paragraph{\textit{Exercise 2.}} \textit{Retarded and advanced \stuck tricks} \\

The goal of this Exercise is to highlight a few subtleties about the difference between retarded and advanced \stuck tricks. Let us consider the coordinate transformation $x^\mu \rightarrow x^{\prime\mu} = x^\mu + \epsilon^\mu$. A scalar $\phi$ transforms as 
\begin{align}
	\phi(x) \quad \rightarrow \quad \phi'(x + \epsilon) = \phi(x) \quad \leftrightarrow \quad \phi'(x)=\phi(x-\epsilon)\,.
\end{align}
Let us consider the flat spacetime action 
\begin{align}
	S = \int \dd^4 x \mathcal{L}[\phi(x);x] ,
\end{align}
where $\mathcal{L}[\phi(x);x]$ is a Lagrangian density that depends on both the scalar field $\phi(x)$ and the spacetime coordinate $x$. Under coordinate transform, the action becomes
\begin{align}
	S = \int \dd^4 x \mathcal{L}[\phi(x);x]  \rightarrow \int \dd^4 x \mathcal{L}[\phi'(x);x] &= \overbrace{\int \dd^4 x \mathcal{L}[\phi(x- \epsilon);x]}^{\text{Method I}} 
	= \overbrace{\int \dd^4 \widetilde{x} \mathcal{L}[\phi(\widetilde{x});\widetilde{x} + \epsilon]}^{\text{Method II}}. 
\end{align}
The last step, from Method I to Method II is a change of coordinates in the integral, as opposed to a transformation of the fields.
\begin{enumerate}
	\item Consider the advanced diff $x^\mu \rightarrow x^\mu + \epsilon^\mu_a$, which transform the two branches of the path integral in opposite directions,
	\begin{align}
		\phi_+(x) \rightarrow \phi_+'(x + \epsilon_a) = \phi_+(x) \quad \leftrightarrow \quad \phi_+'(x ) = \phi_+(x-\epsilon_a)\,, \\
		\phi_-(x) \rightarrow \phi_-'(x - \epsilon_a) = \phi_-(x)  \quad \leftrightarrow \quad \phi_-'(x ) = \phi_-(x+\epsilon_a) \,.
	\end{align}
	Discuss Method II in this case. 
	\item Reproduce the discussion about Method I and Method II for the case of the metric that transforms under $x^\mu \to x^{\prime\mu}=x^\mu+\epsilon^\mu$ as
	\begin{align}\label{eq:gtransfov0}
		g_{\mu\nu}(x) \quad \rightarrow \quad g'_{\mu\nu}(x')= \frac{\partial x^{\alpha} }{\partial x^{\prime \mu}} \frac{\partial x^{\beta} }{\partial x^{\prime\nu}} g_{\alpha \beta}(x)\,.
	\end{align}
	You should find that the action remains invariant for 
	\begin{align}
		 \text{Method I: }& \qquad \Delta g_{\mu\nu}=- 2 \nabla_{(\mu} \epsilon_{\nu)} \,,  \qquad \Delta g^{\mu\nu}=+ 2 \nabla^{(\mu} \epsilon^{\nu)} \\
		 \text{Method II: }& \qquad \Delta g_{\mu\nu}=- g_{\mu\alpha} \partial_{\nu} \epsilon_{\alpha}- g_{\nu\alpha} \partial_{\nu} \epsilon_{\alpha} \,,  \qquad \Delta g^{\mu\nu}=+ 2 \partial^{(\mu} \epsilon^{\nu)}.
	\end{align}
	\item While Method II is the most convenient for retarded diffs, it falls short when one considers advanced diffs for the reason presented in 1. For this reason, we use Method I for the latter. Give $g^{00}$ and $a^{00}$, $g^{0i}$ and $a^{0i}$, $g^{ij}$ and $a^{ij}$ under both retarded and advanced diffs.
\end{enumerate}


\begin{center}
	\noindent\rule{8cm}{0.4pt}
\end{center}

\paragraph{\textit{Exercise 3.}} \textit{Universal part in the Keldysh basis} \\

The goal of this Exercise is to express the universal part of the EFT of Inflation and Dark Energy in the Keldysh basis, at linear order in the advanced fields. The theory being unitary, the effective functional separates into
\begin{align}
	S_{\mathrm{univ}}[g_+,g_-] = S_{\mathrm{univ}}[g_+] - S_{\mathrm{univ}}[g_-]\,,
\end{align}
with
\begin{align}\label{eq:unitpmv2}
	&S_{\mathrm{univ}}[g]= \int \dd^4 x \sqrt{-g} \left[ \frac{M^2_{\mathrm{\mathrm{Pl}}}}{2}R - \Lambda(t) - c(t )g^{00}\right].
\end{align}
\begin{enumerate}
	\item Expand the effective functional at linear order in the advanced metric $a^{\mu\nu}$. A useful expression is
	\begin{align}
		\sqrt{-g_\pm} &= \sqrt{- \mathrm{det}\left( g \pm \frac{a}{2}\right)}=\sqrt{-g}\left[1 \mp \frac{1}{4} g_{\mu\nu} a^{\mu\nu} + \cdots \right]. \label{eq:sqrt1v2}
	\end{align}
	\item Neglecting the perturbation of the Ricci tensor \cite{Salcedo:2025ezu}, you should reach
	\begin{align}\label{eq:univfinv2}
		S_{\mathrm{univ}}=\int d^{4}x\sqrt{-g}&\bigg[\frac{\Mpl^{2}}{2}G_{\mu\nu}a^{\mu\nu}+\frac{\Lambda(t)}{2} g_{\mu\nu}a^{\mu\nu}+\frac{c(t)}{2}g^{00}g_{\mu\nu}a^{\mu\nu}-c(t)a^{00} \bigg].
	\end{align}
	Perform an advanced \stuck trick on this expression and an integration by part to obtain 
	\begin{align}
		S_{\mathrm{univ}} \rightarrow S_{\mathrm{univ}} + \int \dd^4 x \sqrt{-g} \left[ - \dot{\Lambda}(t) \pia  - 2 c(t) g^{0\mu}\partial_\mu \pia  - \dot{c}(t) \gr \pia \right].
	\end{align}
	\item Vary with respect to $\pia$ to recover the usual continuity equation of the EFT of Inflation and Dark Energy.
	\item Perform a retarded \stuck trick and identify the usual kinetic term of the scalar field. 
\end{enumerate}


\section{Conclusion}

Cosmology presents phenomena that go beyond the familiar framework of flat-space quantum field theory in the vacuum. It compels us to broaden our theoretical toolkit to address non-equilibrium and open-system dynamics. These lecture notes aim to provide a starting point for those interested in applying the Schwinger–Keldysh formalism to cosmological settings. Above all, we hope to have conveyed that this is a vibrant and accessible area of research — rich with open questions at nearly every level — well within reach for researchers trained in general relativity and quantum field theory. 

Beyond cosmology, SK-EFTs hold exciting potential in areas such as the quark–gluon plasma, black hole and neutron-star physics, holography, and dissipative hydrodynamics. We hope these notes will contribute to building bridges among these fields. The study of SK-EFTs is still in its infancy, and cosmology stands to benefit greatly from cross-disciplinary exchanges on topics as diverse as non-equilibrium renormalization group or non-Abelian strong-to-weak symmetry breaking.

The \textit{Disordered Universe Summer School 2025} has been a fertile ground for stimulating discussions among the participants. Far from being exhaustive, the following is a selection of topics highlighted during the courses that may inspire future investigations:
\begin{itemize}
	\item \textbf{Beyond correlators:} Can the SK-EFT formalism be extended to compute information-spreading measures such as out-of-time-ordered correlators (OTOCs), entropy measures, or off-diagonal density-matrix elements? 
	\item \textbf{Open scalar theory in Minkowski:} What is the single-exchange, tree-level trispectrum in the dissipative theory? How does the framework simplify when local thermodynamic equilibrium is imposed through the KMS symmetry? 
	\item \textbf{Symmetries in the SK contour:} How can we extend the discussion to non-linear realisations of diffeomorphisms? Can the in-in coset construction be adapted to cosmology? Is there a differential-geometric formulation of gauge symmetries within the SK framework? 
	\item \textbf{EFT construction:} Can we develop non-equilibrium renormalization-group techniques for open EFTs to clarify their power counting and cutoff structure? Is it possible to constrain EFT coefficients using entropy bounds, stability requirements, or microcausality? 
	\item \textbf{Phenomenology:} How can the Open EFT of Inflation be matched to gauge-inflation or warm-inflation models? Could the theory of gravity in a medium help describe dissipative dark sectors? Can similar methods model the dissipative propagation of gravitational waves in the interstellar medium? What observational limits can be placed on EFT coefficients using current data? 
\end{itemize}
I hope these lecture notes provide an initial step toward acquiring the tools needed to address these exciting questions and to develop a new generation of EFTs, capable of extracting novel physics from the wealth of forthcoming data.

\paragraph{Acknowledgements.} First and foremost, I would like to thank Paolo Benincasa for organising the \textit{Disordered Universe Summer School 2025}. It was one of the most inspiring weeks I have experienced in academia and will continue to shape my aspirations as a physicist. I am also grateful to all the students and participants of the summer school for their enthusiasm and dedication in exploring this subject. Finally, I wish to thank S.~Ag\"u\'i Salcedo, Sebastian Cespedes, Lennard Dufner, Enrica Lausdei, Enrico Pajer, Zhehan Qin and Xi Tong for sharing with me this journey that led us to develop bottom-up Open EFTs for gravity and cosmology. This work has been supported by STFC consolidated grant ST/X001113/1, ST/T000694/1, ST/X000664/1, ST/Y509127/1 and EP/V048422/1.

	\bibliographystyle{JHEP}
	\bibliography{biblio}

\end{document}